\documentclass[letterpaper,11pt]{article}
\usepackage[utf8]{inputenc}

\pdfoutput=1 
\usepackage{jheppub}
\usepackage{graphicx,verbatim}
\usepackage{blindtext}
\usepackage[T1]{fontenc}
\usepackage{epsf}
\usepackage{mathtools}
\usepackage{lmodern}
\usepackage[mathscr]{euscript}
\usepackage{textpos}
\makeatletter
\newcommand{\extp}{\@ifnextchar^\@extp{\@extp^{\,}}}
\def\@extp^#1{\mathop{\bigwedge\nolimits^{\!#1}}}
\makeatother
\usepackage{amsmath}
\usepackage{amsfonts}
\usepackage{amssymb}
\usepackage{mathrsfs}
\usepackage{slashed}
\usepackage{xcolor}

\def\bE{\mathbb{E}}
\def\qe{\mathfrak{q}}
\def\kq{\mathfrak{q}}
\def\fq{\mathfrak{q}}

\def\fA{\mathfrak{A}}

\def\fa{\mathfrak{a}}

\def\rx{\mathrm{x}}

\def\ri{\mathrm{i}}
\def\rj{\mathrm{j}}

\def\bs{\mathbf{s}}
\def\bz{\mathbf{z}}

\def\ba{\mathbf{a}}

\def\bm{\mathbf{m}}
\def\BC{\mathbb{C}}
\def\BR{\mathbb{R}}
\def\BE{\mathbb{E}}
\def\BZ{\mathbb{Z}}

\def\BG{\mathbb{G}}
\def\BP{\mathbb{P}}
\def\BL{\mathbb{L}}
\def\CalN{\mathcal{N}}
\def\CalA{\mathcal{A}}

\def\CalR{\mathcal{R}}
\def\CalZ{\mathcal{Z}}
\def\CalH{\mathcal{H}}

\def\CalO{\mathcal{O}}
\def\CalM{\mathcal{M}}

\def\CalT{\mathcal{T}}
\def\qe{\mathfrak{q}}
\def\CalL{\mathcal{L}}

\newcommand{\mf}{\rm{m}}
\def\Tr{{\rm Tr}}

\def\ve{{\varepsilon}}

\def\ta{\mathtt{a}}
\def\tb{\mathtt{b}}
\def\tc{\mathtt{c}}
\def\td{\mathtt{d}}
\def\tm{\mathtt{m}}
\def\tx{\mathtt{x}}

\def\tU{\mathtt{U}}
\newcommand{\ssl}{\mathfrak{sl}}
\def\fz{\mathfrak{z}}

\def\sK{\mathsf{K}}
\def\sJ{\mathsf{J}}

\def\sS{\mathsf{S}}

\def\sw{\mathsf{w}}

\def\sT{\mathsf{T}}
\def\sz{\mathsf{z}}

\def\sk{\mathsf{k}}
\def\sc{\mathsf{c}}

\def\sH{\mathscr{H}}

 \def\O{\Omega}
 \def\o{\omega }
 \def\U{\Upsilon}

\def\EN{\EuScript{N}}
\def\EY{\EuScript{Y}}
\def\EX{\EuScript{X}}

\def\EZ{\EuScript{Z}}
\def\EG{\EuScript{G}}
\def\EH{\EuScript{H}}
\def\EA{\EuScript{A}}
\def\EV{\EuScript{V}}

 \def\p{\partial}
 
 \def\a{\alpha}
 \def\b{\beta}
 \def\g{\gamma}
 \def\d{\delta}

 \def\th{\theta}

 \def\l{\lambda}
 \def\m{\mu}

 \def\r{\rho}

 \def\s{\sigma}
 \def\t{\tau}
 \def\th{\theta}
 \def\z{\zeta }
 
 \def\G{\Gamma}

\def\spch{XXX_{{\ssl}_2}}
\def\beq{\begin{equation}}
\def\eeq{\end{equation}}

\title{{Intersecting defects in gauge theory, \\quantum spin chains,\\and Knizhnik-Zamolodchikov equations}}
\author[a]{Saebyeok Jeong}
\author[b]{, Norton Lee}
\author[b,c,d]{, and Nikita Nekrasov}

\affiliation[a]{New High Energy Theory Center, Rutgers University, \\ 136 Frelinghuysen Road, Piscataway, NJ 08854-8019, USA}
\affiliation[b]{C. N. Yang Institute for Theoretical Physics,\\ Stony Brook University, Stony Brook, NY, 11794-3636, USA}
\affiliation[c]{Simons Center for Geometry and Physics, \\ Stony Brook University,Stony Brook, NY, 11794-3636, USA}
\affiliation[d]{{\rm on leave from}: Center for Advanced Studies, Skoltech, and Kharkevich IITP RAS, Moscow, Russia}
\emailAdd{saebyeok.jeong@physics.rutgers.edu}
\emailAdd{norton.lee@stonybrook.edu}
\emailAdd{nnekrasov@scgp.stonybrook.edu}

\vspace{2cm}
\abstract{We propose an interesting BPS/CFT correspondence playground: 
the correlation function of two intersecting half-BPS surface defects in four-dimensional $\EN=2$ supersymmetric $SU(N)$ gauge theory with $2N$ fundamental hypermultiplets. We show it satisfies a difference equation, 
the \textit{fractional quantum T-Q relation}.  
Its Fourier transform is the $5$-point conformal block
of the $\widehat{\mathfrak{sl}}_N$ current algebra with one of the vertex operators corresponding to the $N$-dimensional $\mathfrak{sl}_N$ representation, which we demonstrate with the help of the
Knizhnik-Zamolodchikov equation.  
We also identify the correlator with a state of the $XXX_{\mathfrak{sl}_2}$ spin chain of $N$ Heisenberg-Weyl modules over $Y({\ssl}_2)$. We discuss
the associated quantum Lax operators, and connections to isomonodromic deformations.}

\begin{document}

\maketitle

\section{Introduction}
Apparently distinct realms of theoretical physics find themselves connected through supersymmetric field theories. A certain amount of supersymmetry facilitates both qualitative and quantitative understanding of these correspondences. If the theory can be embedded in string/M-theoretic setup then these somewhat obscure relations become more visible in the form of dualities. Meanwhile, the availability of exact computations of relevant physical quantities makes the suggested links more tangible, and serves as a verification of string/M-theory dualities. 
Certainly, being able to make different branches of physics talk to each other via any intermediary is a useful ability. 

One of such striking relations is between the four-dimensional supersymmetric gauge theories and (non-supersymmetric) two-dimensional conformal field theories. It was first observed in the ${\EN}=4$ case in \cite{Vafa:1994tf}, then more generally in the ${\EN}=2$ case in \cite{Nekrasov:2002qd,NN2004ii,nekrasov2006seiberg}, stated as the BPS/CFT correspondence in \cite{NN2004i, Nikita:I}. In \cite{Alday:2009aq} the correspondence was given a very large class of precise (conjectural) examples,  with Nekrasov partition functions of $A_{1}$-type $S$-class theories on one side, and
conformal blocks of Liouville conformal field theory, on the other side. The celebrated AGT correspondence was further extended \cite{Alday:2009aq,Wyllard:2009hg,AT,KPPW,Wyllard:2010a,Wyllard:2010b,Tachikawa2011,K-T} to conformal field theories possessing various infinite-dimensional symmetry algebras. The relevant symmetry algebras were conjectured to be quantum Drinfeld-Sokolov reductions of affine Lie algebras, depending on the constituents of the gauge theory counterpart. The realization of the $\EN=2$ supersymmetric gauge theories as the effective field theory on a stack of fivebranes, compactified on a Riemann surface, or a type II string on local Calabi-Yau geometry \cite{Klemm:1996bj}, provides a physical reason for the emergent relation, by e.g. interpreting the fields originating from the six dimensional tensor multiplet in the presence of $\Omega$-deformation 
as the localized two dimensional chiral fields \cite{Nekrasov:2002qd}, or by duality arguments \cite{gai1,Gaiotto:2009hg}. 

Another related connection, between the supersymmetric gauge theories and integrable systems, reveals itself in the identification of the geometry of the low-energy states of the four-dimensional $\EN=2$ supersymmetric field theory  with the phase space of an algebraic integrable systems \cite{Gorsky:1995zq,DW1}. In the case of the ${\EN}=2^{*}$ case with $SU(N)$ gauge group the comparison of the construction \cite{DW1} and \cite{Gorsky:1994dj} suggests a relation to Hitchin systems. Indeed, 
for $S$-class $\EN=2$ theories \cite{gai1} the associated integrable systems can be argued to be precisely Hitchin integrable systems \cite{Gaiotto:2009hg}. Another vast class of theories, namely the ones corresponding to the quivers with unitary gauge groups, are demonstrated \cite{Nekrasov:2012xe} to be dual to algebraic integrable systems on the moduli spaces of double-periodic instantons or periodic monopoles. 
The classical spin chains are connected through the equivalence of the spectral curves \cite{Gorsky:1996hs,Gorsky:1997a}. The correspondence was uplifted to the quantum spin chains in \cite{Nekrasov:2009ui,Nekrasov:2009uh}, in the context of two-dimensional $\EN=(2,2)$ gauge theories (with the restriction that the spin representations are highest-weight). Here, the quantum Hamiltonians of the spin chain are identified with the twisted chiral ring of the $\EN = (2,2)$ gauge theory, while the common eigenstates are associated to the supersymmetric vacua. Such a connection between supersymmetric gauge theories and quantum integrable systems is called the Bethe/gauge correspondence \cite{Nekrasov:2009ui,Nekrasov:2009uh,Nikita-Shatashvili,Dorey:2011pa}. 

Therefore, it turns out that conformal field theories and spin chain integrable systems are associated in a nontrivial manner by sharing the same counterpart in the BPS/CFT correspondence and the Bethe/gauge correspondence. Consequently, it is expected that analytic properties of the correlation functions of conformal field theories and the spectral properties of spin chain systems can be explored on the same footing. Of the goals of this work is to manifestly realize such a framework by studying relevant gauge theory objects.

The ingredients on the gauge theory side are the half-BPS codimension-two (surface) defects in four-dimensional $\EN=2$ supersymmetric gauge theories transversally intersecting each other. The surface defects relevant to our study descend from the gauge origami configuration defined on an orbifold, e.g. intersecting stacks of D3-branes in the IIB string theory on an orbifold \cite{Nikita:II,Nikita:III,Nikita:IV,Nikita:V}. These surface defects are divided into two classes. One is the
regular surface defect \cite{NN2004ii}, also known as the Gukov-Witten monodromy type surface defect \cite{gukwit,gukwit2} defined by the singular boundary conditions along a surface, which could be modelled by an orbifold construction \cite{K-T,Nikita:IV}.\footnote{In the language of the six-dimensional $\EN= (2,0)$ theory, these surface defects are identical to \textit{codimension-two} defects \cite{K-T,frenguktes}, which are constructed by introducing M5-branes intersecting the worldvolume of the original stack of M5-branes with a four-dimensional intersection \cite{Gaiotto:2009hg}.} The other type of surface defect is realized by adding the \textit{folded} branes in the gauge origami language \cite{Nikita:III,Nikita:IV}.\footnote{In the language of the six-dimensional $\EN=(2,0)$ theory, these surface defects correspond to \textit{codimension-four} defects \cite{frenguktes} realized by M2-branes ending on a stack of M5-branes along a two-dimensional surface \cite{Gaiotto:2009hg}.} We consider the configuration of intersecting surface defects on $\BC^{2}_{12}$, in which an orbifold surface defect extends along the $\BC_1$-plane at $z_2 = 0$,  while a folded brane surface defect extends along the $\BC_2$-plane at $z_1 = 0$. The partition function of the generic gauge origami configuration is calculable by supersymmetric localization. The correlation function of the intersecting surface defects is a special case of the gauge origami partition function. It should be emphasized that the current setting is similar to, but slightly different from the intersecting surface defects considered in \cite{JeongNekrasov}, where both surface defects were of the vortex string type.

The correlators of the gauge origami system obey Dyson-Schwinger relations which express the compactness of the moduli space of spiked instantons \cite{Nikita:II,NaveenNikita}, implying nontrivial equations on the gauge theory correlation functions \cite{Nikita:V,Jeong:2017pai,Jeong:2018qpc,Jeong2017,Jeong2019b,JeongNekrasov,NekrasovVI,Lee:2020hfu,NT,Chen:2019vvt}, including the chiral ring relations of the gauge theory \cite{Jeong2019b}, both with and without the surface defects in the $\O$-background. Some of these equations are identified with the Belavin-Polyakov-Zamolodchikov (BPZ) equations \cite{bpz} and the Knizhnik-Zamolodchikov
(KZ) equations \cite{Knizhnik:1984nr} satisfied by the CFT correlation functions, leading to a direct proof of various incarnations of the BPS/CFT correspondence \cite{Nikita:V,Jeong2019b,Jeong:2018qpc,NT}.

In the present work, we find the non-perturbative Dyson-Schwinger equations obeyed by the correlation function of the intersecting surface defect observables. We identify them  with a set of functional difference equations, which we call the  \textit{fractional quantum T-Q equations}. With the help of these T-Q equations we clarify the link between the conformal field theory and the spin chain system. Also, 
as the additional evidence for the BPS/CFT correspondence, the fractional quantum T-Q equation is the Fourier transform of the KZ equations for the $4$-point conformal block with additional insertion of a degenerate field.
It is an extension of the statement that the vacuum expectation value of the regular orbifold surface defect in the $SU(N)$ gauge theory with $2N$ fundamental hypermultiplets obeys the KZ equation also obeyed by the $4$-point $\widehat{{\ssl}}_{N}$  conformal block \cite{NT}. We show that the insertion of a vortex-string type surface defect transverse to the regular monodromy defect
on the BPS side amounts to the insertion of the $N$-dimensional representation of $\mathfrak{sl}_N$ on the CFT side. 

At the same time, in support of the Bethe/gauge correspondence the fractional quantum T-Q equation provides a fractionalization of a refinement of the Baxter T-Q equation \cite{bax} for the $XXX_{\mathfrak{sl}_2}$ spin chain system. In particular, it can be expressed through the action of Lax operators on the $N$ spin sites. By concatenating the Lax operators we get the monodromy matrix of the spin chain. We note that the construction generalizes both the setup of \cite{Nekrasov:2009uh,Nekrasov:2009ui} by incorporating unbounded weight representations, the so-called HW-modules \cite{NT}, and the setup of \cite{Lee:2020hfu}, by quantization. We also show that the higher-rank $qq$-characters yield non-perturbative Dyson-Schwinger equations which express the spin chain transfer matrix in gauge theory language.

Our results imply, 
in agreement with Bethe/gauge correspondence, that the NS limit \cite{Nikita-Shatashvili} ${\ve}_{2} \to 0$ translates, on the CFT side, to the critical level limit of the genus zero KZ equations for $\mathfrak{sl}_N$, which indeed becomes the spectral problem for the $\mathfrak{sl}_N$ Gaudin system,  generalizing   \cite{Feigin:1994in}. Thus, the nontrivial connection between the KZ equations for $\mathfrak{sl}_N$ and the Lax operators of the $XXX_{\mathfrak{sl}_2}$ spin chain we found through the four dimensional gauge theory provides a refinement of the bispectral duality \cite{mironov2013spectral,Chen:2013jtk}. Its implications will be presented in a separate work \cite{JLN}.

The paper is organized as follows: 
We first review the gauge origami construction which leads to the folded brane surface defect and the $qq$-characters in section~\ref{sec:qq-char}. We then introduce the monodromy type surface defect via orbifolds in the section~\ref{sec:orb}, along with the folded brane surface defects. We also discuss the local observables, the $qq$-characters,  in the presence of the surface defects. The $qq$-characters are used to derive the (fractional) quantum T-Q equation in section~\ref{sec:TQ-DS}. In section \ref{sec:vortex}, we show that the folded brane surface defect is related to the surface defect constructed by a vortex string through Fourier transformation. In section~\ref{sec:KZ}, we will give a brief review of the constructions of $\mathfrak{sl}_N$-modules and the KZ equations for $\mathfrak{sl}_N$, and verify that they are satisfied by the correlation function of intersecting surface defect observables. 
In section~\ref{sec:spin} we turn our focus to the correspondence between the $\spch$ spin chain and $\EN=2$ supersymmetric gauge theories. We construct the Lax operators, the generators of the Yangian $Y({\ssl}_{2})$ and the monodromy matrix of $\spch$ spin chain. It is pleasing to recognize that the trace of the monodromy matrix shares an identical structure with the higher-rank $qq$-character in the NS limit, as it becomes a (Yangian version) of the $q$-character, in agreement with \cite{EFR}.
We end with the discussion of our results and future directions in the section~\ref{sec:discussion}. The appendices contain various computational details.

\acknowledgments
The authors are grateful to Gregory Moore, Alexander Tsymbaliuk, Fei Yan, and Sasha Zamolodchikov for discussions. The work of SJ was supported by the US Department of Energy under grant DE-SC0010008. The work of NL was supported by the Simons Center for Geometry and Physics.

\section{Intersecting branes, $Q$-observables, and $qq$-characters} \label{sec:qq-char}

We study special classes of codimension-two ($Q$-observables) and codimension-four ($qq$-characters) defects in four-dimensional $\EuScript{N}=2$ supersymmetric gauge theories. These defects are special since they originate from the configuration of intersecting D-branes in string theory, whose low energy description is the \textit{gauge origami} of \cite{Nikita:III}. In this section, we briefly review the gauge origami and the gauge theory defects it provides. For more details on the subject, see \cite{Nikita:I,Nikita:II,Nikita:III,NaveenNikita}.

\bigskip \noindent $\bullet$ 
The starting point is a Calabi-Yau fourfold $Z$. We consider 
a configuration of intersecting $D3$-branes in the type IIB string theory on $Z \times {\BC}_{\ba}$. We are being sloppy with the signature of the metric. In one setup ${\BC}_{\ba}$ stands for the two dimensional Minkowski space ${\BR}^{1,1}$, so that the 
$D3$-branes are actually the $S$-branes. In another setup ${\BC}_{\ba}$ is Euclidean, while $Z = {\BR}^{1,1} \times B$ with a local Calabi-Yau threefold $B$, so that some of the $D3$-branes are the usual physical branes wrapping ${\BR}^{1,1} \times \Sigma$,
with $\Sigma \subset B$ a complex curve, while others are  
euclidean $D3$-branes wrapping complex surfaces inside $Z$, all of them localized in the \emph{Coulomb} 
${\BC}_{\ba}$-factor.

\bigskip \noindent $\bullet$
Below, the subscripts $a \in \underline{\mathbf{4}} = \{ 1, 2,3, 4 \} $ denote the coordinates $\mathbf{z}_{a}$ on ${\BC}^{4}$, or, 
in case we describe some orbifolds of ${\BC}^4$, the coordinates $\hat{\mathbf{z}}_{a}$ on the covering space, which we shall also denote using a hat,  $\widehat{\BC}^{4}$, in order to avoid confusion.
We also use 
\begin{align}
     \underline{\mathbf{6}}=\{12, 13,14,23,24,34 \},
\end{align}
and, for $A = ab \in  \underline{\mathbf{6}}$, denote by $\mathbb{C}^2_A = \mathbb{C}_a \times \mathbb{C}_b \subset Z$ two
corresponding complex two-plane. In toric origami there can be at most six stacks $S_{A}, A \in \underline{\mathbf{6}}$, of $D3$-branes, with the multiplicities $n_A \in \mathbb{Z}_{\geq 0}$ and the worldvolumes $\mathbb{C}^2 _A$. The union  
\beq
S \equiv \bigcup_A \, n_A\, \mathbb{C}^2 _A \subset Z
\eeq
is the \textit{origami worldvolume}.

\bigskip \noindent $\bullet$ In this paper we shall be only looking at the cases $Z = {\BC}^{4} \equiv {\BC}^{4}_{1234}$, the local K3 orbifold
$Z = {\BC}^{2}_{12} \times {\BC}^{2}_{34}/{\Gamma}_{34}$, with the cyclic group ${\Gamma}_{34} = {\BZ}_{p}$, and 
the local CY3 orbifold
$Z = {\BC}^{1}_{1} \times {\BC}^{3}/{\Gamma}_{24} \times {\Gamma}_{34}$, with another cyclic group ${\Gamma}_{24} = {\BZ}_{n}$. 
The group ${\Gamma}_{ab}$ acts on the two-dimensional factor ${\BC}^{2}_{ab}$ via
\beq
\left( \mathbf{z}_{a}, \mathbf{z}_{b} \right) \mapsto \left( {\varpi} \mathbf{z}_{a}, {\varpi}^{-1} \mathbf{z}_{b} \right)
\eeq
with $\varpi$ being the corresponding roots of unity. We use the notations
\begin{align}
\begin{split}
& {\CalR}_{\o} \in {\Gamma}^{\vee}_{34} \, , \qquad {\omega} = 0, 1, \ldots, p-1 \equiv -1  \\
& {\mathfrak{R}}_{\o'} \in {\Gamma}_{24}^{\vee}\, , \qquad {\o'} = 0 , 1, \ldots, n-1 \equiv -1
\label{eq:rep1}
\end{split}
\end{align}
for the irreducible representations of the groups ${\Gamma}_{A}$, $A = 24, 34$.

\bigskip \noindent $\bullet$ 
 We shall further stick to the case $p=3$ to produce the $SU(N)$ theory with $2N$ fundamental hypermultiplets on the ${\BC}^{2}_{12}$-plane.

\bigskip \noindent $\bullet$ 
Now we study the effective field theory on the origami wolrdvolume $S$. At each one of six stacks of branes, the effective theory is locally the $\EuScript{N}=2$ supersymmetric gauge theory with the gauge group $\bigtimes_{i=0} ^{p-1} U(n_{A,i})$. These gauge theories are interacting with each other in an intricate manner through the couplings at the intersections of their worldvolumes. As a whole, the effective theory on the origami worldvolume $X$ defines what is called the \textit{generalized gauge theory}. Among the six intersecting worldvolumes, we differentiate $\mathbb{C}^2 _{12}$ as the support of the main affine $\hat{\Gamma}$-quiver gauge theory with the gauge group $\bigtimes_{i=0} ^{p-1} U(n_{12,i})$. Then the fields associated to other gauge theories on $\mathbb{C}^2 _A$, $A\in \underline{\mathbf{6}}\setminus\{12\}$, can be integrated out, realizing codimension-two or codimension-four defects in the gauge theory on $\mathbb{C}^2 _{12}$, depending on whether $A \cap \{12\}$ is empty, $\{1\}$, or $\{2\}$. In the path integral formulation of the four-dimensional gauge theory, these local and non-local defects would result in observable insertions, and we are interested in recovering those observables from the partition functions of the gauge origami.

\bigskip \noindent $\bullet$
Supersymmetry localizes the path integral for the generalized gauge theory on $S$ onto the BPS configurations of $D(-1)$-instantons dissolved into $S$.\footnote{To be precise, we turn on the appropriate B-field to push the $D(-1)$-instantons dissolved into the worldvolume $S$. See \cite{Nikita:II,NaveenNikita}.} The point-like BPS objects in the generalized gauge theory constructed in such a way is called \textit{spiked instantons}. The path integral reduces to a finite-dimensional integration over the moduli space $\mathfrak{M}_S$ of spiked instantons on $S$, which can be constructed as the Higgs branch of the matrix theory supported by the collection of $D(-1)$-instantons.

\bigskip \noindent $\bullet$  The symmetry 
\beq
\mathsf{H} \, = \ \bigtimes_{A \in  \underline{\mathbf{6}}} \, \bigtimes_{r \in {\Gamma}_{24}^{\vee} \times {\Gamma}_{34}^{\vee}} \ U(n_{A,r}) \times U(1)^{3}_{\ve}
\label{eq:origamisymm}
\eeq
of the gauge origami setup, with its maximal torus  $\mathsf{T}_\mathsf{H} \subset \mathsf{H} $ naturally act on the moduli space $\mathfrak{M}_S$ of spiked instantons, allowing for further equivariant localization of the finite-dimensional integral. As a result, 
the partition function is computed to be  a rational function in the equivariant parameters $\xi \in \text{Lie} (\mathsf{T_H})$,
\begin{align}
    \mathcal{Z}_S (\xi) = \int_{\mathfrak{M}_S} ^{\mathsf{T_H}} 1.
\end{align}
The equivariant localization reduces the partition function $\mathcal{Z}_S$ to a sum over the fixed points $\mathfrak{M}_S ^{\mathsf{T_H}}$ with respect to $\mathsf{T_H}$. The fixed points are classified by a set $\boldsymbol\l = \{\boldsymbol\lambda_{A,i}\}$, $A\in\underline{6}$, $i=0,\dots,p-1$, of partitions. 
The partition function becomes that of a statistical mechanical model defined on $\boldsymbol\l$. The general formula for gauge origami partition function is derived in \cite{Nikita:III}. See appendix~\ref{sec:parti-func} for an illustration

Let us briefly explain the notation used in writing the partition function $\CalZ_S$. For more details, see appendix \ref{sec:parti-func}. We use the same letters for both the vector spaces themselves and for their characters. For example,
\begin{subequations}
\begin{align}
    N_{A} & = 
    \sum_{i=1}^{p-1} \left(\sum_{\alpha=1}^{n_{A,i}} e^{a_{A,i,\alpha}} \right) \CalR_{i}, 
    \\
    K_{A} & = 
    \sum_{i=1}^{p-1} \left( \sum_{\alpha=1}^{n_{A,i}}\sum_{\Box\in\boldsymbol\lambda_{A,i}} e^{a_{A,i,\alpha}}e^{c_{A,\Box}} \right) \CalR_{i},
\end{align}
\end{subequations}
where $c_{A,\Box}=(\ri-1)\ve_a+(\rj-1)\ve_b$ for $A=ab\in\underline{\mathbf{6}}$. The character $S_A$ of the universal bundle is
\begin{align}
    S_A = N_A - P_A K_A.
\end{align}
The exponentiated $\Omega$-background parameters are 
\begin{align}
    q_a = e^{\ve_a}, \ P_a = 1 - q_a, \ P_A = P_aP_b, \quad \prod_{a=1} ^4 q_a =1.
\end{align}
Given a virtual character $X = \sum_{\ta} \tm_\ta e^{\tx_\ta}$ we denote by $X^* = \sum_{\ta} \tm_\ta e^{-\tx_\ta}$ the dual virtual character. 

The pseudo-measure associated to the instanton configuration $\boldsymbol\lambda$ is defined through the 
\textit{plethystic exponent} operator $\BE$ (also related to Adams operations in $K$-theory) converting the additive Chern characters to the multiplicative classes
\begin{align}
    \BE \left[ \sum_\ta \tm_\ta e^{\tx_\ta}  \right] = \prod_\ta \tx_\ta^{\tm_\ta}.
\end{align}
The above brane construction of spiked instantons suggests a projection of the moduli space of spiked instantons to the moduli space of ordinary (noncommutative) instantons on $\mathbb{C}^2 _{12}$, $\mathfrak{M}_S \longrightarrow \mathfrak{M}_{\mathbb{C}^2_{12}}$. Integration along the fibers of this projection casts the gauge origami partition function in a form of  a correlation function of the associated codimension-two and codimension-four defects in the four-dimensional bulk gauge theory. Schematically,
\begin{align}
    \mathcal{Z}_S  (\xi) = \int_{\mathfrak{M}_{\mathbb{C}^2 _{12}} } ^{\mathsf{T_H}} \prod_{A \in \underline{\mathbf{6}} \setminus \{12\} }\EuScript{O}_{A} = \Bigg\langle \prod_{A \in \underline{\mathbf{6}} \setminus \{12\} }\EuScript{O}_{A}  \Bigg\rangle \mathcal{Z}_{\mathbb{C}^2_{12}},
\end{align}
where the bracket denotes the vacuum expectation value in the gauge theory on $\mathbb{C}^2 _{12}$.

\subsection{Surface defects from folded branes}
We consider a specific class of half-BPS surface (codimension-two) defects in the four-dimensional $\EuScript{N}=2$ supersymmetric gauge theories. This type of the surface defects can be constructed by introducing an additional stack of D-branes in the gauge origami construction, on top of the original stack of D-branes engineering the bulk four-dimensional gauge theory. The worldvolume of the additional stack of branes has a two-dimensional intersection with the worldvolume of the bulk gauge theory. Thus, from the point of view of the original bulk theory observer, one has a codimension two defect. 

\subsubsection{The bulk gauge theory}
The four-dimensional gauge theory that we will mainly consider is the $\EuScript{N}=2$ supersymmetric $U(N)$ gauge theory with $N$ fundamental and $N$ anti-fundamental hypermultiplets. We can engineer this particular gauge theory from the simplest gauge origami configuration, composed of single stack of branes on $\mathbb{C}^2 _{12}$ with $\Gamma = A_2 = \mathbb{Z}_3$, as follows. We set $n_{12,0} = n_{12,1}=n_{12,2} \equiv N$ and $n_A = 0$ for $A \in \underline{\mathbf{6}}\setminus\{ 12\}$. Also we assign the $\mathbb{Z}_3$-charge as
\begin{align}
    N_{12} = \sum_{\a=1} ^N e^{a_\a} \mathcal{R}_0 + \sum_{\a=1} ^N e^{m_\a ^- - \ve_4} \mathcal{R}_1 + \sum_{\a=1} ^N e^{m^+ _\a - \ve_3} \mathcal{R}_2.
\end{align}
The $\EuScript{N}=2$ gauge theory on $\mathbb{C}^2 _{12}$ constructed in this way is the affine $\hat{A}_2$-quiver gauge theory, with the gauge group $\bigtimes_{i=0} ^2 U(N)_i$. The fixed points on the moduli space of spiked instantons are classified, a priori, by three $N$-tuples of partitions, $\boldsymbol\l_i=\boldsymbol\lambda_{A,i}$, $i=0,1,2$.

Throughout this work, our main consideration in the four-dimension side would be the $A_1$-quiver gauge theory, which can be obtained by freezing two of the gauge nodes in the above affine $\hat{A}_2$-quiver gauge theory, making the corresponding $U(N)$-factors a flavor symmetry (which can be enhanced to $SU(2N)$). 

\bigskip \noindent $\bullet$ In what follows we use on several occasions the trick of taking the limit ${\qe}_{1} = {\qe}_{2}$ thereby killing all instantons in the nodes $1$ and $2$,   $\boldsymbol\l_{1} = \boldsymbol\l_{2} = \varnothing$. We call this procedure the
\emph{freezing}  ${\hat A}_2 \to A_1$, or simply freezing, for short.  We denote the remaining coupling by $\qe  \equiv \qe_{0}$, while the remaining $N$-tuples of partitions as $\boldsymbol\l \equiv \boldsymbol\l_{0}$. The gauge origami partition function reduces to
\begin{align}
    \mathcal{Z}_S = \sum_{\boldsymbol\l} \qe^{\vert \boldsymbol\l \vert} \mathbb{E} \left[ \frac{- S S^* +M^+ S^* + M^{-} S^*}{P_{12}^*}  \right] \equiv \mathcal{Z}_{\mathbb{C} ^2 _{12}} =  \sum_{\boldsymbol\l} \qe^{\vert \boldsymbol\l \vert} \mathbb{E} \left[ \mathcal{T}_{12} \right] \ , 
\end{align}
where we have defined
\begin{align}
\begin{split}
    &N \equiv \sum_{\a=1} ^N e^{a_\a} , \quad M^\pm \equiv \sum_{\a=1} ^N e^{m^\pm _\a}, \quad M \equiv M^+ + M^-\\
    &S \equiv N - P_{12} K_{12}.
\end{split}
\end{align}
This is the partition function of the rank $N$ $A_1$ theory \cite{Nekrasov:2012xe}, i.e. $U(N)$ gauge theory with $2N$ fundamentals. We emphasize that the ${\Gamma}_{34}$-orbifold plays an auxiliary role of reducing the theory with adjoint
fields to the theory with fundamental matter. 
The ${\Gamma}_{24}$ orbifold, introduced in section \ref{sec:orb}, is another auxiliary tool, 
generating the surface defects.

\subsubsection{Introducing surface defects from folded branes: the $Q$-observables} \label{sec:qobs}
Having engineered the \emph{bulk four-dimensional gauge theory} of our interest, 
we move on to a class of surface defects, called the $Q$-observables, constructed by the next-to-the simplest gauge origami configuration. Namely, take now  two stacks of branes, one on $\mathbb{C}^2_{12}$ and another on $\mathbb{C}^2_{23}$, with 
the Chan-Paton spaces carrying the $\Gamma_{34}$ representations, decomposing into the irreps ${\CalR}_{\o}$ with 
the multiplicities $n_{12,0} = n_{12,1}=n_{12,2} = N$ and $n_{23,1} = 1$, respectively. Namely, we assign the $\mathbb{Z}_3$-charges in the following way:
\begin{align}
\begin{split}
    &N_{12} = \sum_{\a=1} ^N e^{a_\a} \mathcal{R}_0 + \sum_{\a=1} ^N e^{m_\a ^- - \ve_4} \mathcal{R}_1 + \sum_{\a=1} ^N e^{m^+ _\a - \ve_3} \mathcal{R}_2\\
    &N_{23} = e^{x+ \ve_2+\ve_3} \, \mathcal{R}_1.
\end{split}
\end{align}
The gauge origami partition function is computed as
\begin{align}
    \mathcal{Z}_S = \sum_{\boldsymbol\l} \prod_{i= 0,1,2} \qe_i ^{\vert \boldsymbol\l_{12,i} \vert + |\boldsymbol\lambda_{23,i}|} \mathbb{E}\left[ - \frac{P_3 S_{12} S_{12} ^*}{P_{12} ^*} - \frac{ P_1  S_{23} S_{23}^* }{P_{23} ^*} + q_3 P_{4} \frac{S_{12} S_{23}^*}{P_{2} ^*} \right] ^{\mathbb{Z}_3}
\end{align}
In the freezing limit $\qe_1=\qe_2 =0 $, only $K_{23} = \varnothing$  gives non-zero contribution. The partition function reduces to a sum over $N$-tuples of partitions which we still denote as $\boldsymbol\l$. It is given by
\begin{align}
    \mathcal{Z}_S = \sum_{\boldsymbol\l} \qe^{\vert \l \vert} \mathbb{E} \left[ \frac{-SS^* +M S^* }{P_{12}^* }- \frac{e^{x} (S^* - M^{-*})}{P_2 ^*}   \right].
\end{align}
Note that the first term gives the usual measure for the $A_1$-quiver gauge theory partition function. The second term is the contribution obtained by integrating out the fields on the brane on $\mathbb{C}^2 _{23}$. Thus, in the four-dimensional point of view it is interpreted as a surface defect on the $\mathbb{C}_2$-plane, which we call the \textit{$Q$-observable}. By discarding the $\G$-function involving the hypermultiplet mass, which can be absorbed into the 1-loop part of the partition function $\CalZ_S$, the $Q$-observable is represented on the partitions $\boldsymbol\l$ as
\begin{align} \label{eq:qexp}
\begin{split}
Q(x) [\boldsymbol\l] &= \mathbb{E} \left[ - \frac{e^x S^* [\boldsymbol\l]}{P_2 ^*} \right].
\end{split}
\end{align}
It is important to note that $Q(x)$ is in fact regular in $x$, i.e.,
\begin{align} \label{eq:qreg}
\begin{split}
    Q(x)[\boldsymbol\l]  &= \prod_{\a=1} ^N \left[ \frac{ \left( -\ve_2  \right)^{\frac{ x  -a_\a}{\ve_2}} }{\Gamma \left( - \frac{ x-a_\a}{\ve_2} \right)} \prod_{\Box \in \l ^{(\a)}} \frac {x-c_\Box - \ve_1}{x-c_\Box} \right] \\
    &= \prod_{\a=1} ^N \left[ \frac{(-\ve_2) ^{\frac{x - a_\a}{\ve_2}}}{\G\left( \l ^{(\a)} _1  - \frac{x-a_\a}{\ve_2} \right)} \prod_{j=1} ^{\l^{(\a)} _1 } \left( j-1 - \frac{x-a_\a -\l^{(\a)t} _j \ve_1}{\ve_2} \right)  \right].
\end{split}
\end{align}
Also, it should be noted that the $Q$-observable is related to the $\EuScript{Y}$-observable by
\begin{align}
    \frac{Q(x)}{Q(x-\ve_2)} = \mathbb{E} \left[ -e^x S^*  \right] =\EuScript{Y}(x).
\end{align}
The gauge origami partition function thus provides the vacuum expectation value of the $Q$-observable,
\begin{align}
    \mathcal{Z}_S = \Big\langle Q(x) \Big\rangle \mathcal{Z}_{\mathbb{C}^2 _{12}}.
\end{align}

Similarly, we can construct the $Q$-observable from intersecting branes on $\mathbb{C}^2 _{12}$ and $\mathbb{C}^2 _{24}$ instead. Namely, we consider the gauge origami configuration with 
\begin{align}
\begin{split}
    &N_{12} = \sum_{\a=1} ^N e^{a_\a} \mathcal{R}_0 + \sum_{\a=1} ^N e^{m_\a ^- - \ve_4} \mathcal{R}_1 + \sum_{\a=1} ^N e^{m^+ _\a - \ve_3} \mathcal{R}_2\\
    &N_{24} = e^{x+ \ve_2+\ve_4} \, \mathcal{R}_2.
\end{split}
\end{align}
A computation similar to the one above shows
\begin{align}
\begin{split}
    \mathcal{Z}_S &= \sum_{\boldsymbol\l} \prod_{i= 0,1,2} \qe_i ^{\vert \boldsymbol\l_i \vert} \mathbb{E}\left[ - \frac{P_3 S_{12} S_{12} ^*}{P_{12} ^*} - \frac{ P_1  S_{24} S_{24}^* }{P_{24} ^*} + q_4 P_{3} \frac{S_{12} S_{24}^*}{P_{2} ^*} \right] ^{\mathbb{Z}_3}\\
    &=\sum_{\boldsymbol\l} \qe^{\vert \l \vert} \mathbb{E} \left[ \frac{-SS^* +M S^* }{P_{12}^* }- \frac{e^{x} (S^* - M^{+*})}{P_2 ^*}   \right].
\end{split}
\end{align}
Hence we obtain the same $Q$-observable, up to the $\G$-function involving the hypermultiplet mass which can be absorbed into the 1-loop part of the partition function $\CalZ_S$.

In the following discussion, it will be convenient to use a redefined version of the $Q$-observable, which is still regular in $x$ but dressed with the $\G$-function involving the hypermultiplet masses:
\begin{align}
    \tilde{Q}(x) [\boldsymbol\l] \equiv \frac{Q(x) [\boldsymbol\l]}{\prod\limits_{\a=1} ^N \ve_2 ^{\frac{x-m^+ _\a}{\ve_2}} \G \left( 1 + \frac{x-m^+ _\a}{\ve_2} \right)}
\end{align}
Note that the $\G$-function produces a polynomial in hypermultiplet masses by taking the ratio:
\begin{align}
    \frac{\tilde{Q} (x)}{\tilde{Q} (x-\ve_2) } =  \frac{\EY(x)}{P^+ (x)}.
\end{align}

\subsection{$qq$-characters from crossed branes} \label{sec:crbr}
Now we study two transversal stacks of branes. Such a configuration defines local BPS operators at the intersection of the components of two braneworlds. From the point of view of either four-dimensional gauge theory, integrating out the degrees of freedom on the other produces the local observable called the $qq$-character.

\subsubsection{Bulk theory with just one $qq$-character }
Consider the gauge origami with only two orthogonal stacks of branes, one on $\mathbb{C}^2 _{12}$ and another on $\mathbb{C}^2 _{34}$.
As a minimal modification of rank $N$  $A_1$ theory on ${\BC}^{2}_{12}$ we start with a single brane on $\mathbb{C}^2 _{34}$. 
We still have a choice of a ${\BZ}_{p}$ representation to assign to that brane. It turns out the only interesting choice is to 
assign it a singlet representation. 
So, we assign $n_{12,0} = n_{12,1} =n_{12,2} = N$ and $n_{34,0} = 1$:
\begin{align}
\begin{split}
    &N_{12} = \sum_{\a=1} ^N e^{a_\a} \mathcal{R}_0 + \sum_{\a=1} ^N e^{m_\a ^- - \ve_4} \mathcal{R}_1 + \sum_{\a=1} ^N e^{m^+ _\a - \ve_3} \mathcal{R}_2\\
    &N_{34} = e^{x} \, \mathcal{R}_0.
\end{split}
\end{align}
The corresponding gauge origami partition function is computed as
\begin{align}
    \mathcal{Z}_S = \sum_{\boldsymbol\l} \kq_0^{|K_{34}|} \prod_{i= 0,1,2} \qe_i ^{\vert \boldsymbol\l_i \vert} \mathbb{E}\left[ - \frac{P_3 S_{12} S_{12} ^*}{P_{12} ^*} - \frac{P_1 S_{34} S_{34} ^*}{P_{34} ^*} - q_{12}^{-1} S_{12} S_{34} ^* \right] ^{\mathbb{Z}_3}.
    \label{eq:gopf1}
\end{align}
Once again, in order to obtain the $A_1$-quiver gauge theory we take the freezing limit $\qe_1 = \qe_2 = 0$. Then $\vert K_{34} \vert = 0 \text{ or }1$ and 
\eqref{eq:gopf1} can be cast as the expectation value of a surface defect in $A_1$ theory. Therefore, the gauge origami partition function can be split as the sum of the expectation values of two observables in $A_1$ theory, one with $|K_{34}|=0$ and the another with $|K_{34}|=1$:
\begin{align}
\begin{split}
    \mathcal{Z}_S &= \sum_{\boldsymbol\l} \qe^{\vert \boldsymbol\l \vert} \mathbb{E} \left[ \frac{-S S^* +M S^* }{P_{12} ^*} \right] \left( \mathbb{E} \left[ - e^{x+\ve} S^* \right] + \qe \mathbb{E} \left[ e^x S^* - e^x M^{*}   \right] \right) \\
    &=  \sum_{\boldsymbol\l} \qe^{\vert \boldsymbol\l \vert} \mathbb{E}  \left[ \mathcal{T}_{12} \right] \left( \EuScript{Y}(x+\ve) +\qe \frac{ P (x) }{\EuScript{Y}(x)} \right) .
\end{split}
\end{align}
In the language of the four-dimensional gauge theory, this is the vacuum expectation value 
\begin{align}
    \mathcal{Z}_S = \Big\langle \EuScript{X} (x) \Big\rangle \mathcal{Z}_{\mathbb{C}^2 _{12}} \, , 
\end{align}
of an observable called the fundamental $qq$-character,
\begin{align}
    \EuScript{X} (x) = \EuScript{Y}(x+ \ve) + \qe \frac{P (x)}{\EuScript{Y}(x)}\ .
\end{align}

\subsubsection{Bulk theory with the surface defect $Q$-observable and the $qq$-character}

Now we are interested in a more complicated example. We want to study the correlation functions of several local and non-local observables. In our setup these are induced by simultaneous insertions of additional branes. Specifically, we consider three stacks of branes: $N$ regular branes wrapping ${\BC}^2 _{12}$, one $\CalR_1$-type brane on ${\BC}^2 _{23}$, and one ${\CalR}_{0}$-brane on  ${\BC}^2 _{34}$. In other words, the ${\Gamma}_{34}^{\vee}$-multiplicities are: $n_{12,0}=n_{12,1}=n_{12,2} = N$, $n_{23,1}= 1$ and $n_{34,0} = 1$. This assignment translates to the following Chan-Paton characters:
\begin{align}\label{eq:gosetup}
\begin{split}
    {N}_{12} & = \sum_{\alpha} e^{a_{\alpha}} \cdot \CalR_{0} + \sum_{\alpha} e^{m_{\alpha}^- -\ve_4} \cdot \CalR_{1} + \sum_{\alpha} e^{m_{\alpha}^+ -\ve_3} \cdot \CalR_{2} \\
    {N}_{23} & = e^{x' + \ve_2 + \ve_3} \cdot \CalR_{1} \\
    {N}_{34} & = e^{x} \cdot \CalR_{0}
\end{split}.
\end{align}
The gauge origami partition function is then given by the sum of plethystic exponents \cite{Nikita:III} :
\begin{align} \label{eq:gopf}
\begin{split}
    \CalZ_{S} = \sum_{\vec{\boldsymbol\lambda}}  \prod_{i=0,1,2} \qe_i ^{\vert \boldsymbol\l_i \vert} \BE & \left[ -\frac{{P}_3 {S}_{12}{S}_{12}^*}{{P}_{12}^*} - \frac{{P}_{1} {S}_{23}{S}_{23}^*}{{P}_{23} ^*} - \frac{{P}_{1}{S}_{34}{S}_{34}^*}{{P}_{34} ^*} \right. \\ 
    & \left. -{q}_{12} ^{-1} {S}_{12}{S}_{34} ^* + {q}_{3} {P}_4 \frac{{S}_{12}{S}_{23}^*}{{P}_{2}^*} + {q}_{4} {P}_1 \frac{{S}_{23}{S}_{34}^*}{{P}_{3}^*}   \right] ^{\BZ_3}.
\end{split}
\end{align}
Again, freeze ${\hat A}_{2} \to A_{1}$. Let us address the last term in the plethystic exponent. It comes from the interaction between the components $\mathbb{C}^2 _{23}$ and $\mathbb{C}^2 _{34}$ of the origami braneworld. Thanks to our
choice \eqref{eq:gosetup} of $\mathbb{Z}_3$-charges ${K}_{23}=\varnothing$. 

\bigskip \noindent $\bullet$ We now argue, that the perturbative pseudo-measure contribution
\beq
   {q}_{4} {P}_1 \frac{{N}_{23}{N}_{34}^*}{{P}_{3}^*} 
   \label{eq:origpert}
\eeq
can be replaced by 
\beq
    \BE\left[-{q}_{23}{N}_{34}{N}_{23}^* - {q}_{34}{N}_{23}{N}_{34}^* \right] , 
    \label{eq:effpert}
\eeq
without introducing additional poles or zeroes in the $x$-variable. 

Indeed \eqref{eq:origpert}
can be interpreted as the effect of imposing an infinite sequence of equations on the ADHM data, cf. \cite{Nikita:III}:
\begin{subequations}\label{eq:nocrossing}
\begin{align}
    & J_{23} B_3^{k} I_{34} = 0 \, , \qquad
     J_{34} B_{3}^{k} I_{23} = 0 \, , \\
    &  \qquad\qquad k \geq 0\ .
\end{align}
\end{subequations}
Now we argue that the matrix $B_3$ vanishes in the setup of the Eq.~\eqref{eq:gosetup}.  Firstly, $B_3 I_{12}=0$ by the standard  stability condition of gauge origami. Secondly, the instantons cannot move onto the $\BC_{23}^2$ subspace, $K_{23} = 0$,  therefore:
\beq
    B_3 I_{23}=0 \ .
\eeq
Finally, the would-be vector $B_3 I_{34} (N_{34})$ belongs to the $\CalR_1$-component of $K_{34}$, which is zero in the frozen limit, so 
\beq
    B_3 I_{34} = 0 \ .
\eeq
Therefore
all the constraints \eqref{eq:nocrossing} are automatically satisfied for all $k > 0$. The only remaining constraints imposed to the gauge origami data are
\begin{align}
    J_{23} I_{34} = 0, \quad J_{34} I_{23} = 0,
\end{align}
whose combined contributions to the pseudo-measure read exactly
\begin{align}
    \BE\left[-{q}_{23}{N}_{34}{N}_{23}^* - {q}_{34}{N}_{23}{N}_{34}^* \right] .
\end{align}
Also, the non-perturbative piece of ${S}_{34}$ contributes
\begin{align}
     \mathbb{E} \left[ {P}_{14} {N}_{23} {K}_{34}^* \right]
\end{align}
Consequently, the last term of \eqref{eq:gopf} is replaced by the sum of the two contributions in the decoupling limit.

\bigskip \noindent $\bullet$ After these preparations
the gauge origami partition function reads as follows:
\begin{align}
      \CalZ_{S} = \sum_{\vec{\boldsymbol\lambda}}   \qe ^{\vert \boldsymbol\l \vert} \BE & \left[ -\frac{{P}_3 {S}_{12}{S}_{12}^*}{{P}_{12}^*} - \frac{{P}_{1} {S}_{23}{S}_{23}^*}{{P}_{23} ^*} - \frac{{P}_{1}{S}_{34}{S}_{34}^*}{{P}_{34} ^*} \right. \\ 
    & \left. -{q}_{12} ^{-1} {S}_{12}{S}_{34} ^* + {q}_{3} {P}_4 \frac{{S}_{12}{S}_{23}^*}{{P}_{2}^*} -q_{23} N_{34} N_{23} ^* -{q}_{34} {N}_{23} {N}_{34} ^* + {P}_{14} {N}_{23} {K}_{34} ^*   \right] ^{\BZ_3}. \nonumber
\end{align}
Using the $Q$-observables and $\EuScript{Y}$-observables that we have studied, we notice that the partition function can be written as the following
\begin{align}
\begin{split}
    \CalZ_{S} &= \sum_{\boldsymbol{\lambda}} \qe^{\vert \boldsymbol\l \vert} \mathbb{E} \left[ \mathcal{T}_{12} \right]  \left[ (x'-x) Q (x' )[\boldsymbol{\lambda}] \EuScript{Y}(x + \ve)[\boldsymbol{\lambda}]  + \fq (x'-x-\ve_1) Q (x')[\boldsymbol{\lambda}] \frac{P (x) }{\EuScript{Y}(x)[\boldsymbol{\lambda}]} \right] \\
    &= -\Big\langle T_{N+1} (x) Q(x') \Big\rangle \mathcal{Z}_{\mathbb{C}^2 _{12}},
\end{split}
\end{align}
where we have defined the $qq$-character in the presence of the $Q$-observable
\begin{align}\label{eq:qyeq}
     {T}_{N+1}(x) Q(x')=(x-x')Q(x')\EuScript{Y}(x+\ve) + \fq (x-x'+\ve_1) P(x) \frac{Q(x')}{\EuScript{Y}(x)}. 
\end{align}
By the compactness of the moduli space of spiked instantons, the vacuum expectation value of \eqref{eq:qyeq} has no poles in the variable  $x$. Thus the observable $T_{N+1} (x)$ is a polynomial in $x$ of degree $N+1$.

\bigskip \noindent $\bullet$ This is our main new tool. 

\section{Intersecting surface defects from branes on orbifold} \label{sec:orb}
As reviewed above,  the gauge origami construction produces correlators of both local and non-local defects in the four-dimensional gauge theory.  As the field theory effects of these defects arise from integrating out the elementary degrees of freedom of bi-fundamental strings connecting distinct components of the worldvolumes, these defects are \textit{electric} in nature.

In this section, we introduce another type of half-BPS codimension-two, \emph{monodromy type} defect in four-dimensional supersymmetric gauge theory. In a sense, it is \textit{magnetic} in nature. This surface defect \cite{NN2004ii, gukwit,gukwit2} is defined by prescribing a specific singular behavior of the fundamental fields along a two-dimensional surface. We shall sometimes call it the Gukov-Witten type defect. 

For practical purposes the monodromy defect can be represented by an orbifold construction \cite{biquard, biswas, Bruzzo:2010fk, K-T, Nikita:IV}. The main use of orbifold construction is that it allows for the straightforward localization computation of its partition function \cite{K-T, Nikita:IV, Jeong:2018qpc}. Below we show that the correlation function of the  $Q$-observables that we have constructed in the previous section \textit{fractionalize} in the presence of the orbifold surface defect, also is computable explicitly. Since the orbifold can be replaced by a Gukov-Witten type (monodromy) surface defect, the fractional $Q$-observables on an orbifold can thus be interpreted as intersecting the surface defects in the bulk gauge theory. We remark that the current setting of intersecting surface defects is related to, although is explicitly different from the one in \cite{JeongNekrasov}, cf. also \cite{gfpp,pp}. In those papers the surface defects were introduced in the form of vortex strings. These can be viewed as a dual description \cite{frenguktes,Jeong:2018qpc} of a $\BZ_2$-orbifold surface defect, as opposed to the regular $\BZ_N$-orbifold considered in \cite{Nikita:IV, Nikita:V, Jeong:2017pai, NT} and in the present work. 

In the limit $m_f \to \infty$, ${\qe}\to 0$ defining the pure super-Yang-Mills theory the instantons in the presence of monodromy surface defect
can be related to the two dimensional sigma model instantons valued in the infinite-dimensional K\"ahler manifold $LG/T$ \cite{NN2004ii}. Their enumeration can be then analyzed using intersection homology \cite{Braverman:2004vv}, leading to a degenerate version of
the KZ equation which we discuss in full generality below. 

\subsection{Orbifold surface defect as the disorder operator}

We start by recalling the map \cite{Nikita:IV} of gauge theory on orbifold to gauge theory with a monodromy type surface defect. We also give the equivariant integral expression for the surface defect observable from the orbifold projection of the gauge theory measure.

\subsubsection{Construction of the surface defect via an orbifold}\label{sec:cons-defect}

Let us view the worldvolume of the four-dimensional gauge theory of interest as the one immersed at the ${\hat {\mathbf{z}}}_{3} = {\hat{\mathbf{z}}}_{4} = 0$ locus of the  $\mathbb{Z}_n$-orbifold $\hat{\BC}_1 \times \left( \hat{\BC}_2 /\mathbb{Z}_n \right) \subset {\hat{\BC}}^{4}/{\Gamma}_{24}$. 
On the quotient space the worldvolume can be identified with $\mathbb{C}^2_{12}$ as complex manifolds, via
\begin{align} \label{eq:map}
\begin{split}
    \hat{\mathbb{C}}_1 \times \left( \hat{\mathbb{C}}_2 / \mathbb{Z}_n \right) &\longrightarrow \mathbb{C}^2 _{12} \\
    (\hat{\mathbf{z}}_1,\hat{\mathbf{z}}_2) \quad &\longmapsto ( \mathbf{z}_1 = \hat{\mathbf{z}}_1 ,
     \, \mathbf{z}_2 = \hat{\mathbf{z}}_2 ^n).
\end{split}
\end{align}
This map has a branching locus at the  plane $\{ \, \mathbf{z}_2 = 0\, \}$, which is a copy of the complex line ${\BC}_{1}$. 

A more rigorous treatment, e.g. a K{\"a}hler quotient construction of the resolution of singularities $\widetilde{{\BC}^{4}/({\Gamma}_{24} \times {\Gamma}_{34})}$ adds to that ${\BC}^{2}_{12}$ a chain of two-spheres. 
The fractional instanton charges we are about to discuss below can be interpreted as the fluxes of Chan-Paton bundles on those spheres.

At any rate, the four-dimensional gauge theory on the orbifold $\hat{\BC}_1 \times \left( \hat{\mathbb{C}}_2 /\mathbb{Z}_n \right)$ is equivalent to gauge theory on the smooth space ${\BC}^2_{12}$ supplemented with specific singular boundary conditions imposed on its fields along $\mathbb{C}_1$ at $\mathbf{z}_2 = 0$. Since the orbifold reduces the isometries of the spacetime, the supersymmetry
preserved by the surface defect is at least a half of the original supersymmetry. 

\bigskip \noindent $\bullet$ To distinguish the orbifold and the ordinary space with the defect, we use the notation of putting hats above the symbols on the orbifold side, as in $\hat{\mathbb{C}}_1 \times \left( \hat{\mathbb{C}}_2 / \mathbb{Z}_n \right)$. This rule will also apply to gauge theory parameters such as Coulomb moduli and the masses of hypermultiplets.

The singular boundary condition prescribed by the orbifold breaks the global symmetry. The boundary condition along $\{\hat{\bz}_2 = 0 \}$ can be written as
\begin{align}
    A_\mu dx^\mu \sim \text{diag}(\a_1, \a_2, \cdots, \a_{N}) d\theta,
\end{align}
where $(r,\th)$ are local radial coordinates near the surface $\{\hat{\bz}_2 = 0\}$. Then the singularity $(\a_1,\cdots \a_N)$ is assumed to have the structure
\begin{align}
    (\a_1, \cdots, \a_N) = (\underbrace{\a_{(0)}, \cdots, \a_{(0)}}_{N_0}, \underbrace{\a_{(1)}, \cdots \a_{(1)}}_{N_1}, \cdots, \underbrace{\a_{(n-1)},\cdots, \a_{(n-1)}} _{N_{n-1}}), 
\end{align}
explicitly breaking the global gauge symmetry to $U(N_0) \times \cdots \times U(N_{n-1}) \subset U(N)$. This is a choice that characterizes the surface defect, which can be conveniently encoded in the \textit{coloring function} $c:[N]\longrightarrow \mathbb{Z}_n$ that assigns a representation $\mathfrak{R}_{c(\alpha)}$ of $\BZ_n$ to each Coulomb modulus $\hat{a}_\alpha$, $\a= 1,\dots,N$.  

In the presence of $N$ (anti-)fundamental hypermultiplets, the singular boundary condition breaks the flavor symmetry in a similar way. The choice of the remnant flavor symmetry characterizes the surface defect, encoded in additional coloring functions $\sigma^\pm:[N]\to\mathbb{Z}_n$ which assign a representation $\mathfrak{R}_{\sigma^{\pm}(f)}$ to each (anti-)fundamental hypermultiplet mass $\hat{m}_f^{\pm}$, $f=1,\dots,N$.

By localization the path integral of the four-dimensional $\EuScript{N}=2$ gauge theory on the orbifold reduces to a finite-dimensional integral over the moduli space $\mathfrak{M}_{\mathbb{C}^2 _{12}}  ^{\text{orb}} $ of instantons on the orbifold. There is a natural projection $\r$ of this moduli space to the moduli space of instantons on the ordinary $\mathbb{C}^2 _{12}$, $\r: \mathfrak{M}_{\hat{\mathbb{C}}^2 _{12}} ^{\text{orb}} \longrightarrow \mathfrak{M}_{\mathbb{C}^2 _{12}}$, induced by the map \eqref{eq:map}. Accordingly, the equivariant integration over $\mathfrak{M}_{\hat{\mathbb{C}}^2 _{12}} ^{\text{orb}}$ can be broken into the integration over the bulk $\mathfrak{M}_{\mathbb{C}^2 _{12}}$ and the integration over the fiber of the projection. Namely,
\begin{align}
    \hat{\mathcal{Z}}_{\hat{\mathbb{C}}^2 _{12}} = \int_{\mathfrak{M}_{\hat{\mathbb{C}}^2 _{12}} ^{\text{orb}}} ^{\mathsf{T_H}} 1 =   \int_{\mathfrak{M}_{\mathbb{C}^2 _{12}} } ^{\mathsf{T_H}} \EuScript{S} = \Big\langle \EuScript{S} \Big\rangle \mathcal{Z}_{\mathbb{C}^2 _{12}}.
\end{align}
Hence the integration over the fiber of the projection gives rise to the surface defect observable $\EuScript{S}$ supported on the $\mathbf{z}_1$-plane, so that the path integral of the $\EuScript{N}=2$ gauge theory on the orbifold is identified with the vacuum expectation value of the surface defect observable $\EuScript{S}$.

Having established the concept of the surface defect in interest, we now turn to the actual computation of the partition function $ \hat{\mathcal{Z}}_{\hat{\mathbb{C}}^2 _{12}}$. The $\mathbb{Z}_n$-orbifold used in this construction can be embedded into the gauge origami, and a slight generalization of the gauge origami setup described earlier provides a systematic way to obtain the partition function, as we explain below.  

\subsubsection{Vacuum expectation value of the surface defect observable}
In the gauge origami setup where our bulk four-dimensional gauge theory is embedded as the effective field theory on the wolrdvolume of D3-branes on $\mathbb{C}^2 _{12} \subset S \subset Z$, so that the $4_{\BC}$-dimensional orbifold is $\hat{Z}= \hat{\mathbb{C}}^4_{1234} / \left( \mathbb{Z}_p \times \mathbb{Z}_n \right)$, where the ${\Gamma}_{24} \times {\Gamma}_{34}$ action is given by
\begin{align}
\begin{split}
    (\hat{\bz}_1, \hat{\bz}_2, \hat{\bz}_3,\hat{\bz}_4) \longmapsto (\hat{\bz}_1, \eta \hat{\bz}_2, \zeta  \hat{\bz}_3, \zeta^{-1} \eta^{-1} \hat{\bz}_4) ,\quad \eta^n = \zeta^p = 1.
\end{split}
\end{align}
Note that even though the $SU(4)$ symmetry is broken by the  presence of the branes and orbifolds, 
its maximal torus $U(1)^3 _\ve \subset SU(4)$ still acts preserving the whole structure. The gauge origami partition function can still be computed by equivariant localization applied to the $\mathbb{Z}_p \times \mathbb{Z}_n$-invariant locus. Thus the constructions of codimension-two and codimension-four defects in four-dimensional gauge theory investigated in the previous section generalize to the current setup with the additional orbifold. Again, we focus on the $U(N)$ gauge theory with $N$ fundamental and $N$ anti-fundamental hypermultiplets, which is obtained by choosing $p=3$ and taking the decoupling limit removing the instantons with nonzero $\mathbb{Z}_3$-charges. 

Specifically, we consider a stack of $3N$ parallel branes extended along $\hat{\mathbb{C}}^{2}_{12}$, with $N$ branes of every  ${\BZ}_3$-charge, as above. Furthermore, we assign the ${\BZ}_{n}$ charges to colors and flavors with the help of the functions $c:[N] \longrightarrow \mathbb{Z}_n$ and $\s^\pm : [N] \longrightarrow \mathbb{Z}_n$. Namely, cf. \eqref{eq:rep1}
\begin{align}
    \hat{N}_{12} = \sum_{\o \in \mathbb{Z}_n} \left( \sum_{\a \in c^{-1} (\o)  } e^{\hat{a}_\a} \mathcal{R}_0 \otimes \mathfrak{R}_\o 
    + \sum_{f \in \left(\s^- \right) ^{-1} (\o)} e^{\hat{m} _f ^- - {\ve}_4} \mathcal{R}_{1} \otimes \mathfrak{R}_\o 
    + \sum_{f \in \left(\s^+ \right) ^{-1} (\o)} e^{\hat{m} _f ^+ - {\ve}_3} \mathcal{R}_{2} \otimes \mathfrak{R}_\o  \right) \ .
\end{align}
 Then the partition function is given by
\begin{align}
    \hat{\mathcal{Z}}_{X;c,\s^\pm} = \sum_{\hat{\boldsymbol\l}} \prod_{\substack{i \in \mathbb{Z}_3 \\ \o \in \mathbb{Z}_n}} \hat{\qe}_{i,\o} ^{\vert \hat{\boldsymbol\l}_{i,\o} \vert} \mathbb{E} \left[ - \frac{\hat{P}_3 \hat{S}_{12} \hat{S}_{12} ^*}{\hat{P}_{12} ^*} \right]^{\mathbb{Z}_3 \times \mathbb{Z}_n}.
\end{align}
As above, we take the freezing limit whereby the instantons with nonzero $\mathbb{Z}_3$-charges are eliminated. The remaining couplings account for the instantons with zero $\mathbb{Z}_3$-charge. There are $n$ such couplings, which we denote by $\left(\hat{\qe}_\o \right)_{\o\in \mathbb{Z}_n}$,  to account
for the ${\Gamma}_{24}$-charges. 
In the language of string theory, the bulk instantons fractionalize into $n$ types by getting onto the surface of the surface defect. The coupling ${\hat\qe}_{\o}$ counts the number of fractional instantons of type $\o$, so that $\hat{\qe}_{\o+ n} \equiv \hat{\qe}_\o$. The bulk coupling is recovered from:
\begin{align}
    {\qe} = \hat{\qe}_0 \hat{\qe}_1\cdots \hat{\qe}_{n-1} \quad .
\end{align}
We define useful variables $\{z_{\omega}\}_{\omega=0}^{n-1}$ via
\begin{align}\label{def:z_w}
    \hat{\kq}_{\omega} = \frac{z_{\omega+1}}{z_{\omega}}\, ,  
\end{align}
which we extend to all integers by $z_{\omega+n} = {\qe} z_{\omega}$.
Thus, upon freezing the partition function becomes
\begin{align} \label{eq:orbpart}
    \hat{\mathcal{Z}}_{X;c,\s^\pm} = \sum_{\hat{\boldsymbol\l}}\  \prod_{\o \in \mathbb{Z}_n} \hat{\qe}_\o ^{\vert \hat{\boldsymbol\l}_\o \vert} \ \mathbb{E} \left[ \frac{-\hat{S} \hat{S}^* + \hat{M}^+ \hat{S}^* + \hat{M}^{-} \hat{S}^*}{\hat{P}_{12} ^*} \right]^{\mathbb{Z}_n}.
\end{align}

The projection onto the $\mathbb{Z}_n$-invariant piece can be performed in a way that reveals the structure of $\hat{\mathcal{Z}}_S$ as the vacuum expectation value of the surface defect observable. First, note that the projection $\rho : \mathfrak{M}_{\mathbb{C}^2 _{12}} ^{\text{orb}} \longrightarrow \mathfrak{M}_{\mathbb{C}^2 _{12}}$ descends to the projection at the level of fixed points which we still denote as $\rho : \hat{\boldsymbol\l}  \longrightarrow \boldsymbol\l$. This projection between partitions is given by $\r(\hat{\boldsymbol\l}) = \boldsymbol\l$, where \cite{Nikita:IV,Jeong:2018qpc}
\begin{align}
    \l_i ^{(\a)} \equiv \left\lfloor\frac{\hat{\l}_i ^{(\a)} + c(\a)}{n} \right\rfloor, \quad 1\leq i \leq l\left( \hat{\l} ^{(\a)} \right), \quad \a = 1, \cdots, N.
\end{align}
Next, the two descriptions of the gauge theory, the orbifold defect and Gukov-Witten surface defect, are connected by the mapping \eqref{eq:map} accompanied by the appropriate redefinition of the gauge theory parameters. It is immediate to see that the $\Omega$-background parameters are related by
\begin{align} \label{eq:shiftomega}
    n\hat{\ve}_2 \equiv \ve_2,
\end{align}
corresponding to the $z_2 = \hat{z}_2 ^n$ mapping (the parameter ${\hat\ve}_2$ was denoted by ${\tilde\ve}_{2}$ in \cite{NT}). We further define shifted Coulomb moduli and the shifted hypermultiplet masses by
\begin{align}\label{shifted moduli}
{\hat{a}_\alpha}-{c(\alpha)}\hat{\ve}_2 \equiv {{a}_\alpha};\quad {\hat{m}_f^\pm}-{{\sigma^\pm(f)}}\hat{\ve}_2 \equiv {{m}_f^\pm}.
\end{align}
(again, ${\hat a}_{\alpha}$ and $\hat{m}_f^{\pm}$ correspond to ${\tilde a}_{\alpha}$ and ${\tilde m}_{f}^{\pm}$ of \cite{NT}, respectively). 
These shifted moduli will be the relevant parameters of the gauge theory after the mapping onto the ordinary $\mathbb{C}^2 _{12}$ with the surface defect. Note that these parameters are neutral under the $\mathbb{Z}_n$-action due to the shifts by the right amount of $\hat{\ve}_2$.

Keeping the new parameters in mind, we can write out all the relevant characters decomposed according to the $\mathbb{Z}_n$-representations as
\begin{subequations} \label{eq:rechar}
\begin{align}
    &\hat{N}=\sum_{\omega=0}^{N-1} {N}_\omega \hat{q}_2^{\omega}{\mathfrak{R}}_\omega,\quad {N}_\omega=\sum_{c(\alpha)=\omega}e^{{a}_\a},\quad {N} \equiv \sum_{\omega=0}^{N-1} {N}_\omega; \\
    &\hat{M} ^\pm =\sum_{\omega=0}^{N-1} {M} ^\pm _{\omega}\hat{q}_2^{\omega}{\mathfrak{R}}_\omega,\quad {M} ^\pm _\omega=\sum_{\sigma^\pm (f)=\omega}e^{{m}_{f}^{\pm}},\quad {M} ^\pm \equiv \sum_{\omega=0}^{N-1}M ^\pm _\omega; \\
    &\hat{K}=\sum_{\omega=0}^{N-1}{K}_\omega \hat{q}_2^{\omega}{\mathfrak{R}}_\omega,\quad {K}_\omega=\sum_{\alpha}e^{{a}_\alpha} \sum_{\underset{c(\alpha)+j-1 \equiv \omega \text{ mod } n }{(i,j)\in\hat{\lambda}^{(\alpha)}}}q_1^i\hat{q}_2^j, \quad K \equiv K_{n-1} ; \\
    & \hat{S}=\hat{N}-P_1\left(1-\hat{q}_2{\mathfrak{R}}_1\right)\hat{K}= \sum_\o \hat{S}_\o \mathfrak{R}_\o =\sum_\omega {S}_\omega \hat{q}_2^{\omega}{\mathfrak{R}}_\omega.
\end{align}
\end{subequations}
Note that the ADHM data for the ordinary instantons on $\mathbb{C}^2 _{12}$ are realized as linear maps on the spaces $N$ and $K$. In particular, we have
\begin{align}
\begin{split}
    K &= K_{n-1} = \sum_{\a=1} ^N e^{a_\a} \sum_{\underset{c(\alpha)+j-1 \equiv n-1 \text{ mod } n }{(i,j)\in\hat{\lambda}^{(\alpha)}}}q_1^i\hat{q}_2^j
    = \sum_{\a=1} ^N e^{a_\a} \sum_{(i,j) \in \l^{(\a)}} q_1 ^i q_2 ^j,
\end{split}
\end{align}
where we used $\rho(\hat{\boldsymbol\l}) = \boldsymbol\l$ in the last equality. This implies that the linear maps defined on the vector spaces $N$ and $K$ are indeed the ADHM data for the moduli space $\mathfrak{M}_{\mathbb{C}^2 _{12}}$ of instantons on the ordinary $\mathbb{C}^2 _{12}$. In particular, the number of instantons on $\mathbb{C}^2 _{12}$ is determined by the number of instantons on the $\mathbb{Z}_n$-orbifold with the $\mathbb{Z}_n$-charge $n-1$: $\vert \boldsymbol\l \vert = \vert \hat{\boldsymbol\l}_{n-1} \vert$.

We have defined the fractional characters
\begin{subequations}
\begin{align}
{S}_\omega
&={N}_\omega-P_1 {K}_\omega+P_1 {K}_{\omega-1}\, , \qquad \omega=1,\dots,n-1; \\
{S}_0
&={N}_0-P_1 {K}_0+q_2P_1 {K}_{n-1}.
\end{align}
\end{subequations}
Then the character of the universal sheaf $S$ is obtained by summing over the fractional characters $S_\o$,
\begin{align}
    \quad {S}=\sum_{\omega=0}^{N-1}{S}_\omega = N - P_1 P_2 K. \label{S}
\end{align}

We define fractional $\EY$-function:
\begin{align}
    \EY_{\omega}(x) = \BE \left[- e^x S_{\omega}^*\right]
\end{align}
so that the bulk $\EY(x)$ is a product of all fractional $\EY(x)$ by virtue of \eqref{S}:
\begin{align}
    \EY(x) = \prod_{\omega\in\BZ_n} \EY_{\omega}(x).
\end{align}
Finally, using the characters with shifted parameters, the partition function \eqref{eq:orbpart} can be reorganized as
\begin{align}
    \hat{\CalZ}_{X;c,\sigma^\pm} = \sum_{\boldsymbol\lambda} \kq^{|\boldsymbol\lambda|} \EZ_{\rm bulk}[\boldsymbol\lambda]  \sum_{\hat{\boldsymbol\lambda} \in \rho ^{-1} ( \boldsymbol\lambda ) } \prod_{\omega=0}^{n-1} z_\omega^{k_{\omega-1}-k_{\omega}} \EZ_{\rm defect}[\hat{\boldsymbol\lambda}]  .
\end{align}
The bulk and regular surface defect contributions to the grand canonical ensemble are 
\begin{subequations}\label{def:defect ensemble weights}
\begin{align}
     & \EZ_{\rm bulk}[\boldsymbol\lambda] = \BE \left[ \frac{-SS^*+M S^* }{P_{12}^*}\right], \\
     & \EZ_{\rm defect}[\hat{\boldsymbol\lambda}] = \BE \left[ \frac{1}{P_1^*} \sum_{0 \leq \omega<\omega' \leq N-1} \left( S_{\omega}S_{\omega'}^* - M_{\omega}^+S_{\omega'}^* - M_{\omega}^{-}S_{\omega'}^* \right) \right].
\end{align}
\end{subequations}
Now, the partition function can be expressed in the following form by treating the defect contributions as observable:
\begin{align}
    \hat{\CalZ}_{X;c,\s^\pm} = \sum_{\boldsymbol\l} \qe^{\vert \boldsymbol\l \vert} \; \EuScript{S}_{c,\s^\pm} [\boldsymbol\l] \; \EZ_{\rm bulk}[\boldsymbol\lambda] = \Big\langle \EuScript{S}_{c,\s^\pm} \Big\rangle \mathcal{Z}_{\mathbb{C}^2 _{12}},
\end{align}
where:
\begin{align} \label{eq:surfobs}
    \EuScript{S}_{c,\s^\pm}[\boldsymbol\l] = \sum_{\hat{\boldsymbol\l} \in \r^{-1} (\boldsymbol\l)} \prod_{\o=0} ^{n-1} z_\o ^{k_{\o-1} -k_\o} \EZ_{\rm defect}[\hat{\boldsymbol\lambda}]
\end{align}
The $P_1 ^* = 1- q_1 ^{-1}$ in the denominator signifies this contribution is indeed coming from integrating out the degrees of freedom on the $\mathbf{z}_1$-plane. Therefore, the partition function is interpreted as the vacuum expectation value of the surface defect observable supported on the $\mathbf{z}_1$-plane upon the mapping \eqref{eq:map}.

The expression \eqref{eq:surfobs} of the surface defect observable in fact suggests a dual description of the defect: a two-dimensional sigma model coupled to the bulk four-dimensional gauge theory. Indeed, it can be shown that \eqref{eq:surfobs} gives the partition function of the two-dimensional supersymmetric sigma model on a bundle over flag variety in the decoupling limit $\qe \to 0$ \cite{Nikita:IV,Jeong:2018qpc}. The coupling between the two-dimensional sigma model and the four-dimensional gauge theory produces additional terms at nonzero $\qe$.

Although the orbifold construction produces the description of all Gukov-Witten type defects, our main interest will be the special case referred to as the \textit{regular} surface defect. The latter breaks the gauge group down to its maximal torus along the surface of the surface defect. It is defined by choosing $n = N$ and the coloring functions $c(\alpha)$ and $\sigma^\pm(f)$ as the one-to-one functions,
\begin{align}\label{eq:std-color}
    c(\alpha)=\alpha-1, \ \alpha=1,\dots,N; \quad \sigma^\pm(f) = f-1, \ f=1,\dots,N.
\end{align}
unique up to the $S(N) \times S(2N)$ permutations. 

\subsection{Folded branes on orbifold and fractional $Q$-observables} \label{sec:fracqobs}
Now we turn to the case where we insert additional stacks of branes on top of the stack on $\hat{\mathbb{C}}^2_{12} \subset \hat{Z}$, and on top of the regular surface defect. Again, the ${\BZ}_3$-charges are assigned as $n_{12,0} = n_{12,1} = n_{12,2} = N$ and $n_{23,1} = 1$. The ${\BZ}_N$-charges are assigned so as to produce the regular surface defect, namely, 
\begin{align}
\begin{split}
  \hat{N}_{12} 
   & = \sum_{\o'' \in \mathbb{Z}_N} \left(  e^{\hat{a}_{\o''+1}} \mathcal{R}_0 \otimes \mathfrak{R}_{\o''} +  e^{\hat{m} _{\o''+1} ^- - \ve_4} \mathcal{R}_{1} \otimes \mathfrak{R}_{\o''} + e^{\hat{m} _{\o''} ^+ - \ve_3} \mathcal{R}_{2} \otimes \mathfrak{R}_{\o''}  \right) \\
   & = \sum_{\omega''=0}^{N-1} 
   \left(  e^{{a}_{\o''+1}} \hat{q}_2^{\omega''} \mathcal{R}_0 \otimes \mathfrak{R}_{\o''} + e^{{m} _{\o''} ^- - \ve_4} \hat{q}_2^{\omega''} \mathcal{R}_{1} \otimes {\mathfrak{R}}_{\o''} + e^{{m} _{\o''+1} ^+ - \ve_3} \hat{q}_2^{\omega''} \mathcal{R}_{2} \otimes {\mathfrak{R}}_{\o''}   \right) \\
   \hat{N}_{23} 
   & = e^{x+(\o+1)\hat{\ve}_2+\hat{\ve}_3} \mathcal{R}_1 \otimes {\mathfrak{R}}_{\o+1} \ .
\end{split}
\end{align}
Here, given a choice of the surface defect data, there are additional $N$ choices of $\o \in \mathbb{Z}_N$. The gauge origami partition function is written as
\begin{align}
    \hat{\mathcal{Z}}_{X;\o} = \sum_{\hat{\boldsymbol\l}} \prod_{\substack{i= 0,1,2 \\ \o' \in \mathbb{Z}_N }} \hat{\qe}_{i,\o'} ^{\vert \hat{\boldsymbol\l}_{i,\o'} \vert} \mathbb{E}\left[ - \frac{\hat{P}_3 \hat{S}_{12} \hat{S}_{12} ^*}{\hat{P}_{12} ^*} - \frac{ \hat{P}_1  \hat{S}_{23} \hat{S}_{23}^* }{\hat{P}_{23} ^*} + \hat{q}_3 \hat{P}_{4} \frac{\hat{S}_{12} \hat{S}_{23}^*}{\hat{P}_{2} ^*} \right] ^{\mathbb{Z}_3 \times \mathbb{Z}_N}.
\end{align}
Removing the instantons with ${\CalR}_{\pm 1}$ $\mathbb{Z}_3$-charge by setting the corresponding fractional couplings to zero, we get:
\begin{align} \label{eq:partfracq}
    \hat{\mathcal{Z}}_{X;\o} = \sum_{\hat{\boldsymbol\l}} \prod_{\o' \in \mathbb{Z}_N} \hat{\qe}_{\o'} ^{\vert \hat{\boldsymbol\l}_{\o'} \vert}  \mathbb{E} \left[ \frac{-\hat{S}\hat{S}^* +\hat{M}^+ \hat{S}^* +\hat{M}^{-} \hat{S}^*}{\hat{P}^* _{12}} - \frac{e^{x+\o \hat{\ve}_2} (\hat{S}^* - \hat{M}^{-*})}{\hat{P}_2 ^*}  \right]^{\mathbb{Z}_N}.
\end{align}
The first term is precisely the measure that defines the partition function of the gauge theory on the orbifold, \eqref{eq:orbpart}. The second term can be interpreted as the surface defect observable supported on the $\hat{\mathbf{z}}_2$-plane, obtained by integrating out the degrees of freedom on $\hat{\mathbb{C}}^2 _{23}$. 

\bigskip \noindent $\bullet$
We define the \textit{fractional $Q$-observables} (with the \emph{mostly entire} convention, to be made more precise below) by
\begin{align}\label{def:hatQ}
    {Q}_\o (x) & \equiv \mathbb{E} \left[ - e^{x+ \o \hat{\ve}_2}  \frac{\hat{S}^* }{\hat{P}_2 ^*}  \right]^{\mathbb{Z}_N}, \quad \o=0, \cdots, N-1. \\
    & = \BE \left[ - \frac{e^x}{1-q_2^{-1}}\sum_{\omega'\leq\omega}S_{\omega'} ^*  - \frac{e^x q_2^{-1}}{1-q_2^{-1}}\sum_{\omega'>\omega} S_{\omega'} ^*  \right] \nonumber
\end{align}
Note that the ratio of fractional $Q$-observables produces the fractional $\EY$-observables,
\begin{align}
    \frac{{Q}_\o (x)}{{Q}_{\o-1} (x)} = \mathbb{E} \left[ - e^{x }  {S}^* _\o \right]=\EuScript{Y}_\o (x).
\end{align}
It will be convenient to define the $\G$-dressed fractional $Q$-observables:
\begin{align} \label{def:tilq}
    \tilde{Q}_{\o} (x) \equiv \frac{{\ve}_{2}^{-\o} \, Q_{\o}(x)}{\prod\limits_{\o'=0} ^{N-1}  \G \left( 1+ \frac{x- m^+ _{\o-\o'}}{\ve_2} \right)}.
\end{align}
Then the ratio of the redefined fractional $Q$-observables is
\begin{align}
    \frac{\tilde{Q}_\o (x)}{\tilde{Q}_{\o-1} (x) } =  \frac{\EY_\o (x)}{ x- m^+ _\o} 
\end{align}
This identity will be useful in deriving the fractional quantum T-Q equation in section~\ref{sec:TQ-DS-frac}.

In these notations, the partition function can be expressed as expectation value of an observable on the colored partitions $\hat{\boldsymbol\l}$ as
\begin{align}
    \hat{\mathcal{Z}}_{X;\o} = \Big\langle {Q}_\o (x ) \Big\rangle_{\mathbb{Z}_N} \hat{\mathcal{Z}}_{\hat{\mathbb{C}}^2 _{12}},
\end{align}
where the subscript $\mathbb{Z}_N$ is to distinguish from the vacuum expectation value in the bulk gauge theory without
defects.

\bigskip \noindent $\bullet$
Using the map \eqref{eq:map} the same partition function \eqref{eq:partfracq} is interpreted as the correlation function of the intersecting surface defect observables, supported on the $\mathbf{z}_1$-plane and the $\mathbf{z}_2$-plane respectively, in the absence of the $\mathbb{Z}_N$-orbifold. More pedantically, one could distinguish three operators whose correlation function represents \eqref{eq:partfracq}: the two surface defects and one bi-local operator, inserted at their intersection.

By the redefinition of parameters and the rearrangement of the characters \eqref{eq:shiftomega}, \eqref{shifted moduli}, \eqref{eq:rechar}, the partition function becomes
\begin{align}
    \hat{\mathcal{Z}}_{X;\o} = \sum_{\boldsymbol\l} \qe^{\vert \boldsymbol\l \vert} \; \EuScript{O}_\o (x) [\boldsymbol\l] Q (x) [\boldsymbol\l]\; \mathbb{E}\left[ \frac{-SS^* + M  S^* }{P_{12} ^*} \right] =  \Big\langle \EuScript{O}_\o (x) Q (x) \Big\rangle \mathcal{Z}_{\mathbb{C}^2 _{12}},
\end{align}
where the $Q$-observable supported on the $\mathbf{z}_2$-plane is still given by
\begin{align}
    Q(x)[\boldsymbol\l] = \mathbb{E} \left[ - \frac{e^x S^* [\boldsymbol\l] }{P_2 ^*} \right],
\end{align}
and the observable $\EuScript{O}_\o$ is obtained as
\begin{align} \label{eq:oo}
    \EuScript{O}_\o (x) [\boldsymbol\l] &=  \sum_{\hat{\boldsymbol\l} \in \r^{-1} (\boldsymbol\l)} \prod_{\o'=0} ^{N-1} z_{\o'} ^{k_{\o'-1} -k_{\o'}} \EuScript{Z}_{\rm defect} [\hat{\boldsymbol\l} ] \, \Xi_\o ^{(0)} (x) [\hat{\boldsymbol\l}],
\end{align}
where $\EuScript{Z}_{\rm defect}$ is the contribution from the regular surface defect \eqref{def:defect ensemble weights} and $\Xi^{(0)} _\o (x)$ is the contribution from interaction between the two defects through the intersection point (the origin in our setup) given by
\begin{align} \label{def:inter}
    \Xi ^{(0)} _\o (x) [\hat{\boldsymbol\l}] \equiv \BE\left[ e^x \sum_{ \o < \o' \leq N-1}  S_{\o'} ^* [\hat{\boldsymbol\l}] \right]. 
\end{align}

It should be noted that the interaction term \eqref{def:inter} vanishes for $\o = N-1$. In this case we simply get
\begin{align}
    {Q}_{N-1} (x) = Q(x).
\end{align}
Namely, the bulk $Q$-observable is identified with the last component of the fractional $Q$-observables. This also agrees with the definition of the fractional $Q$-observables in \eqref{def:hatQ}.

\subsection{Fractional $qq$-characters}
We top up by another stack of branes on the orbifold $\hat{Z} = \hat{\mathbb{C}}^4 _{1234} /\left( \mathbb{Z}_3 \times \mathbb{Z}_N \right)$. This is the generalization of the construction of the $qq$-characters from the crossed instantons that we have seen in the section \ref{sec:crbr}. In the presence of the orbifold, these branes fractionalize according to the ${\BZ}_N$-representations. 
Accordingly, integrating out the degrees of freedom of these additional fractional branes 
 produces the \textit{fractional} $qq$-characters. Technically this is done by computing the orbifold version of the gauge origami partition function. It can be cast in the form of the correlation function of a point-like observable and the intersecting surface defects.

\subsubsection{$qq$-character and the regular surface defect}
We consider the case where there are two stacks of intersecting branes on $\hat{\mathbb{C}} ^2 _{12}$ and $\hat{\mathbb{C}}^2 _{34}$. The corresponding Chan-Paton spaces are the representations of the orbifold group:
\begin{align}
\begin{split}
   \hat{N}_{12} 
   & = \sum_{\o'' \in \mathbb{Z}_N} \left(  e^{\hat{a}_{\o''+1}} \mathcal{R}_0 \otimes \mathfrak{R}_{\o''} +  e^{\hat{m} _{\o''} ^- - \ve_4} \mathcal{R}_{1} \otimes \mathfrak{R}_{\o''} + e^{\hat{m} _{\o''+1} ^+ - \ve_3} \mathcal{R}_{2} \otimes \mathfrak{R}_{\o''}  \right) \\
   & = \sum_{\omega''=0}^{N-1} 
   \left(  e^{{a}_{\o''+1}} \hat{q}_2^{\omega''} \mathcal{R}_0 \otimes \mathfrak{R}_{\o''} + e^{{m} _{\o''} ^- - \ve_4} \hat{q}_2^{\omega''} \mathcal{R}_{1} \otimes \mathfrak{R}_{\o''} + e^{{m} _{\o''+1} ^+ - \ve_3} \hat{q}_2^{\omega''} \mathcal{R}_{2} \otimes \mathfrak{R}_{\o''}   \right) \\
   %%%%%%%%%%%%%%%%%%%%%%%%%%%%%%%%%%%%
%   \hat{N}_{23} 
%   & = e^{x'+\hat{\ve}_2 + \ve_3} \hat{q}_2^{\omega'} \ \mathcal{R}_1 \otimes \mathcal{R}_{\o'+1}  \nonumber\\
   %%%%%%%%%%%%%%%%%%%%%%%%%%%%%%%%%%%%
   \hat{N}_{34} 
   & = e^x \hat{q}_2^{\omega} \ \mathcal{R}_0 \otimes \mathfrak{R}_\o \ .
\end{split}
\end{align} 
Again, taking decoupling limit, the gauge origami partition function reduces to 
\begin{align} \label{eq:fracqqpart}
\begin{split}
    \hat{\mathcal{Z}}_{X;\o} &=    \sum_{\hat{\boldsymbol\l}} \prod_{\o' \in \mathbb{Z}_n} \hat{\qe}_{\o'} ^{\vert \hat{\boldsymbol\l}_{\o'} \vert} \EZ_{\rm bulk}[\boldsymbol\lambda] \EZ_{\rm defect} [\hat{\boldsymbol\lambda}] \left( \mathbb{E} \left[ - e^{x+{{\ve}_1}} {S}^* _{\o+1} \right] + \hat{\qe}_{\o} \mathbb{E} \left[ e^x {{S}}^* _\o - e^x e^{-m_{\omega}^+} _\o - e^{x} e^{-m_{\omega}^-}  \right] \right) \\
    &=    \sum_{\hat{\boldsymbol\l}} \prod_{\o' \in \mathbb{Z}_n} \hat{\qe}_{\o'} ^{\vert \hat{\boldsymbol\l}_{\o'} \vert} \EZ_{\rm bulk}[\boldsymbol\lambda] \EZ_{\rm defect} [\hat{\boldsymbol\lambda}] \left( {{\EuScript{Y}}}_{\o+1} (x+{{\ve}_1}) + \hat{\qe}_\o \frac{{{P}}_\o  (x)}{{{\EuScript{Y}}}_\o (x)}  \right) \\
    & = \left\langle  {{\EuScript{Y}}}_{\o+1} (x+{{\ve}_1}) + \hat{\qe}_\o \frac{{{P}} _\o  (x) }{{{\EuScript{Y}}}_\o (x)} \right\rangle_{\mathbb{Z}_N} \hat{\mathcal{Z}}_{\hat{\mathbb{C}}^2 _{12}},
\end{split}
\end{align}
where the subscript indicates it is a vacuum expectation value in the gauge theory on the $\mathbb{Z}_N$-orbifold. The point-like observables defined in this way are called the \emph{fractional $qq$-characters},
\begin{align} \label{eq:fracqq}
    {\EuScript{X}}_\o (x) \equiv {{\EuScript{Y}}}_{\o+1} (x+{{\ve}_1}) + \hat{\qe}_\o \frac{{{P}} _\o  (x) }{{{\EuScript{Y}}}_\o (x)}.
\end{align}
Using the map \eqref{eq:map}, the partition function \eqref{eq:fracqqpart} can also be viewed as the correlation function of the regular surface defect observable supported on the $\mathbf{z}_1$-plane and a point-like observable at the origin.

\subsubsection{With fractional $Q$-observables} \label{sec:fracqqq}
At last, we consider the three stacks of branes on $\hat{\mathbb{C}}^2 _{12}$, $\hat{\mathbb{C}}^2 _{23}$, and $\hat{\mathbb{C}}^2 _{34}$. The orbifold group representations for the Chan-Paton spaces are:
\begin{align}
   \hat{N}_{12} 
   & = \sum_{\o'' \in \mathbb{Z}_N} \left(  e^{\hat{a}_{\o''+1}} \mathcal{R}_0 \otimes {\mathfrak{R}}_{\o''} +  e^{\hat{m} _{\o''} ^- - \ve_4} \mathcal{R}_{1} \otimes {\mathfrak{R}}_{\o''} + e^{\hat{m} _{\o''+1} ^+ - \ve_3} \mathcal{R}_{2} \otimes {\mathfrak{R}}_{\o''}  \right) \nonumber\\
   & = \sum_{\omega''=0}^{N-1} 
   \left(  e^{{a}_{\o''+1}} \hat{q}_2^{\omega''} \mathcal{R}_0 \otimes {\mathfrak{R}}_{\o''} + e^{{m} _{\o''} ^- - \ve_4} \hat{q}_2^{\omega''} \mathcal{R}_{1} \otimes \mathcal{R}_{\o''} + e^{{m} _{\o''+1} ^+ - \ve_3} \hat{q}_2^{\omega''} \mathcal{R}_{2} \otimes {\mathfrak{R}}_{\o''}   \right) \nonumber\\
   %%%%%%%%%%%%%%%%%%%%%%%%%%%%%%%%%%%%
   \hat{N}_{23} 
   & = e^{x'+\hat{\ve}_2 + \ve_3} \hat{q}_2^{\omega'} \ \mathcal{R}_1 \otimes {\mathfrak{R}}_{\o'+1}  \nonumber\\
   %%%%%%%%%%%%%%%%%%%%%%%%%%%%%%%%%%%%
   \hat{N}_{34} 
   & = e^x \hat{q}_2^{\omega} \ \mathcal{R}_0 \otimes {\mathfrak{R}}_\o
\end{align}
Here, notice that there are $N$ choices for both $\omega, \omega'\in\BZ_N$, of total $N^2$ such configurations. The gauge origami partition function is written as 
\begin{align}
    \hat{\CalZ}_{X;\omega,\omega'}
    = \sum_{\hat{\boldsymbol\lambda}} \prod_{\substack{ i\in \mathbb{Z}_3 \\\omega''\in\BZ_N}} \hat{\kq}_{i,\omega''}^{|\hat{\boldsymbol\lambda}_{i,\omega''}|} \BE & \left[ -\frac{\hat{P}_3 \hat{S}_{12}\hat{S}_{12}^*}{\hat{P}_{12}^*} - \frac{\hat{P}_{1} \hat{S}_{23}\hat{S}_{23}^*}{\hat{P}_{23} ^*} - \frac{\hat{P}_{1}\hat{S}_{34}\hat{S}_{34}^*}{\hat{P}_{34} ^*} \right. \\ 
    & \left. -\hat{q}_{12} ^{-1} \hat{S}_{12}\hat{S}_{34} ^* + \hat{q}_{3} \hat{P}_4 \frac{\hat{S}_{12}\hat{S}_{23}^*}{\hat{P}_{2}^*} + \hat{q}_{4} \hat{P}_1 \frac{\hat{S}_{23}\hat{S}_{34}^*}{\hat{P}_{3}^*}   \right] ^{\BZ_3 \times \BZ_N}. \nonumber
\end{align}
We again take the decoupling limit where the instantons with nonzero $\BZ_3$-charges are prohibited. 
Using the same argument as the case without the $\BZ_N$-orbifold, we modify the last term coming from the interaction between $\hat{\BC}^2_{23}$ and $\hat{\BC}_{34}^2$ to 
\begin{align}
    \BE \left[ -\hat{q}_{23}\hat{N}_{34}\hat{N}_{23}^* - \hat{q}_{34} \hat{N}_{23}\hat{N}_{34}^* + \hat{P}_{14} \hat{N}_{23}\hat{K}_{34}^* \right]^{\BZ_3 \times \BZ_N}
\end{align}
With some decent but tedious calculation, we find the gauge origami partition function can be organized into the following form
\begin{align}
    \hat{\CalZ}_{X; \o,\o'} &= -\sum_{\hat{\boldsymbol\l}} \prod_{\o'' \in \mathbb{Z}_N}  \hat{\qe}_{\o''} ^{k_{\omega''}} \EZ_{\rm bulk}[\boldsymbol\lambda] \EZ_{\rm defect} [\hat{\boldsymbol\lambda}] \times \hat{T}_{N+1,\omega} (x)[\hat{\boldsymbol\lambda}] {{Q}}_{\omega'}(x') [\hat{\boldsymbol\lambda}] \nonumber\\
    % & \times \left[ (x-x')^{\d_{\o\o'}} \EY_{\omega+1}(x+\ve_1){Q}_{\o'} (x')+ \hat{\fq}_\o (x-x'+\ve_1)^{\d_{\o\o'}} \frac{{P}_\o  (x) }{{\EuScript{Y}}_\o (x)} {Q}_{\o'} (x') \right]. \\
    & =- \left \langle \hat{T}_{N+1,\omega} (x) {{Q}}_{\omega'}(x') \right \rangle_{\mathbb{Z}_N}  \hat\CalZ_{\hat{\mathbb{C}}^2 _{12}} .
\end{align}
The fractional $qq$-character, which is a correlation function of gauge theory observables, consists of fractional ${Q}$ and $\EY$-observables:
\begin{align}\label{eq:qq-orbid}
    \hat{T}_{N+1,\omega} (x){Q}_{\omega'}(x')
    & = (x-x')^{\d_{\o\o'}} {\EY}_{\omega+1}(x+{\ve}_1){{Q}}_{\o'} (x')+ \hat{\fq}_\o (x-x'+\ve_1)^{\d_{\o\o'}} \frac{ {P}_\o(x) }{{{\EuScript{Y}}}_\o (x)} {{Q}}_{\o'} (x')
\end{align}
Due to the compactness of the moduli space of spiked instantons on orbifold \cite{Nikita:II,NaveenNikita}, the vacuum expectation value of the fractional $qq$-character
\begin{align}
    \Big\langle \hat{T}_{N+1,\omega}(x){{Q}}_{\omega'}(x') \Big\rangle_{\mathbb{Z}_N} =
    & (x-x')^{\delta_{\omega\omega'}} \left \langle {\EY}_{\o+1} (x+{\ve}_1) {{Q}}_{\omega'}(x') \right \rangle_{\mathbb{Z}_N} \nonumber\\
    & + \hat\kq_{\omega} (x-x'+\ve_1)^{\delta_{\omega\omega'}} {P} _{\omega}(x) \left \langle \frac{{{Q}}_{\omega'}(x')}{{\EY}_{\omega}(x)} \right \rangle_{\mathbb{Z}_N} 
\end{align}
is regular in $x$. In particular, it is a degree 2 polynomial in $x$ when $\omega=\omega'$, and a degree 1 polynomial in $x$ when $\omega\neq\omega'$.

% \begin{align}
% \begin{split}
%     \hat{\CalZ}_{X; \o,\o'} &= \sum_{\hat{\boldsymbol\l}} \prod_{\o'' \in \mathbb{Z}_N}  \hat{\qe}_{\o''} ^{\vert \boldsymbol\l_{\o''} \vert} \mathbb{E} \left[ \frac{-\hat{S} \hat{S}^* + \hat{M}^+ \hat{S}^* + \hat{M} ^{-*} \hat{S} }{\hat{P}^* _{12}} \right]^{\mathbb{Z}_N} \times  \\
%     &\left[ (x'-x)^{\d_{\o\o'}} \hat{Q}_{\o'} (x' ) \hat{\EuScript{Y}}_\o (x + \hat{\ve})+ \hat{\fq}_\o (x'-x-\ve_1)^{\d_{\o\o'}} \hat{Q}_{\o'} (x') \frac{\hat{P}_\o ^+ (x) \hat{P}_{\o+1} ^- (x+\hat{\ve})}{\hat{\EuScript{Y}}_\o (x)} \right].
% \end{split}
% \end{align}

\section{Quantum T-Q equations as Dyson-Schwinger equations}\label{sec:TQ-DS}
The $qq$-characters in the $\EuScript{N}=2$ gauge theory contain nontrivial analytic information on the gauge theory correlation functions. Their crucial property is, as introduced in the previous sections, the regularity of their vacuum expectation value \cite{Nikita:I} following from the compactness theorem for the moduli space of spiked instantons \cite{Nikita:II}. The regularity constrains relevant gauge theory correlation functions by requiring the vanishing conditions for their singular parts, yielding nontrivial equations that they have to satisfy. These equations are called the non-perturbative Dyson-Schwinger equations \cite{Nikita:I}.

Generally, the non-perturbative Dyson-Schwinger equations encode the chiral ring relations \cite{Jeong2019b}. With the insertion of surface defects, the chiral ring is subject to nontrivial relations between the observables from the bulk gauge theory and the defect. The $\Omega$-background uplifts these relations to differential equations in coupling constants obeyed by the vacuum expectation value of the defect observable. The non-perturbative Dyson-Schwinger equation can effectively used to exactly derive such differential equations, as shown in \cite{Nikita:V,Jeong:2017pai,Jeong2017,Jeong:2018qpc,JeongNekrasov,Lee:2020hfu}.

In this section, rather, we focus on a set of \textit{difference} equations satisfied by the vacuum expectation values of (fractional) $Q$-observables that we introduced in section \ref{sec:qobs} and section \ref{sec:fracqobs}. We refer to the associated difference equation as the \textit{(fractional) quantum T-Q equation}. We present the relation of these difference equations to the Baxter T-Q equation for the $XXX_{\mathfrak{sl}_2}$ spin chain, which explains how these names are coined.

\subsection{Quantum T-Q equations}

% The strategy of using \eqref{eq:qyeq} is the following: first, compute ${\hat T}_{N+1}(x)$
% by expanding at $x \to \infty$ till $x^{0}$ terms.
% \begin{align}
%     Y(x) = \prod_{\alpha=1}^N(x-a_\alpha) \exp \left[  \frac{{\ve}_1 {\ve}_2}{x^2}k + \sum_{i=2}^\infty\frac{{\ve}_1 {\ve}_2}{x^{i+1}}D^{(i)} \right]
% \end{align}
% In this way ${\hat T}_{N+1}$ is expressed as a polynomial in $k (= D^{(1)})$ and $D^{(i)}$'s with $i$ going up to $N$.
We consider the $qq$-character in the presence of the $Q$-observable, without the regular surface defect. Recall that the $qq$-character in this case is given by \eqref{eq:qyeq}. Now we set $x=x'$ and  to $x = x' - {\ve}_{1}$ in \eqref{eq:qyeq}, yielding 
\begin{subequations}\label{eq:twotq}
\begin{align}
    & \Biggl\langle \EY(x+{\ve}_2) Q(x) \Biggr\rangle = \left \langle Q(x+{\ve}_{2})  \right \rangle
    = - \frac{1}{{\ve}_{1}} \left \langle {\hat T}_{N+1}(x - {\ve}_{1})  Q(x) \right \rangle \ , \\
    %%%%%%%%%%%%%%%%%%%%%%%%%%%%%%%%
    & {\qe} P(x) \Biggl\langle \frac{Q(x)}{\EY(x)} \Biggr\rangle \equiv  {\qe} P(x)\left \langle Q(x-{\ve}_{2}) \right \rangle
    = \frac{1}{{\ve}_{1} } \left \langle {\hat T}_{N+1}(x)  {Q}(x) \right \rangle \ 
\end{align}
\end{subequations}
Let us define a degree $N$ polynomial $T_N (x)$ by,
\begin{align}
     T_N(x) =  \frac{1}{\ve_1} \left (  \hat{T}_{N+1} (x)  -   \hat{T}_{N+1} (x-\ve_1) \right).
\end{align}
The $T_N (x)$ can be obtained explicitly by expanding \eqref{eq:qyeq} in large $x$ and taking the difference at two values. Since the $\EY$-observable is the generating function of the $\EN=2$ chiral observable $\Tr\phi^k$, $k\in \BZ_{>0}$, the coefficients of $T_N (x)$ are given by combinations of them.

Then we obtain the following difference equation for the vacuum expectation value of the $Q$-observable, called the \textit{quantum T-Q equation}:
\begin{align}\label{eq:qTQ}
     \Big\langle Q(x+\ve_2)  \Big\rangle + \fq P (x)  \Big\langle Q(x-\ve_2) \Big\rangle =  \Big\langle T_N(x) Q(x)  \Big\rangle.
\end{align}
In terms of the redefined $Q$-observables, the quantum T-Q equation is written as
\begin{align}
    P^+ (x+\ve_2) \Big\langle \tilde{Q} (x+\ve_2) \Big\rangle + \qe P^- (x) \Big\langle \tilde{Q}(x-\ve_2 ) \Big\rangle = \Big\langle T_N (x) \tilde{Q} (x) \Big\rangle.
\end{align}

Note that the quantum T-Q equation is valid with two non-zero $\Omega$-background parameters $\ve_1$ and $\ve_2$; and also that the vacuum expectation value of the $Q$-observable involves nontrivial ensemble average over partitions. In the limit $\ve_1 \to 0$, the ensemble average is dominated by the evaluation on the limit shape \cite{Nekrasov:2003rj, Poghossian:2010pn}.\footnote{See also \cite{Nikita:IV, Poghosyan:2016mkh, Jeong:2017mfh, Jeong:2017pai} for other studies of the relation between the Baxter TQ equation and the null-vector decoupling equation for degenerate vertex operator in vertex algebra.} In particular, the vacuum expectation value of the $Q$-observable becomes a regular function in $x$, which we may call the Baxter $Q$-operator \cite{bax}. The quantum T-Q equation reduces to the Baxter T-Q equation for $XXX_{\mathfrak{sl}_2}$ spin chain satisfied by the $Q$-operator. The spectra of Hamiltonians, encoded in the coefficients of the polynomial $T_N(x)$, are given by the vacuum expectation values of the chiral observables $\Tr \phi ^k$ in the NS limit. The Baxter T-Q equation obtained in this way is identical to the one in \cite{Nikita-Pestun-Shatashvili} arising from the $q$-characters of the same $\EN=2$ gauge theory.

\subsection{Fractional quantum T-Q equations}\label{sec:TQ-DS-frac} 
We consider the $qq$-characters in the presence of the fractional $Q$-observable, constructed in section \ref{sec:fracqqq}. Recall that the $qq$-characters are given by
\begin{align}
    \hat{T}_{N+1,\omega} (x){Q}_{\omega'}(x')
    & = (x-x')^{\d_{\o\o'}} \EY_{\omega+1}(x+\ve_1){Q}_{\o'} (x')+ \hat{\fq}_\o (x-x'+\ve_1)^{\d_{\o\o'}} \frac{{P}_\o  (x) }{{\EuScript{Y}}_\o (x)} {Q}_{\o'} (x').
\end{align}
The vacuum expectation value is regular in $x$. Namely,
\begin{align}\label{eq:qyeq-frac}
    \left \langle \hat{T}_{N+1,\omega} (x){Q}_{\omega'}(x') \right \rangle _{\mathbb{Z}_N} 
    =
    & (x-x')^{\delta_{\omega\omega'}} \left \langle \EY_{\omega+1}(x+\ve_1) {Q}_{\omega'}(x') \right \rangle _{\mathbb{Z}_N} \nonumber\\ 
    & + \hat\kq_{\omega} (x-x'+\ve_1)^{\delta_{\omega\omega'}} P_{\omega}(x) \left \langle \frac{{Q}_{\omega'}(x')}{\EY_{\omega}(x)} \right \rangle _{\mathbb{Z}_N}  
\end{align}
has no singularities in $x$. We can compute the left hand side by explicitly expanding the right hand side in large $x$. The building block $\EY_{\omega}(x)$ in large $x$ behaves as
\begin{align}\label{eq:Y-large-x}
    {\EY}_{\omega}(x) = (x-{a}_{\omega}) \exp \left[ \frac{\ve_1}{x}\nu_{\omega-1} + \frac{\ve_1}{x^2} D^{(1)}_{\omega-1} + \cdots \right]
\end{align}
with $\nu_{\omega}=k_{\omega}-k_{\omega+1}$ and
$$
    D_{\omega}^{(1)} = \ve_2 k_{\omega} + \sum_{\Box\in {\sK}_{\omega}} \hat{c}_\Box - \sum_{\Box\in {\sK}_{\omega+1}} \hat{c}_\Box = \ve_2 k_\omega + \hat{c}_\omega - \hat{c}_{\omega+1}.
$$
$\hat{T}_{N+1,\omega}(x)$ is a degree 2 polynomial when $\omega=\omega'$
\begin{align}\label{eq:T_N+1,=}
    \hat{T}_{N+1,\omega}(x)
    = & (x-x')(x-{a}_{\omega+1} + \ve_1\nu_{\omega} + \ve_1 ) + \hat\qe_{\omega} (x-x'+\ve_1)(x-{m}_{\omega} + {a}_{\omega} - \ve_1\nu_{\omega-1}) \nonumber\\
    & + \ve_1 D_{\omega}^{(1)} - \hat\qe_{\omega} \ve_1 D_{\omega-1}^{(1)} + \frac{\ve_1^2}{2}\nu_{\omega}^2 - \ve_1 {a}_{\omega+1}\nu_{\omega} \nonumber\\ 
    & + \hat\qe_{\omega} \left( \frac{\ve_1^2}{2}\nu_{\omega-1}^2 + ({m}_{\omega}-{a}_{\omega})\ve_1\nu_{\omega-1} + {P}_{\omega}({a}_{\omega}) \right).
\end{align}
where ${m}_\omega : = \sum_{\pm} {m}_\omega ^\pm$.
$\hat{T}_{N+1,\omega}(x)$ is a degree 1 polynomial when $\omega\neq\omega'$:
\begin{align}
    \hat{T}_{N+1,\omega}(x) = x - {a}_{\omega+1} + \ve_1 + \ve_1 \nu_{\omega} + \hat\kq_{\omega} ( x - {m}_\omega + {a}_{\omega} - \ve_1 \nu_{\omega-1} ).
\end{align}

Now let us repeat the steps deriving the quantum T-Q equation in eq.~\eqref{eq:qTQ}: We set $\omega=\omega'$ and take the difference between $x=x'$ and $x'=x+\ve_1$ cases in \eqref{eq:T_N+1,=}:
\begin{align} \label{eq:fractqint}
    &  (x-m^+ _{\o+1}) \Biggl\langle {\tilde{Q}} _{\omega+1} (x) \Biggl\rangle _{\mathbb{Z}_N}  +  \hat{\qe}_\omega (x-m^- _\o) \Biggl\langle {\tilde{Q}}  _{\omega-1} (x) \Biggl\rangle _{\BZ_N}  \\
    &=\Biggl\langle \left(x -{a}_{\omega+1} +\ve_1 \nu_{\omega} \right) {\tilde{Q}} _\omega (x) \Biggl\rangle _{\mathbb{Z}_N}  + \hat{\qe}_\omega \Biggl\langle \left(x+{a}_\omega-{m}_\omega - \ve_1 \nu_{\omega-1} \right) {\tilde{Q}} _\omega (x)  \Biggl\rangle _{\mathbb{Z}_N} , \nonumber
\end{align}
By multiplying the perturbative prefactor, we define the full vacuum expectation value of the regular surface defect observable as
\begin{align}\label{eq:pert}
    {\Psi} (\qe,z) = \prod_{\omega=0} ^{N-1} z_\omega ^{\frac{m_{\omega} ^+ -a_{\omega}}{\ve_1}} \hat{\CalZ}_{\hat{\mathbb{C}}^2 _{12}}.
\end{align}
Now we can replace the vacuum expectation values of the fractional instanton charges in \eqref{eq:fractqint} by differentials in fractional couplings acting on the correlation function of intersecting surface defects. This leads to the following difference equation for the fractional $Q$-observables, which we call the \textit{fractional quantum T-Q equation}:
\begin{align} 
\begin{split}
    &\Biggl\langle (x-m^+ _{\o+1}) {\tilde{Q}} _{\omega+1} (x) +\hat{\qe}_\omega (x-m^- _\o)  {\tilde{Q}} _{\omega-1} (x) \Biggl\rangle _{\BZ_N} {\Psi} \\
    &=\Big[x-{m}_{\omega +1} ^+ +\ve_1 z_{\omega+1} \partial_{z_{\omega+1}} + \hat{\qe}_\omega (x- {m}_\omega ^- -\ve_1 z_{\omega} \partial_{z_{\omega}}) \Big] \Biggl\langle {\tilde{Q}} _{\omega} (x)  \Biggl\rangle _{\BZ_N} {\Psi} . \\
    & := \Big[(1+\hat{\qe}_{\omega})x+\rho_\omega \Big] \Biggl\langle {\tilde{Q}} _{\omega} (x)  \Biggl\rangle _{\BZ _N} {\Psi}  
\end{split}
\end{align}
For notational convenience, we will define 
\begin{align}
    T_{N,\omega}(x) := (1+\hat{\qe}_{\omega})x+\rho_\omega
\end{align}
as a differential operator. The fractional quantum T-Q equation is then simply written as
\begin{align}\label{eq:qTQvev}
   (x-m^+ _{\o+1}) \Big\langle  {\tilde{Q}} _{\omega+1} (x) \Big\rangle_{\BZ_N} \Psi + \hat{\qe}_\omega (x-m^- _\o) \Big\langle  {\tilde{Q}} _{\omega-1} (x) \Big\rangle _{\BZ_N} {\Psi}  =  T_{N,\o} (x) \Big\langle \tilde{Q}_\o (x) \Big\rangle_{\BZ_N} \Psi.
\end{align}

In section \ref{sec:spin}, we show that the fractional quantum T-Q equation can be reorganized into a matrix equation valued in an auxiliary two-dimensional space $V_{\text{aux}}$. More specifically, it is translated to $\mathfrak{sl}_2$-homomorphisms ${L}^{\text{XXX}}_\o (x) \in \text{End} \left( \EuScript{V} _{\mathbf{s}_\o, \mathfrak{a}_\o} \otimes V_{\text{aux}} \right)$, which are identified as the Lax operators of the $XXX_{\mathfrak{sl}_2}$ spin chain with $N$ sites, with particular $\mathfrak{sl}_2$-modules $\left( \EH_{\mathbf{s}_\o, \mathfrak{a}_\o} \right)_{\o=0} ^{N-1}$ (see section \ref{sec:spin} for the definition of $\EH_{\mathbf{s}_\o,\mathfrak{a}_\o}$). By concatenating the Lax operators, we produce the monodromy matrix of the spin chain, as a $2 \times 2$ matrix in $\text{End}(V_{\text{aux}})$ with its entry valued in $\text{End}\left( \bigotimes_{\o=0} ^{N-1} \EH_{\mathbf{s}_\o ,\mathfrak{a}_\o} \right)$, represented as differential operators in the fractional couplings $(z_\o)_{\o=0} ^{N-1}$. The transfer matrix is obtained by taking the trace in $V_{\text{aux}}$, yielding a degree $N$ polynomial in $x$ whose coefficients are quantum Hamiltonians represented on $\bigotimes_{\o=0} ^{N-1} \EH_{\mathbf{s}_\o , \fa_\o}$ as differential operators. In this sense, as the name suggests, the fractional quantum T-Q equation can indeed be regarded as the fractionalization of the quantum T-Q equation \eqref{eq:qTQ}.

More interestingly, in section \ref{sec:KZ} the fractional quantum T-Q equation will be shown to be the Fourier transform of the degenerate $5$-point KZ equation for ${\ssl}_N$. Accordingly, the solutions to the KZ equation is given by the Fourier transform of the vacuum expectation value of the fractional $Q$-observable. Thus, the fractional quantum T-Q equation plays a fundamental role in connecting the $\EN=2$ gauge theory to the system of the $\mathfrak{sl}_N$ KZ equation and the system of $\mathfrak{sl}_2$ spin chain simultaneously, establishing an intricate spectral relation between the two systems. We will explain the details of this correspondence in a separate work \cite{JLN}.

\section{The vortex string defect} \label{sec:vortex}
In this section we Fourier transform the folded brane induced observables to define another surface defect, which is the analogue
of the vortex string defect studied in \cite{Nikita:IV, Jeong:2018qpc}, with a contact term arising at the intersection with the
regular surface defect. For a discussion of contact terms between $2$-observables in Donaldson theory and its generalizations, 
see \cite{Moore:1997pc,Losev:1997tp}. Unlike those infrared contact terms, our contact term is an ultraviolet observable, which we discuss below. 

\subsection{Fourier transform to vortex string defect}

\bigskip \noindent $\bullet$ 
Let us define the new observable ${\boldsymbol\Upsilon}(y)$ as a 
Fourier transform of the vector of vacuum expectation values of fractional ${Q}$-observables \eqref{def:tilq}:
\begin{align}\label{def:upsilon}
    \boldsymbol\Upsilon (y) \equiv \Upsilon^{\text{pert}} (y) \sum_{x \in L} \Big( \langle \tilde{Q}_\omega (x) \rangle_{\BZ_N} \Psi \Big)_{\omega=0} ^{N-1} y^{-\frac{x}{\ve_2}} =\Upsilon^{\text{pert}} (y) \sum_{x \in L} \begin{pmatrix} \langle \tilde{Q}_0 (x) \rangle_{\BZ_N} \\ \langle \tilde{Q}_1 (x) \rangle_{\BZ_N} \\ \vdots \\ \langle \tilde{Q}_\omega (x) \rangle_{\BZ_N} \\ \vdots \\ \langle \tilde{Q}_{N-1} (x) \rangle_{\BZ_N} \end{pmatrix} \Psi y^{-\frac{x}{\ve_2}} ,
\end{align}
where $L = L +  \mathbb{Z} \ve_2 \subset {\BC}$ is a lattice of complex numbers with step $\ve_2$ chosen so that the above expression converges. 

\bigskip \noindent $\bullet$
The physical meaning of ${\boldsymbol\Upsilon}(y)$ is that it is essentially a vortex string surface defect \cite{Nikita:IV, Jeong:2018qpc}. One can interpret it as a partition function of an $A$-type model (specifically, in a gauged linear sigma model realization) on the total space of a sum of $N$ copies of ${\CalO}(-1)$ line bundle over the projective space ${\BP}^{N-1}$.

The parameter $y$
plays the r{\^o}le of the complexified K{\"a}hler modulus. Depending on the domain in which $y$ is, this projective space is either the projectivization ${\BP}(N)$ of the color Chan-Paton space, or the projectivization ${\BP}(M^{\pm})$ of the half of the flavor spaces. 
In that sense the original $Q$-observable could be thought of as the analytic continuation of a path integral in the two dimensional theory living on the vortex string
to the complex values $x/{\ve}_2$ of the instanton charge. It is remarkable that the latter can be identified with the Coulomb
modulus of a theory living on the ${\BC}_{23}^2$-plane in the folded construction.

\bigskip \noindent $\bullet$ 
The perturbative contribution $\Upsilon^{\text{pert}} (y)$ is a simple function of $y$ to be determined. Each individual component of $\boldsymbol\Upsilon$ can be referred to as
\begin{align} \label{eq:Yo}
    \Upsilon_\omega (y) \equiv \Upsilon^{\text{pert}} (y) \sum_{x \in L } \langle \tilde{Q}_\omega (x) \rangle_{\BZ_N} \Psi y^{-\frac{x}{\tilde\ve_2}},
\end{align}
which possesses a twisted periodicity
\begin{align}
    \Upsilon_{\omega+N} (y) = y \Upsilon_\omega (y).
\end{align} 

\bigskip \noindent $\bullet$
We stress here that the vacuum expectation value above is taken in the gauge theory in the presence of the $\mathbb{Z}_N$-orbifold, which can be converted to the vacuum expectation value in the ordinary gauge theory with additional insertion of the regular surface defect. As a result, ${\boldsymbol\Upsilon}(y)$ is \emph{the pair correlator of intersecting surface defect observables}.

\bigskip \noindent $\bullet$ 
In defining the Fourier transform $\boldsymbol{\Upsilon}(y)$ \eqref{def:upsilon}, we should require that the series converges. The convergence is guaranteed only with appropriate choices of the lattice $L$, and, moreover, different choices of the lattice lead to series with different convergence domains. This can be shown as follows.

From the definition \eqref{def:tilq}, we notice that the fractional $Q$-observable $\tilde{Q}_\o (x)$ has simple zeros at $x = m^+_{\o-\o'} -(n+1)\ve_2$, $\o' =0,\cdots, N-1$, $n \in \BZ_{\geq 0}$. Thus we choose the lattices
\begin{align}
L_\o \equiv \{ m^+ _\o + n \ve_2\; \vert \; n \in \BZ \}, \quad \o=0, \cdots, N-1.
\end{align}
Then the infinite summation in \eqref{def:upsilon} terminates to the left due to the zeros of the fractional $Q$-observables, giving
\begin{align} \label{eq:latchoice}
\begin{split}
        \boldsymbol\Upsilon^{(\o')} (y)  &\equiv \Upsilon^{\text{pert}} (y) \sum_{x \in L_{\o'}} \Big( \langle \tilde{Q}_\omega (x) \rangle_{\BZ_N} \Psi \Big)_{\omega=0} ^{N-1} y^{-\frac{x}{\ve_2}} \\
        & = \Upsilon^{\text{pert}} (y) \sum_{n = 0} ^\infty  \Big( \langle \tilde{Q}_\omega (m^+ _{\o'} + n \ve_2) \rangle_{\BZ_N} \Psi \Big)_{\omega=0} ^{N-1} y^{-\frac{m^+ _{\o'}}{\ve_2} -n}.
\end{split}
\end{align}
Then the series converges in the domain $0<\vert \qe \vert < 1< \vert y \vert$. In other words, with the specific choices for the lattice $L$ above, the solutions to the KZ equations expressed as series \eqref{eq:latchoice} are valid only inside the particular domain $0<\vert \qe \vert < 1< \vert y \vert$.

We can continuously vary the parameter $y$ to other convergence domains, where the solutions have to be properly analytically continued. Such analytic continuations of correlation functions of surface defect observables was studied in depth in \cite{Jeong:2018qpc, JeongNekrasov}. For instance, in the domain $0<\vert y \vert < \vert \qe \vert <1$, we can take the Fourier transform similar to \eqref{def:upsilon} to construct the solutions as a series,
\begin{align}
    \boldsymbol\Upsilon' {}^{(\o')} (y)  \equiv {\Upsilon'}^{\text{pert}} (y) \sum_{x \in L_{\o'}} \Big( \langle \tilde{Q}_\omega (x) \rangle_{\BZ_N} \Psi \Big)_{\omega=0} ^{N-1} \left(\frac{y}{\qe} \right) ^{\frac{x}{\ve_2}}  ,
\end{align}
for which we can derive the $5$-point KZ equations by repeating the same computations. Note that even though we are seemingly using the same lattices, the convergence domains are distinct so that we have to redefine $m^+ _\a \leftrightarrow m^- _\a$ in the latter solution to properly patch the two solutions together through analytic continuation. Such analytic continuations across convergence domains lead to nontrivial connection formulas between these solutions \cite{Jeong:2018qpc,JeongNekrasov}.

In the language of the supersymmetric gauged linear sigma model with the target space being the total space of the vector bundle $\mathcal{O}(-1) \otimes \BC^N \rightarrow \BP^{N-1}$ over the projective space, the analytic continuation corresponds to the flop transition initiated by the variation of the K\"{a}hler modulus $y$, which roughly exchanges the base and the fiber of the target space \cite{Jeong:2018qpc}.

\subsection{On the intersection of surface defects}

\bigskip \noindent $\bullet$
Consider two $A$-twisted topological sigma models on K\"{a}hler manifolds $X_{1,2}$ with the worldsheets $\Sigma_{1,2}$, respectively. 
Let $L \subset X_{1} \times X_{2}$ be a subvariety (a correspondence), and $p_1 \in \Sigma_1$, $p_2 \in \Sigma_2$ be a couple of points. Then one can define a bi-local observable ${\varpi}_{L}$ in the combined theory which is a condition for the holomorphic maps ${\phi}_{i} : {\Sigma}_{i} \to X_i$, $i=1,2$, to agree at $p_i$ in the sense of the $L$-correspondence: $({\phi}_{1}(p_1), {\phi}_{2}(p_2)) \in L$. Using K{\"u}nneth decomposition, the Poincar\'{e} dual ${\delta}_{L} \in H^{*}(X_{1} \times X_{2})$ to $L$ (assuming compactness)
decomposes as:
\beq
{\delta}_{L} = \sum_{a,b} N^{ab} e^{(1)}_{a} \otimes e^{(2)}_{b}
\eeq
where $e^{(i)}_{a}$, $a = 1, \ldots, \dim H^{*}(X_{i})$ are the bases of the respective cohomology groups. Using this decomposition, the
bi-local observable ${\varpi}_{L}$ can be expanded in the basis of the ordinary $0$-observables of respective sigma models
(this is similar to the Eq. (3.3) in \cite{Aganagic:2002qg}). 

\bigskip \noindent $\bullet$
In our case, the surface defects support the sigma models on the (total spaces of certain equivariant vector bundles) over the complete flag variety $F(N)$ for the regular defect, and the projective space ${\BP}(N)$ for the vortex string. 
Define the correspondence 
\beq
L_{\omega} \subset F(N) \times {\BP}(N) 
\eeq
as the variety of pairs
$(V_{0} \subset V_{1} \subset \ldots  \subset V_{N-1} \subset V_{N} \equiv N , {\ell} \subset N)$, with  ${\rm dim} N_{i} = i$, ${\rm dim}{\ell} = 1$, 
such that ${\ell} \subset V_{\omega+1}$. It is a nontrivial correspondence for $\omega = 0, \ldots, N-2$. 

We expect the local observable defined by the evaluation of the ${\Xi}^{(0)}$ \eqref{def:inter} at $x \in L$ to be a localization of
the observable ${\varpi}_{L_{\omega}}$. It would be nice to work this out in detail.

\section{Knizhnik–Zamolodchikov equations}\label{sec:KZ}
In this section, we verify that the correlation function of the intersecting surface defect observables in the $\EuScript{N}=2$ supersymmetric gauge theory introduced above satisfies the KZ equations associated with affine Lie algebra $\widehat{\ssl}_N$.

Let $\mathfrak{g}$ be a simple Lie algebra over $\mathbb{C}$. The KZ equations were originally derived for the correlation functions of primaries in the WZNW model, in which an affine Lie algebra $\widehat{\mathfrak{g}}$ is the conserved current algebra \cite{Knizhnik:1984nr}. The level of the relevant lowest-weight $\widehat{\mathfrak{g}}$-modules is identified with the level $k \in \mathbb{Z}$ of the WZNW model, and therefore is constrained to be an integer. The KZ equations were later reformulated in a representation theoretical manner \cite{TK,FR}, where the correlation functions are defined as matrix elements of products of intertwining operators between lowest-weight $\widehat{\mathfrak{g}}$-modules of level $k\in \mathbb{C}$ and evaluation $\widehat{\mathfrak{g}}$-modules of level $0$. 

Let us only briefly recall the formulation of Knizhnik-Zamolodchikov equations here, without going into details of representation theory of affine Lie algebras. We consider lowest-weight $\mathfrak{g}$-modules $\EuScript{V}_0$, $\EuScript{V}_\infty$ and any $r+1$ $\mathfrak{g}$-modules $\left(\EH_\ri \right)_{\ri=0} ^r$. To each lowest-weight $\mathfrak{g}$-module ${\EuScript{V}}_0$ and ${\EuScript{V}}_\infty$, we associate the induced lowest-weight $\widehat{\mathfrak{g}}$-modules ${\EuScript{V}}_{0,k}$ and ${\EuScript{V}}_{\infty,k}$ of level $k \in \mathbb{C}$, considering them to be located at $\sz_{r+1} = 0$ and $\sz_{-1} = \infty$, respectively. For the rest of the $\mathfrak{g}$-modules $\EH_\ri$, we construct the evaluation modules $\EH_\ri (\sz_\ri)$ with complex parameters $\sz_\ri \in \mathbb{P}^1$. The intertwining operator is defined as a $\widehat{\mathfrak{g}}$-homomorphism between a lowest-weight $\widehat{\mathfrak{g}}$-module and the product of a lowest-weight $\widehat{\mathfrak{g}}$-module and an evaluation $\widehat{\mathfrak{g}}$-module. By taking a consecutive product of the intertwining operators, we can construct a $\widehat{\mathfrak{g}}$-homomorphism between $\EuScript{V}_{\infty,k}$ and $\bigotimes_{\ri =0} ^r \EH_\ri (\sz_\ri) \otimes \EuScript{V} _{0,k}$. The correlation function $\psi(\sz) = \psi(\sz_0, \cdots, \sz_r)$ is defined as the matrix element of this product of intertwining operators, valued in $\EuScript{V}_\infty ^*  \otimes \bigotimes_{\ri=0} ^r \EH_\ri \otimes \EuScript{V}_0 $. Here, $\EuScript{V}_\infty ^*$ is the restricted dual of $\EuScript{V}_\infty$, i.e., the direct sum of the duals of weight subspaces of $\EuScript{V}_\infty$.

Now let us denote the basis of the Lie algebra $\mathfrak{g}$ by $\{\sT^\sk\}$. We also denote $\sT^\sk_\ri$ the representation of $\{\sT^\sk\}$ on the module at $\sz_\ri$. Then the KZ equations read
\begin{align}
    \left[ (k+ h^\vee) \frac{\partial}{\partial\sz_\ri} - \sum_{\substack{\rj = 0 \\\rj\neq \ri}} ^{r+1} \frac{\sT^\sk_\ri \otimes \sT_\rj^\sk }{\sz_\ri-\sz_\rj} \right] \psi (\sz) = 0, \quad \ri = 0, \cdots, r.
\end{align}
where $h^\vee$ is the dual Coxeter number. As evident from the equations, the space of solutions is $\mathfrak{g}$-invariant. Thus we may restrict our attention to the correlation function valued in the space of $\mathfrak{g}$-invariants, $\psi(\sz) \in \left(\EuScript{V}_\infty^*  \otimes \bigotimes_{\ri=0} ^r \EH_\ri \otimes \EuScript{V}_0 \right)^{\mathfrak{g}}$, at our interest toward the correspondence with the $\EuScript{N}=2$ gauge theory.

In connection to the $\EuScript{N}=2$ gauge theory, the subject of our study, the relevant simple Lie algebra will be $\mathfrak{g}=\mathfrak{sl}_N$. Our main example will be the Riemann sphere with five punctures ($r=2$),
\begin{align}
    \sz_{-1} = \infty, \ \sz_0 = y, \ \sz_1 = 1, \ \sz_2 = \qe, \ \sz_3 = 0,
\end{align}
at each of which we attach an $\mathfrak{sl}_N$-module as we just described. In particular, we associate the lowest-weight Verma modules at $0$ and $\infty$, the Heisenberg-Weyl modules (HW modules) at $\qe$, $1$, and finally the standard $N$-dimensional representation at $y$. It turns out that, as we will see below, the $N$-dimensional representation can be realized as a submodule of the HW module with specialized weights. From the point of view of the current algebra, this corresponds to inserting a degenerate primary field at $y$. 

We show that the Fourier transform of the correlation function of the intersecting surface defect observables that we studied in section \ref{sec:fracqobs} solves the degenerate $5$-point KZ equations. In the view of the BPS/CFT correspondence, there have been earlier conjectures relating the four dimensional gauge theory correlation functions  to the analytically continued
WZNW conformal blocks \cite{NN2004i, AT,KPPW,K-T}. Our proof of the KZ equations for the gauge theory correlation function is an explicit confirmation of some of these conjectures.

\subsection{Knizhnik-Zamolodchikov equations for ${\ssl}_N$}
We first introduce the construction of relevant ${\ssl}_N$-modules that compose the degenerate 5-point genus-$0$ correlation function by flag varieties. Then we will describe how the degenerate 5-point KZ equations are expressed with these representations.

\subsubsection{Some representations of ${\ssl}_N$}
As we have briefly mentioned, the relevant $\mathfrak{sl}_N$-modules are
the lowest-weight (highest-weight) Verma modules, the Heisenberg-Weyl (HW) modules, and the standard $N$-dimensional representation. For the Verma modules and the HW modules, we shall only review how these modules are constructed without providing the proofs for their desired properties. For the details of the proofs, we refer to \cite{NT}. Then we construct the $N$-dimensional representation as a submodule of a special HW module.

To make the notations concise, we first present the constructions of $\mathfrak{gl}_N$-modules. The representations of $\mathfrak{sl}_N$ are defined on the same spaces by properly redefining the Cartan generators.

Let $W = \BC^N$ be the complex vector space of dimension $N$. We choose a basis $\{e_\ta\}_{\ta=1} ^N$ in $W$, with the dual basis in $W^*$ by $\{\tilde{e}^\tb\}_{\tb=1} ^N$, so that the Lie algebra $\mathfrak{gl}_N$ is represented by the linear maps
\begin{align}
    \sT_\ta ^{\tb} = e_\ta \otimes \tilde{e}^\ta \in \text{End}(W)
\end{align}
with the commutation relations
\begin{align}
    \left[ \sT_\ta ^\tb , \sT_\tc ^\td \right] = \d_\tc ^\tb \sT_\ta ^\td - \d_\ta ^\td \sT_\tc ^\tb.
\end{align}
The Lie algebra $\mathfrak{sl}_N$ is spanned by $\sT_\ta ^\tb$ with $\ta \neq \tb$, and the Cartan generators 
\begin{align}
    \mathfrak{h}_i \equiv \sT_i ^i - \sT_{i+1} ^{i+1}, \quad i = 1, \cdots, N-1.
\end{align}

\paragraph{Verma modules} \begin{comment} The highest-weight (lowest-weight) Verma modules  $V_{\mathbf{s}}^\pm$, for a given weight $\mathbf{s}=(s_1,\dots,s_N)\in\BC^{N}$, are generated by the highest-weight (lowest-weight) vector $\Omega^\pm$ satisfying the relations
\begin{align}
\begin{split}
    &J^\ta_\tb \Omega^+ = 0, \quad \ta<\tb \quad\quad (J^\ta_\tb \Omega^- = 0, \quad \ta>\tb) \\
    & J^\ta_\ta \Omega^\pm = s_\ta \Omega^\pm, \quad s_\ta \in \mathbb{C} \quad \left( \sum_{\ta=1} ^N s_\ta  = 0 \right) , \quad \ta = 1,\dots,N,
\end{split}
\end{align}
where we denoted by $J^\ta_\tb$, $\ta,\tb=1,\dots,N$ the generators of $\mathfrak{sl}_N$, $\sT^\ta_\tb|_{ V^\pm _{\mathbf{s}}} = J^\ta_\tb$, obeying
\begin{align}
    \left[ J^\ta_\tb, J^\tc_\td \right] = \delta_\tb^\tc J^\ta_\td - \delta^\ta_\td J^\tc_\tb.
\end{align}
Namely, as vector spaces the highest-weight (lowest-weight) Verma modules are given as $V^+ _{\mathbf{s}} \equiv \BC[J^\ta_\tb]_{\ta>\tb} \Omega^+$ and $V^- _{\mathbf{s}} \equiv \BC[J^\ta_\tb]_{\ta<\tb} \Omega^-$. \end{comment}

For the purpose of our study, it is convenient to construct Verma modules by using flag variety. Let us consider complete flags of $W=\mathbb{C}^N$,
\begin{align}
    \{0\} = V_0 \subset V_1 \subset  \cdots \subset V_{N-1} \subset V_N = W, \quad \dim V_i = i,
\end{align}
with the embeddings $\tU_i : V_i \rightarrow V_{i+1}$. The action of $ \BG= GL(V_1) \times GL(V_2) \times \cdots \times GL(V_{N-1})$ on the embeddings is simply given by
\begin{align}
    &g : \ (\tU_i)_{i=1}^{N-1} \mapsto (g_{i+1}\tU_{i} g_i^{-1})_{i=1}^{N-1}, \quad g_i \in GL(V_i).
\end{align}
Then the flag variety is given by the GIT quotient 
\begin{align}\label{def:flag}
    F(W) = \{ (\tU_i) \ | \ \tU_i : \ V_i \to V_{i+1}, \ i=1,\dots,N-1 \}^{\rm stable} / \BG,
\end{align}
where the stability condition requires all the embeddings $(\tU_i)_{i=1} ^{N-1}$ to be injective.

Let us also define the exterior power of the product of embeddings,
\begin{align}
    \pi_ i \equiv \extp^i \left( \tU_{N-1} \cdots \tU_i \right) : \extp^i V_i \longrightarrow \extp^i W.
\end{align}
Now we choose a basis $(e_\ta)_{\ta=1}^{N}$ of $W$, and denote its dual basis by $(\tilde{e}^{\ta})_{\ta=1}^{N}$ with $\tilde{e}^{\ta}(e_\tb) =\d^ \ta _\tb$. We consider an open patch $F(W)^\circ \subset F(W)$ associated to the chosen basis, defined by
\begin{align}\label{def:open flag}
    F (W)^\circ =\{ (\tU_i) | \ \tilde{\pi}_ \circ ^i (\pi_i) \neq 0, \;\; \forall i \} \subset F(W),
\end{align}
where we defined the poly-covector $\tilde{\pi}_ \circ ^i \equiv \tilde{e}^{1} \wedge \cdots \wedge \tilde{e}^{i} \in \extp^i W^*$. 

We construct a lowest-weight Verma module as follows. Let $\boldsymbol\z = (\z_1, \cdots \z_{N-1}) \in \mathbb{C}^{N-1}$ be given. Let us define $\Omega_{\boldsymbol\z} \equiv \prod_{i=1} ^{N-1} \left( \tilde{\pi} _\circ ^i (\pi_i) \right) ^{\z_i}$, and freely generate the space $\EuScript{V}_{\boldsymbol\z} = \Omega_{\boldsymbol\z} \; \mathbb{C} [u^{(i)} _\b ]$ as a space of polynomial in
\begin{align}
    u^{(i)} _\b \equiv \frac{ \tilde{\pi}_\circ ^{i+1} (e_\b \wedge \pi_i)}{\tilde{\pi}_\circ ^i (\pi_i)  }, \quad\quad \begin{split} &i = 1, \cdots, N-1 \\ &\b=1, \cdots, i \end{split}.
\end{align}
We will only work in the patch $F(W)^\circ$, so that the space is well-defined. Then the generators of $\mathfrak{gl}_N$ are represented on $\EuScript{V}_{\boldsymbol\z}$ by
\begin{align} \label{eq:jv}
    \sT^\tb _\ta \vert_{\EV_{\boldsymbol{\z}}} \equiv J^\tb _\ta  = - \sum_{\tm=1} ^{N-1} \left(\tU_{N-1} \right)^\tb _\tm \frac{\p}{\p \left( \tU_{N-1}\right)^\ta _\tm }.
\end{align}
We can show that $\EuScript{V}_{\boldsymbol\z}$ is a lowest-weight Verma module with the lowest-weight vector $\Omega_{\boldsymbol\z}$:
\begin{align}
\begin{split}
  &J_\ta ^\tb \Omega_{\boldsymbol\z} = 0, \quad \ta > \tb, \\
  &\mathfrak{h}_i  \vert_{\EV_{\boldsymbol{\z}}} \Omega_{\boldsymbol\z} = -\z_i \Omega_{\boldsymbol\z} , \quad\quad i=1, \cdots N-1.\
\end{split}
\end{align}
See \cite{NT} for the details of the proof.

A highest-weight Verma module can be constructed in a similar manner. We consider complete flags of the dual space
\begin{align}
    \{0\} = \tilde{V}_0 \subset \tilde{V}_1 \subset \cdots \subset \tilde{V}_{N-1} \subset \tilde{V}_N = W^*, \quad \dim \tilde{V}_i =i,
\end{align}
with the forgetful maps $\tilde\tU_i : \tilde{V}_{i+1} \rightarrow \tilde{V}_i$. The action of $\tilde{\mathbb{G}} = GL(\tilde{V}_1) \times \cdots GL(\tilde{V}_{N-1})$ on these maps is simply
\begin{align}
    g: (\tilde\tU_i)_{i=1} ^{N-1} \mapsto (g_{i} \tilde\tU _i g_{i+1} ^{-1} )_{i=1} ^{N-1} ,\quad g_i \in GL(\tilde{V}_i).
\end{align}
Then the flag variety is given by
\begin{align}
    \tilde{F}(W^*) = \{ (\tilde{\tU}_i) \ | \ \tilde{\tU}_i : \ \tilde{V}_{i+1} \to \tilde{V}_{i}, \ i=1,\dots,N-1 \}^{\rm stable} / \tilde{\BG},
\end{align}
where the stability condition requires all the forgetful maps $(\tilde{\tU})_{i=1} ^{N-1} $ are surjective.

We define the exterior power of the product of duals of the forgetful maps,
\begin{align}
    \tilde{\pi}^i \equiv \extp^i (\tilde{\tU}_{N-1} ^* \cdots \tilde{\tU}_i ^* ) : \extp^i \tilde{V}_i ^* \longrightarrow \extp^i W^*.
\end{align}
Also, with the chosen basis of $W$, we define the polyvector $\pi_i ^\circ \equiv e_1 \wedge \cdots \wedge e_i \in \extp^i W$. The associated open patch is
\begin{align}
    \tilde{F}(W^*) ^\circ =\{ (\tilde{\tU}_i) | \ \tilde{\pi} ^i (\pi_i ^\circ) \neq 0, \;\; \forall i \} \subset \tilde{F}(W^*).
\end{align}
Then, for a given $\tilde{\boldsymbol\z} \in \mathbb{C}^{N-1}$, we define a vector $\tilde{\Omega}_{\tilde{\boldsymbol\z}} = \prod_{i=1} ^{N-1} \left( \tilde\pi ^i (\pi_i ^\circ) \right)^{\tilde{\z}_i} $. We construct the space $\tilde{\EV}_{\tilde{\boldsymbol\z }} = \tilde{\Omega}_{\tilde{\boldsymbol\z}} \, \mathbb{C}[ \tilde{u}^\b _{(i)}]$ as the space of polynomials in 
\begin{align}
    \tilde{u} ^\b _{(i)} \equiv \frac{\tilde{e}^\b \wedge \tilde\pi ^i (\pi_{i+1} ^\circ)}{\tilde\pi ^i (\pi_i ^\circ)},  \quad \begin{split} &i = 1, \cdots, N-1 \\  &\b = 1, \cdots, i. \end{split}
\end{align}
We will only work on the patch $\tilde{F}(W^*)^\circ$ so that the space is well-defined. The generators of $\mathfrak{gl}_N$ are represented on $\tilde{\EV}_{\tilde{\boldsymbol\z}}$ by
\begin{align} \label{eq:jv'}
    \sT^\tb _\ta \vert_{\tilde{\EV}_{\tilde{\boldsymbol{\z}}}} \equiv \tilde{J}^\tb _\ta =  \sum_{\tm=1} ^{N-1} \left(\tilde{\tU}_{N-1} \right)^\tm _\ta \frac{\p}{\p  \left(\tilde{\tU}_{N-1} \right)^\tm _\tb }.
\end{align}
It can be shown that $\tilde{\EV}_{\tilde{\boldsymbol\z}}$ thus defined is a highest-weight Verma module with the highest-weight vector $\tilde{\Omega}_{\tilde{\boldsymbol\z}}$:
\begin{align}
\begin{split}
    &\tilde{J}_\ta ^\tb \tilde{\Omega}_{\tilde{\boldsymbol\z}} = 0 , \quad \ta < \tb, \\
    &\mathfrak{h}_i \vert_{\tilde{\EV}_{\tilde{\boldsymbol{\z}}}}  \tilde{\Omega}_{\tilde{\boldsymbol\z}} = \tilde\z_i \tilde{\Omega}_{\tilde{\boldsymbol\z}}, \quad i = 1, \cdots, N-1.
\end{split}
\end{align}
Details can be found in \cite{NT}.

\paragraph{Heisenberg-Weyl modules} The Heisenberg-Weyl module (HW module) is constructed from the projective space ${\BP}^{N-1}$. Consider an abstract one-dimensional space $\mathbb{L}= \mathbb{C}^1$ and the space of linear maps $\fz : \mathbb{L} \rightarrow W$. The symmetry group $\mathbb{C}^\times$ of $\mathbb{L}$ acts on the space of such maps by
\begin{align}
    \fz \longmapsto t^{-1} \fz, \quad t \in \mathbb{C}^\times.
\end{align}
The space of injective embeddings up to the $\mathbb{C}^\times$-equivalence is the projective space $\mathbb{P}^{N-1}$.

Let us trivialize the sections of the line bundle $\BL \rightarrow \mathbb{P}^{N-1}$ on the open patch near $(\fz^1,\cdots,\fz^N)= (1,\cdots,1)$. Let us be given with $\mu \in \BC$ and $\t \in \BC^{N-1}$. Then we construct the space $\EH_\mu ^{\boldsymbol\t}$ as a space of degree-zero Laurent polynomials (with a multiplicative prefactor):
\begin{align}
\begin{split}
    \EuScript{H}_\mu ^{\boldsymbol\t} \equiv \prod_{\ta=1} ^N \left(\fz^\ta \right)^{\b_\ta} \mathbb{C} \left[ (\fz^1)^\pm, \cdots (\fz^N)^\pm \right]^{\BC^\times},
\end{split}
\end{align}
where $\boldsymbol\b = (\b_1 , \cdots, \b_N) \in \BC^N$ is determined by
\begin{align}
    \mu = \sum_{\ta=1} ^N \b_\ta, \quad \t_i = \b_i - \b_{i+1}, \quad i = 1, \cdots, N-1.
\end{align}
We will only work in the patch near $(\fz^1, \cdots, \fz^N) = (1, \cdots, 1)$ so that $\fz^\ta \neq 0$ for all $\ta= 1, \cdots, N$. Now the generators of $\mathfrak{gl}_N$ are represented on $ \EuScript{H}_\mu ^{\boldsymbol\t}$ by
\begin{align} \label{eq:tz}
    \sT^\tb_\ta|_{\EuScript{H}_\mu ^{\boldsymbol\t}} = -\fz^\tb \frac{\partial}{\partial \fz^\ta} .
\end{align}
Note that the $\mathfrak{sl}_N$ weights of the vectors in $\EuScript{H}_\mu ^{\boldsymbol\t}$ form a lattice including $-\boldsymbol\t$, and each weight subspace is one-dimensional. For instance, $\mathfrak{h}_i \vert_{\EuScript{H}_\mu ^{\boldsymbol\t}} \left( \prod_{\ta=1} ^N \left(\fz^\ta \right)^{\b_\ta}  \right) = -\t_i \left( \prod_{\ta=1} ^N \left(\fz^\ta \right)^{\b_\ta} \right)$, $i = 1, \cdots, N-1$.

We call $\EuScript{H}_\mu ^{\boldsymbol\t}$ the (twisted) Heisenberg-Weyl module. It is the space of degree-zero Laurent polynomials in $(\fz^\ta)_{\ta=1} ^N$, multiplied by the prefactor determining the weight $\boldsymbol\t$ and $\m$.

% \NL{How about if we change the logic a little bit. We say the HW module $\CalH^{\boldsymbol\beta}_\mu$ is defined bulkd on $N$ parameters $\boldsymbol\beta=(\beta_1,\dots,\beta_N)$ satisfying (5.20). And we highlight the Casimirs of such $\CalH^{\boldsymbol\beta}_\mu$ depends only on $\mu$. All follow up notation will be modified accordingly. This way we don't need to deal with the infinite direct sum.}

The same module can be presented in a slightly different way. Let us again consider one-dimensional space $\tilde{\mathbb{L}}= \mathbb{C}^1$ with the symmetry group $\tilde{\mathbb{C}}^\times$. Then the space of the forgetful maps
\begin{align}
    \tilde\fz : W \longrightarrow \tilde{\mathbb{L}},
\end{align}
up to the $\tilde{\mathbb{C}}^\times$ action,
\begin{align}
    \tilde\fz \longmapsto \tilde{t} \tilde\fz , \quad \tilde{t} \in \tilde{\mathbb{C}}^\times,
\end{align}
defines the projective space $\BP^{N-1}$. Then, for given $\tilde\m \in \mathbb{C}$ and $\tilde{\boldsymbol{\t}} \in \BC^{N-1}$, we define the space $\tilde{\EH}_{\tilde\m} ^{\tilde{\boldsymbol\t}}$ by degree-zero Laurent polynomials multiplied by a prefactor,
\begin{align}
 \tilde{\EH}_{\tilde{\mu}}^{\tilde{\boldsymbol\t}} = \prod_{\ta=1} ^N (\tilde{\fz}_\ta)^{\tilde{\beta}_\ta} \mathbb{C} [ \tilde{\fz}_1 ^\pm, \cdots, \tilde{\fz}_N^\pm ]^{\mathbb{C}^\times},
\end{align}
with $\tilde\mu = \sum_{\ta=1} ^N \tilde{\beta}_\ta$ and $\tilde{\t}_i = \tilde{\b}_i - \tilde\b_{i+1}$, $i=1, \cdots, N-1$. The generators of $\mathfrak{gl}_N$ are represented by
\begin{align} \label{eq:tz'}
    \sT^\tb _\ta | _{\tilde{\EH}_{\tilde{\mu}}^{\tilde{\boldsymbol\t}}} =  \tilde{\fz}_\ta \frac{\p}{\p \tilde{\fz}_\tb} .
\end{align}

An important feature of the HW module $\EH_{\mu}^{\boldsymbol\t}$ is that all the Casimirs depend only on $\mu$ ($\tilde\mu$ for $\tilde{\EH}_{\tilde\mu}^{\tilde{\boldsymbol\t}}$). Also, the HW module is neither highest-weight nor lowest-weight generically. Moreover, for generic $\mu$ and $\boldsymbol\t$ the HW module $\EH_\mu ^{\boldsymbol\t}$ is irreducible. At special values of $\mu$ and $\boldsymbol\t$, however, it is reducible into highest-weight and lowest-weight submodules.

\paragraph{$N$-dimensional representation} The standard $N$-dimensional representation of $\mathfrak{sl}_N$ is simply given by the $N$-dimensional vector space $\mathbb{C}^N$, on which the generators of $\mathfrak{gl}_N$ act by the single-entry $GL(N,\mathbb{C})$ matrices,
\begin{align}
    \sT^\tb_\ta|_{\mathbb{C}^N} =E^\tb _\ta.
\end{align}
Note that the $N$-dimensional representation can also be obtained as a finite-dimensional submodule of the Heisenberg-Weyl module with the specialized $\tilde{\boldsymbol\b} = (0,0,\cdots,0,1)$, or equivalently $\tilde\mu=1$ and $\tilde{\boldsymbol\t} = (0,\cdots, 0, -1)$. Namely, it is the $N$-dimensional submodule of $\tilde{\EH}_1^{(0,\cdots, 0,-1)} = \tilde{u}_N \, \mathbb{C}[\tilde{u}_1 ^\pm, \cdots, \tilde{u}_N ^\pm ]^{\mathbb{C}^\times}$, spanned by $\{\tilde{u}_1, \cdots, \tilde{u}_N\} \subset \tilde{\EH}_1^{(0,\cdots,0,-1)}$. The correlation function with an insertion of the $N$-dimensional representation is said to be degenerate in this sense.

\subsubsection{4-point correlation function}
Before proceeding to the 5-point KZ equations, let us consider the 4-point KZ equation for $\mathfrak{sl}_N$ where the $N$-dimensional representation at $y$ is not present (namely, $r=1$). It was verified in \cite{NT} that the vacuum expectation value of the regular surface defect observable in the $U(N)$ gauge theory with $N$ fundamentals and $N$ antifundamentals provides solutions to the 4-point KZ equation. We will give a brief review here; details can be found in appendix \ref{sec:4-point KZ}.

For the 4-point KZ equation for $\mathfrak{sl}_N$, we consider the four punctures on the Riemann sphere located at
\begin{align}
    \sz_{-1} = \infty, \ \sz_0 = 1, \ \sz_1 = \kq, \ \sz_2 = 0.
\end{align}
As we described earlier, we assign the lowest-weight Verma modules $\EuScript{V}_{0} ( \equiv \EuScript{V}_{\boldsymbol{\z}})$ and $\EuScript{V}_\infty$ to  the points $0$ and $\infty$. We assign the  HW-modules ${\EH}_{\mu ^{(4)}}^{\boldsymbol\t - {\boldsymbol\z}}$ and $\tilde{\EH}_{\tilde{\mu}}^{{\boldsymbol\t} - \tilde{\boldsymbol\z}}$ to $\qe$ and $1$, respectively. 

In this paper we only study the generic weights modules, so that they are irreducible. In particular, since the lowest-weight Verma module $\EuScript{V}_\infty$ is irreducible, its restricted dual $\EuScript{V}_\infty^*$ is a highest-weight Verma module. Thus, we can just replace $\EuScript{V}_\infty^* $ by a highest-weight Verma module that we will denote by $\tilde{\EuScript{V}}_{\tilde{\z}}$.

The 4-point correlation function $\Psi(\qe)$ is valued in
\begin{align}\label{def:4pt-fun}
   \Psi(\qe) \in \left( \EuScript{V}_{\boldsymbol{\z}} \otimes {\EH}_{\mu ^{(4)}}^{\boldsymbol\t - {\boldsymbol\z}} \otimes \tilde{\EH}_{\tilde{\mu}}^{\boldsymbol\t -\tilde{\boldsymbol\z}} \otimes   \tilde{\EuScript{V}} _{\tilde{\boldsymbol{\z}}}  \right)^{\mathfrak{sl}_N}.
\end{align}

\bigskip
$\bullet$
\bigskip

\noindent
Let $\EG = \BG \times \tilde\BG \times \BC^\times \times \tilde\BC^\times$. 
Using the constructions of the $\mathfrak{sl}_N$-modules from flag varieties, we can present
the correlation function as a product of the  ${\rm Lie}({\EG})$-equivariant piece $\Psi_0$, , and a $\EG$-invariant factor $\chi$:
\begin{align} \label{eq:correp}
    \Psi(\qe) = \Psi_0 \cdot \chi (v_1, \cdots, v_{N-1};\qe),
\end{align}
where $\chi(v_1, \cdots, v_{N-1};\qe)$ is a Laurent polynomial in
\begin{align} \label{eq:vdef}
    v_\ta = \frac{ \left( \tilde{\fz} \wedge \tilde\pi^{\ta-1} \right)(\pi_\ta) \cdot \tilde\pi ^\ta \left( \fz \wedge \pi_{\ta-1} \right)}{\tilde\fz (\fz) \cdot \tilde{\pi}^{\ta-1} (\pi_{\ta-1}) \cdot \tilde\pi ^\ta (\pi_\ta) }, \quad \ta=1,\cdots, N.
\end{align}
It can be shown that $\sum_{\ta=1} ^N v_\ta =1$, making only $N-1$ variables independent. Also,
\begin{align} \label{eq:psi0}
    \Psi_0 = \prod_{\ta=1} ^N \left( (\tilde{\fz} \wedge \tilde\pi ^{\ta-1} ) (\pi_\ta) \right)^{\tilde\b _\ta} \left( \tilde\pi ^\ta (\fz \wedge \pi_{\ta-1} ) \right)^{\b_\ta} \cdot \prod_{i=1} ^{N-1} \left( \tilde\pi^i (\pi_i) \right)^{\a_i}.
\end{align}
Note that all the $\mathfrak{sl}_N$ indices in $(v_\ta)_{\ta=1} ^N$ and $\Psi_0$ are contracted, so that $\Psi(\qe)$ is invariant under $\mathfrak{sl}_N$. The ${\rm Lie}({\EG})$-equivariance of $\Psi_0$ reads as: 
\begin{align}
    \Psi_{0}[g_{i+1}\tU_ig_i^{-1},\tilde{g}_i\tilde\tU_i\tilde{g}_{i+1}^{-1},t\fz_\ta,\tilde{t}^{-1}\tilde\fz_\ta] \times
    \prod_{i=1}^{N-1} (\det g_i)^{\zeta_i} (\det \tilde{g}_i)^{-\tilde\zeta_i} \times t^{\mu} \tilde{t}^{-\tilde\mu} = \Psi_{0}[\tU_i,\tilde\tU_i,\fz^\ta,\tilde\fz_\ta]. 
\end{align}
with $g_{i} = {\exp} ( \, h\ {\xi}_{i} ) $, etc. and $h$ nilpotent. 
The $3N-1$ undetermined variables $(\b_\ta,\tilde\b_\ta, \a_i)$ in the exponents of $\Psi_0$ are determined by the $3N-1$ weight parameters $(\z_i,\tilde{\z}_i,  \t_i, \mu,\tilde\m ^{(4)})$ by
\begin{align} \label{eq:4pmat}
    \begin{split} &\tilde\beta_i + \beta_{i+1} + \alpha_i = \zeta_i, \\
    & \tilde\beta_{i+1} + \beta_i + \alpha_i = \tilde\zeta_i, \\
    &\t_i - {\z}_i = \b_i - \b_{i+1}, \\ 
    &\t_i - \tilde{\z}_i = \tilde{\b}_i - \tilde{\b}_{i+1}\end{split}
    \quad\quad i=1,\dots,N-1, \nonumber \\
    \begin{split} & \sum_\tb \beta_\tb  = \mu, \\
    & \sum_\tb \tilde\beta_\tb = \tilde{\mu}^{(4)}.\end{split}
\end{align}
Note that among the first four equations only three of them are mutually independent. Hence we have $3N-1$ equations in total, which completely determine $3N-1$ undetermined variables $(\b_\ta,\tilde\b_\ta, \a_i)$ in terms of $3N-1$ weight parameters $(\z_i,\tilde{\z}_i,  \t_i, \mu,\tilde\m ^{(4)})$.

The 4-point correlation function \eqref{eq:correp} constructed from flag varieties provides a particular representation of the 4-point KZ equation as a differential operator. The 4-point KZ equation reads
\begin{align} \label{eq:4kz}
    \left[-(k+N) \frac{\partial}{\partial \kq} + \frac{\hat{\rm H}_0^{(4)}}{\kq} + \frac{\hat{\rm H}_1^{(4)}}{\kq-1} \right] \Psi(\qe) = 0.
\end{align}
Here, we recall that the generator $\sT^k$ of $\mathfrak{sl}_N$ is represented on the respective modules by \eqref{eq:jv}, \eqref{eq:jv'}, \eqref{eq:tz}, and \eqref{eq:tz'}, yielding
\begin{align}
    \hat{\rm H}_0^{(4)} = - \sum_{\ta,\tb=1}^N J^\ta_\tb \fz^\tb \frac{\partial}{\partial \fz^\ta}, \quad \hat{\rm H}_1^{(4)} = - \tilde\fz(\fz) \sum_{\ta=1}^N \frac{\partial^2}{\partial\fz^\ta\partial\tilde\fz_\ta},
\end{align}
where the superscript was used to distinguish from the 5-point case that will appear later. Using the definition \eqref{eq:vdef} of the variables $(v_i)_{i=0} ^{N-1}$, these differential operators can be rewritten as differential operators in $(v_i)_{i=0} ^{N-1}$. See appendix \ref{sec:4-pt,flag} for the details of the computation.

On the other hand, in the gauge theory side we have the non-perturbative Dyson-Schwinger equation for the vacuum expectation value of the regular surface defect, which follows from the regularity of the vacuum expectation value of the fractional $qq$-characters \eqref{eq:fracqq},
\begin{align}
     [x^{-I}] \Big\langle  \EX_\o (x) \Big\rangle_{\mathbb{Z}_N} = 0, \quad \quad \begin{split} \o \in \BZ_N \\ I \in \BZ_{>0} \end{split}.
\end{align}
The Dyson-Schwinger equations can be organized into differential equations in the gauge coupling $\qe$ and the fractional couplings $(z_\o)_{\o=0} ^{N-1}$ \cite{Nikita:V}. As proven in \cite{NT}, these differential equations induce the 4-point KZ equation \eqref{eq:4kz} that we have discussed so far, when accompanied with a proper matching of the parameters on two sides.
See appendix~\ref{sec:4-point KZ} for the review on the derivation. This correspondence can be regarded as an explicit verification of the equivalence of the gauge theory correlation function and the WZNW correlation function \cite{AT,KPPW,K-T}, for the case at hand.

\subsubsection{Degenerate 5-point correlation function}
Let us now consider $\mathfrak{sl}_N$-modules associated to five points
\begin{align}
    \sz_{-1} = \infty, \ \sz_0 = y, \ \sz_1 = 1, \ \sz_2 = \kq, \ \sz_3 = 0
\end{align}
on the Riemann sphere; lowest-weight Verma modules $\EuScript{V}_0  (=\EuScript{V} _{\boldsymbol{\z}})$ and $\EuScript{V}_\infty$ at $0$ and $\infty$, Heisenberg-Weyl modules ${\EH}_{{\mu}}^{{\boldsymbol\t} - {\boldsymbol\z}}$ and $\tilde{\EH}_{\tilde\mu} ^{\boldsymbol\t - \tilde{\boldsymbol\z}+(0,\cdots, 0 ,1)}$ at $\kq$ and 1, and finally an $N$-dimensional representation $\mathbb{C}^N$ at $y$. Note the HW module with the shifted $\mathfrak{sl}_N$-weights can be written in $N$ equivalent ways: $\tilde{\EH} _{\tilde\m} ^{\boldsymbol\t - \tilde{\boldsymbol\z}+\boldsymbol\d_{N-1}} =\tilde{\EH}_{\tilde\mu} ^{\boldsymbol\t - \tilde{\boldsymbol\z}+\boldsymbol\d_{\ta-1} - \boldsymbol\d_{\ta}}$, $\ta=1, \cdots, N$, where $\boldsymbol\d_\ta \equiv \left( \d_{i,\ta} \right)_{i=1} ^{N-1} \in \BZ^{N-1}$.

The standard $N$-dimensional representation $\BC^N$ is an $N$-dimensional submodule of a special HW module $\tilde{\EH}_{1}^{-\boldsymbol\d_{N-1}} = \tilde{u}_{N}\BC[\tilde{u}_1^\pm,\dots,\tilde{u}_N^\pm]^{\BC^\times}$, which is spanned by $\{\tilde{u}_1, \cdots , \tilde{u}_N\} \subset \tilde{\EH}_1^{-\boldsymbol\d_{N-1}}$. The generators of $\mathfrak{gl}_N$, which act as single-entry $GL(N,\BC)$ matrices $E^\tb_\ta$, can be represented by differentials:
\begin{align}
    \sT^\tb _\ta \vert_{\mathbb{C}^N} = E^\tb_\ta = \tilde{u}_\ta \frac{\partial}{\partial \tilde{u}_\tb}. 
\end{align}

Again, we choose all the weights to be generic, so that all the modules are irreducible. In particular, we can replace the restricted dual $\EuScript{V}_\infty ^*$ by a highest-weight Verma module $\tilde{\EuScript{V}} _{\tilde{\boldsymbol{\z}}}$.

The corresponding degenerate $5$-point correlation function $\boldsymbol{\U}( {\qe},y)$ is valued in
\begin{align} \label{eq:5ptup}
    \boldsymbol{\U}( {\qe},y) \in \left( \EuScript{V} _{ \boldsymbol{\z}} \otimes {\EH}_{\mu} ^{{\boldsymbol\t} - {\boldsymbol\z}} \otimes \tilde{\EH}_{\tilde{\mu}} ^{\boldsymbol\t -\tilde{\boldsymbol\z} + \boldsymbol\d_{N-1}} \otimes   \mathbb{C}^N \otimes \tilde{\EuScript{V}} _{\tilde{\boldsymbol{\z}}} \right)^{\mathfrak{sl}_N} .
\end{align}
Note that we have the following decomposition for the last tensor product,
\begin{align} \label{eq:decom}
    \tilde{\EV}_{\tilde{\boldsymbol\z}} \otimes \BC^N \simeq \bigoplus_{\ta=1} ^N
    \tilde{\EV}_{\tilde{\boldsymbol\z} - \boldsymbol\d_{\ta-1} + \boldsymbol\d_{\ta} },
\end{align}
which follows from the identification of the $N$-dimensional representation as a submodule of a HW module, $\BC^N \subset\tilde{\EH} _{1} ^{-\boldsymbol\d_{N-1}} =\tilde{\EH}_{1} ^{-\boldsymbol\d_{\ta-1} + \boldsymbol\d_\ta}$. Thus the 5-point correlation function $\boldsymbol\Upsilon(\qe,y)$ can be expressed as an $N$-tuple of 4-point correlation functions that we have studied earlier. Note that the shift in the weights of the HW module $\tilde{\EH} _{\tilde\m} ^{\boldsymbol\t - \tilde{\boldsymbol\z}+\boldsymbol\d_{N-1}} =\tilde{\EH}_{\tilde\mu} ^{\boldsymbol\t - \tilde{\boldsymbol\z}+\boldsymbol\d_{\ta-1} - \boldsymbol\d_{\ta}}$ is introduced precisely to account for the shift in the weights of the Verma modules after the decomposition \eqref{eq:decom}, making the space \eqref{eq:5ptup} nonempty.

The 5-point KZ equations satisfied by $\boldsymbol\U (\qe,y)$ are
\begin{subequations}\label{eq:5-ptKZeq}
\begin{align}
    & \left[ -(k+N)\frac{\partial}{\partial \kq} + \frac{\hat{\rm H}_0}{\kq} + \frac{\hat{\rm H}_1}{\kq-1} + \frac{\hat{\rm H}_y}{\kq-y} \right]\boldsymbol\Upsilon=0 \\
    & \left[ -(k+N)\frac{\partial}{\partial y} + \frac{\hat{\rm A}_0}{y} + \frac{\hat{\rm A}_1}{y-1} + \frac{\hat{\rm A}_\kq}{y-\kq} \right]\boldsymbol\Upsilon = 0
\end{align}
\end{subequations}
The operators in the numerators in the expression are the symmetric product of generators $\{\sT^\sk\}$ of $\mathfrak{sl}_N$ represented on the respective modules,
\begin{subequations}
\begin{align}
    & \left( \hat{\rm H}_0 \right) = - \sum_{\ta,\tb=1}^N \fz^\tb J^\ta_\tb \frac{\partial}{\partial \fz^\ta}, \quad 
    \left( \hat{\rm H}_1 \right) = -\tilde\fz(\fz) \sum_{\ta=1}^N \frac{\partial^2}{\partial \fz^\ta \partial \tilde \fz_\ta}, \quad
    \left( \hat{\rm H}_y \right)_{\ta\tb} = - E^\tb_\ta \fz^\ta \frac{\partial}{\partial \fz^\tb}, \\
    %%%%%%%%%%%%%%%%%%%%%%%%%%%%%%%%%%%%%%%%
    & \left( \hat{\rm A}_0 \right)_{\ta\tb} = - E^\tb_\ta J_\tb^\ta, \quad
    \left( \hat{\rm A}_1 \right)_{\ta\tb} = - E^\tb_\ta \tilde{\fz}_\tb\frac{\partial}{\partial\tilde{\fz}_\ta}, \quad
    \left( \hat{\rm A}_\kq \right)_{\ta\tb} =  - \left( \hat{\rm H}_y \right)_{\ta\tb} =   E^\tb_\ta \fz^\ta \frac{\partial}{\partial \fz^\tb}.
\end{align}
\end{subequations}

The degenerate 5-point correlation function $\boldsymbol\Upsilon(\kq,y)$ is a vector in $\BC^N$, which can be constructed by
\beq
\boldsymbol{\Upsilon}(\kq,y) = \sum_{\ta=1}^{N} \Upsilon^{(\ta)}_0 \cdot (\tilde{u} \wedge \tilde{\pi}^{\ta-1}) (\pi_{\ta}) \cdot  \chi_\ta (v_{1}, \ldots , v_{N-1}; \qe,y)\, , 
\eeq
where $\chi_{\ta} (v_1 , \cdots, v_{N-1};\qe ,y)$ is a Laurent polynomial in $(v_i)_{i=1} ^{N-1} $ defined in \eqref{eq:vdef}, and $\Upsilon_{0}^{(\ta)}$ takes a similar form with $\Psi_{0}$ of the 4-point case \eqref{eq:psi0}, only that the exponents are now dependent on the index $\ta = 1, \cdots, N$: 
\begin{align}
    \Upsilon_{0}^{(\ta)} = 
    & \prod_{\tb=1}^{N} ((\tilde{\fz}\wedge \tilde{\pi}^{\tb-1})(\pi_\tb))^{\tilde{\beta}_\tb^{(\ta),*}}  (\tilde\pi^\tb (\fz\wedge\pi_{\tb-1}) )^{\beta_\tb^{(\ta),*}} \prod_{i=1}^{N-1} ( \tilde{\pi}^i(\pi_i) )^{\alpha_i^*}.
\end{align}
As in the 4-point case, the variables $(v_\ta)_{\ta=1} ^N$ and $\Upsilon_0 ^{(\ta)}$ are defined to be $\mathfrak{sl}_N$-invariant. For each $\ta=1, \cdots, N$, there are $3N-1$ undetermined variables $(\b_\tb ^{(\ta),*}, \tilde{\b}_\tb ^{(\ta),*} , \a_i ^*  )$. In a way similar to the 4-point case \eqref{eq:4pmat}, we have:
\begin{align}\label{eq:bbg}
    \begin{split} & \tilde\beta_i^{(\ta),*} + \beta_{i+1}^{(\ta),*} + \alpha_i^* + { \delta_{i,\ta}} = \zeta_i   \\
    & \tilde\beta_{i+1}^{(\ta),*} + \beta_i^{(\ta),*} + \alpha_i^* + {\delta_{i+1,\ta}} = \tilde\zeta_i \\& \t_i - \z_i = \b_i ^{(\ta),*} - \b_{i+1} ^{(\ta),*} \\ &\t_i - \tilde\z_i +\d_{\ta,i+1} - \d_{\ta,i} = \tilde\b_i ^{(\ta),*} - \tilde\b_{i+1} ^{(\ta),*} \end{split} \quad\quad i = 1, \ldots , N-1 \nonumber \\
    \begin{split} & \sum_\tb \beta_\tb^{(\ta),*}  = \mu \\
    & \sum_\tb \tilde\beta_\tb^{(\ta),*} = \tilde{\mu}.\end{split}
\end{align}
Compared to the 4-point case, there are simple shifts by $\d_{i,\ta}$ and $\d_{i+1,\a}$ in the first four equations, among which only three of them are mutually independent. Also the parameter $\tilde\mu$ of the HW module $\tilde{\EH}_{\tilde\mu}^{\boldsymbol\t - \boldsymbol\b + \boldsymbol\d_{N-1}}$ is shifted from its 4-point counterpart \eqref{eq:4pmat} by
\begin{align}
    \tilde\mu = \tilde\mu^{(4)} - 1.
\end{align}
Such shifts in parameters reflect the effect of the additional $N$-dimensional representation insertion. See \cite{Jeong:2018qpc,JeongNekrasov} for instance.

Just as in the 4-point case, we have $3N-1$ independent equations that determine $3N-1$ undetermined variables $(\b_\tb ^{(\ta),*}, \tilde{\b}_\tb ^{(\ta),*} , \a_i ^*  )$ in terms of $3N-1$ weight parameters $(\z_i,\tilde\z_i, \t_i, \m,\tilde\m)$. In particular, given a set of $\{\beta_\ta,\beta_\ta,\alpha_i\}$, $\ta=1,\dots,N$, $i=1,\dots,N-1$, that solves the matching in the 4-point case \eqref{eq:4pmat}, the solutions to \eqref{eq:bbg} are given by
\begin{align}\label{eq:sol5ptpara}
    \beta_\tb^{(\ta),*} = \beta_\tb, \quad \tilde\beta_\tb^{(\ta),*} = \tilde\beta_\tb - \delta_{\ta\tb}, \quad \alpha_i^* = \alpha_i. 
\end{align}

For the comparison with the equations on the gauge theory side, it is convenient to reorganize the KZ equations into differential equations acting on the Laurent polynomial part $\chi_\ta(v_1,\dots,v_{N-1};\qe, y)$, by commuting the differential operators through the prefactors $\Upsilon_0 ^{(\ta)}$. Let us denote $\boldsymbol\chi (v_1,\dots,v_{N-1};\kq,y) = \sum_{\ta=1}^N \tilde{u}^\ta \chi_\ta(v_1,\dots,v_{N-1};\kq,y)$ as a vector in $\BC^N$. Then the KZ equations \eqref{eq:5-ptKZeq} becomes
\begin{subequations}
\begin{align}
    & \left[ -(k+N) \frac{\partial}{\partial \kq} + \frac{\hat{\mathscr{H}}_0}{\kq} + \frac{\hat{\mathscr{H}}_1}{\kq-1} + \frac{\hat{\mathscr{H}}_y}{\kq-y} \right] \boldsymbol\chi = 0, \\
    & \left[ -(k+N) \frac{\partial}{\partial y} + \frac{\hat\EA_0}{y} + \frac{\hat\EA_1}{y-1} + \frac{\hat\EA_y}{y-\kq} \right] \boldsymbol\chi = 0.
\end{align}
\end{subequations}
With some decent calculation,
the residues $\hat{\mathscr{H}}_{0,1,y}$ in the $\qe$-component of the KZ equation are found as
\begin{subequations}\label{eq:KZHs}
\begin{align}
    \left( \hat{\mathscr{H}}_0 \right)_{\ta\tb}
    & =  \hat{\mathscr{H}}_0^{(4)}\delta_{\ta\tb} - \frac{v_{\ta+1}+\cdots+v_{N-1}}{v_\ta} (\nabla^\fz_\ta + \beta_\ta) \delta_{\ta\tb} + \frac{\fz^\ta\tilde\fz_\ta}{\fz^\tb\tilde\fz_\tb} (\nabla^\fz_\tb +\beta_\tb) \theta_{\ta>\tb} \\
    %%%%%%%%%%%%%%%%%%%%%%%%%%%%%%%%%%%%%%
    \left(\hat{\mathscr{H}}_1 \right)_{\ta\tb}
    & = \hat{\mathscr{H}}_1^{(4)} \delta_{\ta\tb} + \frac{1}{v_\ta} (\nabla^\fz_\ta + \beta_\ta) \delta_{\ta\tb}, \\
    %%%%%%%%%%%%%%%%%%%%%%%%%%%%%%%%%%%%%%%
    \left( \hat{\mathscr{H}}_y\right)_{\ta\tb} 
    & = -\frac{\fz^\ta\tilde\fz_\ta}{\fz^\tb\tilde\fz_\tb} (\nabla^\fz_\tb + \beta_\tb), 
\end{align}
\end{subequations}
where variables $\{v_\ta\}_{\ta=1}^N$ are 
\begin{align}
    v_\ta = \frac{\fz^\ta\tilde\fz_\ta}{\fz^1\tilde\fz_1+\cdots+\fz^{N-1}\tilde\fz_{N-1}}.
\end{align}
$\hat{\mathscr{H}}_{0,1}^{(4)}$ are the coefficients in the KZ equation satisfied by the Laurent polynomial part $\chi$ of the 4-point correlation function \eqref{eq:KZ-4p}, whose exact forms can be found in \eqref{eq:KZ-4p-conn}.

The residues of the $y$-component of the KZ equation $\hat{\EA}_{0,1,\kq}$, as differential operators acting on $\chi(v_1,\dots,v_{N-1};\qe,y)$, are found by
\begin{subequations}\label{eq:KZAs}
\begin{align}
    %%%%%%%%%%%%%%%%%%%%%%%%%%%%%%%%%%%%%%
    \left( \hat\EA_0 \right)_{\ta\tb}
    & = v_\ta  \theta_{\tb > \ta} \frac{\partial }{\partial v_\ta} - v_\tb \frac{\fz^\ta}{\fz^\tb}\frac{\tilde{\fz}_\ta}{\tilde{\fz}_\tb} \theta_{\tb > \ta} \frac{\partial }{\partial v_\tb} - \beta_\tb \frac{\fz^\ta}{\fz^\tb}\frac{\tilde{\fz}_\ta}{\tilde{\fz}_\tb} \theta_{\tb> \ta} + \tilde{\beta}_\ta \theta_{\tb > \ta} - \delta_{\ta\tb} \xi_\ta, \\
    %%%%%%%%%%%%%%%%%%%%%%%%%%%%%%%%%%%%%%
    \left( \hat\EA_1 \right)_{\ta\tb}
    & = - (\nabla^{\tilde\fz}_\ta + \tilde\beta_\ta - \delta_{\ta\tb}), \\
    %%%%%%%%%%%%%%%%%%%%%%%%%%%%%%%%%%%%%%
     \left( \hat\EA_\kq\right)_{\ta\tb} 
    & = \frac{\fz^\ta\tilde\fz_\ta}{\fz^\tb\tilde\fz_\tb} (\nabla^\fz_\tb + \beta_\tb). 
\end{align}
\end{subequations}
We use a short hand notation
\begin{align}
    \xi_\ta = \delta^\ta_\tb + \delta^\ta_\tb \sum_{i=1}^N \beta_i^{(\tb),*} \theta_{i > \ta} + \tilde\beta_i^{(\tb),*}\theta_{i\geq \ta} + \delta^\ta_\tb \sum_{i=1}^{N-1} \alpha_i^* \theta_{i\geq \ta} = \zeta_\ta + \cdots + \zeta_{N-1}.
\end{align}
The detailed derivation of the differential operators \eqref{eq:KZHs} and \eqref{eq:KZAs} can be found in appendix~\ref{sec:5pt derive}.

\subsection{Knizhnik-Zamolodchikov equations from the T-Q equations}\label{sec:KZ from TQ}
Now we shall verify that the correlation function of intersecting surface defect observables in the $\EuScript{N}=2$ gauge theory satisfies the degenerate 5-point KZ equations discussed so far. In fact, the KZ equations are shown to be the Fourier transform of the fractional quantum T-Q equation obeyed by the fractional $Q$-observables, which we derived in section \ref{sec:TQ-DS-frac}. Correspondingly, the Fourier transform $\boldsymbol\Upsilon(\qe,y)$ of the correlation function of intersecting surface defects provides the solutions to the KZ equation, and thereby gets identified with the genus $0$ degenerate $5$-point conformal block of the $\widehat{\ssl}_N$ current algebra.

\subsubsection{The $y$-component}
First we show that the correlation function $\boldsymbol\Upsilon(\qe,y)$ of intersecting surface defects \eqref{def:upsilon} satisfies the $y$-component of the 5-point KZ equation. This follows from performing Fourier transformation to the fractional quantum T-Q equations \eqref{eq:qTQvev}.

\bigskip \noindent $\bullet$
The $N$ fractional quantum T-Q equations \eqref{eq:qTQvev} can be expressed into a single $N \times N$ matrix equation:
\begin{align} \label{eq:mateq}
    % & \left[ {\bf U} (x-{\bf M}^+) + \boldsymbol{\qe} ( x - {\bf M}^- ) {\bf U}^{-1} - ( {\bf I}_N+\boldsymbol{\qe} )x +{\boldsymbol{\rho}} \right] \boldsymbol\Upsilon (y) = 0 
    & \left[ ({\bf U} + \boldsymbol{\kq} {\bf U}^{-1} - {\bf I}_N - \boldsymbol{\kq}) \left( \ve_2 y \frac{\partial }{\partial y}  - \ve_2 y\frac{\partial \log \Upsilon^{\text{pert}} (y) }{\partial y}  \right)  + \left({\bf U}{\bf M}^+ + \boldsymbol{\qe} {\bf M}^- {\bf U}^{-1} - {\boldsymbol{\rho}}\right) \right] \boldsymbol\Upsilon(y) = 0
\end{align}
Let us explain the matrices notations appearing above, ${\bf U}$ is given by:
\begin{align}\label{def:U-matrix}
    {\bf U} := \begin{pmatrix}
    0 & 1 & 0 & \cdots & 0 & 0 \\
    0 & 0 & 1 & \cdots & 0 & 0 \\
    & \vdots & & \ddots & & \vdots \\
    0 & 0 & 0 & \cdots & 0 & 1 \\
    y & 0 & 0 & \cdots & 0 & 0
    \end{pmatrix}, \quad 
    {\bf U}^{-1} = \begin{pmatrix}
    0 & 0 & 0 & \cdots & 0 & \frac{1}{y} \\
    1 & 0 & 0 & \cdots & 0 & 0 \\
    0 & 1 & 0 & \cdots & 0 & 0 \\
    & \vdots & & \ddots & & \vdots \\
    0 & 0 & 0 & \cdots & 1 & 0
    \end{pmatrix}, \quad  {\bf U}^N = y {\bf I}_N.
\end{align}
We also have 4 diagonal matrices built up by fundamental matter masses, fractional couplings, and zeros of $T_{N,\omega}$:
\begin{align}\label{def:diagmatrices}
    {\bf M}^\pm = {\rm diag} ({m}_0^\pm,\dots,{m}_{N-1}^\pm), \ \boldsymbol{\qe} = {\rm diag}(\qe_0,\dots,\qe_{N-1}), \ {\boldsymbol{\rho}} = {\rm diag}({\rho}_0,\dots,{\rho}_{N-1}).
\end{align}
% The matrix ${\bf L}$ is defined as
% $$
%     {\bf L} = ({\bf U} + \boldsymbol{\kq} {\bf U}^{-1} - {\bf I}_N - \boldsymbol{\kq}).
% $$

For direct comparison with the KZ equations in the form written in the previous section, we have to express the matrix equation \eqref{eq:mateq} in a different basis. Let us consider the following change of basis,
\begin{align}\label{def:Pi-vector}
    \boldsymbol\Pi = ({\bf U} - {\bf I}_N ) \boldsymbol\Upsilon.
\end{align}
It will be justified in a moment. In terms of $\boldsymbol\Pi$, the matrix equation \eqref{eq:mateq} becomes
\begin{align} \label{eq:ycomp}
    0  
    & = \left[ \frac{\ve_2}{\ve_1} \frac{\partial}{\partial y} - \frac{\ve_2}{\ve_1} \frac{\partial \log \Upsilon^{\text{pert}} (y) }{\partial y} + \frac{1}{\ve_1} \left( \frac{1}{y}\left ({\bf I}_N - {\boldsymbol\kq}{\bf U}^{-1} \right)^{-1} \left({\bf U}{\bf M}^+ + \boldsymbol{\qe} {\bf M}^- {\bf U}^{-1} - {\boldsymbol{\rho}}\right) \left( {\bf U} - {\bf I}_{N} \right)^{-1} \right) \right]\boldsymbol\Pi \nonumber\\
    & := \left[ \frac{\ve_2}{\ve_1} \frac{\partial}{\partial y} + \frac{\hat\CalA_0}{y} + \frac{\hat\CalA_1}{y-1} + \frac{\hat\CalA_\kq}{y-\kq} \right]\boldsymbol\Pi. 
\end{align}
Note that this equation is in the form of the $y$-component of the KZ equation, with the level determined by
\begin{align}
k+N = - \frac{\ve_2}{\ve_1}.
\end{align}
We shall show now that the coefficients $\hat{\CalA}_{0,\qe,1}$ are indeed identical to the ones appearing in the $y$-component of the 5-point KZ equation \eqref{eq:KZAs}, with certain identification of the parameters on two sides.

By an explicit computation, the residues $\hat{\CalA}_{0,1,\kq}$ can be determined as
\begin{subequations}\label{eq:CalA2-new}
\begin{align}
    \left( \hat{\CalA}_0 \right)_{\ta\tb} 
    = & \ \frac{z_{\ta}}{z_{\tb}} \left[ \frac{m_{\tb}^+}{\ve_1}\theta_{\ta<\tb} - \frac{m_{\tb}^-}{\ve_1}\theta_{\ta\leq\tb}\right] + z_{\ta} \left( \frac{\partial}{\partial z_{\ta}} - \frac{\partial}{\partial z_{\tb}} \right)  \theta_{\ta<\tb} - \frac{\ve_2}{\ve_1}\delta_{\ta\tb} \text{Res}_{y=0} \frac{\partial \log \Upsilon^{\text{pert}} (y)}{\partial y},\\
    %%%%%%%%%%%%%%%%%%%%%%%%%%%%%%%%%%%%%%%%%%
    \left( \hat{\CalA}_1 \right)_{\ta\tb} 
    % = & \frac{1}{1-\kq} \left[ u_{\omega'-1}m_{\omega'}^+ + u_{\omega'+1}\kq_{\omega'+1} m_{\omega'+1}^- + u_{\omega'} \left( -m_{\omega'+1}^+ - \kq_{\omega'}m_{\omega'}^- + \ve_1\nabla^z_{\omega'} - \ve_1\kq_{\omega'}\nabla^z_{\omega'-1} \right) \right] \nonumber \\
    = & \ - z_{\ta}\frac{\partial}{\partial z_{\ta}} - \frac{\ve_2}{\ve_1}\delta_{\ta\tb} \text{Res}_{y=1} \frac{\partial \log \Upsilon^{\text{pert}} (y)}{\partial y}, \\
    %%%%%%%%%%%%%%%%%%%%%%%%%%%%%%%%%%%%%%%%%%
    \left( \hat{\CalA}_\kq \right)_{\ta\tb} 
    % = & \frac{\gamma_{\omega}}{z_{\omega'-1}}m_{\omega'}^ + +\frac{\gamma_{\omega}}{z_{\omega'}} m_{\omega'+1}^- + \frac{\gamma_{\omega}}{z_{\omega'}} (-m_{\omega'+1}^+ - \kq_{\omega'}m_{\omega'}^- + \ve_1\nabla^z_{\omega'} - \ve_1\kq_{\omega'}\nabla^z_{\omega'-1}) \nonumber\\
    = & \ \frac{z_{\ta}}{z_{\tb}} \left[ z_{\tb} \frac{\partial }{\partial z_{\tb}} + \frac{m_{\tb}^- - m_{\tb}^+}{\ve_1}  \right] - \frac{\ve_2}{\ve_1}\delta_{\ta\tb} \text{Res}_{y=\qe} \frac{\partial \log \Upsilon^{\text{pert}} (y)}{\partial y},
\end{align}
\end{subequations}
where $\ta,\tb=1,\dots,N$. Here, we have defined the Boolean function $\theta$:
\begin{align}
    \theta_S = 
    \begin{cases}
    1 & \text{if $S$ is true} \\
    0 & \text{otherwise}
    \end{cases}.
\end{align}
See appendix~\ref{sec:y-KZ derive} for details of the calculations.
To determine $\Upsilon^{\text{pert}} (y)$, we consider the traces of the coefficients:
\begin{subequations} \label{eq:traceas}
\begin{align}
    & \Tr \hat\CalA_0 = \sum_{\tb=1}^{N} -\frac{m_{\tb}^-}{\ve_1} - \frac{N\ve_2}{\ve_1} \underset{y=0}{\text{Res}} \frac{\p \log \Upsilon^{\text{pert}} (y)}{\p y} \\
    & \Tr \hat\CalA_1 = \sum_{\tb=1}^{N} -\nabla^z_\tb - \frac{N}{\ve_1} \text{Res}_{y=1} \frac{\p \log \Upsilon^{\text{pert}} (y)}{\p y} = \sum_{\omega=0}^{N-1} -\nabla^z_{\omega} - \frac{N\ve_2}{\ve_1} \underset{y=1}{\text{Res}} \frac{\p \log \Upsilon^{\text{pert}} (y)}{\p y}  \\
    & \Tr \hat\CalA_\kq = \sum_{\tb=1}^{N} \frac{m_{\tb}^- - m_{\tb}^+}{\ve_1} + \nabla^z_\tb - \frac{N\ve_2}{\ve_1} \underset{y=\kq}{\text{Res}} \frac{\p \log \Upsilon^{\text{pert}} (y)}{\p y} 
\end{align}
\end{subequations}
Dependence of $(z_{\omega})_{\omega=0}^{N-1}$ in $\boldsymbol\Upsilon$ comes from the perturbative factor \eqref{eq:pert} and $(\hat\kq_{\omega})_{\omega=0}^{N-1}$ in $\langle Q_{\omega} \rangle_{\BZ_N} \Psi$. For any function $f=f(\hat\kq_0,\dots,\hat\kq_{N-1})$:
\begin{align}
    \sum_{\omega=0}^N z_{\omega}\frac{\partial}{\partial z_{\omega}} f(\hat\kq_0,\dots,\hat\kq_{N-1}) = \sum_{\omega=0}^{N-1} \left( \hat\kq_{\omega+1} \frac{\partial}{\partial \hat\kq_{\omega+1}} - \hat\kq_{\omega}\frac{\partial}{\partial\hat\kq_{\omega}} \right) f(\hat\kq_0,\dots,\hat\kq_{N-1}) = 0.
\end{align}
Hence the only contribution of center of momentum $\sum_{\omega} \nabla^z_{\omega}$ acting on $\boldsymbol\Upsilon$ comes from the perturbative factor \eqref{eq:pert}:
\begin{align}
    \sum_{\omega=0}^{N-1} \nabla^z_{\omega} \boldsymbol\Upsilon = \left( \sum_{\omega=0}^{N-1} \frac{m_{\omega}^+ - a_{\omega}}{\ve_1} \right) \boldsymbol\Upsilon.
\end{align}
Then the prefactor $\Upsilon^{\text{pert}} (y)$ is determined by the condition that $\hat\CalA_{0,1,\kq}$ are traceless, i.e., requiring \eqref{eq:traceas} to vanish,
\begin{align}
   \Upsilon^{\text{pert}} (y) = y^{-\frac{m^-}{N\ve_2}} (y-\kq)^{\frac{m^--a}{N\ve_2}} (y-1)^{-\frac{m^+-a}{N\ve_2}}
\end{align}
with the short handed notation
\begin{align}
    m^\pm = \sum_{\omega=0}^{N-1}m_{\omega}^\pm,\quad a = \sum_{\omega=0}^{N-1}a_{\omega}.
\end{align}
We recall that there is perturbative factor \eqref{eq:pert} in the expectation value of $\langle Q_{\omega}(x) \rangle_{\BZ_{N}} \Psi$. We hence modify derivatives terms 
\begin{align}
    \nabla^z_\tb \mapsto \nabla^z_\tb + \frac{m_\tb^+ - a_\tb}{\ve_1}
\end{align}
when the operators $\hat{\mathcal{A}}_{0,\qe,1}$ act solely on the non-perturbative terms in $\boldsymbol\Pi$. 
As a result, we see that the coefficients in the equation are identical to the ones appearing in the $y$-component of the 5-point KZ equation \eqref{eq:KZAs}:
\begin{align}
    \hat\CalA_{0} = \hat\EA_0, \ \hat\CalA_{1} = \hat\EA_1, \ \hat\CalA_{\kq} = \hat\EA_\kq
\end{align}
with the following identification of parameters on two sides. Namely, the parameters $\boldsymbol\beta$ and $\tilde{\boldsymbol\beta}$ of the 4-point KZ equation are identified by the Coulomb moduli  and hypermultiplet masses:
\begin{align}
    \beta_{\ta} = \frac{m^+_\ta-a_\ta}{\ve_1}, \quad \tilde\beta_{\ta} = \frac{m^-_\ta - a_\ta}{\ve_1}, \quad \ta=1, \cdots N.
\end{align}
Relations between 5-point parameters $(\beta^{(\ta),*}_\tb,\tilde{\beta}^{(\ta),*}_\tb)_{\ta,\tb=1}^N$ and Coulomb moduli and hypermultiplet masses can be obtained through \eqref{eq:sol5ptpara}:
\begin{align}
    & \beta_{\tb}^{(\ta),*} = \frac{m^+_\tb-a_\tb}{\ve_1}, \quad \tilde\beta_\tb^{(\ta),*} = \frac{m^-_\tb - a_\tb}{\ve_1} - \delta_{\ta\tb}, \quad \ta,\tb=1,\dots,N.
\end{align}
Correspondingly, the weights of the $\mathfrak{sl}_N$-modules are determined by the Coulomb moduli and the hypermultiplet masses:
\begin{align} \label{eq:dic1}
\begin{split}
    &\zeta_i = \frac{m^-_{i+1}-m^-_{i}}{\ve_1}, \quad \tilde\zeta_i = \frac{m^+_{i+1}-m^+_{i}}{\ve_1}, \ i=1,\dots,N-1;\\
    &\mu = \sum_{\ta=1}^N \frac{m^-_\ta - a_\ta}{\ve_1}, \quad \tilde\mu = -1 + \sum_{\ta=1}^N \frac{m_\ta^+ - a_\ta}{\ve_1}. 
\end{split}
\end{align}
Also, the fractional couplings $\hat\qe_\o = \frac{z_{\o+1}}{z_\o}$ are identified with the components of the maps $\fz$ and $\tilde{\fz}$ by
\begin{align} \label{eq:dic2}
\begin{split}
    &z_\ta = \tilde\fz_\ta \fz^\ta.
\end{split}
\end{align}
It should be noted that the $N-1$ degrees of freedom  $\boldsymbol\t \in \BC^{N-1}$ for the correlation function $\boldsymbol\Upsilon(\qe,y)$, which determine the $\mathfrak{sl}_N$-weights of the HW modules, precisely correspond to the $N-1$ Coulomb moduli $\left(a_\ta - \frac{1}{N} a \right)_{\ta=1} ^N$ of the $\EN=2$ gauge theory through the matching \eqref{eq:dic1}.

The solutions to the KZ equation also give the equivalence
\begin{align}
   \boldsymbol\Pi(z,\qe,y)= \prod_{\omega=0}^{N-1}z_{\omega}^{\frac{m_{\omega}^+ - a_{\omega}}{\ve_1}} \boldsymbol\chi (v;\qe,y)   ,
\end{align}
with \eqref{eq:dic1} and \eqref{eq:dic2} understood.

\subsubsection{The $\kq$-component}
Next, we show that the correlation function $\boldsymbol\Upsilon(\qe,y)$ satisfies the $\qe$-component of the 5-point KZ equation. So far, we have only used the $\o=\o'$ part of the $qq$-character. Let us consider the non-perturbative Dyson-Schwinger equation
\begin{align}
    \left \langle [x^{-I}] \hat{T}_{N+1,\omega}(x) {Q}_{\omega'}(x') \right \rangle_{\BZ_N} = 0, \ I = 1,2,\dots. 
\end{align}
for any combination of $\omega,\omega'=0,\dots,N-1$.
Using ${Q}_{\omega}$ defined in \eqref{def:hatQ}, we consider the following linear combination
\begin{align}\label{eq:u-comb}
    & \Biggl \langle [x^{-1}] \sum_{\omega\neq \omega'}   u_{\omega}\hat{T}_{N+1,\omega}(x) {Q}_{\omega'}(x') \Biggr \rangle_{\BZ_N} + \left \langle u_{\omega'} \hat{T}_{N+1,\omega'}(x) {Q}_{\omega'}(x') \right \rangle_{\BZ_N} \\
    & = u_{\omega'} (x - x') \left \langle P^+_{\omega'+1}(x) {Q}_{\omega'}(x') \right \rangle_{\BZ_N} + \hat\kq_{\omega'}u_{\omega'} (x-x'+\ve_1)  \left \langle P_{\omega'}^-(x) {Q}_{\omega'-1}(x') \right \rangle_{\BZ_N}. \nonumber
\end{align}
Coefficients $u_{\omega}$ are chosen as
\begin{align}\label{def:u_w}
    & u_{\omega} = 1 + \kq_{\omega+1} + \kq_{\omega+1}\kq_{\omega+2} + \cdots + \kq_{\omega+1}\cdots\kq_{\omega+N} \nonumber\\
    & \implies u_{\omega} - \kq_{\omega+1}u_{\omega+1} = 1 - \kq, \ {}^\forall \omega = 0,1,\dots,N-1.
\end{align}
The Fourier transform \eqref{eq:Yo} of \eqref{eq:u-comb} with $x=x'$ and $x'=x+\ve_1$ yields:
\begin{subequations}
\begin{align}
    & (1-\kq)\ve_1\ve_2\left(\kq\frac{\partial}{\partial\kq} - \kq\frac{\partial\log \Upsilon ^{\text{pert}}(y)}{\partial \kq}  \right) \Upsilon_{\omega'} + \hat{\rm H} \Upsilon_{\omega'} \nonumber\\
    & = -\ve_1 u_{\omega'} \left( -\ve_2 y\frac{\partial}{\partial y} - m_{\omega'+1}^+ \right)\Upsilon_{\omega'+1} + \ve_1u_{\omega'} \left( -\ve_2y\frac{\partial}{\partial y} - m^+_{\omega'+1} + \ve_1\nabla^z_{\omega'+1} \right)\Upsilon_{\omega'}, \\
    %%%%%%%%%%%%%%%%%%%%%%%%%%%%%%%%%%%%%%%%%%%%
    & (1-\kq)\ve_1\ve_2\left(\kq\frac{\partial}{\partial\kq} - \kq\frac{\partial\log \Upsilon ^{\text{pert}}(y)}{\partial \kq}  \right) \Upsilon_{\omega'} + \hat{\rm H} \Upsilon_{\omega'} \nonumber\\
    & = \ve_1 \hat\kq_{\omega'}u_{\omega'} \left( -\ve_2 y\frac{\partial}{\partial y} - m_{\omega'}^- \right)\Upsilon_{\omega'-1} - \ve_1 \hat\kq_{\omega'}u_{\omega'} \left( -\ve_2y\frac{\partial}{\partial y} - m^-_{\omega'} - \ve_1\nabla^z_{\omega'} \right)\Upsilon_{\omega'}.
    \end{align}
\end{subequations}
The operator $\hat{\rm H}$ is defined by
\begin{align}\label{eq:H2}
    \hat{\rm H} :=  \sum_\omega
    & \frac{1-\kq}{2} \left(\ve_1\nabla^z_{\omega}-{{m}^+_{\omega}}\right)^2 + \hat\kq_{\omega} u_{\omega}(\ve_1\nabla^z_{\omega})(\ve_1\nabla^z_{\omega}-m^+_{\omega}+m_{\omega}^-).
\end{align}
See appendix~\ref{sec:q-KZ derive} for details of the computations.

The $y$-derivative terms can be canceled with a proper linear combination of the two. By denoting ${\bf u} = {\rm diag}(u_0,u_1,\dots,u_{N-1})$, and ${\bf G} = \text{diag}(G_0,\dots,G_{N-1})$ with
$$
    G_{\omega} = \frac{u_{\omega}+\kq-1}{u_{\omega}} = \frac{\kq_{\omega+1}u_{\omega+1}}{u_{\omega}},
$$
the $N$ differential equations can be rewritten as one matrix equation:
\begin{align}\label{eq:KZ-q-raw}
    & (1-\kq)\ve_1\ve_2 \left({\bf G} - {\bf U} \right) \left(\kq\frac{\partial}{\partial\kq} - \kq\frac{\partial\log\Upsilon ^{\text{pert}}(y)}{\partial \kq}  \right) \boldsymbol\Upsilon - \left({\bf G} - {\bf U} \right) \hat{\rm H} \boldsymbol\Upsilon \nonumber\\
    & = \ve_1 {\bf G}{\bf u} \left[ {\bf U}({\bf M}^+ - {\bf M}^-){\bf U}^{-1} - \boldsymbol\nabla \right]  ({\bf U} - {\bf I}_N)\boldsymbol\Upsilon
\end{align}
with matrix ${\bf U}$ defined in \eqref{def:U-matrix}. Let us again consider a change of basis $\boldsymbol\Pi$ defined in \eqref{def:Pi-vector}. In terms of $\boldsymbol\Pi$, the matrix equation becomes
\begin{align} \label{eq:qcomp}
    0 & = \left[ \frac{\ve_2}{\ve_1} \frac{\partial}{\partial\kq} + \frac{1}{(1-\kq)\kq} \frac{1}{\ve_1 ^2} \hat{\rm H} - \frac{\ve_2}{\ve_1} \frac{\partial\log \Upsilon ^{\text{pert}}(y)}{\partial \kq} \right. \nonumber \\
    & \qquad \left. - \frac{1}{\ve_1 (1-\kq)\kq} ({\bf U}-{\bf I}_{N})({\bf G}-{\bf U})^{-1} {\bf G}{\bf u} \left( {\bf U}({\bf M}^+ - {\bf M}^-){\bf U}^{-1} - \boldsymbol\nabla \right) \right]\boldsymbol\Pi \nonumber\\
    & :=\left[ \frac{\ve_2}{\ve_1} \frac{\partial}{\partial\kq} + \frac{\hat{\fA}_{0}}{\kq} + \frac{\hat{\fA}_{1}}{\kq-1} + \frac{\hat{\fA}_{y}}{\kq-y} \right] \boldsymbol\Pi = 0.
\end{align}
Note that this is precisely in the form of the $\qe$-component of the 5-point KZ equation, with $k+N = -\frac{\ve_2}{\ve_1}$. We shall show the coefficients $\hat{\fA}_{0,1,y}$ are indeed identical to the ones appearing in the $\qe$-component of the 5-point KZ equation \eqref{eq:KZHs}, with the identification of the parameters given by \eqref{eq:dic1} and \eqref{eq:dic2}.

With some decent computation, we find each individual $\hat{\fA}_{0,1,y}$ as:
\begin{subequations}\label{eq:fA2-new}
\begin{align}
    \left( \hat{\fA}_0 \right)_{\ta\tb} = \
    & \frac{1}{\ve_1^2}\hat{\rm H}|_{\kq=0} \delta_{\ta\tb} - \frac{z_{\ta+1}+\cdots+z_{N-1}}{z_{\ta}}\left( \nabla^z_{\ta} + \frac{m^-_{\ta} - m^+_{\ta}}{\ve_1} \right) \delta_{\ta\tb} \nonumber\\
    & + \frac{z_\ta}{z_\tb} \left( \nabla^z_{\tb} + \frac{m^-_{\tb} - m^+_{\tb}}{\ve_1} \right) \theta_{\ta>\tb},  \\
    % & - \theta_{\omega+1>\omega'} \left( {\bf M}^+_{\omega'+1} - {\bf M}^-_{\omega'} - \ve_1 \nabla^z_{\omega'} - \ve_1 \right) \\
    % & -u_{\omega} {\theta_{\omega>\omega'}} \left( {\bf M}^+_{\omega'+1} - {\bf M}^-_{\omega'} - \nabla^z_{\omega'} - 1 \right) \\
    %%%%%%%%%%%%%%%%%%%%%%%%%%%%%%%%%%%%%%%%%%
    \left( \hat{\fA}_{1} \right)_{\ta\tb} = \
    & -\frac{1}{\ve_1^2} \hat{\rm H}|_{\kq=1} \delta_{\ta\tb} + \frac{z_0+\cdots+z_{N-1}}{z_{\ta}} \left( \frac{m^+_{\ta} - m^-_{\ta}}{\ve_1} - \nabla^z_{\ta} \right) \delta_{\ta\tb}, \\
    % = & -\frac{1}{\ve_1}\hat{\rm H}_2|_{\kq=1} \delta_{\omega,\omega'} \\
    % & + \frac{1}{y-1} \kq_{\omega'}u_{\omega'} \left({y}\right)^{\theta_{\omega>\omega'-1}} (m_{\omega'}^+ - m_{\omega'}^- - \ve_1 \nabla^z_{\omega'-1}) \\
    %%%%%%%%%%%%%%%%%%%%%%%%%%%%%%%%%%%%%%%%%%
    \left( \hat{\fA}_y \right)_{\ta\tb} = \
    & \frac{z_\ta}{z_\tb} \left( \nabla^z_{\tb} + \frac{m^-_{\tb} - m^+_{\tb}}{\ve_1}  \right) - \frac{\delta_{\ta\tb}}{N} \frac{m^--a}{\ve_1},
    % = & - \left( \hat{\CalA}_\kq \right)_{\omega,\omega'}
\end{align}
\end{subequations}
where $\ta,\tb=1,\dots,N$. See appendix~\ref{sec:q-KZ derive} for details of the calculations.
We again find \eqref{eq:fA2-new} agrees with the coefficients appearing in the $\qe$-component of the 5-point KZ equation \eqref{eq:KZHs}: 
\begin{align}
    \hat{\fA}_{0} = \hat{\mathscr{H}}_0, \ \hat{\fA}_{1} = \hat{\mathscr{H}}_{1}, \ \hat{\fA}_{y} = \hat{\mathscr{H}}_{y},
\end{align}
after taking care the perturbative factor in \eqref{eq:pert}:
\begin{align}
    \nabla^z_{\tb} \mapsto \nabla^z_{\tb} + \frac{m_\tb^+ - a_\tb}{\ve_1},
\end{align}
provided that the variables on two sides are related by \eqref{eq:dic1} and \eqref{eq:dic2}.

Therefore, we arrive at our conclusion: The vector $\boldsymbol\Pi$, which is the Fourier transformation of the correlation function of intersecting surface defects, solves the degenerate 5-point KZ equations \eqref{eq:5-ptKZeq} for $\mathfrak{sl}_N$ at the level $k+ N = -\frac{\ve_2}{\ve_1}$. Namely,
\begin{align}
   \boldsymbol\Pi(z;\kq,y) = \prod_{\omega=0}^{N-1}z_{\omega}^{\frac{m_{\omega}^+ - a_{\omega}}{\ve_1}}\boldsymbol\chi(v;\kq,y)
\end{align}
provided that the parameters on two sides are identified by \eqref{eq:dic1} and \eqref{eq:dic2}.

\section{$\spch$ spin chain} \label{sec:spin}
The connection between spin chain systems and supersymmetric gauge theories is one of the most well-known examples of the Bethe/gauge correspondence. It was firstly observed in \cite{Gorsky:1996hs, Gorsky:1997a} that the spectral curves of the classical spin chains are identical to the Seiberg-Witten curves of four-dimensional $\EN=2$ gauge theories. The correspondence was uplifted to the quantum level in \cite{Nekrasov:2009uh,Nekrasov:2009ui}. The corresponding gauge theories were, however, two-dimensional $\EN=(2,2)$ gauged linear sigma models instead of being four-dimensional. Also, the comparison was made within the context of the algebraic Bethe ansatz, restricting the spin representations to be highest-weight (or lowest-weight). With the same restriction on the spin representations, the IR duality discovered in \cite{Dorey:2011pa} between four-dimensional $\EN=2$ theories and two-dimensional $\EN=(2,2)$ theories gave a four-dimensional account for the quantum spin chains. Finally, it was shown in \cite{Lee:2020hfu} that the classical $\spch$ spin chain arises in the Seiberg-Witten geometry of the four-dimensional $\EN=2$ theory \cite{Nekrasov:2012xe,Nikita-Pestun-Shatashvili}, as well as a relation between the $\spch$ spin chain coordinate systems and the defect gauge theory parameters, with more general $\mathfrak{sl}_2$-representations which are neither highest-weight nor lowest-weight. The extension to the supergroup gauge theories is also discussed in \cite{Nek18,Kimura:2019msw,Chen:2020rxu}.

In this section, we generalize these constructions and explain how four-dimensional $\EN=2$ gauge theories give rise to the quantum $\spch$ spin chain systems with non-height-weight infinite-dimensional $\mathfrak{sl}_2$-modules, by the fractional quantum T-Q equations and the higher-rank $qq$-characters.

Let us briefly review the quantum $\spch$ spin magnet and its Lax operators and monodromy matrix. Let $x \in \mathbb{C}$ be a complex number. Also we consider a two-dimensional auxiliary space $V_{\text{aux}} = \BC^2$. The Lax operators are defined as a $2 \times 2$ matrix in $\text{End}(V_{\text{aux}})$ with operator-valued entries:
\begin{align}
    L_{\omega}^{\rm XXX}(x) = x - \theta_{\omega} + \hbar\CalL_{\omega}, \quad \omega=0,1,\dots,N-1
\end{align}
where $\CalL_{\omega} = \bs^0_{\omega}\sigma_0 + \bs^+_{\omega}\sigma_+ + \bs^-_{\omega}\sigma_-$ are $\mathfrak{sl}_2$ matrices. 
The $N$ complex numbers $\theta_\omega \in \mathbb{C}$ are called the \emph{inhomogeneities}. The generators of $\mathfrak{sl}_2$ obey the standard commutation relation:
\begin{align}
    [\bs^0_{\omega},\bs^\pm_{\omega'}] = \pm \bs^\pm_{\omega} \delta_{\omega\omega'}, \quad [\bs_\omega^+,\bs_{\omega'}^-] = 2 \bs^0_{\omega} \delta_{\omega\omega'}.
\end{align}
For each $\o=0,\cdots, N-1$, we construct an $\mathfrak{sl}_2$-module $\EH_{\mathbf{s}_\o,\mathfrak{a}_\o}$ from the space of Laurent polynomials in a complex variable $\g_\o$, namely, 
\beq
\EH_{s_{\omega},\fa_{\omega}} = \gamma_{\omega}^{\fa_{\omega}} \BC[\gamma_{\omega},\gamma_{\omega}^{-1}],
\eeq 
where $\mathfrak{a}_\o$ is a complex number that characterizes the module. Note that $\mathfrak{a}_\o$ is defined up to integer shifts, i.e.,
$$
    \EH_{s_{\omega},\fa_{\omega}} \simeq \EH_{s_{\omega},\fa_{\omega}+n}, \quad \ n\in\BZ.
$$
The generators of $\mathfrak{sl}_2$ are represented by differential operators on this space:
\begin{align}
    \bs_\omega^0 = \gamma_{\omega}\frac{\partial}{\partial \gamma_{\omega}} - s_\omega, \quad \bs_{\omega}^- = \frac{\partial}{\partial \gamma_{\omega}}, \quad \bs^+_{\omega} = 2 s_{\omega}\gamma_{\omega} - \gamma_{\omega}^2 \frac{\partial}{\partial \gamma_{\omega}}.
\end{align}
The space $\EH_{s_{\omega},\fa_{\omega}}$, called the local Hilbert space, constructed in this way is an infinite dimensional module of $\mathfrak{sl}_2$ Lie algebra which are neither highest nor lowest-weight representation.\footnote{It should be noted that the method of algebraic Bethe ansatz does not generally apply for the spin chain with non-height-weight representations such as the ones considered here. There are other methods such as functional Bethe ansatz to solve the spin chain with generic representations. See \cite{sklyanin92} for instance.} Such representation is characterized by generic complex numbers $s_\o$ and $\fa_{\omega}$, for which $\bs^0_{\omega} +s_\o - \fa_{\omega} \in \BZ$. Then, the Lax operator $L_\o ^{\rm XXX} (x)$ assigned to the $(\omega+1)$-th site of $\spch$ spin chain lattice is regarded as a $\mathfrak{sl}_2$-homomorphism, $L_\o ^{\rm XXX} (x) \in \text{End} \left( \EH_{{s}_\o,\mathfrak{a}_\o} \otimes V_{\text{aux}} \right)$.
\begin{comment}
The Lax operator $L_\omega^{\rm XXX}(x)$ is assigned to the $(\omega+1)$-th site of $\spch$ spin chain lattice with the spin operators obeying the standard commutation relation:
\begin{align}
    [\bs^0_{\omega},\bs^\pm_{\omega'}] = \pm \bs^\pm_{\omega} \delta_{\omega\omega'}, \quad [\bs_\omega^+,\bs_{\omega'}^-] = 2 \bs^0_{\omega} \delta_{\omega\omega'}.
\end{align}
Each spin variables act on the local Hilbert space $\EH_{s_{\omega},\fa_{\omega}}$. $\EH_{s_{\omega},\fa_{\omega}}$ is an infinite dimensional module of $\mathfrak{sl}_2$ Lie algebra which are neither the highest nor the lowest weight representation. Such representation is characterized by a generic complex number $\fa_{\omega}$, for which $\bs^0_{\omega} - \fa_{\omega} \in \BZ$, via differential operators
\begin{align}
    \bs_\omega^0 = \gamma_{\omega}\frac{\partial}{\partial \gamma_{\omega}} - s_\omega, \quad \bs_{\omega}^- = \frac{\partial}{\partial \gamma_{\omega}}, \quad \bs^+_{\omega} = 2 s_{\omega}\gamma_{\omega} - \gamma_{\omega}^2 \frac{\partial}{\partial \gamma_{\omega}}.
\end{align}
The module $\EH_{s_{\omega},\fa_{\omega}}$ is isomorphic to the space of Laurent polynomial in $\gamma_{\omega}$, namely, 
\beq
\EH_{s_{\omega},\fa_{\omega}} = \gamma_{\omega}^{\fa_{\omega}} \BC[\gamma_{\omega},\gamma_{\omega}^{-1}].
\eeq
\end{comment}
The full Hilbert space is the tensor product of all the local Hilbert spaces
\begin{align}
    \CalH = \EH_{s_0,\fa_0} \otimes \EH_{s_1,\fa_1} \otimes \cdots \otimes \EH_{s_{N-1},\fa_{N-1}}.
\end{align}

For generic values of $s$ and $\fa$, these modules are irreducible. However, for special, quantized values of $\fa$ and $s$ these modules contain $\mathfrak{sl}_2$-invariant submodules, allowing to take quotients. For example, we have Verma modules in $\EuScript{V}^-_s\subset\EH_{s,0}$ and $\EuScript{V}^+_s \subset \EH_{s,2s}$; moreover, for integer $2s\in \BZ_{>0}$, we have $\EH_{s,0} \approx \EH_{s,2s}$ so that taking quotients leads to the familiar finite dimensional representations.

% Let us briefly recall the infinite dimensional representation of $\mathfrak{sl}_2$ Lie algebra: There are Verma modules $V_h^\pm$ of the lowest or highest weight, in which the spectrum of the operator of the operator $\bs^0$ (in the usual basis $\bs^0,\bs^+,\bs^-$) belongs to the set $h+n$ with $n\in\BZ_{\geq0}$ or $n\in\BZ_{\leq0}$, respectively, with $h \in \BC$ being the eigenvalue of $\bs^0$ of the vacuum vector, annihilated by $\bs^-$, or $\bs^+$, respectively. For this modules the spin $s$ of the system, defined through the value of $s(s+1)$ of the Casimir operator $\bs^+\bs^-+\bs^-\bs^++2(\bs^0)^2$, is determined by $h$.
% However, there are modules $\EuScript{V}_{s,\fa}$ that is neither the highest or the lowest weight, for which $s^0 - \fa \in \BZ$. 

% For generic value $s$ and $\fa$ the representation $\EuScript{V}_{s,\fa}$ is irreducible. However, for special, quantized value of $s$ and $\fa$ these modules contains $\mathfrak{sl}_2$-invariant submodules, allowing to take quotient. For instance, $V^+_s \subset \EuScript{V}_{s,0}$, $V_s^- \subset \EuScript{V}_{s,2s}$, and for half integer $2s \in \BZ_+$, $\EuScript{V}_{s,2s} \sim \EuScript{V}_{s,0}$ allowing to take quotient leading to the familiar representation of finite dimension.

% In particular, we shall use verma module $\EuScript{V}_{s_\omega,\fa_{\omega}}$ for the $\spch$ spin chain:

% with $N$ complex spins $s_{\omega}$ and $N$ complex numbers $\fa_\omega$.
The monodromy matrix is an ordered product of Lax operators
\begin{align}\label{def:monoT}
    \mathbf{T}_{\rm SC}(x) = K(\kq) L_{N-1}^{\rm XXX}(x)\cdots L_0^{\rm XXX}(x) \in \text{End}\left(  \bigotimes_{\o=0} ^{N-1} \EH_{s_\o,\fa_\o} \otimes V_{\text{aux}} \right).
\end{align}
The \textit{twist matrix} $K(\kq)$ is a constant matrix in $\text{End}(V_{\rm aux})$. $K(\kq)$ satisfies
\begin{align}
    \Tr \ K(\kq) = 1+\kq, \quad \det K(\kq) = \kq.
\end{align}

\subsection{Construction of Lax operators}\label{sec:spinLax}

We will demonstrate how one may recognize \eqref{def:monoT} in $\EN=2$ supersymmetric gauge theory in 4-dimension. We start with the fractional T-Q equation \eqref{eq:qTQvev}:
\begin{align}
    & P_{\omega+1}^+(x) \langle \tilde{Q}_{\omega+1}(x) \rangle_{\BZ_N} \Psi + \hat\kq_{\omega} P_{\omega}^-(x) \langle \tilde{Q}_{\omega-1}(x) \rangle_{\BZ_N} \Psi
    = \hat{T}_{N,\omega}(x) \langle \tilde{Q}_{\omega}(x) \rangle_{\BZ_N} \Psi.
\end{align}
Let us define $2\times 1$ vector
\begin{align}
    \Xi_{\omega}(x) = 
    \begin{pmatrix}  \langle \tilde{Q}_{\omega} ( x )  \rangle_{\BZ_N} \\ \langle \tilde{Q}_{\omega-1} ( x ) \rangle_{\BZ_N} \end{pmatrix} \Psi 
\end{align}
to translate fractional TQ equation to degree one matrix equation:
\begin{align}
    \Xi_{\omega+1}(x) = \frac{1}{P^+_{\omega+1}(x)} \begin{pmatrix} \hat{T}_{N,\omega}( x ) & -\hat\kq_{\omega} P_{\omega}^- ( x ) \\ P_{\omega+1}^+ ( x ) & 0 \end{pmatrix} \Xi_{\omega}(x) = \frac{1}{P^+_{\omega+1}(x)}\tilde{L}_{\omega}(x)\Xi_{\omega}(x).
    \label{eq:qBXi}
\end{align}
Matrix $\tilde{L}_{\omega}$ is of the form:
\begin{align}
    \tilde{L}_{\omega}(x) = & (x - m_{\omega+1}^+) \begin{pmatrix} 1+\hat\kq_{\omega} & -\hat\kq_{\omega} \\ 1 & 0 \end{pmatrix} +
    \begin{pmatrix} (1+\hat\kq_{\omega}) (\rho_{\omega}+m_{\omega+1}^+ ) & \hat\kq_{\omega} (m_{\omega}^- - m_{\omega+1}^+) \\
    0 & 0 \end{pmatrix}.
\end{align}
We take gauge transformation $\Theta_{\omega}(x) = g_{\omega}\Xi_{\omega}(x)$ satisfying
\begin{align}\label{def:gaugetrans}
    g_{\omega+1} \begin{pmatrix} 1+\hat\kq_{\omega} & -\hat\kq_{\omega} \\ 1 & 0 \end{pmatrix}
    g_{\omega}^{-1} = \begin{pmatrix} 1 & 0 \\ 0 & 1 \end{pmatrix}.
\end{align}
The twisted matrix $K$ defined based on the gauge transformation
\begin{align}
    g_{N}^{-1} = g_0^{-1}  K.
\end{align}
% We denote 
% \begin{align}
%     g_{\omega} = h_{\omega} \begin{pmatrix}
%     1 & -1 \\ 0 & 1
%     \end{pmatrix}
% \end{align}
% so that
% \begin{align}
%     h_{\omega+1} \begin{pmatrix} \kq_{\omega} & 0 \\ 1 & 1 \end{pmatrix} h_{\omega} ^{-1} = 
%     \begin{pmatrix} 1 & 0 \\ 0 & 1 \end{pmatrix}
% \end{align}
% The twist matrix $K$ is defined via
% \begin{align}
%     K = h_0 h_{N}^{-1} = h_0 \begin{pmatrix} \kq & 0 \\ (\kq-1)\frac{\gamma_{-1}}{z_{-1}} & 1 \end{pmatrix} h_0^{-1}
% \end{align}
When $\kq\neq1$, we choose gauge $g_0$ by picking the twisted matrix $K$ to have the following form 
\begin{align}
    K = \begin{pmatrix} \kq & 0 \\ 0 & 1 \end{pmatrix} \implies g_0 = \frac{1}{\gamma_{0}-\gamma_{-1}} \begin{pmatrix} 1 & -1 \\ -\gamma_{-1} & \gamma_{0} \end{pmatrix}
    \implies g_\omega = \frac{1}{\gamma_{\omega} - \gamma_{\omega-1}} \begin{pmatrix} 1 & -1 \\ -\gamma_{\omega-1} & \gamma_{\omega}. \end{pmatrix}
\end{align}
We define a new set of parameters $\{\gamma_{\omega}\}$:
\begin{align}
    \gamma_{\omega} = \frac{z_{\omega+1} + z_{\omega+2}+\cdots+z_{\omega+N}}{\kq-1}, \quad \gamma_{\omega} - \gamma_{\omega-1} = z_{\omega}, \quad \gamma_{\omega+N} = \kq\gamma_{\omega}.
\end{align}

% \begin{align}
%     h_n=g_n\begin{pmatrix} 1 & -1 \\ 0 & 1 \end{pmatrix}
% \end{align}
% \begin{align}
%     h_{\omega+1} = h_{0}\frac{1}{u_{\omega+1}^{\vee}-u_{\omega}^{\vee}} 
%     \begin{pmatrix}
%     1 & 0 \\ -u_{\omega}^{\vee} & u_{\omega+1}^{\vee}-u_{\omega}^{\vee}
%     \end{pmatrix}
% \end{align}
In terms of $\Theta_{\omega}(x)$, \eqref{eq:qBXi} becomes
\begin{align}
    \Theta_{\omega+1}(x) = \frac{1}{P^+_{\omega+1}(x)} \left( x - m_{\omega+1}^+ + L_{\omega} \right) \Theta_{\omega}(x) = \frac{1}{P^+_{\omega+1}(x)} L_{\omega}^{\rm XXX}(x), \Theta_{\omega}(x).
\end{align}
where gauge transformed $L_{\omega}$ is given by
\begin{align}
    L_{\omega} 
    % & = h_{\omega+1} \begin{pmatrix} -\ve_1(\gamma_{\omega+1}-\gamma_\omega)\frac{\partial}{\partial\gamma_{\omega}}  + \kq_{\omega}(m_{\omega+1}^+ - m_{\omega}^-) & -\ve_1(\gamma_{\omega+1}-\gamma_\omega)\frac{\partial}{\partial\gamma_{\omega}} \\ 0 & 0 \end{pmatrix} h_{\omega}^{-1},  \nonumber\\
    % & = \frac{1}{\gamma_{\omega+1}-\gamma_{\omega}} \begin{pmatrix} 1 & 0 \\ -\gamma_{\omega} & \gamma_{\omega+1}-\gamma_{\omega} \end{pmatrix} 
    % \begin{pmatrix} -\ve_1(\gamma_{\omega+1}-\gamma_\omega)\frac{\partial}{\partial\gamma_{\omega}} + \kq_{\omega}(m_{\omega+1}^+ - m_{\omega}^-) & -\ve_1(\gamma_{\omega+1}-\gamma_\omega)\frac{\partial}{\partial\gamma_{\omega}} \\
    % 0 & 0 \end{pmatrix}
    % \begin{pmatrix}
    % \gamma_{\omega}-\gamma_{\omega-1} & 0 \\ \gamma_{\omega-1} & 1
    % \end{pmatrix} \nonumber\\
    & = 
    \begin{pmatrix}
    -\ve_1\frac{\partial}{\partial\gamma_{\omega}}\gamma_{\omega} + (m_{\omega+1}^+ - m_{\omega}^-) & -\ve_1\frac{\partial}{\partial\gamma_{\omega}} \\ -\gamma_{\omega} (-\ve_1\frac{\partial}{\partial\gamma_{\omega}}\gamma_{\omega} + (m_{\omega+1}^+ - m_{\omega}^-)) & \gamma_{\omega}\ve_1\frac{\partial}{\partial\gamma_{\omega}}
    \end{pmatrix}.
\end{align}
The trace of the Lax operator $L_{\omega}$ is
\begin{align}
    \Tr L_{\omega} & = -\ve_1\frac{\partial}{\partial\gamma_{\omega}}\gamma_{\omega} + m_{\omega+1}^+ - m_{\omega}^- + \gamma_{\omega}\ve_1\frac{\partial}{\partial\gamma_{\omega}} = m_{\omega+1}^+ - m_{\omega}^- - \ve_1 = 2s_{\omega} \ve_1,
\end{align}
where we identified the spin $s_\o$ of the $\mathfrak{sl}_2$-representations and the inhomogeneity $\th_\o$ with hypermultiplet masses by
\begin{align}
\begin{split}
    &s_\o = \frac{m^+ _{\o+1} -m^- _\o -\ve_1}{2 \ve_1},\quad \th_\o = \frac{m^+ _{\o+1} +m^- _\o +\ve_1}{2}
\end{split}
\end{align}
% We multiply the defect instanton partition function by the following perturbative factor
% \begin{align}
%     \prod_{\omega=0}^{N-1} z_{\omega}^{-\frac{a_{\omega+1}^+ - m_{\omega+1}^+}{\ve_1}}
% \end{align}
% so that the differential operator $\tP_{\omega}$ becomes the conjugate momentum of coordinate $\gamma_{\omega}$:
% \begin{align}
%     \frac{1}{\gamma_{\omega+1}-\gamma_{\omega}}\tP_{\omega} = \ve_1 \frac{\partial}{\partial z_{\omega}} - \ve_1 \frac{\partial}{\partial z_{\omega-1}} = -\ve_1 \frac{\partial}{\partial \gamma_{\omega}}
% \end{align}
We may now denote the individual Lax operator as
\begin{align}
    L^{\rm XXX}_{\omega}(x) 
    & = x - \theta_\omega - \ve_1
    \begin{pmatrix} \gamma_{\omega}\frac{\partial}{\partial\gamma_{\omega}} - s_\omega & \frac{\partial}{\partial\gamma_{\omega}} \\ 2 \gamma_{\omega} s_{\omega}-\gamma_{\omega}^2 \frac{\partial}{\partial\gamma_{\omega}} & -\gamma_{\omega}\frac{\partial}{\partial\gamma_{\omega}} + s_{\omega}
    \end{pmatrix} \nonumber\\
    & = x-\theta_{\omega} - \ve_1\CalL_{\omega}.
\end{align}
We see that the $\O$-background parameter is identified with the Planck constant,
\begin{align}
    \ve_1 = -\hbar. 
\end{align}

Let us specify the $\mathfrak{sl}_2$-modules that comprise the spin chain. Each of the two entries of $\Theta_\o (x)$, on which the Lax operator $L^{\rm XXX} _\o (x)$ acts, is a Laurent polynomials in $(z_\o)_{\o=0} ^{N-1} = \left( \g_\o -\g_{\o-1} \right)_{\o=0} ^{N-1}$ of degree $0$ multiplied by the prefactor \eqref{eq:pert},
\begin{align} \label{eq:pertt}
    \prod_{\o=0} ^{N-1} z_\o^{\frac{m_\o ^+ -a_\o}{\ve_1}} = \prod_{\o=0} ^{N-1} (\g_\o - \g_{\o-1})^{\frac{m_\o ^+ -a_\o}{\ve_1}}.
\end{align}
Depending on the relative norm between $|\gamma_{\omega}|$ and $|\gamma_{\omega+1}|$, this prefactor is expanded differently as a Laurent series. Note that such expansion always respects the hierarchy of $(z_{\omega})_{\omega=0}^{N-1}$:
\begin{align}
    |\hat{\kq}_\omega| < 1 \implies |z_{\omega+1}| < |z_{\omega}|.
\end{align}
Thus, we have $2^N$ domains in the $\BC_\g ^N$-space, specified by an $N$-tuple of ``spins'' ${\bf t} = (t_0,t_1,\dots,t_{N-1})$ (not to be confused with the actual spins ${\bf s}=(s_0,\dots,s_{N-1})$),
$$
    t_{\omega}= \frac{1}{2}  \text{sgn}(|\gamma_{\omega}|-|\gamma_{\omega-1}|)  \in \left\{ - \frac 12 , \frac 1 2 \right\}.
$$
Then let us define $\fa_{\omega}$ on the domain labeled by ${\bf t}$ is given by
\begin{align}
    \fa_{\omega} = \left( t_{\omega} + \frac 12 \right) \times \frac{m_{\omega}^+ - a_{\omega}}{\ve_1} + \left( \frac 12 -t_{\omega+1} \right) \times \frac{m_{\omega+1}^+-a_{\omega+1}}{\ve_1}.
\end{align}
Then we can identify the $\mathfrak{sl}_N$-module that the Lax operator $L^{\rm XXX} _\o (x)$ acts on as $\EH_{s_\o, \fa_\o}$. More precisely, the $\Theta _\o (x)$ resides in a particular weight subspace in the completed tensor product:
\begin{align}
     \Theta_\o (x) \in \left( {\EH}_{s_0,\fa_0} {\hat\otimes} {\EH}_{s_1, \fa_1} \hat\otimes \ldots \hat\otimes {\EH}_{s_{N-1}, \fa_{N-1}} \right) \left[ \sum_{\omega = 0} ^{N-1} {\fa}_{\omega } \right] \otimes V_{\rm aux},
\end{align}
where we defined the auxiliary space $V_{\rm aux} = \BC ^2$.

For an illustration, let us consider the domain of ``all spins down'' labeled by ${\bf t}=(-\frac 12, \dots, - \frac 12)$, which corresponds to expanding \eqref{eq:pertt} in the domain $|\gamma_{-1}|=\frac{|\gamma_{N-1}|}{|\kq|}>|\gamma_0|>|\gamma_1|>\dots>|\gamma_{N-1}|$. In this domain $\Theta_{\omega}$ is of the form
\begin{align}
     \prod_{\omega=0}^{N-1} \gamma_{\omega}^{\frac{m_{\omega+1}^+ - a_{\omega+1}}{\ve_1}} \BC[(\gamma_0)^\pm,\dots,(\gamma_{N-1})^\pm]^{\BC^\times} \otimes \BC^2. \nonumber
\end{align}
The parameters $\fa_\omega$ of local Hilbert spaces $\EH_{s_\omega,\fa_\omega}$ are identified as:
\begin{align}
    \fa_{\omega} = \frac{m_{\omega+1}^+ - a_{\omega+1}}{\ve_1}, \quad \omega=0,\dots,N-1.
\end{align}

The spin chain monodromy matrix is defined as a ordered product over the Lax operators
\begin{align}\label{def:monodromy}
    {\bf T}_{\rm SC}(x) = K(\kq) L_{N-1}^{\rm XXX}(x)\cdots L_{0}^{\rm XXX}(x).
\end{align}
When acting on the first state $\Pi_0(x)$, the monodromy transforms
\begin{align} \label{eq:monodromy}
    % & \tilde{L}_{N-1}(y) \tilde{L}_{N-2}(y) \cdots \tilde{L}_{0}(y)\Xi_0(y) = \Xi_{N}(y+\kappa) = \Xi_0(y+\kappa). \nonumber\\
    % & \implies 
    \frac{1}{P^+(x)} {\bf T}_{\rm SC}(x)\Theta_{0}(x) = \Theta_{N}(x) = \Theta_{0}(x+{\ve}_2) 
\end{align}
% Here we take the aligned surface defect.
% Alternatively, one may consider the intersecting surface defect:
% \begin{align}
%     \Xi_{\omega}^{(2)}(x) := 
%     \begin{pmatrix}  \langle \tilde{Q}_{\omega} ( x + \omega \tilde\ve_2) \widetilde\Psi \rangle \\ P_{\omega}^+ ( x ) \langle \tilde{Q}_{\omega-1} ( x + ( \omega - 1 )\tilde\ve_2 ) \widetilde\Psi \rangle \end{pmatrix}. 
% \end{align}
% The fact that \eqref{eq:qTQvev} and \eqref{eq:qTQvev} have the same structure means all calculation of $\Xi^{(2)}_\omega$ follows that of $\Xi_\omega$. We take the gauge transformation $\Pi^{(2)}_{\omega}(x) = g_\omega \Xi^{(2)}_\omega(x) $. The action of monodromy matrix ${\bf T}_{\rm SC}(x)$ in \eqref{def:monodromy} on $\Pi^{(2)}$: 
% \begin{align}
%     % & \tilde{L}_{N-1}(y) \tilde{L}_{N-2}(y) \cdots \tilde{L}_{0}(y)\Xi_0(y) = \Xi_{N}(y+\kappa) = \Xi_0(y+\kappa). \nonumber\\
%     % & \implies 
%     {\bf T}_{\rm SC}(x)\Pi^{(2)}_{0}(x) = \Pi^{(2)}_{0}(x+{\ve}_2) 
% \end{align}
Hence, the monodromy matrix ${\bf T}_{\rm SC}(x)$ is an operator on the completed tensor product with of all the local Hilbert spaces $\EH_{s_\omega,\fa_{\omega}}$ and the auxiliary space $V_{\rm aux}$  
\begin{align}
    {\bf T}_{\rm SC}(x) \in \text{End} \left(  {\EH}_{s_0,\fa_0} {\hat\otimes} {\EH}_{s_1, \fa_1} \hat\otimes \ldots \hat\otimes {\EH}_{s_{N-1}, \fa_{N-1}}  \otimes V_{\rm aux} \right).
\end{align}
The spin chain constructed is quantum integrable by the fact that local Lax operators $L_{\omega}^{\rm XXX}$ satisfy the $RLL$-relation (train track relation) \eqref{eq:RLL}. 
The $R$-matrix, defined in $V_{\rm aux} \otimes V_{\rm aux} = \BC^2 \otimes \BC^2$ space, is given in \eqref{def:R-matrix}. See appendix~\ref{sec:Bethe} for details.

We remark that there exists a curious $\mathfrak{sl}_N$-action on the Hilbert space defined by 
\beq \label{eq:curiousslN}
{\bf J}_{b}^{a} = {\gamma}_{a-1} \frac{\partial}{\partial\gamma_{b-1}}  - c {\delta}_{a}^{b}\, , \ a, b = 1, \ldots, N
\eeq
with $c = \frac 1N \sum_{\omega} {\fa}_{\omega}$. The relation of this $\mathfrak{sl}_N$-action to the one in section \ref{sec:KZ} is not obvious.

% varies according to which domain of $\{\gamma_{\omega}\}$ we expand the perturbative factor \eqref{eq:pert}.  

% We may take alternative expansion in the domain $\gamma_{N-1}>\gamma_{N-2}>\cdots>\gamma_{1}>\gamma_{0}$. $\Pi_{\omega}$ in this domain resides in vector space 
% \begin{align}
%     \prod_{\omega=0}^{N-1} \gamma_{\omega}^{\frac{m_{\omega}^+ - a_{\omega}}{\ve_1}} \BC[(\gamma_0)^\pm,\dots,(\gamma_{N-1})^\pm] \otimes \BC^2.
% \end{align}
% The character of the modules $\fa_{\omega}$'s in such domain are identified as:
% \begin{align}
%     \fa_\omega = \frac{m_\omega^+ - a_{\omega}}{\ve_1}.
% \end{align}
% We are allows to expand in any domain in $\gamma$-variables as long as $\{z_{\omega}\}$ obeys the hierarchy condition $\vert z_{\omega+1} \vert < \vert z_{\omega} \vert$ for instanton partition function to be properly defined. Module characters $\fa_{\omega}$ shall change accordingly to different choice of domain.

\paragraph{Remark} We emphasize again that the $\mathfrak{sl}_2$-modules at the $N$ spin sites are generically neither highest-weight nor lowest-weight. At the special values of $\left(\fa_\o \right)_{\o=0} ^{N-1}$, however, these modules contain highest-weight or lowest-weight submodules. For example, we may simply set $\fa \in \BZ$ or $\fa - 2s \in \BZ$. Then it is straightforward that $\EuScript{V}^- _s \subset \EH_{s,0}$ and $\EuScript{V}^+ _{s} \subset \EH_{s,2s}$ where $V^- _s = \BC[\g]$ and $V^+ _s = \g^{2s} \BC[\g^{-1}]$ are a lowest-weight and a highest-weight Verma module, respectively. 

Note that the condition $\fa_\o = \frac{m_{\o} ^+ - a_{\o}}{\ve_1} \in \BZ$ (or $\fa_\o = \frac{m_{\o+1} ^+ - a_{\o+1}}{\ve_1} \in \BZ$ if $t_{\omega}=0$)  
% or $\frac{a_{\omega+1}-m ^- _{\omega}}{\ve_1} \in \BZ \Leftrightarrow \fa_{\omega} = \frac{m^+_{\o+1} - m^-_{\o}}{\ve_1} \in \BZ + 2s_{\o}$,  
, which gives rise to $\EH_{s_{\omega},0}$ containing the lowest-weight Verma module $V_{s_\omega}^- \subset \EH_{s_{\omega},0}$ in all domain of expansion of $(\gamma_{\omega})_{\omega=0}^{N-1}$,
% or $\EH_{s_{\omega},2s_{\omega}}$, 
is precisely the restriction considered in \cite{Dorey:2011pa,HYC:2011}, as a particular example of the quantization condition \cite{Nikita-Shatashvili}. It is convenient to adopt the type IIA D-brane picture \cite{Witten1997} to illustrate what happens physically under this condition. We can realize the $U(N)$ gauge theory with $N$ fundamental and $N$ anti-fundamental hypermultiplets by three stack of $N$ D4-branes, stretched between two NS5-branes, stretched from the left NS5-brane to the infinity, and finally stretched from the right NS5-brane to the infinity. Now upon imposing the above condition, the two D4-branes across one of the NS5-brane meet each other. When all the $N$ D4-branes meet in such a way, the NS5 brane can be pulled out transversally to trigger Hanany-Witten brane transition, creating $\fa_\o \in \BZ$ D2-branes stretched between the NS5-brane and the $(\o+1)$-th D4-branes.

At the level of the effective field theory, this brane transition corresponds to the Higgsing of the four-dimensional gauge theory. The field configurations are squeezed into the $\BC_1$-plane, described by the effective two-dimensional $\EN=(2,2)$ theory on the non-compact part of the worldvolume of the D2-branes. The vacuum equation obtained from the two-dimensional effective twisted superpotential evaluated at the locus of the quantization condition $\frac{m^+ _\o -a_\o}{\ve_1} \in \BZ $ is identical to the Bethe equation. In this way, we precisely recover the Bethe/gauge correspondence of \cite{Nekrasov:2009uh,Nekrasov:2009ui}, between the two-dimensional $\EN=(2,2)$ gauged linear sigma model and the $XXX_{\mathfrak{sl}_2}$ spin chain with only lowest-weight $\mathfrak{sl}_2$-representations.

\subsection{Transfer matrix and higher rank {\it qq} -characters}

We consider transfer matrix, i.e., the trace of the monodromy matrix ${\bf T}_{\rm SC}$ over the auxiliary space $V_{\rm aux}$:
\begin{align}
    \Tr_{V_{\rm aux}} {\bf T}_{\rm SC}(x) = \Tr_{V_{\rm aux}}  \left( K L_{N-1}^{\rm XXX}(x) \cdots L_0^{\rm XXX}(x) \right) 
    % = \Tr_{V_{\rm aux}} \left( \tilde{L}_{N-1}(x)\cdots \tilde{L}_{0}(x) \right)
\end{align}
Individual Lax operator $L_{\omega}$ is defined on infinite dimensional representation of $\mathfrak{sl}_2$ algebra. In such representation, the spin operators are denoted by differential operators as we have seen earlier. One should consider the action
\begin{align}
   \Tr_{V_{\rm aux}} {\bf T}_{\rm SC}(x) \Phi, \quad \Phi \in \EH_{s_0,a_0} \otimes \cdots \otimes \EH_{s_{N-1},a_{N-1}}.
\end{align}

Alternatively, we may consider
\begin{align}
    \Tr_{V_{\rm aux}}  \left( K L_{N-1}^{\rm XXX}(x) \cdots L_0^{\rm XXX}(x) \right) = \Tr_{V_{\rm aux}} \left( \tilde{L}_{N-1}(x)\cdots \tilde{L}_{0}(x) \right).
\end{align}
The right hand side is in the context of $\EN=2$ SQCD. More precisely speaking, $\hat{T}_{N,\omega}(x)$ in $\tilde{L}_{\omega}$ \eqref{eq:qBXi} is defined through its action on the expectation value of observable $\langle {Q}_{\omega}(x) \rangle_{\BZ_N} \Psi$ through fractional quantum T-Q equation \eqref{eq:qTQvev}. 
The fact that all $\tilde{L}$ having vanishing lower right component means the trace can be understand as follows: 
We image on each spin lattice site there exists two states: empty or occupied. An empty state at site $j$ contributes 
$$
    \hat{T}_{N,j}(x).
$$
Alternatively when a site $j$ is occupied, it requires its previous site $j-1$ also being occupied. The combined contribution of occupied sites $j$ and $j-1$ is
$$
    -\hat\kq_j P_j(x)
$$
% We image an artificial lattice with $N$ sites. Each artificial lattice site $\omega$ consists exactly two states: empty, which gives contribution
% $$
%     T_{N,\omega}(x)T_{N,\omega-1}(x),
% $$
% or occupied, yielding: 
% $$
%     -\kq_{\omega}P_{\omega}(x).
% $$
% Microscopically each site $\omega$ on the auxiliary lattice is composed by two real spin sites $\omega$ and $\omega-1$ on the actual spin lattice. That is the artificial lattice can be considered as a covering of the real spin lattice. Two neighboring sites on the artificial lattice can not be 
% On the level of the real spin sites, 
The trace of the monodromy matrix is an ensemble over all empty/occupied states on each site:
\begin{align}\label{eq:trmono}
    \Tr_{V_{\rm aux}}  {\bf T}_{\rm SC}(x) 
    & = \sum_{[N] = \sJ\cup\sK\cup\tilde\sK} \prod_{j\in \sJ} \hat{T}_{N,j}(x) \prod_{k\in \sK} (-\kq_kP^-_k(x)P_k^+(x+\delta_{k,0}\ve_2)) 
    % & = \sum+{[N] = \sJ\cup\tilde\sJ \cup \sK\cup\tilde\sK} \prod_{j\in\sJ} \tilde{Y}_{j+1}(x+(j+1)\tilde\ve_2) \prod_{}
\end{align}
where $[N] = \{ 0,1,\dots,N-1\}$. The set $\tilde\sK$ is defined by 
$$
     \tilde\sK = \{k-1\ ({\rm mod} \ N) \ | \ k \in \sK\}.
$$

\paragraph{}
The structure in \eqref{eq:trmono} resembles to rank $N$ $qq$-character of $A_1$ theory, which also has structure of an ensemble over $N$ two level states. We only briefly review the subject here, some details can be found in \cite{Nikita:I,Nikita:V}.

The higher rank $qq$-character is constructed by adding a stack of $D3$-branes transverse to the stack of branes supporting the
 bulk four-dimensional gauge theory of section~\ref{sec:qq-char}. 

Let us study the gauge origami configuration with two orthogonal stacks of branes in $\BC_{12}^2$ abd $\BC_{34}^2$. Stacks of D-branes on $\BC_{12}^2$ is the familiar $n_{12,0}=n_{12,1}=n_{12,2}=N$ in the fundamental case in section~\ref{sec:crbr}. 
On the orthogonal $\BC^{2}_{34}$, we insert $\sw$ stacks of D-branes, all charged neutral under the $\BZ_3$-charge assignment with $n_{34,0}=\sw$, on $\BC^2_{34}$:
\begin{align}\label{def:GO-higherqq}
\begin{split}
    & N_{12} = \sum_{\alpha=1}^{N} e^{a_\alpha} \CalR_{0} + \sum_{\alpha=1}^{N} e^{m^-_\alpha-\ve_4} \CalR_{1} + \sum_{\alpha=1}^{N} e^{m_\alpha^+ - \ve_3} \CalR_{2} \\
    & N_{34} = \sum_{j=1}^{\sw} e^{x+\varrho_j} \CalR_0
\end{split}
\end{align}
with $\sw$-tuples ${\boldsymbol\varrho}$ of complex numbers ${\boldsymbol\varrho} = (\varrho_1,\dots,\varrho_\sw)\in \BC^\sw$ acts as moduli parameters of gauge theory that resides on $\BC^2_{34}$.
The corresponding gauge origami partition function is computed as
\begin{align}
    \CalZ_S = \sum_{\boldsymbol\lambda^{(12)},\boldsymbol\lambda^{(34)}} \prod_{i=0,1,2} \kq_i^{|\boldsymbol\lambda_i^{(12)}|} \kq_0^{|\boldsymbol\lambda^{(34)}|} \BE \left[ -\frac{P_3S_{12}S_{12}^*}{P_{12}^*} - \frac{P_1S_{34}S_{34}^*}{P_{34}^*} -q_{12}^* S_{12}S_{34}^* \right]^{\BZ_3}
\end{align}

Similar to the fundamental $qq$-character case in section~\ref{sec:crbr}, we take the decoupling limit $\kq_1=\kq_2=0$ to obtain the $A_1$-quiver gauge theory. The decoupling limit restricts the instanton configurations that enter the ensemble, each D-brane on $\BC_{34}^2$ can only have 0 or 1 instanton, with a total $2^\sw$ allowed configurations. Therefore, the gauge origami partition function comprises of the usual four dimensional bulk terms and an ensemble over the instanton configurations on $\BC^2_{34}$:
\begin{align}
    \CalZ_S 
    & = \sum_{\boldsymbol\lambda^{(12)}, \boldsymbol\lambda^{(34)}} \kq^{|\boldsymbol\lambda^{(12)}|} \kq^{|\boldsymbol\lambda^{(34)}|} \BE \left[ \frac{ - SS^* + M S^*  }{P_{12}} \right]  \BE\left[ -\frac{P_1S_{34}S_{34}^*}{P_{34}^*} - q_{12}^{-1}S_{12}S_{34}^*
    \right]^{\BZ_3} \nonumber\\
    & = \sum_{\boldsymbol\lambda^{(12)}} \kq^{|\boldsymbol\lambda^{(12)}|} \BE[\CalT_{12}] \EX_{\sw,{\boldsymbol\varrho}}(x) 
\end{align}
Rank $\sw$ $qq$-character of $A_1$ theory is:
\begin{align}\label{eq:higherqq}
    \EX_{\sw,{\boldsymbol\varrho}}(x) = \sum_{[\sw]=\sJ\cup \sK} \kq^{|\sK|} \prod_{j\in \sJ, k\in \sK}\sS(\varrho_j-\varrho_k) \prod_{k\in \sK} \frac{P(x+\varrho_k)}{\EY(x+\varrho_k)} \prod_{j\in \sJ}\EY(x+\ve_++\varrho_j)
\end{align}
where $[\sw]=\{1,\dots,\sw\}$ and
\begin{align}
    \sS= \frac{(x+\ve_1)(x+\ve_2)}{x(x+\ve_1+\ve_2)}.
\end{align}
The $\sS$-factor is not present in the fundamental $qq$-character. It can be viewed as a contribution of the $D3$-$D3$ open strings ending on the $\BC^2_{34}$ (there are ${\sw}^2$ of those).

As we integrate out the degrees of freedom in the $\BC^2_{34}$ space orthogonal to the physical $\BC^2_{34}$, we obtain a local observable which is called the higher rank $qq$-character 
$\EX_{\sw,{\boldsymbol\varrho}}(x)$ in \eqref{eq:higherqq}. The gauge origami partition function is identified as expectation value of $qq$-character:
\begin{align}
    \CalZ_S = \Biggl \langle \EX_{\sw,{\boldsymbol\varrho}}(x) \Biggr \rangle \CalZ_{\BC^2_{12}}.
\end{align}
The expectation value $\langle \EX_{\sw,{\boldsymbol\varrho}}(x)\rangle$ is a degree $N^2$ polynomial in $x$.

Eq.~\eqref{eq:higherqq} can be interpreted as a lattice system of length $N$. On each lattice site $j$ there are exactly two states: empty $j\in\sJ$ or occupied $j\in\sK$. When a site $j$ is empty, it contributes
$$
    \EY(x+\varrho_j+\ve_+)
$$
to the system. While occupied, it gives:
$$
    {\qe} \, \frac{{P}(x+\varrho_j)}{\EY(x+\varrho_j)}.
$$
The described system has a long range interaction between the sites with different occupation status 
$$
    \prod_{j\in \sJ, k\in\sK}\sS(\varrho_j-\varrho_k).
$$

Let us again introduce a $\BZ_n$ orbifold $\hat{\BC}_1 \times \left( \hat\BC_2/\BZ_n \right)$ in the same way as we have done in section~\ref{sec:cons-defect}, with the same coloring function $c(\alpha)$ for moduli parameters $\{a_\alpha\}$ and $\sigma(f)^\pm$ for (anti-)fundamental matters \eqref{eq:std-color}. In the orthogonal direction $\BC^2_{34}$, we assign coloring function $\varsigma:[\sw]\to \BZ_n$ for the orthogonal moduli parameters $\{\hat\varrho_i\}_{i=1}^\sw$.

We are interested in the \textit{regular} surface defect by choosing $\BZ_n=\BZ_N$. Furthermore, we consider $\sw=N$ and choose the coloring function $\varsigma$ as simple one-to-one functions for $\boldsymbol\varrho$,
\begin{align}
    \varsigma(x+\hat\varrho_{j}) = j-1, \ j=1,\dots,N.
\end{align}
Namely, \eqref{def:GO-higherqq} becomes
\begin{subequations}
\begin{align}
   \hat{N}_{12} 
   & = \sum_{\o'' \in \mathbb{Z}_N} \left(  e^{\hat{a}_{\o''+1}} \mathcal{R}_0 \otimes {\mathfrak{R}}_{\o''} +  e^{\hat{m} _{\o''} ^- - \ve_4} \mathcal{R}_{1} \otimes {\mathfrak{R}}_{\o''} + e^{\hat{m} _{\o''+1} ^+ - \ve_3} \mathcal{R}_{2} \otimes {\mathfrak{R}}_{\o''}  \right) \\
   & = \sum_{\omega'' \in \BZ_N} 
   \left(  e^{{a}_{\o''+1}} \hat{q}_2^{\omega''} \mathcal{R}_0 \otimes {\mathfrak{R}}_{\o''} + e^{{m} _{\o''} ^- - \ve_4} \hat{q}_2^{\omega''} \mathcal{R}_{1} \otimes {\mathfrak{R}}_{\o''} + e^{{m} _{\o''+1} ^+ - \ve_3} \hat{q}_2^{\omega''} \mathcal{R}_{2} \otimes {\mathfrak{R}}_{\o''}   \right) \nonumber\\
   %%%%%%%%%%%%%%%%%%%%%%%%%%%%%%%%%%%%
%   \hat{N}_{23} 
%   & = e^{x'+\hat{\ve}_2 + \ve_3} \hat{q}_2^{\Omega} \ \mathcal{R}_1 \otimes \mathcal{R}_{\Omega+1}  \nonumber\\
   %%%%%%%%%%%%%%%%%%%%%%%%%%%%%%%%%%%%
   \hat{N}_{34} 
   & = \sum_{\omega=0}^{N-1} e^{x} e^{\hat\varrho_\omega} \ \mathcal{R}_0 \otimes {\mathfrak{R}}_\o = \sum_{\omega=0}^{N-1} e^{x} e^{\varrho_\omega} \hat{q}_2^{\omega} \ \mathcal{R}_0 \otimes {\mathfrak{R}}_\o.
\end{align}
\end{subequations}
The shifted moduli for $\boldsymbol\varrho$ are defined by
$$
    \hat\varrho_{\omega+1} = \varrho_{\omega+1} - \omega\hat\ve_2.
$$
Gauge origami partition is
\begin{align}
    \hat\CalZ_{X;c,\sigma^\pm,\varsigma} 
    & = \sum_{\hat{\boldsymbol\lambda}_{12},\hat{\boldsymbol\lambda}_{34}} \prod_{\omega\in\BZ_n} \hat{\kq}_{\omega}^{|\hat{\boldsymbol\lambda}_{12,\omega}|+|\hat{\boldsymbol\lambda}_{34,\omega}|} \BE \left[ - \frac{\hat{P}_3\hat{S}_{12}\hat{S}_{12}}{\hat{P}_{12}^*} - \frac{\hat{P}_1 \hat{S}_{34}\hat{S}_{34}^*}{\hat{P}_{34}^*}  - \hat{q}_{12}^*\hat{S}_{12}\hat{S}_{34}\right]^{\BZ_3 \times \BZ_n} \nonumber\\
    & =  \sum_{\hat{\boldsymbol\lambda}_{12}} \prod_{\omega\in\BZ_n} \hat{\kq}_{\omega}^{|\hat{\boldsymbol\lambda}_{12,\omega}|} \EZ_{\rm bulk}[\boldsymbol\lambda_{12}]\EZ_{\rm defect}[\hat{\boldsymbol\lambda}_{12}] \EX_{\rm defect}(x) \nonumber\\
    & = \left \langle \EX_{\rm defect}(x) \right \rangle_{\BZ_N} \hat{\CalZ}_{\hat\BC^2_{12}}.
\end{align}
% in the same way we have done in section \ref{sec:TQDS}. The parameter $x$ is chosen to be in the $\CalR_0$ representation of the $\BZ_N$ orbifolding. The parameters $\varrho_{\omega}$ are assigned to the $\CalR_{\omega-1}$ representation of the $\BZ_N$ orbifold.
The defect rank $N$ $qq$-character is of the form 
\begin{align}\label{def:qq-high-defect}
    \EX_{\rm defect}(x) = \sum_{[0,\dots,N-1]=\sJ\cup\sK} & \prod_{\omega\in \sJ, \omega'\in\sK} \sS_{-}(\hat\varrho_{\omega}-\hat\varrho_{\omega'})^{\delta_N(\omega-\omega'+1)} \nonumber\\
    & \times \prod_{\omega\in\sJ} {\EY}_{\omega+1}(x + \ve_1 + \varrho_{\omega}) \prod_{\omega'\in\sK} \hat\kq_{\omega'} \frac{{P}_{\omega'}(x+\varrho_{\omega'})}{{\EY}_{\omega'}(x+\varrho_{\omega'})}
\end{align}
with  
$$
    \sS_{-}(x) = 1 - \frac{\ve_1}{x+\ve_1+\hat\ve_2} = \frac{x+\hat\ve_2}{x+\ve_1+\hat\ve_2}.
$$
Defect $\EX_{\rm defect}$ satisfies the non-perturbative Dyson-Schwinger equation. The expectation value of defect $qq$-character
\begin{align}
    \left \langle \EX_{\rm defect}(x) \right \rangle_{\BZ_N}
\end{align}
is a degree $N$-polynomial in $x$.

We mentioned that the higher rank $qq$-character can be understand as a lattice of $N$ two-states system. In the absence of orbifold, the described lattice system has long range interaction between any two lattice with opposite occupation status. In the orbifolded version of the story \eqref{def:qq-high-defect}, such interaction becomes local. More precisely speaking, the system consist only a nearest neighbor interaction.

In particular, if we choose 
$$
    \hat\varrho_\omega = (\omega-1)\hat\ve_2 \implies {\varrho}_{\omega} = 0, \ \forall \omega.
$$
Such choice puts a strong restriction on the set of $\sJ$ and $\sK$ that give non-vanishing contribution. More precisely speaking, if $\omega+1$ is not in the same set $\sJ$ or $\sK$ as $\omega$, it obtains a factor of
\begin{align}
    \sS_{-}(-\hat\ve_2) = 0 \nonumber
\end{align}
The defect $qq$-character is greatly simplified to 
\begin{align}
    \EX_{\rm defect}(x) =
    & \prod_{\omega=0}^{N-1} {\EY}_{\omega+1}(x+\ve_1) + \prod_{\omega=0}^{N-1} \hat\kq_{\omega} \frac{{P}_{\omega}(x)}{{\EY}_{\omega}(x)} \nonumber\\
    & + \frac{\ve_2}{\ve_1+\ve_2} \EY_N(x+\ve_1) \frac{\hat\kq_0P_0(x)}{\EY_0(x)} \sum_{\omega=1}^{N-2} \prod_{k=1}^{\omega} \frac{\hat\kq_k P_k(x)}{\EY_{k}(x)} \prod_{j=\omega+1}^{N-2} \EY_{j+1}(x+\ve_1). 
\end{align}
with $\EY_N(x)=\EY_0(x+\ve_2)$. 
The first term represents a full empty state, while the second term corresponds to a full occupied configuration. The remaining $N-2$ terms correspond to having the first $\omega$ site occupied, and the rest empty. 

Defect $qq$-character becomes the bulk fundamental $qq$-character in $\ve_2\to0$ limit:
\begin{align}
    \EX_{\rm defect}(x) 
    & = \prod_{\omega=0}^{N-1} {\EY}_{\omega+1}(x+\ve_1) + \prod_{\omega=0}^{N-1} \hat\kq_{\omega} \frac{{P}_{\omega}(x)}{{\EY}_{\omega}(x)} 
    = \EY(x+\ve_1) + \kq \frac{P(x)}{\EY(x)}
\end{align}

% One interesting observation is that in the NS-limit $\ve_2\to0$, the defect rank $N$ $qq$-character becomes the bulk fundamental $qq$-character
% \begin{align}
%     \lim_{\ve_2\to0}\CalX_{\rm defect}(x) = \prod_{\omega} Y_{\omega}(x+\ve_+) + \prod_{\omega} \kq_{\omega}\frac{P_{\omega}(x)}{Y_{\omega}(x)} = \EY(x+\ve_+) + \kq \frac{P(x)}{\EY(x)} = \EX_{\rm fund}(x).
% \end{align}

\subsubsection{With fractional $Q$-observables}

The building blocks $\hat{T}_{N,j}(x)$ showing in transfer matrix \eqref{eq:trmono} is a differential operator acting on
\begin{align}
    \hat{T}_{N,\omega}(x) \left \langle {Q}_{\Omega}(x) \right \rangle_{\BZ_N} \Psi
    & = \left[ (1+\hat\kq_{\omega}) x + \rho_{\omega} \right] \langle {Q}_{\Omega}(x) \rangle_{\BZ_N} \Psi \\
    & = \left \langle \left[ {\EY}_{\omega+1}(x) + \frac{\hat\kq_{\omega}P_{\omega}(x)}{{\EY}_{\omega}(x)} \right] {Q}_\Omega(x) \right \rangle_{\BZ_N} \Psi \nonumber
\end{align}

That is to say, the proper way to think about the transfer matrix is its action on 
\begin{align}
    \Tr_{V_{\rm aux}} {\bf T}_{\rm SC}(x)  \left \langle {Q}_{\Omega}(x) \right \rangle_{\BZ_N} \Psi
\end{align}
with any $\Omega=0,\dots,N-1$.

Let us consider the higher rank analogue of \eqref{eq:qyeq} by taking the following gauge origami setup similar to the rank one $qq$-character case \eqref{eq:gosetup}. We introduce one additional D-brane on $\BC^{2}_{23}$ in \eqref{def:GO-higherqq}:
\begin{subequations}
\begin{align}
    \hat{N}_{12} & = \sum_{\alpha=1}^N e^{a_{\alpha}} \cdot \CalR_{0} + \sum_{\alpha=1}^N e^{m_{\alpha}^--\ve_4} \cdot \CalR_{1} + \sum_{\alpha=1}^N e^{m_{\alpha}^+-\ve_3} \cdot \CalR_{2}; \\
    \hat{N}_{23} & = e^{x'+\ve_2+\ve_3} \cdot \CalR_{1}; \\
    \hat{N}_{34} & = \sum_{j=1}^N e^{x+\varrho_{j-1}} \cdot \CalR_{0}.
\end{align}
\end{subequations}
We again take the decoupling limit $\kq_{1} = 0 = \kq_{2}$ and $\kq \equiv \kq_{0}$. For later convenience, we slightly modify our notation on the $N$-tuples $\boldsymbol\varrho = (\varrho_0,\dots,\varrho_{N-1})$.
% This setup can be achieved by $\BZ_{3}$ orbifolding on the gauge origami on $\BC^2_{34}$.
The gauge origami instanton partition function is 
\begin{align}
    \CalZ_{S} = \sum_{\boldsymbol{\lambda}^{(12)}}\sum_{\boldsymbol\lambda^{(34)}} \kq^{|\boldsymbol\lambda^{(12)}|+|\boldsymbol\lambda^{(34)}|} \BE & \left[ -\frac{\hat{P}_3^*\hat{S}_{12}\hat{S}_{12}^*}{\hat{P}_{12}^*} - \frac{\hat{P}_{1}^*\hat{S}_{23}\hat{S}_{23}^*}{\hat{P}_{23}} - \frac{\hat{P}_{1}\hat{S}_{34}\hat{S}_{34}^*}{\hat{P}_{34}} \right. \nonumber\\ 
    & \left. -\hat{q}_{12}\hat{S}_{12}^*\hat{S}_{34} + \hat{q}_{1} \hat{P}_4 \hat{S}_{23}\frac{\hat{S}_{12}^*}{\hat{P}_{2}^*} - (\hat{q}_{23}+\hat{q}_{12}) N_{34} N_{23}^* +\hat{P}_{1}\hat{P}_4 N_{23}K_{34}^* \right]_{0}^{\BZ_3}
\end{align}
We modified the interaction between $\BC_{23}^2$ and $\BC_{34}^2$ using the same argument as in  arriving at \eqref{eq:gopf}. The gauge origami instanton partition function can be written as the following form:
\begin{align}
    \CalZ_{S} = \sum_{\boldsymbol{\lambda}^{(12)}} \BE[\CalT_{12}]  \EX_N(x){Q}(x') = \left\langle \EX_N(x){Q}(x') \right\rangle \CalZ_{\BC^2_{12}} 
\end{align}
where the rank $N$ $qq$-character $\EX_N$ is of the form:
\begin{align}
    \EX_N(x){Q}(x') = \sum_{[0,\dots,N-1] = \sJ\cup\sK} & \prod_{j\in \sJ,k\in\sK} \sS(\varrho_j-\varrho_k)  \times \prod_{j\in\sJ}(x-x'+\varrho_j) \EY(x+\ve_++\varrho_j) \nonumber\\
    &\times \prod_{k\in\sK} \kq (x+\varrho_k-x'+\ve_{1}) \frac{P(x+\varrho_k)}{\EY(x+\varrho_k)} {Q}(x')
\end{align}
such that the expectation value 
$$
    \left\langle \EX_N(x) {Q}(x')\right\rangle
$$
is a degree $N^2+N$ polynomial in $x$.

We now introduce regular surface defect in the form of $\BZ_N$-orbifold in the $\BC_{24}^2$ direction in the same way we have done in section~\ref{sec:orb}. The coloring functions $c$ and $\sigma^\pm$ for moduli parameters and (anti-)fundamental matters are the same as \eqref{eq:std-color}. The coloring function for $\boldsymbol\varrho$ are chosen as simple one-to-one function
\begin{align}
    \varsigma(\hat\varrho_\omega) = \omega, \ \omega=0,\dots,N-1.
\end{align}
The representations for the orbifolding action are assigned as
\begin{subequations}
\begin{align}
   \hat{N}_{12} 
   & = \sum_{\o'' \in \mathbb{Z}_N} \left(  e^{\hat{a}_{\o''+1}} \mathcal{R}_0 \otimes {\mathfrak{R}}_{\o''} +  e^{\hat{m} _{\o''} ^- - \ve_4} \mathcal{R}_{1} \otimes {\mathfrak{R}}_{\o''} + e^{\hat{m} _{\o''+1} ^+ - \ve_3} \mathcal{R}_{2} \otimes {\mathfrak{R}}_{\o''}  \right) \nonumber\\
   & = \sum_{\omega'' \in \BZ_N} 
   \left(  e^{{a}_{\o''+1}} \hat{q}_2^{\omega''} \mathcal{R}_0 \otimes {\mathfrak{R}}_{\o''} + e^{{m} _{\o''} ^- - \ve_4} \hat{q}_2^{\omega''} \mathcal{R}_{1} \otimes {\mathfrak{R}}_{\o''} + e^{{m} _{\o''+1} ^+ - \ve_3} \hat{q}_2^{\omega''} \mathcal{R}_{2} \otimes {\mathfrak{R}}_{\o''}   \right), \nonumber\\
   %%%%%%%%%%%%%%%%%%%%%%%%%%%%%%%%%%%%
   \hat{N}_{23} 
   & = e^{x'+\hat{\ve}_2 + \ve_3} \hat{q}_2^{\Omega} \ \mathcal{R}_1 \otimes {\mathfrak{R}}_{\Omega+1},  \nonumber\\
   %%%%%%%%%%%%%%%%%%%%%%%%%%%%%%%%%%%%
   \hat{N}_{34} 
   & = \sum_{\omega=0}^{N-1} e^{x} e^{\hat\varrho_\omega} \ \mathcal{R}_0 \otimes {\mathfrak{R}}_\o = \sum_{\omega=0}^{N-1} e^{x} e^{\varrho_\omega} \hat{q}_2^{\omega} \ \mathcal{R}_0 \otimes {\mathfrak{R}}_\o.
\end{align}
\end{subequations}
With some decent calculation, we find the gauge origami partition function can be organized as the following form:
\begin{align}
    \hat\CalZ_{X,\Omega} 
    & = \sum_{\hat{\boldsymbol\lambda}} \prod_{\omega\in\BZ_N} \hat{\kq}_{\omega}^{k_{\omega}}\EZ_{\rm bulk}[\boldsymbol\lambda] \EZ_{\rm defect}[\hat{\boldsymbol\lambda}] \EX_{N,{\rm defect}}(x) Q_{\Omega}(x') \nonumber\\
    & = \left \langle \EX_{N,{\rm defect}}(x) Q_\Omega(x') \right \rangle_{\BZ_N} \hat{\CalZ}_{X;c,\sigma^\pm}.
\end{align}
The defect $qq$-character reads
\begin{align}
    \EX_{N,{\rm defect}}(x){Q}_{\Omega}(x') = \sum_{[0,\dots,N-1] = \sJ \cup \sK} 
    & \prod_{\omega\in\sJ,\omega'\in\sK} \sS_{-}(\hat\varrho_{\omega}-\hat\varrho_{\omega'})^{\delta_N(\omega-\omega'+1)} \\
    & \times \prod_{\omega\in \sJ} (x-x'+\hat\varrho_\omega - \Omega\hat\ve_2)^{\delta_{N}(\omega-\Omega)} {\EY}_{\omega+1}(x+\ve_1+\varrho_{\omega}) \nonumber\\
    &\times \prod_{\omega'\in\sK} \hat\kq_{\omega'} (x+\hat\varrho_{\omega'}-x' - \Omega\hat\ve_2+\ve_{1})^{\delta_{N}(\omega'-\Omega)} \frac{{P}_{\omega'}(x+\varrho_{\omega'})}{{\EY}_{\omega'}(x+\varrho_{\omega'})} \times {Q}_{\Omega}(x') \nonumber
\end{align}
with $\varrho_{\omega} = \hat\varrho_{\omega} - \omega\hat\ve_2$.
We denote the set $[\hat{\Omega}] = [0,\dots,N-1]\backslash \Omega$ by specifying whether site $\Omega$ is empty or occupied: 
\begin{align}
    & \EX_{N,{\rm defect}}(x){Q}_{\Omega}(x')  \\
    = & \sum_{[\hat{\Omega}] = \sJ \cup \sK} \prod_{\omega\in\sJ,\omega'\in\sK} \sS_{-}(\hat\varrho_{\omega}-\hat\varrho_{\omega'})^{\delta_N(\omega-\omega'+1)} \prod_{\omega'\in K} \sS_{-}(\hat\varrho_{\Omega}-\hat\varrho_{\omega'})^{\delta_N(\Omega-\omega'+1)} \nonumber\\
    & \times \prod_{\omega\in \sJ} {\EY}_{\omega+1}(x+\ve_1+\varrho_{\omega}) \prod_{\omega'\in\sK} \hat\kq_{\omega'}  \frac{{P}_{\omega'}(x+\varrho_{\omega'})}{{\EY}_{\omega'}(x+\varrho_{\omega'})} \times (x-x'+\varrho_{\Omega}) {\EY}_{\Omega+1}(x+\varrho_\Omega+\ve_1) {Q}_{\Omega}(x') \nonumber\\
    %%%%%%%%%%%%%%%%%%%%%%%%%%%%%%%%%%%%%%%%%%%%
    + 
    & \prod_{\omega\in\sJ,\omega'\in\sK} \sS_{-}(\hat\varrho_{\omega}-\hat\varrho_{\omega'})^{\delta_N(\omega-\omega'+1)} \prod_{\omega\in \sJ}\sS_{-}(\hat\varrho_\omega-\hat\varrho_\Omega)^{\delta_{N}(\omega-\Omega+1)} \nonumber\\
    & \times \prod_{\omega\in \sJ} {\EY}_{\omega+1}(x+\ve_1+\varrho_\omega) \prod_{\omega'\in\sK} \hat\kq_{\omega'}  \frac{{P}_{\omega'} (x+\hat\varrho_{\omega'})}{{\EY}_{\omega'}(x+\varrho_{\omega'})}  \times \hat\kq_{\Omega} (x-x' + \varrho_\Omega + \ve_{1}) \frac{{P}_{\Omega}(x+\varrho_\Omega)}{{\EY}_{\Omega}(x+\varrho_\Omega)} {Q}_{\Omega}(x') \nonumber
\end{align}
such that the expectation value 
\begin{align}
    \left \langle \EX_{N,{\rm defect}}(x){Q}_{\Omega}(x') \right \rangle _{\BZ_N  }
\end{align} 
is a degree $N+1$ polynomial in $x$. 
% We consider the following arrangement for the $\nu_{\omega}$ by
% $$
%     \nu_{\omega} = \omega\tilde{\ve}_2.
% $$
% The defect $qq$-character becomes
% \begin{align}
%     \CalX_{N,{\rm defect}}^{(2)}(x) {Q}_{\Omega}(y)
%     = & (x-y+\Omega\tilde{\ve}_2) \prod_{\omega=0}^{N-1} Y_{\omega+1}(x+\omega\tilde{\ve}_2 + \tilde{\ve}) {Q}_{\Omega}(y) \nonumber\\
%     & + (x-y+\Omega\tilde{\ve}_2+\ve_1) \frac{\kq P(x)}{\prod_{\omega=0}^{N-1}Y(x+\omega\tilde{\ve})} {Q}_{\Omega}(y)
% \end{align}

% For our interests, we consider the case
% $$
%     \varrho_{\omega} = \omega\hat\ve_2, \implies \hat\varrho_{\omega} = 0, \ \omega=0,\dots,N-1
% $$
Let us take $x'=x+\varrho_\Omega$ and $x'=x+\varrho_\Omega+\ve_1$ respectively:
\begin{subequations}
\begin{align}
    \EX_{N,{\rm defect}}(x){Q}_{\Omega}(x+\varrho_{\Omega}) 
    = \sum_{[\hat{\Omega}] = \sJ \cup \sK} & \prod_{\omega\in\sJ,\omega'\in\sK} \sS_{-}(\hat\varrho_{\omega}-\hat\varrho_{\omega'})^{\delta_N(\omega-\omega'+1)} \prod_{\omega\in \sJ}\sS_{-}(\hat\varrho_\omega-\hat\varrho_\Omega)^{\delta_{N}(\omega-\Omega+1)} \nonumber\\
    & \times \prod_{\omega\in \sJ} {\EY}_{\omega+1}(x+\ve_1+\varrho_{\omega}) \prod_{\omega'\in\sK} \hat\kq_{\omega'}  \frac{{P}_{\omega}(x+\varrho_{\omega})}{{\EY}_{\omega}(x+\varrho_\omega)} \nonumber\\
    & \times \hat\kq_{\Omega} \ve_{1} {{P}_{\Omega}(x+\varrho_\Omega)} {Q}_{\Omega-1}(x+\varrho_\Omega), \\
    %%%%%%%%%%%%%%%%%%%%%%%%%%%%%%%%%%%%%%%%%
    \EX_{N,{\rm defect}}(x-\ve_1){Q}_{\Omega}(x+ \varrho_{\Omega}) 
    = \sum_{[\hat{\Omega}] = \sJ \cup \sK} & \prod_{\omega\in\sJ,\omega'\in\sK} \sS_{-}(\hat\varrho_{\omega}-\hat\varrho_{\omega'})^{\delta_N(\omega-\omega'+1)} \prod_{\omega'\in \sK}\sS_{-}(\hat\varrho_\Omega-\hat\varrho_{\omega'})^{\delta_{N}(\Omega-\omega'+1)} \nonumber\\
    & \times \prod_{\omega\in \sJ} {\EY}_{\omega+1}(x+\varrho_{\omega}) \prod_{\omega'\in\sK} \hat\kq_{\omega'}  \frac{{P}_{\omega}(x + \varrho_{\omega} - \ve_1)}{{\EY}_{\omega}(x+\varrho_{\omega}-\ve_1)}  \nonumber\\
    & \times (-\ve_{1})  {Q}_{\Omega+1}(x+\varrho_\Omega).
\end{align}
\end{subequations}
The difference between their expectation value 
\begin{align}
    & \frac{1}{\ve_1} \Biggl\langle \EX_{N,{\rm defect}}(x)  {Q}_{\Omega}(x+\varrho_\Omega) \Biggr\rangle_{\BZ_N} 
    - \frac{1}{\ve_1} \Biggl\langle  \EX_{N,{\rm defect}}(x-\ve_1)  {Q}_{\Omega}(x+\varrho_\Omega) \Biggr\rangle_{\BZ_N} \\
    %%%%%%%%%%%%%%%%%%%%%%%%%%%%%%%%%%%%
    & = \sum_{[\hat{\Omega}] = \sJ \cup \sK}  \Biggl\langle \prod_{\omega\in\sJ,\omega'\in\sK} \sS_{-}(\hat\varrho_{\omega}-\hat\varrho_{\omega'})^{\delta_N(\omega-\omega'+1)} \prod_{\omega'\in \sK}\sS_{-}(\hat\varrho_\Omega-\hat\varrho_{\omega'})^{\delta_{N}(\Omega-\omega'+1)} \nonumber\\
    & \qquad \qquad \times \prod_{\omega\in \sJ} {\EY}_{\omega+1}(x+\varrho_\omega) \prod_{\omega'\in\sK} \hat\kq_{\omega'}  \frac{{P}_{\omega}(x +\varrho_\omega - \ve_1)}{{\EY}_{\omega}(x+\varrho_\omega-\ve_1)}  \times   {Q}_{\Omega+1}(x+\varrho_{\Omega}) \Biggr\rangle_{\BZ_N} \nonumber\\
    %%%%%%%%%%%%%%%%%%%%%%%%%%%%%%%%%%%%%%%%%
    & \qquad\quad + \Biggl\langle \prod_{\omega\in\sJ,\omega'\in\sK} \sS_{-}(\hat\varrho_{\omega}-\hat\varrho_{\omega'})^{\delta_N(\omega-\omega'+1)} \prod_{\omega\in \sJ}\sS_{-}(\hat\varrho_\omega-\hat\varrho_\Omega)^{\delta_{N}(\omega-\Omega+1)} \nonumber\\
    & \qquad\qquad\times \prod_{\omega\in \sJ} {\EY}_{\omega+1}(x+\ve_1+\varrho_\omega) \prod_{\omega'\in\sK} \hat\kq_{\omega'}  \frac{{P}_{\omega}(x+\varrho_\omega)}{{\EY}_{\omega}(x+\varrho_\omega)}  \times \hat\kq_{\Omega} {P_{\Omega}(x+\varrho_\Omega)} {Q}_{\Omega-1}(x+\varrho_{\Omega}) \Biggr\rangle_{\BZ_N} 
    \nonumber
    %%%%%%%%%%%%%%%%%%%%%%%%%%%%%%%%%%%%%%%%%
    % & := \langle \CalT_N(x) {Q}_{\Omega}(x+\nu_{\Omega}) \rangle \nonumber
\end{align}
is a degree $N$ polynomial in $x$. 

Of our interest, we consider
\begin{align}
    \hat\varrho_{\omega} = \omega\hat\ve_2, \implies \varrho_\omega=0, \quad \omega=0,\dots,N-1.
\end{align}
As we have seen previously, such choice of $\varrho_{\omega}$ restricts allowed sets of $\sJ$ and $\sK$, which leads to simplification of the $qq$-character:
\begin{align}
    & \frac{1}{\ve_1} \Biggl\langle \EX_{N,{\rm defect}}(x)  {Q}_{\Omega}(x) \Biggr\rangle_{\BZ_N} 
    - \frac{1}{\ve_1} \Biggl\langle  \EX_{N,{\rm defect}}(x-\ve_1)  {Q}_{\Omega}(x) \Biggr\rangle_{\BZ_N} \\
    & = \Biggl \langle \left[ \prod_{\omega=0}^{N-1} \EY_{\omega+1}(x) + \frac{\kq P(x)}{\prod_{\omega=0}^{N-1}\EY_{\omega}(x)} + \frac{\ve_2}{\ve_1+\ve_2} \sum_{\omega'=0}^{N-2} \frac{Q_N(x)Q_{-1}(x)}{Q_{\omega'}(x)Q_{\omega'+1}(x)} \prod_{k=0}^{\omega'} \hat\kq_{k} P_k(x) \right] Q_\Omega(x) \Biggr \rangle_{\BZ_N} \nonumber
\end{align}

In the $\ve_2 \to 0$ limit, it becomes:
\begin{align}
    & \frac{1}{\ve_1} \Biggl\langle \EX_{N,{\rm defect}}(x)  {Q}_{\Omega}(x) \Biggr\rangle 
    - \frac{1}{\ve_1} \Biggl\langle  \EX_{N,{\rm defect}}(x-\ve_1)  {Q}_{\Omega}(x) \Biggr\rangle_{\BZ_N}  \\
    % & = \Biggl \langle \prod_{\omega\in[\hat{\Omega}]} {\EY}_{\omega+1}(x) {Q}_{\Omega+1}(x) \Biggr \rangle
    % + \hat\kq_{\Omega}{P}_\Omega(x)\Biggl \langle \prod_{\omega\in[\hat{\Omega}]}\hat\kq_{\omega}\frac{{P}_{\omega}(x)}{{\EY}_\omega(x)} {Q}_{\Omega-1}(x) \Biggr\rangle \nonumber\\
    % & = \Biggl \langle \left[ {\EY}(x) + \kq\frac{{P}(x)}{{\EY}(x)} \right] {Q}_{\Omega}(x) \Biggr \rangle_{\BZ_N} \nonumber \\
    & = \Biggl \langle \left[ {\EY}(x+\ve_2) + \kq\frac{{P}(x)}{{\EY}(x)} \right] {Q}_{\Omega}(x) \Biggr \rangle_{\BZ_N} \nonumber
\end{align}
In particular, in the case of $\Omega=N-1$, $Q_{N-1}(x)=Q(x)$ is identified as bulk $Q$-observable.
\begin{align}
    & \frac{1}{\ve_1} \Biggl\langle \EX_{N,{\rm defect}}(x)  {Q}_{N-1}(x) \Biggr\rangle _{\BZ_N}
    - \frac{1}{\ve_1} \Biggl\langle  \EX_{N,{\rm defect}}(x-\ve_1)  {Q}_{N-1}(x) \Biggr\rangle_{\BZ_N} \nonumber\\
    & = \left \langle Q(x+\ve_2) \right \rangle + \kq P(x) \left \langle Q(x-\ve_2) \right \rangle_{\BZ_N} \nonumber\\
    & = \left \langle T_N(x) Q(x) \right \rangle_{\BZ_N}
\end{align}
where $T_N(x)$ is a degree $N$ polynomial whose coefficients depend on the bulk instanton configuration \eqref{eq:qTQ}.

\subsubsection{Similarity with the spin chain transfer matrix}

The trace of the spin chain monodromy matrix in \eqref{def:monodromy} can be seen as an effective quiver system which bears some resemblance to the one for the construction of the higher rank $qq$-character. 
The higher rank $qq$-character considers a $A_N$ quiver in the auxiliary space $\BC^2_{34}$ in the context of gauge origami. Each gauge node has exactly two instanton configurations: having no instanton (empty) or exactly one instanton (occupied). Adding one instanton to the gauge node $\omega$ (from empty to occupied state) changes the site's contribution 
$$
{\EY}_{\omega+1}(x) \implies \frac{\hat\kq_{\omega}{P}_{\omega}(x)}{{\EY}_{\omega}(x)}
$$
in the $qq$-character along with interaction contribution (the $\sS$ factor). In the orbifolded situation, the $\sS$ bi-fundamental factor is localized to nearest neighbor as a direct consequence of the $\delta_{N}(\omega-\omega'+1)$ power as seen in previous section.

On the level of the actual spin sites, trace of the monodromy matrix is
\begin{align}
    \Tr_{V_{\rm aux}}  {\bf T}_{\rm SC}(x) 
    & = \sum_{[N] = \sJ\cup\sK\cup\tilde\sK} \prod_{j\in \sJ} \hat{T}_{N,j}(x) \prod_{k\in \sK} (-\kq_kP^-_k(x)P_k^+(x+\delta_{k,0}\ve_2)) 
    % & = \sum+{[N] = \sJ\cup\tilde\sJ \cup \sK\cup\tilde\sK} \prod_{j\in\sJ} \tilde{Y}_{j+1}(x+(j+1)\tilde\ve_2) \prod_{}
\end{align}
where $[N] = \{ 0,1,\dots,N-1\}$. The set $\tilde\sK$ is defined by 
$$
     \tilde\sK = \{k-1\ ({\rm mod} \ N) \ | \ k \in \sK\}.
$$
$\hat{T}_{N,j}(x)$ defined in \eqref{eq:qTQvev} is a differential operation:
\begin{align}
    \hat{T}_{N,\omega}(x) \left \langle {Q}_{\Omega}(x) \right \rangle_{\BZ_N} \Psi
    % & = \frac{1}{\ve_1} \left \langle \hat{T}_{N+1,\omega}^{(2)}(x) {Q}_\Omega(x)  \right \rangle -  \nonumber\\ 
    % & = \left \langle \left[ x - a_{\omega+1} + \ve_1\nu_{\omega} + \kq_{\omega}( x - {m}_{\omega} + {a}_{\omega} - \ve_1\nu_{\omega-1} ) \right] \tilde{Q}_\Omega(x+\Omega\tilde\ve_2) \right \rangle \Psi \nonumber\\
    & = \left \langle \left[ {\EY}_{\omega+1}(x) + \frac{\hat\kq_{\omega}P_{\omega}(x)}{{\EY}_{\omega}(x)} \right] {Q}_\Omega(x) \right \rangle_{\BZ_N} \Psi
\end{align}

At first glance, it seems that there exists no interaction factor in the transfer matrix as a lack of $\sS$ factors. However, when the occupation status of lattice site $\omega$ is flipped from empty to occupied. It is required for one of its nearest neighbor to flip along with it. Hence there is a nearest neighboring interaction in the structure of transfer matrix.
The combined contribution of empty and occupied states at site $\omega$ and $\omega-1$ gives
\begin{align} \label{eq:TT-qP}
    & \left\langle \left[ \left( {\EY}_{\omega+1}(x) + \frac{\hat\kq_{\omega}{P}_{\omega}(x)}{{\EY}_{\omega}(x)} \right) \left( {\EY}_{\omega}(x) + \frac{\hat\kq_{\omega-1}{P}_{\omega-1}(x)}{{\EY}_{\omega-1}(x)} \right) - \hat\kq_{\omega}{P}_{\omega}(x) \right] {Q}_\Omega(x) \right \rangle_{\BZ_N} \Psi \nonumber\\
    & = \left \langle \left[ {\EY}_{\omega+1}(x) {\EY}_{\omega}(x) + \hat\kq_{\omega-1}P_{\omega-1}(x) \frac{{\EY}_{\omega+1}(x)}{{\EY}_{\omega-1}(x)} + \frac{\hat\kq_{\omega}P_{\omega}(x)}{{\EY}_{\omega}(x)}\frac{\hat\kq_{\omega-1}P_{\omega-1}(x)}{{\EY}_{\omega-1}(x)} \right] {Q}_\Omega(x) \right\rangle_{\BZ_N} \Psi 
\end{align}
which can be rewritten in the form similar to the higher rank $qq$ character:
\begin{align}
    & \Biggl\langle \left[ {\EY}_{\omega+1}(x) {\EY}_{\omega}(x) + \frac{\hat\kq_{\omega}{P}_{\omega}(x)}{{\EY}_{\omega}(x)}\frac{\hat\kq_{\omega-1}{P}_{\omega-1}(x)}{{\EY}_{\omega-1}(x)} + \sS_{-}((\omega-1)\hat{\ve}_2 - \omega\hat{\ve}_2)^{\delta_N(\omega-1-\omega+1)} \hat\kq_{\omega}{P}_{\omega}(x) \right. \nonumber\\
    % &  \quad  \\
    & \quad \left. + \sS_{-}(\omega\hat{\ve}_2 - (\omega-1)\hat{\ve}_2)^{\delta_N(\omega-(\omega-1)+1)} \hat{\EY}_{\omega+1}(x+(\omega+1)\hat{\ve}_2) \frac{\hat\kq_{\omega-1}{P}_{\omega-1}(x)}{{\EY}_{\omega-1}(x)} \right] {Q}_\Omega(x)  \Biggr \rangle_{\BZ_N} \Psi \nonumber
\end{align}
The other $\sS$-factor associated to $\omega$ can be obtained by considering the combined contribution of $\omega$ and $\omega+1$.

The action of the transfer matrix on $\langle {Q}_{\Omega}(x) \rangle$ now can be written as a modified form of the higher rank $qq$-character with a special set of complex numbers $\hat\varrho_{\omega}=\omega\hat{\ve}_2$ in the $\ve_2\to0$ limit:
\begin{align}
    & \Tr_{V_{\rm aux}}{\bf T}_{\rm SC}(x) \left \langle {Q}_{\Omega}(x) \right \rangle_{\BZ_N} \Psi \nonumber\\
    & = \Biggl\langle \left[ \sum_{[N]=\sJ\cup\tilde\sJ\cup\sK\cup\tilde{\sK}}  \prod_{j\in\sJ} \hat{T}_{N,j}(x) \hat{T}_{N,j-1}(x) \prod_{k\in\sK} (-\hat\kq_k P_k(x)) \right] {Q}_{\Omega}(x) \Biggr\rangle_{\BZ_N} \Psi \nonumber\\
    & = \Biggl\langle \left[ \sum_{[N] = \sJ\cup \sK} \prod_{j\in\sJ,k\in\sK}\sS_{-}(\varrho_j - \varrho_k)^{\delta_N(j-k+1)} \prod_{j\in\sJ}\EY_{j+1}(x) \prod_{k\in\sK} \frac{\hat\kq_k{P}_k(x)}{\EY_{k}(x)} \right] {Q}_{\Omega}(x) \Biggr\rangle_{\BZ_N} \Psi \nonumber\\
    & = \frac{1}{\ve_1} \Biggl\langle \EX^{(2)}_{N,{\rm defect}}(x){Q}_{\Omega}(x) \Biggr\rangle_{\BZ_N} \Psi
    - \frac{1}{\ve_1} \Biggl\langle  \EX^{(2)}_{N,{\rm defect}}(x-\ve_1)  {Q}_{\Omega}(x) \Biggr\rangle_{\BZ_N} \Psi
\end{align}
The left hand side is a degree $N$ polynomial
\begin{align}
    \Tr_{V_{\rm aux}}{\bf T}_{\rm SC}(x) = (1+\kq) x^{N} + \hat{\rm h}_1(\gamma_\omega,\partial_{\gamma_{\omega}}) x^{N-1} + \hat{\rm h}_2(\gamma_\omega,\partial_{\gamma_{\omega}}) x^{N-2} + \cdots + \hat{\rm h}_{N} (\gamma_\omega,\partial_{\gamma_{\omega}})
\end{align}
whose coefficients are functions of the $\mathfrak{sl}_2$ module coordinates $(\gamma_{\omega},\partial_{\gamma_{\omega}})$.
The right hand side is a degree $N$ polynomial. The choice of $\hat\varrho_{\omega} = \omega \tilde\ve_2$ gives in $\ve_2\to0$ limit
\begin{align}
   \Tr_{V_{\rm aux}}{\bf T}_{\rm SC}(x) \left \langle {Q}_{\Omega}(x) \right \rangle_{\BZ_N} \Psi
    = \left \langle \left[ {\EY}(x+\ve_2) + \kq \frac{{P}(x)}{{\EY}(x)} \right] {Q}_\Omega (x)  \right \rangle_{\BZ_N} \nonumber \nonumber
\end{align}
Let us take $\Omega=N-1$, where ${Q}_{N-1}(x) = {Q}(x)$ is the bulk $Q$-observable. We obtain
\begin{align}
    \Tr_{V_{\rm aux}}{\bf T}_{\rm SC}(x) \left \langle {Q}(x) \right \rangle_{\BZ_N} \Psi 
    % & = \left \langle \left[ {\EY}(x+\ve_2) + \kq \frac{{P}(x)}{{\EY}(x)} \right] \tilde{Q} (x)  \right \rangle \nonumber \nonumber\\
    & =  \left \langle {Q}(x+\ve_2) \right \rangle_{\BZ_N} \Psi + \fq P(x) \left \langle {Q}(x-\ve_2) \right \rangle_{\BZ_N} \Psi \nonumber\\
    & = \left \langle T_N(x) {Q}(x) \right \rangle_{\BZ_N} \Psi
\end{align}
where $T_N(x)$ is a degree $N$ polynomial whose coefficients depends on the bulk instanton configuration defined in \eqref{eq:qTQ}. 
The equivalence of the two equations establishes the Schr\"{o}dinger equations for all the conserving Hamiltonians.

\section{Discussion}\label{sec:discussion}
In this paper, we derived novel difference equations from non-perturbative Dyson-Schwinger equations for the correlation function of the intersecting surface defects in the four-dimensional $\EN=2$ supersymmetric gauge theory. The difference equations, called the fractional quantum T-Q equations, are satisfied by the correlation function of the intersecting surface defect observables, one of which is constructed out of the $\BZ_N$-orbifold and the other is constructed out of folded branes. We showed that the Fourier transform of the non-perturbative Dyson-Schwinger equations induce the $5$-point KZ for $\mathfrak{sl}_N$, where one of the ${\ssl}_N$-modules is the $N$-dimensional representation, with a proper matching of the parameters on two sides. We also constructed the quantum $\spch$ spin chain from the fractional quantum T-Q equations, achieving the Lax operators, the monodromy matrix, and the $\mathfrak{sl}_2$-representations at $N$ spin sites in gauge theoretical terms. The trace of the monodromy matrix is found to be identical to the fractional $qq$-character of rank $N$ in the NS limit $\ve_2\to0$.

We provide a few remarks on further developments of our study:
%\begin{enumerate}
    % \item The $Q(x)$ function was introduced in $N_{2}$ orthogonal to the $\BZ_n$ surface defect, which we called the intersecting case. 
    % One may also consider $Q$-function lying parallel to the surface defect, which we called the aligned case.  
%\end{enumerate}

\subsection{Isomonodromic deformations of higher-rank Fuchsian systems}
It was conjectured in \cite{GIL2012} that the isomonodromic tau function of the $\mathfrak{sl}_2$ Fuchsian system can be expressed as an infinite sum of the $SU(2)$ gauge theory partition functions in the self-dual limit ${\ve}_{1} + {\ve}_{2} \to 0$ of the $\Omega$-background. On the other hand, the isomonodromic tau function is a quasiclassical object, corresponding to the 
${\ve}_{1} \to 0$ or ${\ve}_{2} \to 0$. In \cite{JeongNekrasov,NekrasovVI}, the two approaches to the isomonodromic problem are reconciled by placing the gauge theory in the presence of the surface defect on the blowup $\widehat{\mathbb{C}}^2$ and studying novel blowup formula \cite{ny} for the vacuum expectation value of the surface defect observable \cite{JeongNekrasov,NekrasovVI}. Moreover, the horizontal section of the Fuchsian system was constructed from the correlation function of intersecting surface defects, allowing computation of the monodromy data of the Fuchsian system in gauge theoretical terms \cite{JeongNekrasov}.
Our work completes this circle of ideas by explicitly constructing the meromorphic connection (with special residues at $\qe$ and $1$) for the general $N$ rank case, thereby giving an explicit limit $\ve_1 \to 0$ to the isomonodromic problem. 

Specifically, the $\ve_1 \to 0$ asymptotics of the correlation function constructed in section \ref{sec:KZ from TQ} has the form
\begin{align}
    \boldsymbol\Pi(z;\qe,y) = e^{ \frac{S(\qe,z)}{\ve_1} } \left( \boldsymbol\pi(z;\qe,y) + \mathcal{O}(\ve_1) \right),
\end{align}
corresponding to the geometry of the regular surface defect extended in the $\BC_1$-plane while the vortex string defect is extended along the $\BC_2$-plane. Accordingly, the $S({\qe}, z)$ function in the exponential, being the effective twisted superpotential of the theory on the regular surface defect coupled to bulk gauge theory, is independent of the coupling $y$ of the transverse surface defect, the latter creating only a local disturbance. This is consistent with the limit 
of the $\qe$-component of the $5$-point KZ equation \eqref{eq:qcomp} giving
\begin{align} \label{eq:hid}
   0=\frac{\p S(\qe,z) }{\p \qe}  +  H(z,p;\qe) ,
\end{align}
where the Hamiltonian $H(z,p;\qe)$ is obtained in terms of $\fA_{0,1} \equiv \lim_{\ve_1 \to 0} \frac{\ve_1}{\ve_2} \hat{\fA}_{0,1}$. Here, we have the conjugate momenta $p_\o \equiv \frac{\p S(\qe,z)}{\p z_\o}$ appearing in $H(z,p;\qe)$. This equation is none other than the Hamilton-Jacobi equation for the isomonodromic deformations of $\mathfrak{sl}_N$ Fuchsian system, with the Hamilton-Jacobi potential $S(\qe,z)$. Moreover, the $y$-component of the 5-point KZ equation \eqref{eq:ycomp} becomes
\begin{align}
    0= \left[\frac{\p}{\p y} + \frac{\mathcal{A}_0}{y}+ \frac{\mathcal{A}_\qe}{y-\qe} + \frac{\CalA_1}{y-1} \right] \boldsymbol\pi (z;\qe,y),
\end{align}
where $\CalA_{0,\qe,y} \equiv \lim_{\ve_1 \to 0} \frac{\ve_1}{\ve_2} \hat{\CalA}_{0,\qe,y}$. Hence the regular part $\boldsymbol\pi(z;\qe,y)$ of the correlation function of the intersecting defects is precisely the horizontal section of the $\mathfrak{sl}_N$ Fuchsian system. 

It would be nice to verify that the $\ve_1 \to 0$ limit of the blowup formula for the expectation value of the regular surface defect yields the isomonodromic tau function for higher-rank $\mathfrak{sl}_N$ Fuchsian systems, expressed as an infinite sum of the gauge theory partition functions, generalizing \cite{JeongNekrasov, NekrasovVI}. It is also expected that the monodromy data of the higher-rank $\mathfrak{sl}_N$ Fuchsian systems can be computed in gauge theoretical terms, following \cite{JeongNekrasov}. Just as in \cite{JeongNekrasov,Jeong:2018qpc}, in the Darboux coordinates $(\boldsymbol\a,\boldsymbol\b)$ of the $SL(N)$ monodromy space constructed in \cite{Jeong:2018qpc} (higher-rank analogues of NRS coordinates \cite{NRS2011}; see also \cite{hol}) the monodromy data would be computed as
\begin{align}
    \b_\o = \frac{\p S}{\p \a_\o},\quad \o=0,\cdots, N-1,
\end{align}
where the half of the coordinates $(\a_\o)_{\o=0} ^{N-1}$ are identified with the Coulomb moduli, so that the potential $S(\qe,z)$ is the generating function of the Riemann-Hilbert map between the moduli space of $\mathfrak{sl}_N$ Fuchsian systems and the $SL(N)$ monodromy space, $(z,p) \leftrightarrow (\boldsymbol\a,\boldsymbol\b)$.

The higher-rank isomonodromic deformations are in fact not fully accounted by \eqref{eq:hid}. It is more natural to introduce $N-1$ \textit{higher times}, as opposed to the original \textit{time} $\qe$, along which further monodromy preserving deformations of the Fuchsian system are generated by \textit{higher} Hamiltonians. In the gauge theory side, the higher times can be introduced by explicit coupling terms with the higher Casimirs \cite{nekmar,marnek}. The higher-rank isomonodromic deformations of such kinds are not very well-known, at least to our best knowledge.

\subsection{Separation of variables and KZ/BPZ correspondence}
The separation of variables of the quantum integrable system is deeply involved with our study. It was indeed shown in \cite{Lee:2020hfu} that, in the limit $\ve_2 \to 0$, the vacuum expectation value $\Psi(\qe,z)$ \eqref{eq:pert} of the regular surface defect admits a Mellin-Barnes integral representation. This integral transform led to the expression of the eigenfunction in separated variables for the $XXX_{\mathfrak{sl}_2}$ spin chain. We expect that it would be possible to establish such an integral transform formula without the unrefinement of $\ve_2 \to 0$, both for the vacuum expectation value of the regular surface defect and the correlation function of the intersecting surface defects. In the view of the BPS/CFT correspondence, it would be equivalent to the KZ/BPZ correspondence, in which the solutions to the KZ equation and the BPZ equation are transformed to each other. 

In the rank $1$ case, it has been known that the coordinate transformation connecting the two sides of the KZ/BPZ correspondence is the separation of variable transformation \cite{Fren95,sto}. Physically, the integral transformation was interpreted as a consequence of the Hanany-Witten type M-brane transitions which interchange codimension-two defects (M5-branes) and codimension-four defects (M2-branes) \cite{frenguktes}. See also \cite{tes1}. The integral transform we are looking for would be its higher-rank analogues.

\subsection{Quantization conditions}

The four-dimensional gauge theory construction of the $XXX_{\mathfrak{sl}_2}$ spin chain suggests an application of the quantization scheme  of \cite{Nikita-Shatashvili}. Indeed, in section \ref{sec:spinLax}, we have shown that a specific quantization condition is equivalent to the vacuum equation of the dual two-dimensional gauged linear sigma model, leading to the quantization by algebraic Bethe ansatz \cite{Dorey:2011pa,HYC:2011}.

The quantization conditions in \cite{Nikita-Shatashvili} can be viewed as the boundary condition of the effective two-dimensional gauge theory on a cylinder, obtained by reducing the four-dimensional gauge theory subject to the half $\O$-background \cite{ref:nekwit}. It would be nice to exactly characterize the spectral problems induced by different choices of boundary conditions.

\subsection{Representation theory aspects}
We have shown that the correlation functions of intersecting surface defects give rise to certain $\mathfrak{sl}_N$-representations and $\mathfrak{sl}_2$-representations simultaneously. It would be interesting to further investigate the algebraic meaning of this relation. In particular, a natural conjecture is that the proper surface defect arrangement in the quiver gauge theory based
on a quiver of $ADE$ or ${\hat A}, {\hat D}, {\hat E}$-type, the $\ssl_2$ spin chain would be replaced by the corresponding spin chain based on the Yangian of the corresponding Lie algebra. In the quasiclassical limit this is supported by the identification
\cite{Nekrasov:2012xe} of Seiberg-Witten geometries of these theories with the moduli spaces of $ADE$ monopoles on ${\BR}^2 \times S^1$ or instantons on ${\BR}^{2} \times T^{2}$. The deformation quantization of these spaces produces the corresponding Yangian algebras \cite{Nikita-Pestun-Shatashvili, Elliott:2018yqm}.

The relation of the action in \eqref{eq:curiousslN} to the Heisenberg-Weyl representation of $\mathfrak{sl}_N$ is obscured at this moment. It may provide deeper insight for the connection between $\spch$ spin chain and representations of $\mathfrak{sl}_N$. 

In the algebraic engineering of $\EN=2$ gauge theories \cite{Bourgine2017b}, the gauge theory correlation functions are expressed as correlation functions of intertwining operators of representations of quantum toroidal algebra of $\mathfrak{gl}_1$. The regular orbifold surface defects can be incorporated by a lift to quantum toroidal algebra of $\mathfrak{gl}_N$ \cite{JB2019}. It would be nice if we can account for the duality between the KZ equations and the spin chains in this quantum toroidal algebra context.

Another subtle point concerns the precise definition of the tensor products ( \eqref{def:4pt-fun}, \eqref{eq:5ptup} ) etc. Since our computations involve infinite power series in various fugacities, the generating functions we obtain belong to certain completions of the tensor products. A good handle on the required topology comes from the study of the ${\qe} \to 0$ limit. On the KZ theory side this limit corresponds to diagonalizing a pair Casimir ${\hat {\rm{H}}}_{0}$, meaning decomposing the product of a lowest-weight Verma module and the HW module into irreducibles. On gauge theory side we would be computing the $J$-function of a flag variety valued sigma model, similarly to the computations done in  \cite{Lee:2020hfu}. It would be nice to make the precise match.

\subsection{Categorification of conformal blocks}

\begin{figure}
    \centering
    \includegraphics[scale=0.4]{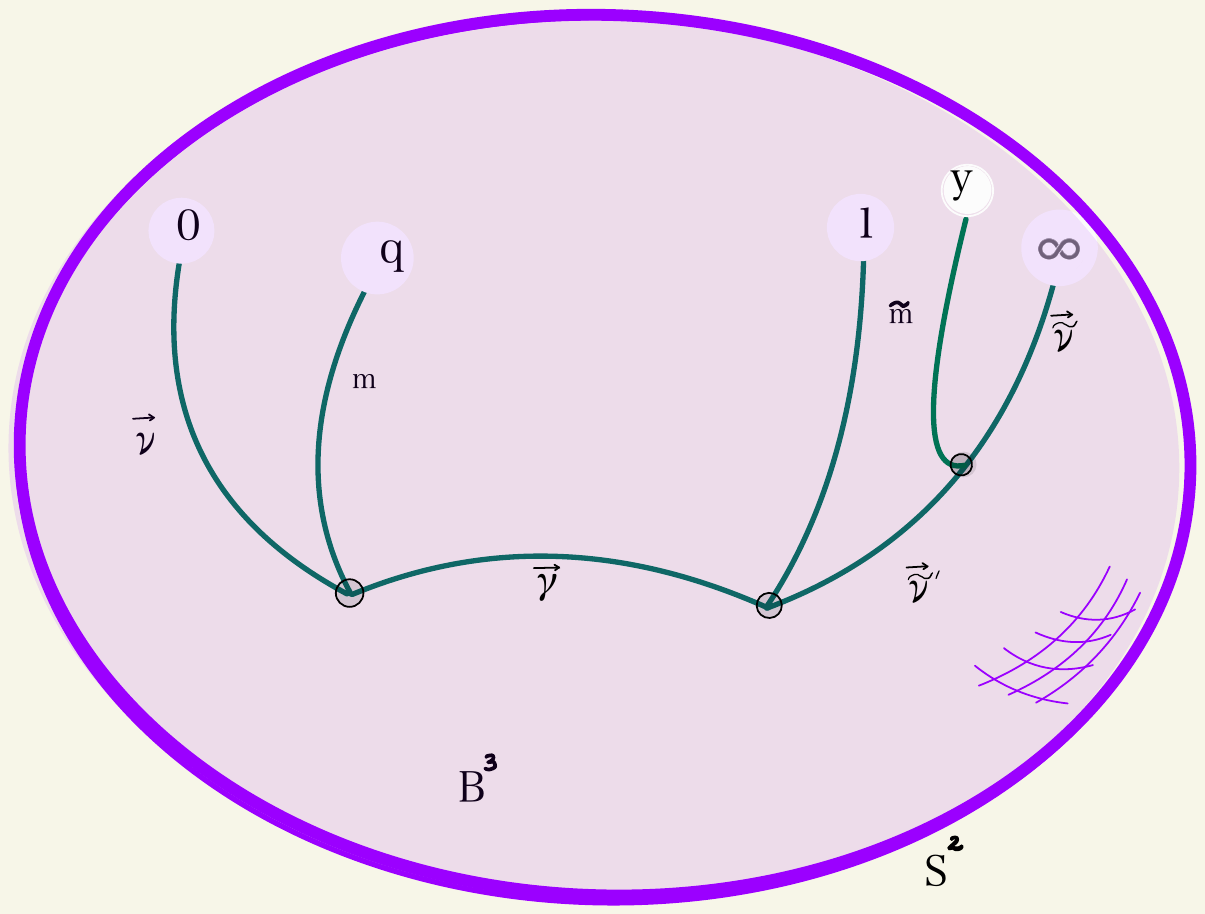}
    \caption{Wilson graph corresponding to the $5$-point conformal block}
    \label{fig:pic1}
\end{figure}

The results of our paper provide a non-trivial check of the BPS/CFT correspondence. 
As in \cite{NT}, it is interesting to recast our statement the language of \cite{ref:nekwit}, as well
as in view of \cite{wit, Witten:2011zz}. Namely, the higher dimensional perspective on the conformal blocks of
current algebra reveals a connection to the mysterious $(0,2)$-theory in six dimensions. The theory relevant to present considerations is of the $A_{N-1}$-type. As the $4$-point block studied in detail in \cite{NT}, the
$5$-point block, for the integral level $k$ and the dominant levels of Verma and HW modules admitting integrable quotients, has an interpretation as a wavefunction of a state in three dimensional Chern-Simons theory on a three-ball $B^3$ with the action 
\beq
\frac{k}{4\pi} \int_{B^3} {\rm Tr} \left( A \wedge d A + \frac{2}{3} A  \wedge A \wedge A \right)
\eeq
with the gauge fields having a curvature singularity along an embedded graph $\Gamma$, as in the Fig.~\ref{fig:pic1}.
Our paper provides an analytic continuation to the
case of complex levels and weights. The paper \cite{wit} offers such a continuation for the Chern-Simons level. 
As explained in \cite{NT} it does not seem to be possible to analytically continue the graph observable as line operators in the analytically continued Chern-Simons theory, as in \cite{wit}. In the present case one leg $l$
of the graph $\Gamma$ corresponds to the $N$-dimensional representation, for which the matrix elements
of the holonomy $T_{{\BC}^{N}} \, P {\exp} \int_{l} A$ are well-defined. Thus, we might expect 
the analytically continued observable to be a junction of a surface defect in the topologically twisted ${\CalN}=4$ theory on a four dimensional manifold
with corners, which locally looks like $B^3 \times I$, and a line operator. 

\begin{figure}
    \centering
    \includegraphics[scale=0.4]{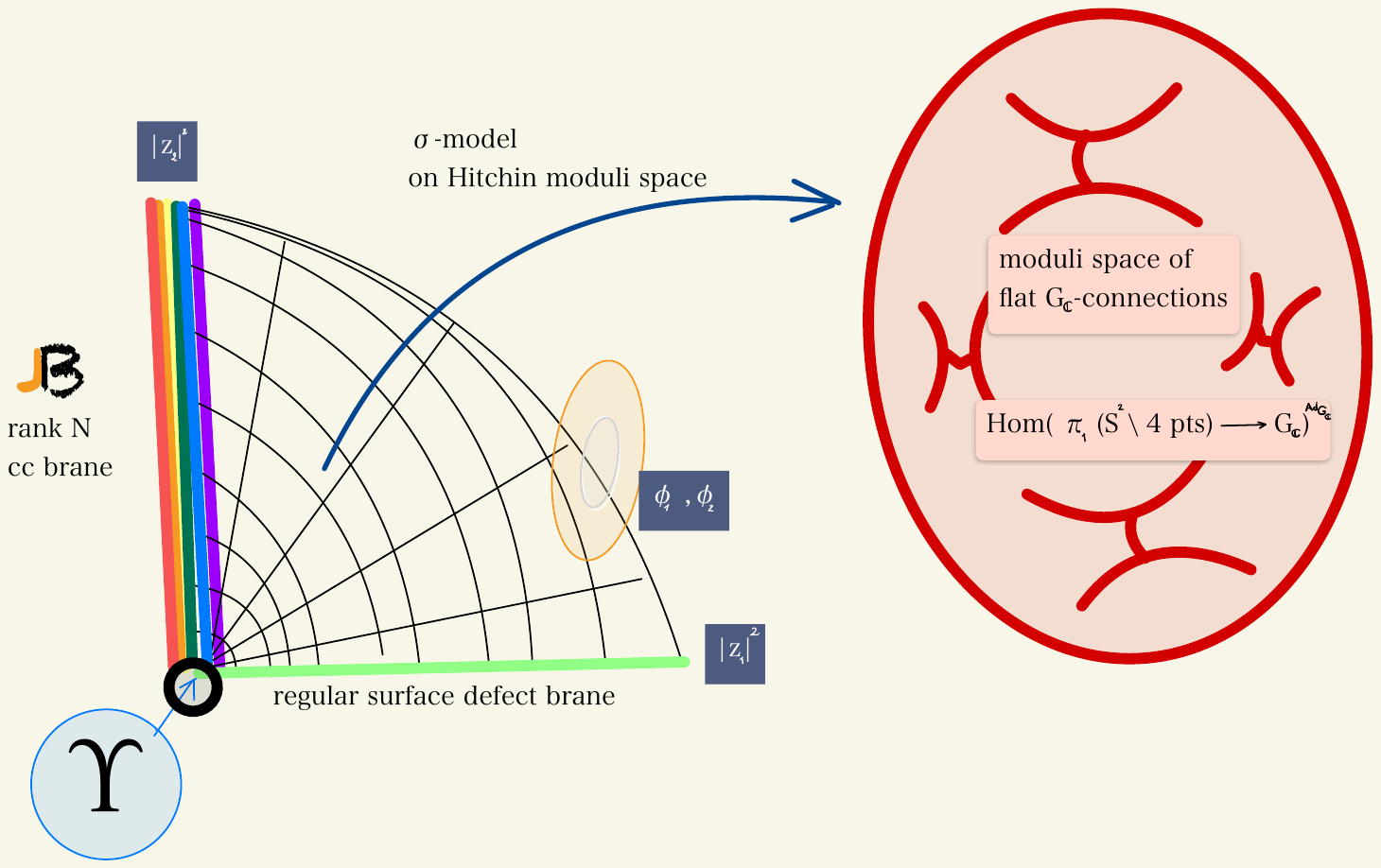}
    \caption{Four dimensional gauge theory in two dimensional presentation}
    \label{fig:pic2}
\end{figure}

{}On the other hand, the surface defect in four dimensions
can be related \cite{ref:nekwit} to boundary conditions in the two dimensional sigma model valued in the moduli space
of vacua of the theory, compactified on a circle, which in the present case is believed to be the moduli space ${\CalM}_{N} \left( S^{2} \backslash 4\, {\rm pts} \, ; {\vec\nu}, {\mf}, {\tilde\mf}, {\vec{\tilde\nu}} \right)$
of $SU(N)$ Higgs pairs on a $4$-punctured sphere with the regular punctures at $0$ and $\infty$, and the
minimal punctures at $\qe$ and $1$, see the Fig.~\ref{fig:pic2}. 
The homotopy between these two representatives of a cohomology class of an intrinsic operator in the six dimensional theory proceeds by viewing the two dimensional sigma model, with the worldsheet $C$ as a long distance limit of the four dimensional ${\CalN}=2$ $\Omega$-deformed theory
compactified on a two-torus $T^2$ as in \cite{ref:nekwit}, which, in turn, is a limit of the $A_{N-1}$ $(0,2)$-theory 
compactified on $\left( S^{2} \backslash \{ 0, {\qe}, 1, {\infty} \}  \right) \times T^2$, which, finally, can be reinterpreted, as
the ${\CalN}=4$ theory on $C \times \left( S^{2} \backslash \{ 0, {\qe}, 1, {\infty} \}  \right)$ with the canonical parameter \cite{kapwit} identified \cite{ref:nekwit} with the ratio ${\ve}_{2}/{\ve}_{1}$. With $C$ having the topology of the corner ${\BR}_{+}^{2}$, as in Fig.~\ref{fig:pic2}, the suitably twisted ${\CalN}=4$ theory 
looks very much like a gradient flow theory of the analytically continued Chern-Simons theory on ${\BR}_{+} \times S^{2}$, with certain boundary conditions. Of special interest is the brane (in the sigma model sense) located at the $z_1 = 0$ component of the boundary on the ~Fig.\ref{fig:pic2}. In the setup of \cite{NT} that brane could be identified \cite{ref:nekwit} with the space-filling canonical coisotropic brane \cite{Kapustin:2001ij, kapwit}. Adding the light surface defect generating the $Q$-observable seems to endow this brane with a rank $N$ Chan-Paton bundle. It is tempting to identify this bundle with the universal Higgs bundle \cite{ref:nekwit} evaluated at the point $y \in S^{2} \backslash \{ 0, {\qe}, 1, {\infty} \}$.

\bigskip
\centerline{***}
\bigskip

\appendix

\section{Partition functions of $\EuScript{N}=2$ supersymmetric gauge theories}\label{sec:parti-func}
We consider $\EN=2$ $A_1$ quiver gauge theory in four dimensions, with gauge group $SU(N)$ and $2N$ fundamental hypermultiplets. The Lagrangian is parametrized by the complex coupling
\begin{align}
    \tau = \frac{4\pi i}{g^2} + \frac{\theta}{2\pi}, \quad \kq = \exp 2 \pi i \tau
\end{align}
and by the choice of $2N$ fundamental matter masses ${\bf m}$, which we splits into $N$ {\it fundamental} ${\bf m}^+ = (m_1^+,\dots,m^+_N)$ and $N$ {\it anti-fundamental} ${\bf m}^- = (m^-_1,\dots,m^-_N)$. The choice of the vacuum is characterized by the $N$ Coulomb moduli parameters ${\bf}=(a_1,\dots,a_N)$ obeying
\begin{align}
    \sum_{\alpha=1}^N a_\alpha = 0.
\end{align}

{}The localization of $\Omega$-deformed theory \cite{Nekrasov:2003rj,Nikita:I} produces the statistical model whose configuration space is $\BP^N$. Each instanton configuration is labeled by $N$-tuples of Young diagrams $\boldsymbol\lambda = (\lambda^{(1)},\dots,\lambda^{(N)})$. Each individual Young diagram $\lambda^{(\alpha)}$, $\alpha=1,\dots,N$, is a collection $\lambda^{(\alpha)} = (\lambda^{(\alpha)}_1,\lambda^{(\alpha)} _2,\dots) $ of non-negative integers satisfying 
\begin{align}
    \lambda^{(\alpha)}_{\ri} \geq \lambda^{(\alpha)}_{\ri+1}, \, \ri=1,2,\dots
\end{align}
where each $\lambda^{(\alpha)}_\ri$ labels the number of squares in the $\ri$-th row of Young diagram $\lambda^{(\alpha)}$. 

The pseudo-measure associated to the instanton configuration $\boldsymbol\lambda$ is defined through 
 \emph{plethystic exponent} $\bE$ operator, which converts the additive Chern characters to the multiplicative classes
\begin{align}
    \mathbb{E}\left[\sum_{a} {\mathtt{m}}_a e^{{\rx}_a}\right] = \prod_a {\rx}_a^{-{\mathtt{m}}_a}
\end{align}
where ${\mathtt{m}}_{a} \in {\BZ}$ is the multiplicity of the Chern root $\rx_a$.
For ${\boldsymbol\lambda}$ the associated pseudo-measure is computed by:
\begin{align}
    \CalZ({\ba},{\bm}^\pm,\vec{\epsilon})[\boldsymbol{\lambda}]= \mathbb{E}\left[-\frac{\hat{S}\hat{S}^*}{P_{12}^*} + \frac{\hat{M}\hat{S}^*}{P_{12}^*} \right]
    \label{def:instparti}
\end{align}
where
\begin{align}
    \hat{N} = \sum_{\alpha=1}^N e^{a_{\alpha}}, \quad
    \hat{K} = \sum_{\alpha=1}^N\sum_{\Box\in \lambda^{(\alpha)}} e^{c_\Box}, \quad \hat{S} = \hat{N} - P_{12} \hat{K}, \quad 
    \hat{M} = \sum_{f=1}^N e^{m_f^+} + e^{m_f^-}. 
\end{align}
We use a short hand notation $c_\Box = a_\alpha + (\ri-1)\ve_1 + (\rj-1)\ve_2$.
$q_i=e^{\ve_i}$ are the exponentiated complex $\Omega$-deformation parameters $\ve_1, \ve_2 \in \BC$ \cite{Nekrasov:2002qd,Nekrasov:2003rj,Pestun:2016zxk}, and
\begin{align}
    P_i = 1 - q_i, \quad P_{12} = (1-q_1)(1-q_2).
\end{align}
Given a virtual character $\hat{X} = \sum_{a} {\mathtt{m}}_{a} e^{{\rx}_{a}}$ we denote by  $\hat{X}^{*} = \sum_{a} {\mathtt{m}}_{a} e^{-{\rx}_{a}}$  the dual virtual character.

Let us define the $\EY$-observable
\begin{align}
    \EY(x) [\boldsymbol\l]
    & = \BE \left[ - e^x \hat{S}^*[\boldsymbol{\lambda}] \right] 
    = \prod_{\alpha=1}^N (x-a_\alpha) \prod_{\Box\in\lambda^{(\alpha)}} \frac{(x-c_\Box-\ve_1)(x-c_\Box-\ve_2)}{(x-c_\Box) (x- c_\Box - \ve_1 - \ve_2)} \nonumber
\end{align}
on the instanton configuration $\boldsymbol\lambda$.

In the $A_1$-type theory, $\EY$-observable can be expressed as ratio of analytic $Q$-observable with a shifted argument:
\begin{align}\label{def:Q-func}
    \EY(x) = \frac{Q(x)}{Q(x-\ve_2)}.
\end{align}
% The zeros of $Q^{(1)}$ are located at
% \begin{align}
%     a_\alpha + \ve_1(\ri-1) + \ve_2\lambda^{(\alpha)}_\ri, \, \alpha=1,\dots,N, \, \ri=1,2,\dots
% \end{align}
The zeros of ${Q}$ are located at
\begin{align}
    a_\alpha + \ve_2(\rj-1) + \ve_1 \lambda^{(\alpha),t}_\rj, \, \alpha=1,\dots,N, \, \rj=1,2,\dots
\end{align}
where $\boldsymbol\lambda^t$ is the transpose of $\boldsymbol\lambda$.

\section{Integrability of $XXX_{\mathfrak{sl}_2}$ spin chain}\label{sec:Bethe}

Consider one-dimensional quantum periodic spin chain with $N$ sites. Each spin site is associated with a local spin operator $\vec{\bs}_n=(\bs_n^+,\bs_n^-,\bs_n^0)$ of general spin. The spin variables act on the Hilbert space $h$. The full Hilbert space is the tensor product of all local Hilbert spaces
\begin{align}
    \CalH = h_0\otimes h_1\otimes\cdots\otimes h_N.
\end{align}
% The Hamiltonian of the system (with no twist imposed) is given by
% \begin{align}
%     H = \sum_{n=1}^N \vec{s}_n \cdot \vec{s}_{n+1}, \quad \vec{s}_{N+n} = \vec{s}_{n}. 
% \end{align}
We shall use the permutation operator 
\begin{align}
    P = \frac{1}{2} (I \otimes I + \vec{\sigma} \otimes \vec{\sigma} ).
\end{align}
% By choosing the basis of $\BC^2$ the vectors
% \begin{align}
%     |+\rangle = \begin{pmatrix} 1 \\ 0 \end{pmatrix}, |-\rangle = \begin{pmatrix} 0 & 1 \end{pmatrix}
% \end{align}
% the permutation operator is given by the matrix
% \begin{align}
%     P = 
%     \begin{pmatrix}
%         1 & 0 & 0 & 0 \\ 
%         0 & 0 & 1 & 0 \\
%         0 & 1 & 0 & 0 \\
%         0 & 0 & 0 & 1
%     \end{pmatrix}
% \end{align}
% in the basis of $|++\rangle$, $|+-\rangle$, $|-+\rangle$, and $|--\rangle$.
% The Hamiltonian can be written in terms of permutation operators
% \begin{align}
%     H = \frac{1}{2}\sum_{n=1}^N (P_{n,n+1} - 1)
% \end{align}

The Lax operator $L_{n,a}$ is defined on the local space $h_n\otimes V_{\rm aux}$. In our example the auxiliary $V_{\rm aux} = \BC^2$ but it can be chosen otherwise. The Lax operator is given by
\begin{align}
    L_{n,a}(x) & = \lambda( I_{h_n} \otimes I_{\rm aux} ) + \hbar ( \vec{\sigma}_n \otimes \vec{\sigma}_{\rm aux} ) \\
    % & = \left( \lambda - \frac{\hbar}{2} \right) I + i P_{n,a} \\
    & = x + \hbar
    \begin{pmatrix} \bs_n^0 & \bs_n^- \\ \bs_n^+ & -\bs_n^0 \end{pmatrix} = x I +\hbar \CalL_{n}
\end{align}

The commutation relation of the matrix elements in $2\times 2$ matrix $L_{n,a}$ is governed by the $RLL$-relation (train track relation):
\begin{align}\label{eq:RLL}
    R_{a_1,a_2}(x - x') L_{n,a_1}(x) L_{n,a_2}(x') = L_{n,a_2}(x) L_{n,a_1}(x') R_{a_1,a_2}(x - x').
\end{align}
This is an equation acting on the space $V_{a_1}\otimes V_{a_2} \otimes h_n$. The indices $a_1$ and $a_2$ and variables $x$ and $x'$ are associated to the auxiliary spaces $V_{a_1}$ and $V_{a_2}$. The $R_{a_1,a_2}$ governing the commutation is given by
\begin{align}\label{def:R-matrix}
    R_{a_1,a_2}(x) 
    & = \frac{1}{x+\hbar} (x I_{a_1,a_2} + \hbar P_{a_1,a_2})
\end{align}
In particular, by choosing $V_{a_1} = V_{a_2} = \BC^2$, $R_{a_1,a_2}$ is a $4\times 4$ matrix
\begin{align}
    R_{a_1,a_2}(z) = \frac{1}{x+\hbar} 
    \begin{pmatrix}
        x+\hbar & 0 & 0 & 0 \\
        0 & x & \hbar & 0 \\
        0 & \hbar & x & 0 \\
        0 & 0 & 0 & x+\hbar 
    \end{pmatrix}
\end{align}
with 
\begin{subequations}
\begin{align}
    L_{n,a_1}(x) & = x (I\otimes I_{a_1} \otimes I_{a_2}) + \hbar (\bs_n^x\otimes\sigma_x\otimes I_{a_2} + \bs_n^y\otimes \sigma_y\otimes I_{a_2} + \bs_n^z\otimes \sigma_z\otimes I_{a_2}) \nonumber\\
    & \equiv x I + \hbar \CalL_1, \\
    L_{n,a_2}(x') & = x' (I \otimes I_{a_1} \otimes I_{a_2}) + \hbar (\bs_n^x\otimes I_{a_1} \otimes \sigma_x + \bs_n^y\otimes I_{a_1} \otimes \sigma_y + \bs_n^z\otimes I_{a_1}\otimes \sigma_z) \nonumber\\
    & \equiv x' I + \hbar \CalL_2.
\end{align}
\end{subequations}
The validity of \eqref{eq:RLL} can be computed via direct calculation
\begin{subequations}
\begin{align}
    R_{a_1,a_2}L_{a_1}L_{a_2} - L_{a_2}L_{a_1}R_{a_1,a_2}
    % = & (\lambda-\mu) \lambda\mu I + \hbar\lambda\mu P + (\lambda-\mu)\hbar (\lambda M_2 + \mu M_1) \nonumber\\
    % & + \hbar^2 P(\lambda M_2 + \mu M_1) + (\lambda-\mu)\hbar^2 M_1M_2 + \hbar^3 PM_1M_2 \\
    % & - \left[ (\lambda-\mu) \lambda\mu I + \hbar\lambda\mu P + (\lambda-\mu)\hbar (\lambda M_2 + \mu M_1)\right. \nonumber\\
    % & +\left. \hbar^2 (\lambda M_2 + \mu M_1)P + (\lambda-\mu)\hbar^2 M_2M_1 + \hbar^3 M_2M_1P\right]\nonumber\\
    = & x \hbar^2 (P\CalL_2-\CalL_2P) + x' \hbar^2 (P\CalL_1-\CalL_1P) \nonumber\\ 
    & + (x-x')\hbar^2(\CalL_1\CalL_2-\CalL_2\CalL_1) + \hbar^3 (P\CalL_1\CalL_2 - \CalL_2\CalL_1P)
\end{align}
\end{subequations}
with each element written as $4 \times 4$ matrix in $V_{a_1}\otimes V_{a_2}$ spaces,
\begin{subequations}
\begin{align}
    & P\CalL_1\CalL_2 
    % \begin{pmatrix}
    % 1 & & & \\
    % & & 1 & \\
    % & 1 & & \\
    % & & & 1
    % \end{pmatrix}
    % \begin{pmatrix}
    % s^0 & s^- & & \\
    % s^+ & -s^0 & & \\
    % & & s^0 & s^- \\
    % & & s^+ & -s^0
    % \end{pmatrix}
    % \begin{pmatrix}
    % s^0 & & s^- & \\
    % & s^0 & & s^- \\
    % s^+ & & -s^0 & \\
    % & s^+ & & -s^0
    % \end{pmatrix}
    =\begin{pmatrix}
    \bs^0\bs^0 & \bs^-\bs^0 & \bs^0\bs^- & \bs^-\bs^- \\
    \bs^0\bs^+ & \bs^-\bs^+ & -\bs^0\bs^0 & -\bs^-\bs^0 \\
    \bs^+\bs^0 & -\bs^0\bs^0 & \bs^+\bs^- & -\bs^0\bs^- \\
    \bs^+\bs^+ & -\bs^0\bs^+ & -\bs^+\bs^0 & \bs^0\bs^0
    \end{pmatrix} \\
    & \CalL_2\CalL_1P  
    % \begin{pmatrix}
    % A & & B & \\
    % & A & & B \\
    % C & & D & \\
    % & C & & D
    % \end{pmatrix}
    % \begin{pmatrix}
    % A & B & & \\
    % C & D & & \\
    % & & A & B \\
    % & & C & D
    % \end{pmatrix}
    % \begin{pmatrix}
    % 1 & & & \\
    % & & 1 & \\
    % & 1 & & \\
    % & & & 1
    % \end{pmatrix}
    =\begin{pmatrix}
    \bs^0s^0 & \bs^-s^0 & \bs^0\bs^- & \bs^-\bs^- \\
    \bs^0\bs^+ & \bs^-\bs^+ & -\bs^0\bs^0 & -\bs^-\bs^0 \\
    \bs^+\bs^0 & -\bs^0\bs^0 & \bs^+\bs^- & -\bs^0\bs^- \\
    \bs^+\bs^+ & -\bs^0\bs^+ & -\bs^+\bs^0 & \bs^0\bs^0
    \end{pmatrix} = P \CalL_1 \CalL_2
\end{align}
\end{subequations}
The remaining terms are
\begin{align}
    (x-x')( \CalL_1\CalL_2-\CalL_2\CalL_1 ) = (x-x')
    \begin{pmatrix}
    0 & [\bs^-,\bs^0] & [\bs^0,\bs^-] & 0 \\
    [\bs^+,\bs^0] & [-\bs^0,\bs^0] & [\bs^+,\bs^-] & [-\bs^0,\bs^-] \\
    [\bs^0,\bs^+] & [\bs^-,\bs^+] & [\bs^0,-\bs^0] & [\bs^-,-\bs^0] \\
    0 & [-\bs^0,\bs^+] & [\bs^+,-\bs^0] & 0
    \end{pmatrix}
\end{align}
and
\begin{align}
    x' (P\CalL_1-\CalL_1P) + x (P\CalL_2-\CalL_2P) & = (x-x')
    \begin{pmatrix}
    0 & -\bs^- & \bs^- & 0 \\
    \bs^+ & 0 & -2\bs^0 & -\bs^- \\
    -\bs^+ & 2\bs^0 & 0 & \bs^- \\
    0 & \bs^+ & -\bs^+ & 0
    \end{pmatrix}
\end{align}
% \begin{subequations}
% \begin{align}
%     \mu (PM_1-M_1P) + \lambda (PM_2-M_2P) & = (\mu - \lambda)
%     \begin{pmatrix}
%     0 & -s^- & s^- & 0 \\
%     s^+ & 0 & -2s^0 & -s^- \\
%     -s^+ & 2s^0 & 0 & s^- \\
%     0 & s^+ & -s^+ & 0
%     \end{pmatrix}\\
%     \lambda (M_2P - PM_2) & = \lambda
%     \begin{pmatrix}
%     0 & B & -B & 0 \\
%     -C & 0 & A-D & B \\
%     C & D-A & 0 & -B \\
%     0 & -C & C & 0
%     \end{pmatrix} = - \lambda(M_1P-PM_1)
% \end{align}
% \end{subequations}

We find \eqref{eq:RLL} holds if the spin chain operators satisfies the commutation relation
\begin{align}
    \left[\bs^0_j,\bs_k^\pm \right] = \pm \bs_j^\pm \delta_{jk}, \quad \left[\bs_j^+,\bs_k^-\right] = 2 \bs_j^0 \delta_{jk}. 
    % \quad (\ell_j^0)^2 + \frac{1}{2}( \ell_j^+\ell_j^- + \ell_j^-\ell_j^+) = s_j(s_j+1) \hbar^2. 
\end{align}
for all the representation.

The monodromy matrix ${\bf T}_{\rm SC}(z)$ is defined as an ordered product of Lax operators 
\begin{align}
    {\bf T}_{a}(x) = K_{a}(\kq) L_{N,a}(x)L_{N-1,a}(x)\cdots L_{1,a}(x),
\end{align}
where $K_a(\kq)$ is a twisted matrix introduced to the system. It is obvious that the monodromy matrix ${\bf T}_{a}(z)$ satisfies the train track commutation relations same as Lax operators, namely
\begin{align}\label{eq:RTT}
    R_{a_1,a_2}(x-x') {\bf T}_{a_1}(x) {\bf T}_{a_2}(x') = {\bf T}_{a_2}(x') {\bf T}_{a_1}(x) R_{a_1,a_2}(x-x')
\end{align}
in the absence of twist, $K_{\kq} = I_a$. When a twist matrix $K(\kq)$ is introduced, one extra condition to check for validity of Eq.~\eqref{eq:RTT} is
\begin{align}\label{eq:RKK}
    R_{a_1,a_2}(x-x')K_{a_1}(\kq)K_{a_2}(\kq) = K_{a_2}(\kq)K_{a_1}(\kq) R_{a_1,a_2}(x-x').
\end{align}
The twist matrix can always be decomposed into 
\begin{align}
    K_a(\kq) = K_{a,1} I + K_{a,x} \sigma_x + K_{a,y} \sigma_y + K_{a,z} \sigma_z.
\end{align}
Eq.~\eqref{eq:RKK} can be verified via direct calculation. We conclude that Eq.~\eqref{eq:RTT} holds for monodromy matrix with general twisted matrix.

\section{Some computational details for $4$-point KZ equation}
\subsection{Representation side} \label{sec:4-pt,flag}
Given the parameters $(\boldsymbol\zeta,\tilde{\boldsymbol\zeta},\boldsymbol\tau,\mu,\tilde\mu)$,
 a way to solve \eqref{eq:4pmat} is to introduce $N-1$ free parameters $\boldsymbol\t=(\t_1,\dots,\t_{N-1})$:
 \begin{align}
     \t_i = \beta_i + \tilde\beta_i + \alpha_i
 \end{align}
 so that \eqref{eq:4pmat} can be reorganized,
 \begin{align}
 \begin{split}
     & \beta_{i+1} - \beta_i = \zeta_i - \t_i, \quad \sum_{\tb} \beta_\tb = \mu, \\
     & \tilde\beta_{i+1} - \tilde\beta_i = \tilde\zeta_i - \t_i, \quad \sum_{\tb}\beta_\tb = \tilde\mu^{(4)}.
 \end{split}
\end{align}
We solve for $(\beta_\ta)_{\ta=1}^{N}$ satisfying \eqref{eq:4pmat} in terms of $(\boldsymbol\zeta,\tilde{\boldsymbol\zeta},\boldsymbol\tau,\mu,\tilde\mu)$:
\begin{align}\label{eq:solbeta}
\begin{split}
    \beta_N & = \frac{\mu}{N} + \sum_{j=1}^{N-1} \frac{j}{N} (\tau_j - \tilde\zeta_j). \\
    \beta_{i} & = \beta_{N} - (\beta_N - \beta_{N-1}+\beta_{N-1}-\beta_{N-2} + \cdots + \beta_{i+1} - \beta_i) \\
    & = \frac{\mu}{N} + \sum_{j=1}^{N-1} \frac{j}{N}(\tau_j - \tilde\zeta_j) + \sum_{j=i}^{N-1} (\tau_j - \tilde\zeta_j), \ i=1,\dots,N-1,
\end{split}
\end{align}
and similarly for $(\tilde\beta_\ta)_{\ta=1}^N$:
\begin{align}\label{eq:soltilbeta}
\begin{split}
    \tilde\beta_N & = \frac{\tilde\mu}{N} + \sum_{j=1}^{N-1} \frac{j}{N} (\tau_j - \zeta_j). \\
    \tilde\beta_{i} & = \tilde\beta_{N} - (\tilde\beta_N - \tilde\beta_{N-1}+\tilde\beta_{N-1}-\tilde\beta_{N-2} + \cdots + \tilde\beta_{i+1} - \tilde\beta_i) \\
    & = \frac{\tilde\mu}{N} + \sum_{j=1}^{N-1} \frac{j}{N}(\tau_j - \zeta_j ) - \sum_{j=i}^{N-1} (\tau_j - \tilde\zeta_j), \ i=1,\dots,N-1.
\end{split}
\end{align}
Finally for $(\alpha_i)_{i=1}^{N-1}$:
\begin{align}\label{eq:solalpha}
\begin{split}
    \alpha_i 
    % & = \t_i - \beta_i - \tilde\beta_i 
    = \t_i - \frac{\mu+\tilde\mu}{N} + \sum_{j=1}^{N-1} \frac{j}{N}(\zeta_j + \tilde\zeta_j) - \sum_{j=\tb}^{N-1} (\zeta_j + \tilde\zeta_j), \quad i=1,\dots,N-1.
\end{split}
\end{align}
 The dependence of $\t_i$'s on $(\beta_\ta,\tilde\beta_\ta,\alpha_i)$ is equivalent to transformation \eqref{eq:transPsi}, with the parameters matching:
 \begin{align}
 \begin{split}
     & \varkappa_{i+1} - \varkappa_i = -\t_i, \ i=1,\dots,N-1, \\
     \implies & \varkappa_\tb = \frac{1}{N}\sum_{j=1}^{N-1}(N-j)\t_j - \sum_{j=\tb}^{N-1} \t_j, \ \tb=1,\dots,N.
 \end{split}
 \end{align}
 \eqref{eq:solalpha}, \eqref{eq:solbeta}, and \eqref{eq:soltilbeta} have an equivalent expression in terms of $\{\varkappa_\tb\}$:
 \begin{subequations}\label{eq:solvarkappa}
 \begin{align}
     &\beta_\tb + \varkappa_\tb = \frac{\mu}{N} + \sum_{j=1}^{N-1}\frac{j}{N}\zeta_j - \sum_{j=\tb}^{N-1} \zeta_j; \\
     & \tilde\beta_\tb + \varkappa_\tb = \frac{\tilde\mu}{N} + \sum_{j=1}^{N-1} \frac{j}{N}\tilde\zeta_j - \sum_{j=\tb}^{N-1} \tilde\zeta_j;\\
     & \alpha_i - \varkappa_i - \varkappa_{i+1} = -\frac{\mu+\tilde\mu}{N} - \sum_{j=1}^{N-1} \frac{j}{N}(\zeta_j + \tilde\zeta_j) + \sum_{j=i}^{N-1} (\zeta_j + \tilde\zeta_j),
 \end{align}
 \end{subequations}
 where $\tb=1,\dots,N$, $i=1,\dots,N-1$. $(\varkappa_\tb)_{\tb=1}^N$ obeys
 $$
     \sum_{\tb=1}^{N} \varkappa_\tb = 0.
 $$
 Hence $N-1$ parameters $(\t_i)_{i=1}^{N-1}$ and $N$ parameters $(\varkappa_\tb)_{\tb=1}^N$ share the same degree of freedoms.

 The 4-point correlation function $\Psi(\kq)$ constructed from the flag varieties provides a particular representation of the 4-point KZ equation as a differential operators. $\Psi(\kq)$ obeys the 4-point Knizhnik-Zamolodchikov equation
 \begin{align}
     \left( -(k+N)\frac{d}{d\kq} + \frac{\hat{\rm H}_0^{(4)}}{\kq} + \frac{\hat{\rm H}_1^{(4)}}{\kq-1} \right)\Psi(\kq) = 0
 \end{align}
 KZ connections are given by the tensor product of $\mathfrak{sl}_N$ generators in the respective modules 
 \begin{align}
     \hat{\rm H}_0^{(4)} = J^\ta_\tb \fz^\tb\frac{\partial}{\partial \fz^a}, \quad \hat{\rm H}_1^{(4)} = - \tilde\fz(\fz) \sum_{\ta=1}^N \frac{\partial^2}{\partial \fz^a \partial \tilde\fz_a}.
 \end{align}
 The superscript is used to distinguish from the 5-point case.

\subsection{Gauge theory side}\label{sec:4-point KZ}
For completeness, let us briefly review the identification of the surface defect partition function with the 4-point conformal block of $\widehat{\ssl}_N$.
The Verma and HW weights $(\boldsymbol\zeta,\tilde{\boldsymbol\zeta}, \mu,\tilde{\mu})$ are related to the fundamental matter masses, 
$\Omega$-deformation parameter $\ve_1$ and the Coulomb moduli of the $\EN=2$ SQCD by
\begin{align}
\begin{split}
    &\zeta_i = \frac{m^-_{i+1}-m^-_{i}}{\ve_1}, \quad \tilde\zeta_i = \frac{m^+_{i+1}-m^+_{i}}{\ve_1}, \ i=1,\dots,N-1;\\
    &\mu = \sum_{\ta=1}^N \frac{m^-_\ta - a_\ta}{\ve_1}, \quad \tilde\mu = \sum_{\ta=1}^N \frac{m_\ta^+ - a_\ta}{\ve_1}.
\end{split}
\end{align}
The 4-point conformal block $\Psi(\kq)$ obeys the KZ equation
\begin{align}
    \left[ - (k+N) \frac{d}{d\kq} + \frac{\hat{\rm H}_0^{(4)}}{\kq} + \frac{\hat{\rm H}_1^{(4)}}{\kq-1} \right] \Psi(\kq) = 0
\end{align}
with the residues of the meromorphic KZ connection given by
\begin{align}
    \hat{\rm H}_0^{(4)} = -\sum_{\ta,\tb=1}^N \fz^\tb J^\ta_\tb \frac{\partial}{\partial \fz^\ta}, \quad
    \hat{\rm H}_1^{(4)} = - \tilde{\fz}(\fz) \sum_{\ta=1}^{N} \frac{\partial^2}{\partial \fz^\ta \partial \tilde{\fz}_\ta}
\end{align}
with $J_\ta^\tb\pi_i = -e_\ta \wedge \tilde{e}^\tb \pi_{i}$.
We focus on the $\EG$-invariant part $\chi(v_1,\dots,v_{N-1};\kq)$ in the 4-point correlation function $\Psi(\kq)$, which satisfies
\begin{align}\label{eq:KZ-4p}
     \left[- (k+N)\frac{d}{d\kq} + \frac{\hat\sH_0^{(4)}}{\kq} + \frac{\hat\sH_1^{(4)}}{\kq-1} \right] \chi(v_1,\dots,v_{N-1};\kq) = 0.
\end{align}

The operators $\hat\sH^{(4)}_{0,1}$ are obtained by commuting the $\text{Lie}(\EG)$-equivariant $\Psi_0$ \eqref{eq:psi0} through operators $\hat{\rm H}_{0,1}^{(4)}$:
\begin{align}
    \hat{\rm H}_{0,1}^{(4)} \Psi(\kq) = \Psi_0 \hat\sH_{0,1}^{(4)} \chi(v_1,\dots,v_{N-1};\kq).
\end{align}
We may choose, by $SL(N)$-transformation, $\pi_i = \pi_i^\circ$ and $\tilde{\pi}^i = \tilde{\pi}^i_\circ$. We consider action of $J^\ta_\tb$:
\begin{subequations}
\begin{align}
    (\tilde{\fz}\wedge \tilde\pi^{i-1})(J^\ta_\tb \pi_i) 
    & = - \left[(\tilde{\fz}_1\tilde{e}^1 + \cdots + \tilde{\fz}_N\tilde{e}^N ) \wedge (\tilde{e}^1 \wedge \cdots \wedge \tilde{e}^{i-1}) \right] (e_\tb\wedge \tilde{e}^\ta (e_1 \wedge \cdots \wedge e_i) ) \nonumber\\
    & = 
    \begin{cases}
   - \tilde{\fz}_\tb  \delta^\ta_i \theta_{\tb\geq i}, & \ta \neq \tb \\
   -\tilde{\fz}_i \theta_{i\geq \ta}, & \ta=\tb
    \end{cases}
    \nonumber \\
    & = -\tilde{\fz}_\tb  \delta^\ta_i \theta_{\tb > i} - \tilde{\fz}_i \theta_{i\geq \ta} \delta_\tb^\ta  \\
    %%%%%%%%%%%%%%%%%%%%%%%%%%%%%%%%%%%%%%%%%%%%%%%%
    \tilde\pi^i (\fz\wedge J^\ta_\tb \pi_{i-1}) 
    & = -(\tilde{e}^1\wedge \cdots \tilde{e}^{i}) [(\fz^1e_1 + \cdots + \fz^Ne_N)\wedge e_\tb \wedge \tilde{e}^\ta ( e_1 \wedge \cdots \wedge e_{i-1} )] \nonumber\\
    & = 
    \begin{cases}
     \fz^\ta \delta_{i\tb} \theta_{i-1\geq \ta}, & \ta \neq \tb \\
   - \fz^i \theta_{i-1\geq \ta}, & \ta=\tb
    \end{cases}
    \nonumber\\
    & =  \fz^\ta \delta_{i\tb} \theta_{i-1\geq \ta} - \fz^i \theta_{i-1\geq \ta} \delta_\tb^\ta \\
    %%%%%%%%%%%%%%%%%%%%%%%%%%%%%%%%%%%%%%%%%%%%%%%%
    \tilde{\pi}^i(J_\tb^\ta \pi_i) 
    & = -(\tilde{e}^1\wedge \cdots \wedge \tilde{e}^i) (e_\tb\wedge \tilde{e}^\ta (e_1 \wedge \cdots \wedge e_i) ) \nonumber\\
    & =- \delta^\ta_\tb \theta_{\ta\leq i}
\end{align}
\end{subequations}
This gives 
\begin{align}
    J^\ta_\tb \Psi_{0}
    & = \Psi_{0} \left[ \beta_\tb \frac{\fz^\ta}{\fz^\tb} \theta_{\tb > \ta} - \tilde{\beta}_\ta \frac{\tilde{\fz}_\tb}{\tilde{\fz}_\ta}\theta_{\tb > \ta} - \delta^\ta_\tb \sum_{i=1}^N \beta_i \theta_{i-1\geq \ta} - \tilde\beta_i \theta_{i\geq \ta} - \delta^\ta_\tb \sum_{i=1}^{N-1} \alpha_i \theta_{i\geq \ta} \right] \nonumber\\
    & = \Psi_{0} \left[ \beta_\tb \frac{\fz^\ta}{\fz^\tb} \theta_{\tb > \ta} -\tilde{\beta}_\ta \frac{\tilde{\fz}_\tb}{\tilde{\fz}_\ta}\theta_{\tb > \ta} - \delta^\ta_\tb \xi_\ta \right]
\end{align}
with a short handed notation for convenience
$$
    \xi_\ta = \sum_{i=1}^N \beta_i \theta_{i-1\geq \ta} + \tilde\beta_i \theta_{i\geq \ta} + \sum_{i=1}^{N-1} \alpha_i \theta_{i\geq \ta}.
$$
$J^\ta_\tb$ acts on $\chi$ through
\begin{align}
    J^\ta_\tb \chi(v_1,\dots,v_{N-1}) 
    & = \sum_{k=1}^{N-1} (J^\ta_\tb v_k) \cdot \frac{\partial \chi}{\partial v_k} 
    = v_\ta \frac{\tilde{\fz}_\tb}{\tilde{\fz}_\ta} \theta_{\tb > \ta} \frac{\partial \chi}{\partial v_a} - v_b \frac{\fz^\ta}{\fz^\tb} \theta_{\tb > \ta} \frac{\partial \chi}{\partial v_\tb}
\end{align}
with
\begin{align}
    J^\ta_\tb v_k 
    & = v_k \left[ -\frac{\tilde{\fz}_\tb}{\tilde{\fz}_k} \delta^\ta_k \theta_{\tb> k} + \frac{\fz^\ta}{\fz^k}\delta_{k\tb} \theta_{k> \ta} \right].
\end{align}

We find $\hat\sH_{0,1}^{(4)}$ acting on $\chi(v_1,\dots,v_{N-1};\kq)$ of the following form:
\begin{subequations}\label{eq:KZ-4p-conn}
\begin{align}
    \hat\sH_0^{(4)}
    = & - \frac{1}{2}\left( \sum_{\ta=1}^N \nabla^\fz_\ta + \beta_\ta \right)^2 \\
    & + \sum_{\ta=1}^N \frac{1}{2} (\nabla^\fz_\ta + \beta_\ta)^2 + \frac{v_{\ta+1}+\dots+v_{N-1}}{v_\ta} (\nabla^\fz_\ta + \beta_\ta) (\nabla^{\tilde{\fz}}_\ta + \tilde\beta_\ta) - \xi_a (\nabla^\fz_\ta + \beta_\ta) \nonumber \\
    %%%%%%%%%%%%%%%%%%%%%%%%%%%%%%%%%%%%
    \hat\sH_1^{(4)} 
    % & = \hat{\rm H}_1 \Upsilon \chi \nonumber\\
    = &  \sum_{\ta=1}^{N} - \frac{\tilde{\fz}(\fz)}{\tilde{\fz}_\ta \fz^\ta} \left( \fz^\ta \frac{\partial}{\partial \fz^\ta} + \beta_\ta \right) 
    \left( \tilde{\fz}_\ta \frac{\partial}{\partial \tilde{\fz}_\ta} + \tilde{\beta}_\ta \right).
\end{align}
\end{subequations}
Variables $(v_\ta)_{i=1}^{N}$ are given by 
\begin{align}
    v_\ta = \frac{(\tilde{\fz}\wedge \tilde{\pi}^{\ta-1})(\pi_\ta) \cdot \tilde{\pi}^\ta (\fz\wedge \pi_{\ta-1})}{\tilde{\fz}(\fz) \cdot \tilde{\pi}^{\ta-1}(\pi_{\ta-1}) \cdot \tilde{\pi}^i(\pi_\ta)} = \frac{\tilde{\fz}_\ta \fz^\ta}{\tilde{\fz}(\fz)}, 
\end{align}
with
$$
    \tilde{\fz}(\fz) = \sum_{\ta=1}^N \tilde{\fz}_\ta \fz^\ta, \ \implies \sum_{i=1}^N v_i = 1.
$$

Differential operators $\nabla^\fz_\ta$ and $\nabla^{\tilde\fz}_\ta$ acting on $\chi$ in terms of variables $\{v_1,\dots,v_{N-1}\}$ via chain rules:
\begin{subequations}
\begin{align}
    \fz^\ta\frac{\partial}{\partial \fz^\ta} & = \fz^\ta \sum_{i=1}^{N-1} \frac{\partial v_i}{\partial \fz^\ta} \frac{\partial}{\partial v_i} 
    = \sum_{i=1}^{N-1} \fz^a \left( \frac{\delta_{i\ta}\tilde{\fz}_\ta}{\tilde{\fz}(\fz)} - v_i \frac{\tilde{\fz}_\ta}{\tilde{\fz}(\fz)} \right)\frac{\partial}{\partial v_i}
    = v_\ta\frac{\partial}{\partial v_\ta} - v_\ta D, \\
    %%%%%%%%%%%%%%%%%%%%%%%%%%%%%%%%%%%%%%%%%%%%%
    \tilde{\fz}_\ta \frac{\partial}{\partial \tilde{\fz}_\ta} 
    & = \tilde{\fz}_\ta \sum_{i=1}^{N-1} \frac{\partial v_i}{\partial \tilde{\fz}_\ta} \frac{\partial}{\partial v_i} 
    = \sum_{i=1}^{N-1} \tilde{\fz}_\ta \left( \frac{\delta_{i\ta}{\fz}^\ta}{\tilde{\fz}(\fz)} - v_i \frac{{\fz}^\ta}{\tilde{\fz}(\fz)} \right)\frac{\partial}{\partial v_i}
    = v_\ta\frac{\partial}{\partial v_\ta} - v_\ta D,
\end{align}
\end{subequations}
where $D = \sum_{i=1}^{N-1} v_i \partial_{v_i}$. 

\paragraph{}

Fractional $qq$-character \eqref{eq:fracqq} satisfies non-perturbative Dyson-Schwinger equation:
\begin{align}
    \left \langle [x^{-I}] \EX_{\omega}(x) \right \rangle _{\BZ_N}, \ \omega\in \BZ_N, \ I \in \BZ_{>0}.
\end{align}
We take large $x$ expansion of \eqref{eq:fracqq} with the building block $\EY_{\omega}(x)$'s behavior in \eqref{eq:Y-large-x}. The case of $I=1$ gives
\begin{align}
    0 = \left \langle \ve_1 D_{\omega}^{(1)} - \hat\kq_{\omega} \ve_1 D_{\omega-1}^{(1)} + \frac{\ve_1^2}{2}\nu_{\omega}^2 - \ve_1 a_{\omega+1}\nu_{\omega} + \hat\kq_{\omega} \left( \frac{\ve_1^2}{2}\nu_{\omega-1}^2 + (m_{\omega} - a_{\omega}) \ve_1\nu_{\omega-1} + P_{\omega}(a_{\omega}) \right) \right \rangle _{\BZ}
\end{align}
We consider linear combination that eliminate the $\hat c_{\omega} - \hat c_{\omega+1}$ coming from $D_{\omega}^{(1)}$. 
The $\EN=2$ SQCD instanton partition function $\hat\CalZ_{X;c,\sigma^\pm}$ then satisfies differential equation
\begin{align}\label{eq:DS-vanilla}
    \left[ (1-\kq)\kq\ve_1\ve_2\frac{\partial}{\partial\kq} + \hat{\rm H} \right] \hat\CalZ_{X;c,\sigma^\pm} = 0
\end{align}
with
\begin{align}
    \hat{\rm H} = \sum_{\omega=0}^{N-1} \frac{1-\kq}{2} \left( \ve_1^2(\nabla^z_{\omega})^2-2a_{\omega}\ve_1\nabla^z_{\omega} \right) + \frac{z_{\omega+1}+\cdots+z_{\omega+N}}{z_{\omega}} (\ve_1\nabla^z_{\omega} - a_{\omega} + m_{\omega}^+) (\ve_1\nabla^z_{\omega} - a_{\omega} + m_{\omega}^-).
\end{align}
We rewrite \eqref{eq:DS-vanilla} into
\begin{align}\label{eq:DS-KZ}
    \left[ \frac{\ve_2}{\ve_1} \frac{d}{d\kq} + \frac{\hat{\rm H}|_{\kq=0}}{\kq} + \frac{-\hat{\rm H}|_{\kq=1}}{\kq-1} \right] \hat\CalZ_{X;c,\sigma^\pm} = 0
\end{align}
We find eq.~\eqref{eq:KZ-4p} agreeing with \eqref{eq:DS-KZ} with matching between parameters $k+N= - \frac{\ve_2}{\ve_1}$ and variables on the gauge and CFT side:
\begin{align}
    z_\ta = \fz^\ta\tilde\fz_\ta, \quad  v_i = \frac{z_{i}}{z_0+z_1+\cdots+z_{N-1}}.
\end{align}
$\ta=1,\dots,N$, $i=1,\dots,N-1$.
We identify defect partition function $\hat\CalZ_{X;c,\sigma^\pm}$ of four dimensional gauge theory as CFT 4-point correlation function
\begin{align}
    \chi(v_1,\dots,v_{N-1};\kq) = \hat\CalZ_{X;c,\sigma^\pm}(z_0,\dots,z_{N-1};\kq) = \prod_{\omega=0}^{N-1}z_{\omega}^{-\frac{m_{\omega}^+ - a_{\omega}}{\ve_1}} \Psi.
\end{align}
with $\Psi$ is defined in \eqref{eq:pert}.

\section{Some computational details for $5$-point KZ equations}

\subsection{Representation side}\label{sec:5pt derive}
Here we demonstrate how equations \eqref{eq:KZHs} and \eqref{eq:KZAs} are obtained. The 5-point correlation function $\boldsymbol\Upsilon(\kq,y)$ is valued in 
\begin{align}
    \boldsymbol\Upsilon(\kq,y) \in \left(\EuScript{V}_{\boldsymbol\z}  \otimes {\EH}_{\mu}^{\boldsymbol\t - \boldsymbol\z} \otimes \tilde{\EH}_{\tilde{\mu}}^{\tilde{\boldsymbol\t} - \tilde{\boldsymbol\z} + \boldsymbol\d_{N-1}} \otimes \tilde{\EuScript{V}}_{\tilde{\z}}  \otimes \BC^N \right)^{\mathfrak{sl}_N}
\end{align}
The 5-point KZ equations are 
\begin{subequations}
\begin{align}
    & \left[ \frac{\ve_2}{\ve_1}\frac{\partial}{\partial \kq} + \frac{\hat{\rm H}_0}{\kq} + \frac{\hat{\rm H}_1}{\kq-1} + \frac{\hat{\rm H}_y}{\kq-y} \right] \boldsymbol\Upsilon=0 \\
    & \left[ \frac{\ve_2}{\ve_1}\frac{\partial}{\partial y} + \frac{\hat{\rm A}_0}{y} + \frac{\hat{\rm A}_1}{y-1} + \frac{\hat{\rm A}_\kq}{y-\kq} \right] \boldsymbol\Upsilon = 0
\end{align}
\end{subequations}
KZ operators in the numerators in the expression are symmetric tensor product of $\mathfrak{sl}_N$ generators $\{\sT^\sk\}$ represented on the respective modules:
\begin{subequations}
\begin{align}
    & \left( \hat{\rm H}_0 \right) =  \sum_{\ta,\tb=1}^N \fz^\tb %J^\ta_\tb \frac{\partial}{\partial \fz^\ta}, \quad 
    \left( \hat{\rm H}_1 \right) = -\tilde\fz(\fz) \sum_{\ta=1}^N \frac{\partial^2}{\partial \fz^\ta \partial \tilde \fz_\ta}, \quad
    \left( \hat{\rm H}_y \right)_{\ta\tb} = - E^\tb_\ta \fz^\ta \frac{\partial}{\partial \fz^\tb}, \\
    %%%%%%%%%%%%%%%%%%%%%%%%%%%%%%%%%%%%%%%%
    & \left( \hat{\rm A}_0 \right)_{\ta\tb} = E^\tb_\ta J_\tb^\ta, \quad
   \left( \hat{\rm A}_1 \right)_{\ta\tb} = E^\tb_\ta \tilde{\fz}_\tb\frac{\partial}{\partial\tilde{\fz}_\ta}, \quad
    \left( \hat{\rm A}_\kq \right)_{\ta\tb} =  - \left( \hat{\rm H}_y \right)_{\ta\tb} =  E^\tb_\ta \fz^\ta \frac{\partial}{\partial \fz^\tb}.
\end{align}
\end{subequations}

$\boldsymbol\Upsilon(\kq,y)$ is a $N$-dimensional vector in the standard $N$-dimensional representation $\BC^N$ of $\mathfrak{sl}_N$. We denote its bulks by $\tilde{u}_\ta = (\tilde{u} \wedge \tilde{\pi}^{\ta-1}) (\pi_{\ta})$, $\ta=1,\dots,N$. Individual component of vector $\Upsilon$ is 
composed by a $\EG$-invariant function $\chi_\ta(v_1,\dots,v_{N-1};\kq,y)$ and  $\EG$-equivariant function $\Upsilon_{0}^{(\ta)}$:
\beq
\label{eq:slnconfblock}
\boldsymbol\Upsilon = \sum_{\ta=1}^{N} \Upsilon^{(\ta)}_0 \cdot (\tilde{u} \wedge \tilde{\pi}^{\ta-1}) (\pi_{\ta}) \cdot  \chi_\ta (v_{1}, \ldots , v_{N-1}). 
\eeq
The $\EG$-equivariant functions $\Upsilon_{0}^{(\ta)}$, $\ta=1,\dots,N$, take a similar form to $\Psi_{0}$ of the 4-point case \eqref{eq:psi0}: 
\begin{align}
    \Upsilon_{0}^{(\ta)} = 
    & \prod_{\tb=1}^{N} ((\tilde{z}\wedge \tilde{\pi}^{\tb-1})(\pi_\tb))^{\tilde{\beta}_\tb^{(\ta),*}}  (\tilde\pi^\tb (z\wedge\pi_{\tb-1}) )^{\beta_\tb^{(\ta),*}} \prod_{i=1}^{N-1} ( \tilde{\pi}^i(\pi_i) )^{\alpha_i^*}.
\end{align}
so that with group transformation $g=(g_i,\tilde{g}_i,t,\tilde{t})\in \EG$, $\Upsilon_{0}^{(\ta)}$ obeys:
\begin{align}
    \Upsilon_{0}^{(\ta)}[g_{i+1}\tU_ig_i^{-1},\tilde{g}_i\tilde\tU_i\tilde{g}_{i+1}^{-1},t\fz_\ta,\tilde{t}^{-1}\tilde\fz_\ta]
    \prod_{i=1}^{N-1} (\det g_i)^{\zeta_i} (\det \tilde{g}_i)^{-\tilde\zeta_i} t^{\mu} \tilde{t}^{-\tilde\mu} = \Upsilon_{0}^{(\ta)}[\tU_i,\tilde\tU_i,\fz^\ta,\tilde\fz_\ta]. 
\end{align}
The coefficients $\{\beta_\tb^{(\ta),*},\tilde\beta_\tb^{(\ta),*},\alpha_i^*\}$ satisfy similar relations as the 4-point KZ equation \eqref{eq:4pmat} but with a slight modification:
\begin{subequations}
\begin{align}
    & \tilde\beta_i^{(\ta),*} + \beta_{i+1}^{(\ta),*} + \alpha_i^* + { \delta_{i,\ta}} = \zeta_i \\
    & \tilde\beta_{i+1}^{(\ta),*} + \beta_i^{(\ta),*} + \alpha_i^* + {\delta_{i+1,\ta}} = \tilde\zeta_i \\
    & \qquad\qquad i = 1, \ldots , N-1 \nonumber \\
    & \sum_\tb \beta_\tb^{(\ta),*}  = \mu \\
    & \sum_\tb \tilde\beta_\tb^{(\ta),*} = \tilde{\mu}
\end{align}
\end{subequations}
In particular, the HW-module weight $\tilde\mu$ of $\EuScript{V}_{\tilde\mu}$ is shifted from its 4-point counterpart by
\begin{align}
    \tilde\mu = \tilde\mu^{(4)} - 1.
\end{align}
Such shift has been observed in \cite{Jeong:2018qpc}. Given a set of solution $\{\beta_\ta,\beta_\ta,\alpha_i\}$, $\ta=1,\dots,N$, $i=1,\dots,N-1$, to 4-point KZ condition \eqref{eq:4pmat}, solutions to \eqref{eq:bbg} are
\begin{align}
  \beta_\tb^{(\ta),*} = \beta_\tb, \quad \tilde\beta_\tb^{(\ta),*} = \tilde\beta_\tb^* - \delta_{\ta\tb}, \quad \alpha_i^* = \alpha_i. 
\end{align}
for $\ta,\tb=1,\dots,N$, $i=1,\dots,N-1$. The operators $\hat\EH_{0,1,y}$ and $\hat\EA_{0,1,\kq}$ are obtained by commuting $\Upsilon_{0}^{(\ta)}\tilde{u}_\ta$ through $\hat{\rm H}_{0,1,y}$ and $\hat{\rm A}_{0,1,\kq}$ respectively.

Here we demonstrate how the equations \eqref{eq:KZHs} and \eqref{eq:KZAs} are obtained. Let us define $\boldsymbol\chi = \sum_{\tb=1}^N \tilde{u}^\tb \chi_\tb(v_1,\dots,v_{N-1};\kq,y)$ to study how the KZ equations are expressed as differential equations annihilating individual $\chi_\tb(v_1,\dots,v_{N-1};\kq)$. The 5-point KZ equations for $\boldsymbol\chi$ are denoted as 
\begin{subequations}
\begin{align}
    & \left[-(k+N) \frac{\partial}{\partial \kq} + \frac{\hat\sH_0}{\kq} + \frac{\hat\sH_1}{\kq-1} + \frac{\hat\sH_y}{\kq-y} \right] \boldsymbol\chi = 0 \\
    & \left[ -(k+N) \frac{\partial}{\partial y} + \frac{\hat\EA_0}{y} + \frac{\hat\EA_1}{y-1} + \frac{\hat\EA_y}{y-\kq} \right] \boldsymbol\chi = 0
\end{align}
\end{subequations}
First we find $\hat\sH_1$:
\begin{align}\label{eq:KZ5-H1}
    \left(\hat\sH_1\right)_{\ta\tb} 
    & = \left[ - \sum_{\tc=1}^{N} \frac{\tilde{\fz}(\fz)}{\fz^\tc\tilde{\fz}_\tc} \left(\fz^\tc\frac{\partial}{\partial \fz^\tc} + \beta_\tc \right) \left( \tilde{\fz}_\tc\frac{\partial}{\partial \tilde{\fz}_\tc}+ \tilde\beta_{\tc} - \delta_{\tc\ta} \right) \right]\delta_{\ta\tb} \\
    & =  \hat\sH_1^{(4)} \delta_{\ta\tb} + \frac{z_0+\cdots+z_{N-1}}{z_\ta} (\nabla^z_\ta+\beta_\ta) \delta_{\ta\tb}. \nonumber
\end{align}

The action of $\hat\sH_y = - \hat\EA_\kq$ acting on $\chi_\tb$ can be found by taking
\begin{align}\label{eq:KZ5-Hq/Ay}
    \left( \hat\sH_y \right)_{\ta\tb} \chi_\tb
    & =\left( \Upsilon_{0}^{(\ta)} \tilde{u}_\ta\right)^{-1} \left( \hat{\rm H}_y \right)_{\ta\tb} \Upsilon_{0}^{(\tb)} \tilde{u}_\tb \chi_\tb \nonumber\\ 
    % = -\left({\rm A}_\kq\right)_{ab} \Phi
    & =- \left( \Upsilon_{0}^{(\ta)} \tilde{u}_\ta\right)^{-1}  E^\tb_\ta \fz^\ta \frac{\partial}{\partial \fz^\tb} \Upsilon_{0}^{(\tb)} \tilde{u}_\tb \chi_\tb \nonumber\\
    % & = \left( \Upsilon_{0}^{(a)} \tilde{u}_a\right)^{-1} \left( z^a T^b_a \frac{\partial}{\partial z^b} \right) \sum_j \Upsilon_{0}^{(j)}\tilde{u}_j\chi_j \nonumber\\
    & = - \left( \Upsilon_{0}^{(\ta)} \tilde{u}_\ta\right)^{-1} \fz^\ta\frac{\partial}{\partial \fz^\tb} \left( \tilde{u}_\ta\frac{\partial}{\partial \tilde{u}_\tb} \right) \Upsilon_{0}^{(\tb)} \tilde{u}_\tb \chi_\tb \nonumber\\
    & = - \frac{\Upsilon_{0}^{(\tb)}}{\Upsilon_{0}^{(\ta)}}  \frac{\fz^\ta}{\fz^\tb} \left( \fz^\tb\frac{\partial}{\partial \fz^\tb} + \beta_\tb^{(\tb),*}  \right) \chi_\tb \nonumber\\
    & = - \frac{\tilde{\fz}_\ta \fz^\ta}{\tilde{\fz}_\tb \fz^\tb} \left( \fz^\tb\frac{\partial}{\partial \fz^\tb} + \beta_\tb \right) \chi_\tb \nonumber\\
    &= - \frac{z_{\ta}}{z_{\tb}} \left( \nabla^z_{\tb} + \b_\tb \right) \chi_\tb.
\end{align}
Similarly the action of $\hat\EA_1$ on $\chi_\tb$ can be found by
\begin{align}\label{eq:KZ5-A1}
    \left( \hat\EA_1 \right)_{\ta\tb}\chi_\tb
    & = \left( \Upsilon_{0}^{(\ta)} \tilde{u}_\ta\right)^{-1} \left( \hat{\rm A}_1 \right)_{\ta\tb} \Upsilon_{0}^{(\tb)} \tilde{u}_\tb \chi_\tb \nonumber\\
    & = \left( \Upsilon_{0}^{(\ta)} \tilde{u}_\ta\right)^{-1} \left( - \tilde{\fz}_\tb \frac{\partial}{\partial \tilde{\fz}_\ta} \tilde{u}_\ta \frac{\partial}{\partial\tilde{u}_\tb} \right) \Upsilon_{0}^{(\tb)} \tilde{u}_\tb \chi_\tb \nonumber\\ 
    & = - \frac{1}{\Upsilon_{0}^{(\ta)}} \left( \tilde{\fz}_\tb \frac{\partial}{\partial \tilde{\fz}_\ta} \right)  \Upsilon_{0}^{(\tb)} \chi_\tb \nonumber\\
    & = - \frac{\Upsilon_{0}^{(\tb)}}{\Upsilon_{0}^{(\ta)}} \frac{\tilde{\fz}_\tb}{\tilde{\fz}_\ta} \left( \tilde{\fz}_\ta\frac{\partial}{\partial\tilde{\fz}_\ta} + \tilde{\beta}_\ta^{(\tb),*} \right) \chi_\tb \nonumber\\
    & = - \left( \tilde{\fz}_\ta\frac{\partial}{\partial\tilde{\fz}_\ta} + \tilde{\beta}_\ta - \delta_{\ta\tb} \right) \chi_\tb \nonumber\\
    & = - \left( \nabla^z_\ta + \tilde\beta_\ta - \delta_{\ta\tb} \right)\chi_{\tb}
\end{align}

Action of $\hat\EA_0$ on $\chi_\tb$ is: 
\begin{align}\label{eq:KZ5-A0}
    \left( \hat\EA_0 \right)_{\ta\tb} \chi_\tb 
    & = \left( \Upsilon_{0}^{(\ta)}  \tilde{u}_\ta\right)^{-1} (\hat{\rm A}_0)_{\ta\tb} \ \Upsilon_{0}^{(\tb)} (\tilde{u} \wedge \tilde\pi^{\tb-1})(\pi_\tb) \chi_\tb \\ 
    & = \left( \Upsilon_{0}^{(\ta)} \tilde{u}_\ta\right)^{-1} (E^\tb_\ta J_\tb^\ta) \psi^{(\tb)}_0 (\tilde{u} \wedge \tilde\pi^{\tb-1})(\pi_\tb) \chi_\tb \nonumber \\
    & = -\frac{1}{\Upsilon_{0}^{(\ta)}} \frac{\partial}{\partial \tilde{u}_\tb}  \left[ ( J^\ta_\tb \psi^{(\tb)}_0 ) \tilde{u}_\tb \chi_\tb + \psi^{(b)}_0 (J^\ta_\tb (\tilde{u} \wedge \tilde\pi^{\tb-1})(\pi_\tb) ) \chi_\tb + \psi^{(j)}_0 \tilde{u}_\tb (J^\ta_\tb \chi_\tb) \right] \nonumber\\
    % & = \frac{1}{\Upsilon_{0}^{(a)}} \left[ \Upsilon_{0}^{(b)} (J^a_b \chi_b) + (J^a_b \psi^{(b)}) \chi_b + \psi \theta_{b > a} \right] \nonumber\\
    & = \frac{\psi^{(\tb)}_0}{\psi^{(\ta)}_0} \left[ v_\ta \frac{\tilde{\fz}_\tb}{\tilde{\fz}_\ta} \theta_{\tb > \ta} \frac{\partial }{\partial v_\ta} - v_\tb \frac{\fz^\ta}{\fz^\tb} \theta_{\tb > \ta} \frac{\partial }{\partial v_\tb} - \beta_\tb^{(\ta),*} \frac{\fz^\ta}{\fz^\tb} \theta_{\tb> \ta} + \tilde{\beta}_\ta^{(\tb)*} \frac{\tilde{\fz}_\tb}{\tilde{\fz}_\ta}\theta_{\tb > \ta}  \right. \nonumber\\
    & \qquad\qquad \left. + \delta^\ta_\tb + \delta^\ta_\tb \sum_{i=1}^N \beta_i^{(\tb),*} \theta_{i > \ta} + \tilde\beta_i^{(\tb),*}\theta_{i\geq \ta} + \delta^\ta_\tb \sum_{i=1}^{N-1} \alpha_i^* \theta_{i\geq \ta} \right]\chi_\tb \nonumber \\
    & = \left[ z_\ta \left( \frac{\partial}{\partial z_\ta} - \frac{\partial}{\partial z_\tb} \right) - \frac{z_\ta}{z_\tb} \beta_\tb \theta_{\tb>\ta} + \tilde\beta_\ta \theta_{\tb>\ta} - \delta^\ta_\tb \xi_\ta \right] \chi_\tb
\end{align}
% We rewrite the differentiation w.r.t. $w$ with
% \begin{align}
%     \frac{\tilde{z}_a}{\tilde{z}_b} \left[ v_a \frac{\tilde{z}_b}{\tilde{z}_a}\frac{\partial}{\partial v_a} - v_b \frac{z^a}{z^b}\frac{\partial}{\partial v_b}  \right] \theta_{b>a} 
%     = \left[ \nabla^w_a + v_a D - \frac{z^a\tilde{z}_a}{z^b \tilde{z}_b} (\nabla^w_b + v_b D) \right] \theta_{b > a}
%     = w_\alpha \left[ \frac{\partial}{\partial w_a} - \frac{\partial }{\partial w_b}\right] \theta_{b > a}.
% \end{align}
% And the diagonal terms are given by
% \begin{align}
%     \sum_{i=1}^N \beta_i^*\theta_{i>a}+\tilde\beta_i^* \theta_{i\geq a} + \sum_{i=1}^{N-1} \alpha_i^* \theta_{i\geq a} = \sum_{i=a}^{N-1} \tilde{\zeta}_i + \tilde\beta_{N}^* 
% \end{align}
% We recover $\hat\CalA_0$ from the gauge theory side in \eqref{eq:CalA2-new}. 
Last but not least is $\hat\sH_0$. We consider action of $\hat{\rm H}_0$ on $\Upsilon_{0}^{(\tc)}\tilde{u}_\tc \chi_\tc$ for some $\tc=1,\dots,N$. When $J^\ta_\tb$ does not act on $\tilde{u}_{\tc} = (\tilde{u}\wedge \tilde\pi^{\tc-1})(\pi_\tc)$, it gives only diagonal element in the $N\times N$ matrix:
\begin{align}\label{eq:KZ5-H0diag}
    % \hat{\rm H}_0  = 
    & -\left(\Upsilon_{0}^{(\tc)}\right)^{-1}\sum_{\ta,\tb} \fz^\tb \frac{\partial}{\partial \fz^\ta} J^a_b \Upsilon_{0}^{(\tc)}\chi_\tc \\
    = & \sum_{\ta,\tb} \frac{\fz^\tb}{\fz^\ta} \left[ \fz^\ta \frac{\partial}{\partial \fz^\ta} + \beta_\ta^{(\tc),*} \right]  \left[ \left( v_\ta \frac{\tilde{\fz}_\tb}{\tilde{\fz}_\ta}\frac{\partial}{\partial v_a} - v_\tb \frac{\fz^\ta}{\fz^\tb} \frac{\partial}{\partial v_\tb} - \beta_\tb^{(\tc),*} \frac{\fz^\ta}{\fz^\tb} + \tilde\beta_\ta^{(\tc),*} \frac{\tilde{\fz}_\tb}{\tilde{\fz}_\ta} \right) \theta_{\tb>\ta} - \delta^\ta_\tb \left(\theta_{\tc\geq\tb} + \xi_\ta \right) \right]\chi_\tc \nonumber\\
    = & \sum_{\tb>\ta} \frac{\fz^\tb}{\fz^\ta} \left[ \fz^\ta \frac{\partial}{\partial \fz^\ta} + \beta_\ta \right] \left[ \frac{\tilde{\fz}_\tb}{\tilde{\fz}_\ta} \left( \tilde{\fz}_\ta\frac{\partial}{\partial \tilde{\fz}_\ta} + v_a D \right) -\frac{\fz^\ta}{\fz^\tb} \left( \fz^\tb \frac{\partial}{\partial \fz^\tb} + v_\tb D \right) - \beta_\tb \frac{\fz^\ta}{\fz^\tb} + (\tilde\beta_\ta - \delta_{\ta\tc}) \frac{\tilde{\fz}_\tb}{\tilde{\fz}_\ta} \right]\chi_\tc  \nonumber\\
    & - \sum_{\ta=1}^N  \left( \fz^\ta \frac{\partial}{\partial \fz^\ta} + \beta_\ta \right) \left(\theta_{\tc\geq\tb} + \xi_\ta \right)\chi_\tc \nonumber\\
    = & \left[\sum_{\tb>\ta} \frac{\fz^\tb\tilde{\fz}_\tb}{\fz^\ta\tilde{\fz}_\ta} (\nabla^\fz_\ta+\beta_\ta) \nabla^{\tilde{\fz}}_\ta - (\nabla^\fz_\ta+\beta_\ta) \nabla^\fz_\tb - \beta_\tb (\nabla^\fz_\ta+\beta_\ta) + ( \tilde\beta_\ta - \delta_{\ta\tc}) \frac{\fz^\tb\tilde{\fz}_\tb}{\fz^\ta\tilde{\fz}_\ta} (\nabla^\fz_\ta+\beta_\ta) \right. \nonumber\\
    & \left. - \sum_{\ta=1}^{N} \left(\theta_{\tc\geq\tb} + \xi_\ta \right) (\nabla^\fz_\ta+\beta_\ta) \right]\chi_\tc \nonumber \\
    = & - \frac{1}{2}\left( \sum_{\ta=1}^N \nabla^\fz_\ta + \beta_\ta \right)^2 \chi_\tc \nonumber\\
    & + \left[ \sum_{\ta=1}^N \frac{1}{2} (\nabla^\fz_a + \beta_\ta)^2 + \frac{v_{\ta+1}+\cdots+v_{N-1}}{v_\ta} (\nabla^\fz_\ta + \beta_\ta) (\nabla^{\tilde{\fz}}_\ta + \tilde\beta_\ta) - \left( \theta_{\tc \geq \ta} + \xi_\ta \right) (\nabla^\fz_\ta + \beta_\ta) \right]\chi_\tc \nonumber
\end{align}

Unlike $\hat\sH_1$, $\hat\sH_0$ may have off-diagonal components when $J^\ta_\tb$ acts on the unit vector $(\tilde{u} \wedge \tilde\pi^{\tc-1})(\pi_\tc)$:
\begin{align}\label{eq:KZ5-H0Off}
    & \sum_{\ta,\tb} \frac{\fz^\tb}{\fz^\ta} \Upsilon^{(\tc)}_0 \left[ \fz^\ta\frac{\partial}{\partial \fz^\ta} + \beta_\ta \right] (J^\ta_\tb (\tilde{u} \wedge \tilde\pi^{\tc-1})(\pi_\tc)) \chi_\tc \nonumber\\
    & = \sum_{\ta,\tb} \frac{\fz^\tb}{\fz^\ta} \Upsilon^{(\tc)}_0 (\nabla^\fz_\ta + \beta_\ta) (\tilde{u}_\tb\delta^\ta_\tc \theta_{\tb>\tc} + \tilde{u}_\tc \delta^\ta_\tb \theta_{\tc\geq \ta}) \chi_\tc \nonumber\\
    & = \sum_{\tb} \frac{\fz^\tb\tilde{\fz}_\tb}{\fz^\tc\tilde{\fz}_\tc} \Upsilon^{(\tb)}_0 \tilde{u}_\tb (\nabla^\fz_\tc + \beta_\tc ) \theta_{\tb>\tc} \chi_\tc + \psi_{0}^{(\tc)} \tilde{u}_\tc (\nabla^\fz_\tb + \beta_\tb) \theta_{\tc\geq \tb} \chi_\tc
\end{align}
We obtain KZ operator $\hat\sH_0$ as an $N \times N$ matrix with the combination of \eqref{eq:KZ5-H0diag} and \eqref{eq:KZ5-H0Off}: 
\begin{align}\label{eq:KZ5-H0}
    \left( \hat\sH_0 \right)_{\ta\tb} =
    & - \delta_{\ta\tb} \frac{1}{2} \left( \sum_{\tc=1}^{N} \nabla^\fz_\tc + \beta_\tc \right)^2  \\
    & + \delta_{\ta\tb} \left[ \sum_{\tc=1}^N \frac{1}{2}(\nabla^\fz_\tc + \beta_\tc)^2 + \frac{v_{\tc+1}+\cdots+v_{N-1}}{v_\tc}(\nabla^\fz_\tc + \beta_\tc)(\nabla^{\tilde{\fz}}_\tc + \tilde\beta_\tc - \delta_{\ta\tc}) - \xi_\tc(\nabla^z_c + \beta_\tc) \right] \nonumber\\
    & + \frac{\fz^\ta\tilde{\fz}_\ta}{\fz^\tb\tilde{\fz}_\tb}(\nabla^\fz_\tb+\beta_\tb)\theta_{\ta>\tb} \nonumber\\
    = & \ \hat\sH_0^{(4)}\delta_{\ta\tb} - \frac{z_{\ta+1}+\cdots+z_{N-1}}{z_\ta} (\nabla^z_\ta + \beta_\ta) \delta_{\ta\tb} + \frac{z_\ta}{z_\tb} (\nabla^z_\tb +\beta_\tb) \theta_{\ta>\tb} \nonumber
\end{align}
% Here we again recover $\hat\fA_0$ in \eqref{eq:fA2-new}. 
Notice that when acting on $\chi_\tb(v_1,\dots,v_{N-1})$, the operator sum
\begin{align}
    \sum_{\ta=1}^{N} \nabla^\fz_\ta \chi_\tb
    & = \left[ \left( \sum_{i=1}^{N-1} v_i \frac{\partial}{\partial v_i} - v_i D \right) + (v_1+\cdots+v_{N-1} - 1)D \right]\chi_\tb = 0.
\end{align}
This also applies to $\sum_\omega\nabla^z_\omega \chi_\tb = \sum_\ta\nabla^{\tilde{\fz}}_\ta \chi_\tb = 0$. 

\subsection{Gauge theory side}
\subsubsection{The $y$-component of the KZ equation}\label{sec:y-KZ derive}

Here are the details of the computation leading to \eqref{eq:CalA2-new}.
Using ${\bf U}$ defined in \eqref{def:U-matrix}
\begin{align}
    {\bf U} := \begin{pmatrix}
    0 & 1 & 0 & \cdots & 0 & 0 \\
    0 & 0 & 1 & \cdots & 0 & 0 \\
    & \vdots & & \ddots & & \vdots \\
    0 & 0 & 0 & \cdots & 0 & 1 \\
    y & 0 & 0 & \cdots & 0 & 0
    \end{pmatrix}, \quad 
    {\bf U}^{-1} = \begin{pmatrix}
    0 & 0 & 0 & \cdots & 0 & \frac{1}{y} \\
    1 & 0 & 0 & \cdots & 0 & 0 \\
    0 & 1 & 0 & \cdots & 0 & 0 \\
    & \vdots & & \ddots & & \vdots \\
    0 & 0 & 0 & \cdots & 1 & 0
    \end{pmatrix}, \quad  {\bf U}^N = y {\bf I}_N.
\end{align}
and 4 diagonal matrices defined as in \eqref{def:diagmatrices}:
\begin{align}
    {\bf M}^\pm = {\rm diag} ({m}_0^\pm,\dots,{m}_{N-1}^\pm), \ \boldsymbol{\qe} = {\rm diag}(\qe_0,\dots,\qe_{N-1}), \ {\boldsymbol{\rho}} = {\rm diag}({\rho}_0,\dots,{\rho}_{N-1}),
\end{align}
$N$ fractional quantum T-Q equations \eqref{eq:qTQvev} can be rewrite into a single $N\times N$ matrix equation in Fourier space as
% \begin{align}
%     & \left[ {\bf U} (x-{\bf M}^+) + \boldsymbol{\qe} ( x - {\bf M}^-_{\ast} ) {\bf U}^{-1} - ( {\bf I}_N+\boldsymbol{\qe} )x +{\boldsymbol{\rho}} \right] \boldsymbol\Upsilon (y) = 0 \nonumber\\
%     \implies & \left[ - x + ({\bf U} + \boldsymbol{\qe}{\bf U}^{-1} - {\bf I}_N -\boldsymbol{\qe})^{-1} \left({\bf U}{\bf M}^+ + \boldsymbol{\qe} {\bf M}^- {\bf U}^{-1} - {\boldsymbol{\rho}}\right) \right] \boldsymbol\Upsilon(y) = 0
% \end{align}
% with
% \begin{align}
%     -x \boldsymbol\Upsilon(y) = \ve_2 y \frac{\partial }{\partial y} \boldsymbol\Upsilon(y) - \left( \ve_2 y\frac{\partial}{\partial y} \log\psi(y) \right) \boldsymbol\Upsilon(y)
% \end{align}
$\boldsymbol\Upsilon$ is the Fourier transformation of the fractional ${Q}$-observables \eqref{def:hatQ}:
\begin{align}
    \left[ ({\bf U} + \boldsymbol{\kq} {\bf U}^{-1} - {\bf I}_N - \boldsymbol{\kq}) \left( \ve_2 y \frac{\partial }{\partial y}  - \ve_2 y\frac{\partial \log\Upsilon^{\text{pert}}(y) }{\partial y}  \right)  + \left({\bf U}{\bf M}^+ + \boldsymbol{\qe} {\bf M}^- {\bf U}^{-1} - {\boldsymbol{\rho}}\right) \right] \boldsymbol\Upsilon(y) = 0
\end{align}
Let us consider the following change of basis
\begin{align}
    \boldsymbol\Pi = ({\bf U} - {\bf I}) \boldsymbol\Upsilon.
\end{align}
Vector $\boldsymbol\Pi$ satisfies 
\begin{align}
    0  
    & = \left[ \frac{\ve_2}{\ve_1} \frac{\partial}{\partial y} - \frac{\ve_2}{\ve_1} \frac{\partial \log \Upsilon^{\text{pert}}(y) }{\partial y} + \frac{1}{\ve_1}\frac{1}{y}\left ({\bf I}_N - {\boldsymbol\kq}{\bf U}^{-1} \right)^{-1} \left({\bf U}{\bf M}^+ + \boldsymbol{\qe} {\bf M}^- {\bf U}^{-1} - {\boldsymbol{\rho}}\right) \left( {\bf U} - {\bf I}_{N} \right)^{-1} \right]\boldsymbol\Pi \nonumber\\
    % & = \left[ \ve_2 \frac{\partial}{\partial y} - \ve_2 \frac{\partial \log\psi(y) }{\partial y} \right. \nonumber\\
    % & \qquad \left.+ \frac{1}{y}\left ({\bf I}_N - {\boldsymbol\kq}{\bf U}^{-1} \right)^{-1} \left({\bf U}{\bf M}^+ + \boldsymbol{\qe} {\bf M}^- {\bf U}^{-1} - {\bf U}{\bf M}^+{\bf U}^{-1} - {\boldsymbol\kq}{\bf M}^- + {\boldsymbol\nabla} - {\boldsymbol\kq}{\bf U}^{-1}{\boldsymbol\nabla}{\bf U}\right) \left( {\bf U} - {\bf I}_{N} \right)^{-1} \right]\boldsymbol\Pi \nonumber\\
    & = \left[ \frac{\ve_2}{\ve_1} \frac{\partial}{\partial y} - \frac{\ve_2}{\ve_1} \frac{\partial \log \Upsilon^{\text{pert}}(y) }{\partial y} + \frac{\CalA(y)}{(y-1)(y-\kq)} \right] \nonumber\\
    & := \left[ \frac{\ve_2}{\ve_1} \frac{\partial}{\partial y} + \frac{\hat\CalA_0}{y} + \frac{\hat\CalA_1}{y-1} + \frac{\hat\CalA_\kq}{y-\kq} \right]\boldsymbol\Pi. 
\end{align}
Matrix $\CalA$ is given by
\begin{align}
    \CalA(y)_{\ta\tb} & = \left[ - \frac{1}{\ve_1} \sum_{j=0}^{N-1}\left( {\boldsymbol\kq}{\bf U}^{-1} \right)^{j} \left({\bf U}{\bf M}^+ + \boldsymbol{\qe} {\bf M}^- {\bf U}^{-1} - {\bf U}{\bf M}^+{\bf U}^{-1} - {\boldsymbol\kq}{\bf M}^- + {\boldsymbol\nabla} - {\boldsymbol\kq}{\bf U}^{-1}{\boldsymbol\nabla}{\bf U}\right) \sum_{j'=0}^{N-1} {\bf U}^{j'}  \right]_{\ta\tb} \nonumber\\
    & = (1-y)\frac{z_\ta}{z_\tb} \left[ \frac{m_\tb^+}{\ve_1} \left(\frac{\kq}{y}\right)^{\theta_{\ta<\tb}} - \frac{m_{\tb}^-}{\ve_1} \left(\frac{\kq}{y}\right)^{\theta_{\ta<\tb+1}}  \right] - z_\ta \sum_{\tc=1}^N \left(\frac{\kq}{y}\right)^{\theta_{\ta<\tc}} y^{\theta_{\tb<\tc}} \left( \frac{\partial}{\partial z_{\tc}} - \frac{\partial}{\partial z_{\tc-1}} \right)
\end{align}
with $\ta,\tb=1,\dots,N$. We find trace of matrix $\CalA(y)$:
\begin{align}
    \Tr \CalA(y) & = (1-y) \frac{1}{\ve_1} \Tr {\bf M}^+ + \frac{y-1}{y} \frac{\kq}{\ve_1} \Tr{\bf M}^- + (\kq-1) \frac{1}{\ve_1} \Tr \boldsymbol\nabla.
\end{align}
Since the dependence of $z_{\omega}$ in the non-perturbative part of $\boldsymbol\Upsilon(y)$ only comes through the fractional couplings $\hat\qe_\o$, it is annihilated by $\Tr \boldsymbol\nabla$. Hence the only contribution comes from the perturbative factor in eq.~\eqref{eq:pert}, we have
\begin{align}
\Tr {\boldsymbol\nabla} \boldsymbol\Upsilon = \ve_1 \sum_{\o=0} ^{N-1} z_\o \p_{z_\o} \boldsymbol\Upsilon = \sum_{\omega=0}^{N-1} m_{\omega}^+ - a_{\omega}.
\end{align}
Let us choose $\Upsilon^{\text{pert}}(y)$ in \eqref{def:upsilon} so that KZ-connection is traceless:
\begin{align}
    \Upsilon^{\text{pert}}(y)= y^{-\frac{m^-}{N\ve_2}} (y-q)^{\frac{m^--a}{N\ve_2}} (y-1)^{-\frac{m^+-a}{N\ve_2}}
\end{align}
with short handed notation
\begin{align}
\begin{split}
   & a := \sum_{\o=0} ^{N-1} a_\o, \quad m^\pm := \sum_{\o=0} ^{N-1} m^\pm_\o. \\
\end{split}
\end{align}
We obtain individual $\hat{\CalA}_{0,1,\kq}$:
\begin{subequations}
\begin{align}
    \left( \hat{\CalA}_0 \right)_{\ta\tb} 
    = & \ \frac{z_{\ta}}{z_{\tb}} \left[ \frac{m_{\tb}^+}{\ve_1}\theta_{\ta<\tb} - \frac{m_{\tb}^-}{\ve_1} \theta_{\ta\leq\tb}\right] + z_{\ta} \left( \frac{\partial}{\partial z_{\ta}} - \frac{\partial}{\partial z_{\tb}} \right)  \theta_{\ta<\tb} + \frac{\delta_{\ta\tb}}{N} \frac{m^-}{\ve_1},\\
    %%%%%%%%%%%%%%%%%%%%%%%%%%%%%%%%%%%%%%%%%%
    \left( \hat{\CalA}_1 \right)_{\ta\tb} 
    % = & \frac{1}{1-\kq} \left[ u_{\omega'-1}m_{\omega'}^+ + u_{\omega'+1}\kq_{\omega'+1} m_{\omega'+1}^- + u_{\omega'} \left( -m_{\omega'+1}^+ - \kq_{\omega'}m_{\omega'}^- + \ve_1\nabla^z_{\omega'} - \ve_1\kq_{\omega'}\nabla^z_{\omega'-1} \right) \right] \nonumber \\
    = & \ - z_{\ta}\frac{\partial}{\partial z_{\ta}}  + \frac{\delta_{\ta\tb}}{N} \frac{m^+ - a}{\ve_1} , \\
    %%%%%%%%%%%%%%%%%%%%%%%%%%%%%%%%%%%%%%%%%%
    \left( \hat{\CalA}_\kq \right)_{\ta\tb} 
    % = & \frac{\gamma_{\omega}}{z_{\omega'-1}}m_{\omega'}^ + +\frac{\gamma_{\omega}}{z_{\omega'}} m_{\omega'+1}^- + \frac{\gamma_{\omega}}{z_{\omega'}} (-m_{\omega'+1}^+ - \kq_{\omega'}m_{\omega'}^- + \ve_1\nabla^z_{\omega'} - \ve_1\kq_{\omega'}\nabla^z_{\omega'-1}) \nonumber\\
    = & \ \frac{z_{\ta}}{z_{\tb}} \left[  z_{\tb} \frac{\partial }{\partial z_{\tb}} + \frac{m_{\tb}^- - m_{\tb}^+}{\ve_1}  \right] + \frac{\delta_{\omega,\omega'}}{N} \frac{a - m^-}{\ve_1},
\end{align}
\end{subequations}
$\ta,\tb=1,\dots,N$. We multiplied perturbative factor \eqref{eq:pert} to expectation value of $\langle Q_{\omega}(x) \rangle_{\BZ_N} \Psi$. We may modify derivative terms 
\begin{align}
    \nabla^z_\tb \mapsto \nabla^z_\tb + \frac{m_\tb^+ - a_\tb}{\ve_1}
\end{align}
when operators $\hat{\CalA}$ acting in the non-perturbative parts in $\boldsymbol\Pi$.
We find the KZ-connections appearing in the $y$-component of 5-point KZ equation agrees with representation theory data:
\begin{align}
    \hat\CalA_0 = \hat\EA_0, \ \hat\CalA_1 = \hat\EA_1, \ \hat\CalA_\kq = \hat\EA_\kq.
\end{align}

\subsubsection{The $\kq$-component of the KZ equation} \label{sec:q-KZ derive}

Here we derive the $\kq$-component of the KZ equation \eqref{eq:fA2-new} from the non-perturbative Dyson-Schwinger equation obeyed by
the fractional $qq$-character \eqref{eq:qyeq-frac} 
\begin{align}
    \left\langle \left[x^{-I} \right] \hat{T}_{N+1,\omega}(x) Q_{\omega'}(x') \right \rangle_{\BZ_N} = 0, \ \forall I = 1,2,\dots
\end{align}

By exploiting the large $x$ expansion of the building block $\EY_{\omega}(x)$ \eqref{eq:Y-large-x},
we can expand fractional Dyson-Schwinger equation in large $x$ and equates the coefficient of $x^{-1}$ in \eqref{eq:qyeq-frac} to zero. One obtains when $\omega\neq\omega'$:
\begin{align}
    \Biggl\langle & \left[\ve_1 D_{\omega}^{(1)} - \hat\qe_{\omega} \ve_1 D_{\omega-1}^{(1)} + \frac{\ve_1^2}{2}\nu_{\omega}^2 - \ve_1 {a}_{\omega+1}\nu_{\omega} \right. \nonumber\\ 
    & \quad \left. + \hat\qe_{\omega} \left( \frac{\ve_1^2}{2}\nu_{\omega-1}^2 - \frac{\ve_1^2}{2}\nu_{\omega-1} + ({m}_{\omega}-{a}_{\omega})\ve_1\nu_{\omega-1} + P_{\omega}(a_{\omega}) \right) \right] {Q}_{\omega'}(x') \Biggr\rangle _{\BZ_N} = 0
\end{align}
In the case $\omega=\omega'$, we simply take \eqref{eq:T_N+1,=}:
\begin{align}
    \hat{T}_{N+1,\omega}(x)
    = & (x-x')(x+{a}_{\omega+1} + \ve_1\nu_{\omega} ) + \hat\qe_{\omega} (x-x'+\ve_1)(x-{m}_{\omega} + {a}_{\omega} - \ve_1\nu_{\omega-1}) \nonumber\\
    & + \ve_1 D_{\omega}^{(1)} - \hat\qe_{\omega} \ve_1 D_{\omega-1}^{(1)} + \frac{\ve_1^2}{2}\nu_{\omega}^2 - \ve_1 {a}_{\omega+1}\nu_{\omega} \nonumber\\ 
    & + \hat\qe_{\omega} \left( \frac{\ve_1^2}{2}\nu_{\omega-1}^2 + ({m}_{\omega}-{a}_{\omega})\ve_1\nu_{\omega-1} + {P}_{\omega}({a}_{\omega}) \right)
\end{align}
which gives, when $x=x'$:
\begin{align}
    & \left\langle \hat\kq_{\omega}{P}_{\omega}(x) {Q}_{\omega-1}(x) \right\rangle _{\BZ_N} \\
    = & \Biggl\langle \Bigg[ \hat\qe_{\omega} \ve_1(x-{m}_{\omega} + {a}_{\omega} - \ve_1\nu_{\omega-1})  + \ve_1 D_{\omega}^{(1)} - \hat\qe_{\omega} \ve_1 D_{\omega-1}^{(1)}  + \frac{\ve_1^2}{2}\nu_{\omega}^2 - \ve_1 {a}_{\omega+1}\nu_{\omega}  \nonumber\\
    & + \hat\qe_{\omega} \left( \frac{\ve_1^2}{2}\nu_{\omega-1}^2 + ({m}_{\omega}-{a}_{\omega})\ve_1\nu_{\omega-1} + P_{\omega}(a_{\omega}) \right) \Bigg] {Q}_{\omega}(x) \Biggr\rangle _{\BZ_N}, \nonumber
\end{align}
also when $x+\ve_1=x'$:
\begin{align}
    & -\ve_1\left\langle {Q}_{\omega+1}(x) \right\rangle _{\BZ_N} \\
    = & \Biggl\langle \Bigg[ -\ve_1(x - {a}_{\omega+1} + \ve_1\nu_{\omega})  + \ve_1 D_{\omega}^{(1)} - \hat\qe_{\omega} \ve_1 D_{\omega-1}^{(1)} + \frac{\ve_1^2}{2}\nu_{\omega}^2 - \ve_1 {a}_{\omega+1}\nu_{\omega} \nonumber\\
    & + \hat\qe_{\omega} \left( \frac{\ve_1^2}{2}\nu_{\omega-1}^2 + ({m}_{\omega}-{a}_{\omega})\ve_1\nu_{\omega-1} + P_{\omega}(a_{\omega}) \right) \Bigg] {Q}_{\omega}(x) \Biggr\rangle _{\BZ_N}. \nonumber
\end{align}

We find that with the linear combination coefficients $u_{\omega}$ in \eqref{def:u_w} satisfying  $u_{\omega}-\kq_{\omega+1}u_{\omega+1}=1-\kq$, the unwanted $\hat{c}_{\omega}-\hat{c}_{\omega+1}$ in $D_{\omega}^{(1)}$ can be canceled, leaving 
\begin{subequations}
\begin{align}
    & \Biggl\langle (1-\kq)\left[\ve_1\ve_2 k_{N-1} +\sum_\omega\frac{1}{2}\left(\ve_1\nu_\omega-{a}_{\omega+1}\right)^2  -\frac{{a}_{\omega+1}^2}{2}\right] \\
    & \qquad + \left[\sum_{\omega}\kq_{\omega+1} u_{\omega+1}(\ve_1\nu_{\omega}-a_{\omega+1}+m_{\omega+1}^+)(\ve_1\nu_{\omega}-a_{\omega+1}+m_{\omega+1}^-)\right] {Q}_{\omega'}(x+\omega'\tilde{\ve}_2) \Biggr\rangle _{\BZ_N} \nonumber\\
    = & -\ve_1u_{\omega'}\left\langle {P}_{\omega'+1}^+(x) {Q}_{\omega'+1}(x) \right\rangle _{\BZ_N} + \Biggl\langle u_{\omega'} \ve_1(x- a_{\omega+1} +  \ve_1\nu_{\omega'}){Q}_{\omega'}(x) \Biggr\rangle _{\BZ_N} \nonumber\\
    %%%%%%%%%%%%%%%%%%%%%%%%%%%%%%%%%%%%%%%%%%%%
    & \Biggl\langle (1-\kq)\left[\ve_1\ve_2 k_{N-1} +\sum_\omega\frac{1}{2}\left(\ve_1\nu_\omega-{a}_{\omega+1}\right)^2-\frac{{a}_{\omega+1}^2}{2}\right] \\
    & \qquad + \left[\sum_{\omega}\kq_{\omega+1} u_{\omega+1}(\ve_1\nu_{\omega}-a_{\omega+1}+m_{\omega+1}^+)(\ve_1\nu_{\omega}-a_{\omega+1}+m_{\omega+1}^-)\right] {Q}_{\omega'}(x) \Biggr\rangle _{\BZ_N} \nonumber\\
    = & \ve_1 \hat\kq_{\omega'}u_{\omega'}\left\langle {P}_{\omega'}^-(x) {Q}_{\omega'-1}(x) \right\rangle _{\BZ_N} - \Biggl\langle \hat\qe_{\omega'}u_{\omega'} \ve_1(x-m_{\omega'} + a_{\omega'} - \ve_1\nu_{\omega'-1}) {Q}_{\omega'}(x) \Biggr\rangle _{\BZ_N}. \nonumber
\end{align}
\end{subequations}
with ${Q}_{\omega}(x)$ defined in \eqref{def:hatQ}. Using definition of expectation value $\left \langle {Q}_{\omega'}(x) \right \rangle _{\BZ_N} \Psi$:
\begin{align}
    \left \langle {Q}_{\omega'}(x) \right \rangle _{\BZ_N} \Psi
    & = \prod_{\omega=0}^{N-1} z_{\omega}^{\frac{m_{\omega}^+ - a_{\omega}}{\ve_1}} \sum_{\hat{\boldsymbol\lambda}} \prod_{\omega=0}^{N-1} z_{\omega}^{\nu_{\omega-1}} \kq^{k_{N-1}} {Q}_{\omega'}(x)[\vec\lambda] \EZ_{\rm surface}[\boldsymbol\lambda] \EZ_{\rm defect}[\hat{\boldsymbol\lambda}].
    % & = \sum_{\vec\lambda} \prod_{\omega=0}^{N-1} z_{\omega}^{\nu_{\omega-1}} \kq^{k_{N-1}} {Q}_{\omega'}(x+\omega')[\vec\lambda] \widetilde\Psi[\vec\lambda]
\end{align}
We may rewrite
$$
    \left \langle (\ve_1\nu_{\omega}-a_{\omega+1}+m_{\omega+1}^+) {Q}_{\omega'}(x) \right \rangle _{\BZ_N} \Psi = \ve_1 z_{\omega+1}\frac{\partial}{\partial z_{\omega+1}} \left \langle {Q}_{\omega'}(x)  \right \rangle _{\BZ_N} \Psi, \quad \omega=0,1,\dots,N-1
$$
and 
$$
    \left \langle k_{N-1} {Q}_{\omega'}(x) \right \rangle _{\BZ_N} \Psi = \kq\frac{\partial}{\partial \kq} \left \langle {Q}_{\omega'}(x) \right \rangle _{\BZ_N} \Psi.
$$
The two equations can be rewrite as second order differential in $\{z_{\omega}\}$ and first order differential in $\kq$:
\begin{align}
    & (1-\kq)\ve_1\ve_2 \kq \frac{\partial}{\partial \kq} \langle {Q}_{\omega'}(x) \rangle _{\BZ_N} \Psi - \hat{\rm H}  \langle {Q}_{\omega'}(x) \rangle _{\BZ_N} \Psi \\
    & = - \ve_1 u_{\omega'} \langle P_{\omega'+1}^+(x) {Q}_{\omega'+1}(x) \rangle _{\BZ_N} \Psi + \ve_1 u_{\omega'} \langle (x-m_{\omega'+1}^+ + \ve_1\nu_{\omega}) {Q}_{\omega'}(x) \rangle _{\BZ_N} \Psi \nonumber
\end{align}
and $x'=x+\ve_1$:
\begin{align}
    & (1-\kq)\ve_1\ve_2 \kq \frac{\partial}{\partial \kq} \langle {Q}_{\omega'}(x) \rangle _{\BZ_N} \Psi - \hat{\rm H}  \langle {Q}_{\omega'}(x) \rangle _{\BZ_N} \Psi \\
    & = \ve_1 \hat\kq_{\omega'}u_{\omega'}  \langle P_{\omega'}^-(x) {Q}_{\omega'-1}(x) \rangle _{\BZ_N} \Psi - \ve_1 \hat\kq_{\omega'} u_{\omega'} \langle (x-m_{\omega'}^- - \ve_1\nu_{\omega-1}) {Q}_{\omega'}(x) \rangle _{\BZ_N} \Psi \nonumber.
\end{align}
Differential operator $\hat{\rm H}$ is defined in \eqref{eq:H2}:
\begin{align}
    \hat{\rm H} :=  \sum_\omega
    & \frac{1-\kq}{2} \left(\ve_1\nabla^z_{\omega}-{{m}^+_{\omega}}\right)^2 + \hat\kq_{\omega} u_{\omega}(\ve_1\nabla^z_{\omega})(\ve_1\nabla^z_{\omega}-m^+_{\omega}+m_{\omega}^-).
\end{align}
The Fourier transform \eqref{eq:Yo} yields:
\begin{subequations}
\begin{align}
    & (1-\kq)\ve_1\ve_2\kq \left[ \frac{\partial}{\partial \kq} - \frac{\partial\log \Upsilon^{\text{pert}}(y)}{\partial \kq} \right] \Upsilon_{\omega'} - \hat{\rm H} \Upsilon_{\omega'} \nonumber\\
    & = -\ve_1 u_{\omega'} \left( -\ve_2 y\frac{\partial}{\partial y} - m_{\omega'+1}^+ \right)\Upsilon_{\omega'+1} + \ve_1u_{\omega'} \left( -\ve_2y\frac{\partial}{\partial y} - m^+_{\omega'+1} + \ve_1\nabla^z_{\omega'+1} \right)\Upsilon_{\omega'} \\
    %%%%%%%%%%%%%%%%%%%%%%%%%%%%%%%%%%%%%%%%%%%%
    & (1-\kq)\ve_1\ve_2\kq\left[ \frac{\partial}{\partial \kq} - \frac{\partial\log \Upsilon^{\text{pert}}(y)}{\partial \kq} \right] \Upsilon_{\omega'} - \hat{\rm H} \Upsilon_{\omega'} \nonumber\\
    & = \ve_1 \hat\kq_{\omega'}u_{\omega'} \left( -\ve_2 y\frac{\partial}{\partial y} - m_{\omega'}^- \right)\Upsilon_{\omega'-1} - \ve_1 \hat\kq_{\omega'}u_{\omega'} \left( -\ve_2y\frac{\partial}{\partial y} - m^-_{\omega'} - \ve_1\nabla^z_{\omega'} \right)\Upsilon_{\omega'} 
    \end{align}
\end{subequations}
The derivatives w.r.t. $y$ can be eliminated by considering a linear combination:
\begin{align}\label{eq:q-diff-Upsilon}
    & (1-\kq)\ve_1\ve_2  \left(\hat\kq_{\omega'+1}u_{\omega'+1}\kq\frac{\partial}{\partial\kq}\Upsilon_{\omega'} - u_{\omega'}\kq\frac{\partial}{\partial\kq}\Upsilon_{\omega'+1}\right) +  \left(u_{\omega'}\hat{\rm H}_2\Upsilon_{\omega'+1} - \hat\kq_{\omega'+1}u_{\omega'+1}\hat{\rm H} \Upsilon_{\omega'}  \right) \nonumber\\
    & = \ve_1 u_{\omega'} \hat\kq_{\omega'+1}u_{\omega'+1} \left[ (m_{\omega'+1}^+ - m_{\omega'+1}^- - \ve_1\nabla^z_{\omega'+1}) \Upsilon_{\omega'+1} + (m_{\omega'+1}^- - m_{\omega'+1}^+ + \ve_1\nabla^z_{\omega'+1})\Upsilon_{\omega'} \right] 
\end{align}
We denote two $N\times N$ diagonal matrices ${\bf u} = {\rm diag}(u_0,u_1,\dots,u_{N-1})$, and ${\bf G} = \text{diag}(G_0,\dots,G_{N-1})$ with
$$
    G_{\omega} = \frac{u_{\omega}+\kq-1}{u_{\omega}} = \frac{\hat\kq_{\omega+1}u_{\omega+1}}{u_{\omega}}.
$$
The $N$ $\kq$-differential equations \eqref{eq:q-diff-Upsilon} can be written as one $N\times N$ matrix equation:
\begin{align}
    & (1-\kq)\ve_1\ve_2 \left({\bf G} - {\bf U} \right) \left(\kq\frac{\partial}{\partial\kq} - \kq\frac{\partial}{\partial \kq} \log \Upsilon^{\text{pert}}(y) \right) \boldsymbol\Upsilon - \left({\bf G} - {\bf U} \right) \hat{\rm H} \boldsymbol\Upsilon \nonumber\\
    & = \ve_1 {\bf G}{\bf u} \left[ {\bf U}({\bf M}^+ - {\bf M}^-){\bf U}^{-1} - \boldsymbol\nabla \right]  ({\bf U} - {\bf I}_N)\boldsymbol\Upsilon.
\end{align}
Matrix ${\bf U}$ is defined in \eqref{def:U-matrix}, and diagonal matrices in \ref{def:diagmatrices}. We again consider change of basis $\boldsymbol\Pi = ({\bf U} - {\bf I}_N)\boldsymbol\Upsilon$. Vector $\boldsymbol\Pi$ obeys KZ equation:
\begin{align}
    0 & = \left[ \frac{\ve_2}{\ve_1} \frac{\partial}{\partial\kq} + \frac{1}{(1-\kq)\kq} \frac{1}{\ve_1^2} \hat{\rm H} - \frac{\ve_2}{\ve_1} \frac{\partial\log \Upsilon^{\text{pert}}(y)}{\partial \kq} \right. \\
    & \qquad \left. - \frac{1}{(1-\kq)\kq} \frac{1}{\ve_1} ({\bf U}-{\bf I}_{N})({\bf G}-{\bf U})^{-1} {\bf G}{\bf u} \left( {\bf U}({\bf M}^+ - {\bf M}^-){\bf U}^{-1} - \boldsymbol\nabla \right) \right]\boldsymbol\Pi \nonumber\\
    & = \left[ \frac{\ve_2}{\ve_1} \frac{\partial}{\partial\kq} + \frac{1}{(1-\kq)\kq} \frac{1}{\ve_1^2} \hat{\rm H} - \frac{\ve_2}{\ve_1} \frac{\partial\log \Upsilon^{\text{pert}}(y)}{\partial \kq} + \frac{\hat\fA(y,\kq)}{(\kq-1)(\kq-y)} \right]\boldsymbol\Pi \nonumber\\
    & := \left[ \frac{\ve_2}{\ve_1} \frac{\partial}{\partial\kq} + \frac{\hat{\fA}_{0}}{\kq} + \frac{\hat{\fA}_{1}}{\kq-1} + \frac{\hat{\fA}_{y}}{\kq-y} \right] \boldsymbol\Pi 
\end{align}
Elements of matrix $\hat\fA(y,\kq)$ are
\begin{align}\label{eq:fA-new}
    \hat{\fA}(y,\kq)_{\ta\tb}
    & = \frac{1}{\ve_1}\left[ ({\bf U} - {\bf I}) \sum_{j=0}^{N-1} ({\bf G}^{-1}{\bf U})^{j} {\bf u} \left( {\bf U}({\bf M}^+ - {\bf M}^-){\bf U}^{-1} - \boldsymbol\nabla \right) \right]_{\ta\tb} \nonumber\\
    & = \frac{z_\ta}{z_\tb} \left(\kq_{\ta}u_{\ta}\left(\frac{y}{\kq}\right)^{\theta_{\ta+1>\tb}} - u_{\ta-1}\left(\frac{y}{\kq}\right)^{\theta_{\ta>\tb}} \right) \left( \frac{m^+_{\tb} - m^-_{\tb}}{\ve_1} - \nabla^z_{\tb} \right)
\end{align}
with $\ta,\tb=1,\dots,N$. We find individual KZ-operators $\hat{\fA}_{0,1,y}$ as:
\begin{subequations}
\begin{align}
    \left( \hat{\fA}_0 \right)_{\ta\tb} = \
    & \frac{1}{\ve_1^2}\hat{\rm H}|_{\kq=0} \delta_{\ta\tb} - \frac{z_{\ta+1}+\cdots+z_{N-1}}{z_{\ta}}\left( \nabla^z_{\ta} + \frac{m^-_{\ta} - m^+_{\ta}}{\ve_1} \right) \delta_{\ta\tb} \nonumber\\
    & + \frac{z_\ta}{z_\tb} \left( \nabla^z_{\tb} + \frac{m^-_{\tb} - m^+_{\tb}}{\ve_1} \right) \theta_{\ta>\tb}  \\
    % & - \theta_{\omega+1>\omega'} \left( {\bf M}^+_{\omega'+1} - {\bf M}^-_{\omega'} - \ve_1 \nabla^z_{\omega'} - \ve_1 \right) \\
    % & -u_{\omega} {\theta_{\omega>\omega'}} \left( {\bf M}^+_{\omega'+1} - {\bf M}^-_{\omega'} - \nabla^z_{\omega'} - 1 \right) \\
    %%%%%%%%%%%%%%%%%%%%%%%%%%%%%%%%%%%%%%%%%%
    \left( \hat{\fA}_{1} \right)_{\ta\tb} = \
    & - \frac{1}{\ve_1^2} \hat{\rm H}|_{\kq=1} \delta_{\ta\tb} - \frac{z_0+\cdots+z_{N-1}}{z_{\ta}} \left( \frac{m^+_{\ta} - m^-_{\ta}}{\ve_1} - \nabla^z_{\ta} \right) \delta_{\ta\tb} \\
    % = & -\frac{1}{\ve_1}\hat{\rm H}_2|_{\kq=1} \delta_{\omega,\omega'} \\
    % & + \frac{1}{y-1} \kq_{\omega'}u_{\omega'} \left({y}\right)^{\theta_{\omega>\omega'-1}} (m_{\omega'}^+ - m_{\omega'}^- - \ve_1 \nabla^z_{\omega'-1}) \\
    %%%%%%%%%%%%%%%%%%%%%%%%%%%%%%%%%%%%%%%%%%
    \left( \hat{\fA}_y \right)_{\ta\tb} = \
    & - \frac{z_\ta}{z_\tb} \left( \nabla^z_{\tb} + \frac{m^-_{\tb} - m^+_{\tb}}{\ve_1}  \right) + \frac{\delta_{\ta\tb}}{N} \frac{m^--a}{\ve_1} 
    % = & - \left( \hat{\CalA}_\kq \right)_{\omega,\omega'}
\end{align}
\end{subequations}
$\ta,\tb=1,\dots,N$. KZ-operators obtained from supersymmetric gauge theory agree with representation theory prediction:
\begin{align}
    \hat\fA_0 = \hat\sH_0, \ \hat\fA_0 = \hat\sH_1, \ \hat\fA_y = \hat\sH_y.
\end{align}
after taking care the perturbative factor $\Psi$ in \eqref{eq:pert}:
\begin{align}
    \nabla^z_\tb \mapsto \nabla^z_\tb + \frac{m_\tb^+ - a_\tb }{\ve_1}.
\end{align}

\newpage
%%%%%% Bibliography %%%%%%
\bibliographystyle{utphys}
\bibliography{5points}

\end{document}